\documentclass[useAMS,usenatbib]{mn2e}

\usepackage{graphicx}
\usepackage{subfigure}
\title[Ionized outflows in luminous type 2 AGNs]{Ionized outflows in luminous type 2 AGNs at $z<$ 0.6: no evidence for significant impact on the host galaxies.\thanks{Based on observations carried out at the European Southern Observatory
(Paranal, Chile) with FORS2 on VLT-UT1 (programmes 
083.B-0381 and 087.B-0034)}}
\author[Villar Mart\'\i n et al.]{M. Villar-Mart\'{i}n$^{1,2}$, S. Arribas$^{1,2}$,  B. Emonts$^{1,2}$, A. Humphrey$^3$, C. Tadhunter$^4$
\newauthor P. Bessiere$^5$, A. Cabrera Lavers$^6$, C. Ramos Almeida$^{6,7}$\\
$^1$Centro de Astrobiolog\'{i}a (INTA-CSIC), Carretera de Ajalvir, km 4, 28850 Torrej\'on de Ardoz, Madrid, Spain\\
$^2$Astro-UAM, UAM, Unidad Asociada CSIC, Facultad de Ciencias, Campus de Cantoblanco, E-28049, Madrid, Spain \\
$^3$Instituto de Astrof\'{i}sica e Ci\^encias do Espa\c{c}o, Universidade do Porto, CAUP, Rua das Estrelas, PT4150-762 Porto, Portugal\\
$^4$Department of Physics and Astronomy, University of Sheffield, Sheffield S3 7RH, UK \\
$^5$Universidad de Concepci\'on, Departamento de Astronom\'\i a, Casilla 160-C, Concepci\'on, Chile \\
$^6$Instituto de Astrof\'\i sica de Canarias (IAC), C/ V\'\i a L\'actea, s/n, E38205  La Laguna,Tenerife, Spain \\
$^7$Departamento de Astrof\'\i sica, Universidad de La Laguna, E-38206, La Laguna, Tenerife, Spain}

\begin{document}

%\date{Accepted ?.
   %   Received ?;
      %in original form ?.}

\pagerange{\pageref{firstpage}--\pageref{lastpage}}
\pubyear{2013}

\maketitle

\label{firstpage}

\begin{abstract}

We  investigate the presence of extended ionized outflows in    18 luminous type 2 AGNs (11  quasars and 7 high luminosity Seyfert 2s) at 0.3$<$z$<$0.6
based on VLT-FORS2 spectroscopy. We infer typical lower limits on the radial sizes of the outflows $R_ {\rm o}\ga$several$\times$100 pc and  upper limits $R_{\rm o}\la$1-2 kpc. Our results are  inconsistent with related studies  which suggest that large scale ($R_{\rm o}\sim$several-15 kpc) 
are ubiquitous in QSO2. We study the possible causes of  discrepancy and propose that seeing smearing is the cause of the large inferred sizes.
  The implications  in our understanding of the feedback phenomenon are important since  the  mass $M_ {\rm o}$ (through the density), mass injection $\dot M_{\rm o}$ and energy injection $\dot E_{\rm o}$  rates  of the outflows  become highly uncertain.   One conclusion seems unavoidable: $M_ {\rm o}$,  $\dot M_{\rm o}$ and  $\dot E_{\rm o}$ are  modest or low compared  with previous estimations. 
We obtain  typically $M_{\rm o}\la$(0.4-22)$\times$10$^{6}$ M$_{\odot}$ (median 1.1$\times$10$^{6}$ M$_{\odot}$) assuming $n=$1000 cm$^{-3}$. These are $\sim$10$^2$-10$^4$ times lower than values reported in the literature.
Even under the most favorable assumptions,   we obtain  $\dot M_{\rm o}\la$10 M$_{\odot}$ yr$^{-1}$ in general,
100-1000 times lower than claimed in related studies. Although the uncertainties are large, it is probable that these are lower than typical star forming rates. In conclusion,  no evidence is found supporting that typical outflows can affect the interstellar medium of the host galaxies accross spatial scales $\ga$1-2 kpc.

\end{abstract}


\section{Introduction}
\label{Sec:intro}

Evidence for an intimate connection between supermassive black hole (SMBH) growth and the evolution of galaxies is nowadays compelling. Not only have SMBHs been found in many galaxies with a bulge component, but correlations also exist between the black hole mass and some bulge properties, such as the stellar mass and velocity dispersion (e.g. Ferrarese \& Merritt \citeyear{fer00}). The origin of this relation is still an open question, but   outflows induced by the nuclear activity in galaxies might play a critical role. Hydrodynamical simulations show that the energy output from active galactic nuclei (AGN)  can regulate the growth and activity of black holes and their host galaxies (e.g. di Matteo et al. \citeyear{dim05}). Such models show that the energy released by the outflows associated with major phases of accretion expel enough gas to quench both star formation and further black hole growth. 
Observational evidence for the presence of AGN induced outflows in luminous active galaxies is accumulating (see Fabian \citeyear{fab12} for a review), although the impact on their host galaxies is far from clear. It may depend on the luminosity of the AGN and  be more efficient at the highest luminosities (Page et al. \citeyear{pag12}, Zakamska et al. \citeyear{zak14}). 

According to the AGN feedback paradigm, the most powerful outflows with potentially the most extreme impact on the environment are expected in quasars (QSO). In spite of the overwhelming evidence for the existence of outflows in  QSO, the effects they have on the evolution of the host galaxy (e.g. via star formation quenching) are far from clear.  The difficulty comes from the great complexity of the feedback phenomenon, which is the interplay between different mechanisms, involving different gaseous phases (coronal, ionized, neutral, molecular), which are distributed across different spatial scales and require different observing techniques. If we want to quantify the galaxy-scale impact of QSO winds, we must be able to understand what triggers them and  how they affect the different gas phases.

Studies of powerful radio galaxies at different $z$ show that the radio structures can induce outflows (Tadhunter et al. \citeyear{tad94}, Sol\'orzano I\~narrea et al. \citeyear{sol01}, Villar Mart\'\i n et al. \citeyear{vm03}, Humphrey et al. \citeyear{hum06}) that may have enormous energies sufficient to eject a large fraction of the gaseous content out of the galaxy (Morganti et al. \citeyear{mor05}, Nesvadba et al. \citeyear{nes06}, \citeyear{nes08}). However, only $\sim$10\% of active galaxies are radio-loud. How AGN feedback works  in radio-quiet counterparts (i.e., radio-quiet quasars) is still an open question. Several mechanisms could be at work, including stellar winds, outflows from weak radio jets and accretion disc winds. 
Type 2 QSO (QSO2) are unique objects for investigating the way feedback works  in the most powerful non radio-loud (i.e. radio-quiet or radio-intermediate) AGN. The active nucleus is occulted by obscuring material, which acts like a convenient Ônatural coronographÕ, allowing a detailed study of many properties of the surrounding medium. Such studies would
be difficult in type 1 QSO (QSO1) due to the dominant contribution of the quasar point spread function. 
	
	In order to asses the role the outflows play on the host galaxy evolution, it is of special relevance to study their impact on the molecular phase, since  this
is the raw material from which stars form.
 Different works suggest that  molecular outflows  exist in AGN, apparently capable of transporting huge amounts of mass
($\ga$10$^8$ M$_{\odot}$) at enormous rates of tens, even hundreds M$_{\odot}$ yr$^{-1}$ (Cicone et al. \citeyear{cic14}).
 In general, the radial sizes are $R_{\rm o}\la$1 kpc, although claims of giant  molecular outflows ($R_{\rm o}\sim$30 kpc)  also exist (Cicone et al. \citeyear{cic15}). Some of the reported outflows can in principle  exhaust a large fraction of the cold molecular reservoir of the host galaxies and quench the star formation activity (see also Alatalo et al. \citeyear{ala15}).
 
 	On the other hand, the most efficient way to identify the imprint of outflows in large samples of objects at any redshift,  and  trace them on galactic scales is to search for them in the warm ionized phase using strong optical emission lines, such as [OIII]$\lambda\lambda$4959,5007.
 Indeed, during the last  few years it has become clear that ionized outflows are a ubiquitous phenomenon in QSO2 at $z\le$ 0.7 (e.g. Villar Mart\'\i n et al. 2011b, \citeyear{vm14}; Greene et al. \citeyear{gre11}, Liu et al. \citeyear{liu13b}, Harrison et al. \citeyear{har14}, McElroy et al. \citeyear{mce15}). Extreme motions are often measured in the ionized gas, with full width at half maximum FWHM$>$ 1000 km s$^{-1}$,  although  modest velocity shifts of $V_{\rm s} <$ few$\times$100 km s$^{-1}$ are more typical. The outflows are triggered by AGN related processes, which may be disk winds and/or radio jets.  
 This is not in contradiction with their radio-quietness (see for instance Tadhunter et al. \citeyear{tadh14}, Harrison et al. \citeyear{har15a}). In fact, the existence of  jets in many (all?) AGN typically defined as radio-quiet  has been proposed by different authors (e.g. Falcke \& Biermann \citeyear{fal95}, Ghisellini et al. \citeyear{ghi04}). Also,  the most extreme outflows are often found in
objects with a  modest degree of jet activity (e.g. Whittle \citeyear{whi92}, Mullaney et al. \citeyear{mul13}, Villar Mart\'\i n et al. \citeyear{vm14}).
 
Evidence for large scale, wide angle AGN driven ionized outflows in quasars with moderate or low radio activity at different redshifts has been reported in a diversity of studies  in recent years. Spectroscopic studies of QSO2 at $z\la$0.7 have suggested that such galaxy outflows are prevalent among
these systems (Greene et al. \citeyear{gre11}, Liu et al. \citeyear{liu13b}, Harrison et al. \citeyear{har14}, McElroy
et al. \citeyear{mce15}; see also studies of individual quasars by Humphrey et al. \citeyear{hum10}, Rupke \& Veilleux \citeyear{rup11}, Greene et al. \citeyear{gre12}, Harrison et al. \citeyear{har15a}). The inferred spatial extents are in the range
$\sim$6-20 kpc. According to these studies , the impact of the
ionized outflows can be exerted across large spatial scales,
even entire galaxies. 
The estimated kinetic energies, outflow
masses and mass outflow rates may be large enough for the observed
quasar winds to have a significant impact on their
host galaxies (e.g. Liu et al. \citeyear{liu13b}).  \cite{kar16} have recently questioned these results and report smaller  outflow radial sizes $\sim$1-2 kpc,
raising doubts about their effectiveness as a mechanism for AGN feedback.

At high redshifts energetic outflows are predicted to be very common. During the peak of cosmic star formation and QSO activity (i.e. $z\sim$2-3), different studies have reported evidence for  spatially extended AGN-driven ionized outflows in systems with signs of AGN activity and moderate or low radio emission. The reported extensions
are  $\sim$2-8 kpc  (Alexander et al. \citeyear{ale10}, Cano D\'\i az et al. \citeyear{can12}, F\"orster Schreiber  et al. \citeyear{for14},  Genzel et al. \citeyear{gen15}, Carniani et al. \citeyear{car15})
and sometimes  $\ga$10 kpc   (Harrison et al. \citeyear{har12}, Cresci et al. \citeyear{cre15}). 
 Ionized outflows extending up to $\sim$10 kpc from the central black hole have also been reported in obscured quasars at intermediate $z\sim$1.5 (Perna et al. \citeyear{per15}).

  We present here a kinematic study of eighteen high luminosity Seyfert 2s and QSO2 at $z\sim$0.3-0.6. The main purpose of this paper is to investigate the spatially extended kinematics in these systems, putting a special emphasis on the potential signatures of the outflows at large extra-nuclear distances ($\ga$several kpc). Previous results on this sample based on the same data
 have  been published in Villar Mart\'\i n et al. 2011a (hereafter \cite{vm11a}),  2011b  (hereafter \cite{vm11b}), \citeyear{vm12} 
and Humphrey et al. 2015 (hereafter \cite{hum15}). The kinematic analysis, which supported the ubiquitousness of  ionized outflows in QSO2, 
 was focussed  on the nuclear, spatially integrated spectra (\cite{vm11b}), providing no information on their spatial extension.

 In Section 2, we describe the sample. The observations and data reduction are summarized in Section 3. 
The kinematic analysis methods and their uncertainties are described in Section 4. The analysis and results are presented in Section 5 and discussed in Section 6. The summary and conclusions  are presented in Section 7. Results for the majority of the objects are explained in detail in Appendix \ref{Sec:app}.

\section{The sample}
\label{Sec:sample}

The sample consists of 18 luminous type 2 AGN   selected from  \cite{rey08} Sloan Digital Sky Survey (SDSS, York et al. \citeyear{york00}) catalogue (Table \ref{tab:sample}). They are objects  with
narrow  ($<$2000 km s$^{-1}$) emission lines with no underlying broad components for the recombination lines which would be suggestive of a broad line region (BLR).
The emission line ratios are  characteristic of non-stellar ionizing radiation (Zakamska et al. \citeyear{zak03}).   The [OIII]$\lambda$5007 ([OIII] hereafter) luminosities  $L_{\rm [OIII]}$ are in
the range $l_{\rm O3}$=log$\Big(\frac{L_{\rm [OIII]}}{L_{\odot}}\Big)$=7.8-9.7 (luminosities taken from
Reyes et al. \citeyear{rey08} available on the Vizier online catalogue).

Our  sample consists of two-subsamples that were observed in 2009 and 2011, and which are described in detail in two separate papers
 (\cite{vm11a} and  \cite{hum15}). For clarity  we will refer to them as the 2009 and 2011 samples.
  Both contain luminous type 2 AGN with redshift $z\sim$0.3-0.6. As explained in \cite{hum15}, the main difference between the two subsamples is that the 2011 objects are somewhat less luminous. While the 2009 subsample  had a range of  $l_{\rm O3}$ with median value 8.84, the  2011 subsample has 8.27. Adopting  $l_{\rm O3}$=8.3 as the threshold value to select QSO2 versus high luminosity Seyfert 2 (Reyes et al. \citeyear{rey08})  the 2009 subsample contains  7 QSO2 and 1 HSy2 and  the 2011 subsample contains 4 QSO2 and 6 HSy2.

In order to classify the objects according to the radio-loudness, we have used the rest-frame radio power at 5 GHz $P_{\rm 5GHz}$
and the [OIII] luminosities $L_{\rm [OIII]}$, following   \cite{xu99}. We calculate $P_{\rm 5GHz} = 4 \pi D_{L}^{2} S_{\rm 5GHz} (1+z)^{-1-\alpha}$ where  $D_{L}$ is the luminosity distance, $S_{\rm 5GHz}$ is the observed   flux density at 5 GHz and $\alpha$ is the spectral index such that $S_{\nu} \propto \nu^{\alpha}$. Only $S_{\rm 1.4GHz}$  is available for most objects (\cite{hum15}) and these have been used to predict $S_{\rm 5GHz}$. 
The  index $\alpha$  was inferred by \cite{lal10} for SDSS J0903+02, SDSS J0950+03, SDSS J1014+02, SDSS J1017+03
and SDSS J1247+01 based of 1.4 and 8.4 GHz flux densities.  For the rest of the objects, we assume $\alpha = +0.094$, 
the median value found by these authors for a sample of QSO2 at similar
$z$ as our objects.  The estimated $P_{\rm 5GHz}$ values and the final classification are shown in Table \ref{tab:sample}. Ten objects are radio
quiet, seven  are radio-intermediate and one is radio-loud. Applying the same classification criteria,
 \cite{lal10} found that $\sim$15$\pm$5\% of QSO2 are radio-loud and that the  majority of the detected
sources are radio-intermediate.

\begin{table*}
\centering
\caption{Sample of type 2 quasars (QSO2, $l_{\rm O3}$=log($\frac{L_{\rm [OIII]}}{L_{\odot}})>8.3$) and high luminosity Seyfert 2 (HSy2, $l_{\rm O3}\le$8.3). 
``Radio class" refers to the radio-loudness classification according to  RQ (radio-quiet), RI (radio-intermediate) or RL (radio-loud) (see the text).} 
\begin{tabular}{llllllllllllll}
\hline
Target & Short name & z & log($\frac{L_{\rm [OIII]}}{L_{\odot}})$ & Seeing  &  log($P_{\rm 5GHz}$) &  AGN & Radio \\
& &   & &  FWHM(")  & erg s$^{-1}$ Hz$^{-1}$ &  class & class    \\
\hline
  &	&	& 2009	sample &	&	& \\ \hline
SDSS J095514.11+034654.2   &  SDSS J0955+03 & 0.422 & 8.51 &  1.6$\pm$0.1& 30.9  & QSO2 & RI \\
SDSS J115314.36+032658.6  &  SDSS J1153+03 & 0.575 & 9.66& 0.8$\pm$0.1  & 31.2    &QSO2 & RQ \\
SDSS J122845.74+005018.7 &  SDSS J1228+00 & 0.575 & 9.31 & 1.75$\pm$0.09 & 31.5   &QSO2 & RI \\
SDSS J130740.55-021455.2 &  SDSS J1307-02 &   0.424 & 9.01 & 0.83$\pm$0.05 & 30.9  & QSO2 & RQ  \\
SDSS J133735.01-012815.6 & SDSS J1337-01 & 0.328 & 8.73 & 1.6$\pm$0.1 & 30.9 & QSO2 & RQ  \\
SDSS J140740.06+021748.3  & SDSS J1407+02 & 0.309 & 9.00 & 0.73$\pm$0.04 & 30.4  & QSO2 & RQ   \\
SDSS J141315.31-014221.0 & SDSS J1413-01 &  0.380 & 9.18 & 1.54$\pm$0.09 &  31.1  & QSO2 & RQ  \\
SDSS J154613.27-000513.5  & SDSS J1546-00 &  0.383 & 8.18 & 0.90$\pm$0.05 & 30.8  & HSy2 & RI \\ \hline
  &	&	&	2011  sample &	&	& \\ \hline
SDSS J090307.83+021152.2 & SDSS J0903+02 &  0.329 & 8.79 &  0.95$\pm$0.05 & 31.4 &  QSO2 & RI \\
SDSS J092318.06+010144.8 & SDSS J0923+01 &  0.386 & 8.78 & 0.66$\pm$0.06 & 30.6 & QSO2 & RQ \\
SDSS J095044.69+011127.2 & SDSS J0950+01 & 0.404 & 8.22 &  1.00$\pm$0.05  & 30.9 &   HSy2 & RI  \\
SDSS J101403.49+024416.4 & SDSS J1014+02 &  0.573 & 8.29 &  0.65$\pm$0.04 & 31.5 &  HSy2 & RI \\
SDSS J101718.63+033108.2 & SDSS J1017+03 &   0.453 & 8.27 & 0.75$\pm$0.05 & 31.2 & HSy2 & RI \\
SDSS J124749.79+015212.6 & SDSS J1247+01 & 0.427 & 8.23 & 1.25$\pm$0.045 & 32.0 & HSy2 & RL \\
SDSS J133633.65-003936.4 & SDSS J1336-00 & 0.416 & 8.64 & 0.64$\pm$0.05 & $\la$30.4  & QSO2 & RQ  \\
SDSS J141611.77-023117.1 & SDSS J1416-02 &  0.305 & 8.03 & 1.0$\pm$0.05 & 30.5 &HSy2 & RQ  \\
SDSS J143027.66-005614.8 & SDSS J1430-00 & 0.318 &  8.44  & 0.73$\pm$0.05 &  $\la$30.1 &  QSO2 & RQ \\
SDSS J145201.73+005040.2 & SDSS J1452+00 &  0.315 & 7.82 & 0.60$\pm$0.04 & 30.1  &HSy2 & RQ \\
\hline
\end{tabular}
\label{tab:sample}
\end{table*}

\section{Observations and data reduction}
\label{Sec:obs}
The data were obtained with the Focal Reducer and Low Dispersion
Spectrograph (FORS2) for the VLT installed on UT1 (Appenzeller
et al. \citeyear{app98}). The observations were performed in two different runs: 2009 April 17 and 18
(8 objects) and 2011 April 25 to 28  (10 objects).
The observations and data reduction  were described in \cite{vm11a} and \cite{hum15}.
The spectral resolution was 5.2$\pm$0.2 \AA\ for the 2009 observations. It was 5.2$\pm$0.2,
7.2$\pm$0.2 \AA\  and 9.6$\pm$0.3 \AA\ or the 2011
spectra, depending on the object (see  \cite{hum15} for details).

\subsection{Seeing effects}
\label{Sec:seeing}

The main objective of this study is to   determine whether spatially extended ionized outflows are present in our sample. For this, a careful characterization of the seeing size (FWHM) and shape, including  the uncertainties, is crucial. The seeing was very variable during both the 2009 and 2011 VLT observations (FWHM in the range $\sim$0.6\arcsec-1.6\arcsec).    To account for the uncertainties,
 the seeing FWHM was measured from the broad band and narrow/intermediate band images observed at similar time as the spectra,
 using several stars in the field.  The average FWHM 
calculated over the exposure time as measured by the Differential Image Motion Monitor (DIMM) station was also  considered. The dispersion in these values 
gives an estimation of  the FWHM uncertainties, which have been carefully taken into account as discussed in \cite{hum15}.

For each object we reconstructed the spatial profile of the seeing disc along the slit using a non-saturated star in images taken immediately before or after the spectroscopic observation of the science target. Because our study  is based primarily on [OIII], whenever possible the
narrow or intermediate band image containing this line was used, in order to minimize the effects of the wavelength dependence of
the seeing profile. In the absence of either a narrow band or intermediate band image, the broad-band image was instead used. In all
cases, the selected star was sufficiently bright to detect and trace adequately the faint wings of the seeing profile. The stellar flux was
extracted from apertures centered on the stellar centroid and mimicking the slit: 4 (1.0 arcsec) or 5 pixels (1.3 arcsec) wide depending
on the slit used for the  spectroscopic observation of the science targets.
Finally, the sky background was removed from the stellar spatial profile.

We showed in \cite{hum15} that the spatial distribution of the ionized gas  is always dominated by a bright compact, usually spatially
unresolved central source. In addition, extended emission
line features are frequently detected, which can be  more than 100 times fainter in surface brightness. This means 
 that even when the emission lines are spatially extended, the contribution from the central unresolved source can be dominant at distances several times  the seeing FWHM.    Thus, seeing smearing must be  carefully accounted for to characterize  the extended kinematics. An  illustrative example in our sample is SDSS J1153+03. Although   faint extended [OIII] emission is detected  up to $\sim$4.5 arcsec from the AGN, the central unresolved source dominates up $\sim$2.5 arcsec or
    $\times$6 times beyond the seeing FWHM.
Another clear example is   SDSS J1336-00.    The [OIII] profiles are spatially unresolved compared with the seeing  and yet  line emission is detected  up to $\sim$7  times beyond the seeing FWHM.

The slit width  was chosen between 1.0\arcsec~ and 1.3\arcsec~ 
to reach a compromise between optimizing  the observing time,  obtaining an adequate spectral resolution  and avoiding  significant flux loses. The slit  was often
wider than the seeing disk as a consequence (\cite{vm11a}, \cite{hum15}).  This introduces additional uncertainties on  the kinematic measurements for spatially unresolved sources.  We refer the reader to   \cite{hum15} for a detailed discussion on all these aspects. 
 The associated  uncertainties have  been carefully taken into account. They do not have an impact on our results.

\section{Kinematic analysis method}
\label{Sec:methods}

\subsection{Methods to isolate the emission from the outflowing gas}
\label{Sec:methods4p1}

The emission line profiles of QSO2  are often complex and show large asymmetries and  deviations from gaussianity (e.g. \cite{vm11b}, Greene et al. \citeyear{gre11}, Liu et al. \citeyear{liu13b}, Harrison et al. \citeyear{har14}, Villar Mart\'\i n et al. \citeyear{vm14}, McElroy et al. \citeyear{mce15}). In order to characterize the ionized gas kinematics two main methods are usually applied:  parametric   and non 
parametric.  

The first method consists of multiple Gaussian fits. It is based on the assumption that the complex emission line profiles are a combination of several kinematic components, each  being described by a Gaussian velocity distribution. The non-parametric measurements consists of measuring velocities at various
fixed fractions of the peak intensity (e.g. Heckman et al. \citeyear{hec81}),  or by measuring velocities at which some fraction of the line flux is accumulated
(Whittle \citeyear{whi85}).  Commonly used  parameters are the FWHM and  the width containing 80 percent
of the line  flux ($W_{80}$).

The second, non-parametric method  has the advantage of not introducing ad-hoc assumptions (namely, Gaussian velocity distributions). Moreover, it  can be applied to low S/N spectra, for which the parametric analysis may fail especially when the lines are very complex (3 or more kinematic components). It helps to characterize global kinematic properties and is  efficient at identifying high velocity (outflowing) gas (e.g. large $W_{80}$, blue asymmetries). This approach, however, is not exempt from uncertainties. For instance, it does not account for the fact that at every spatial location different gaseous components can contribute to the flux  and modify the shape of the  line profiles: e.g. companion objects (e.g. double nuclei, star forming knots), tidal features, ambient NLR gas or outflowing gas.   Parameters such as $W_{80}$ and very particularly the median velocity  of the line profiles are sometimes difficult to interpret, since the apparent spatial velocity fields could be affected by the relative contribution of the different gaseous components varying in space.  

The parametric analysis overcomes this difficulty, making it possible to isolate the contribution of all individual components at every spatial location. It has the disadvantage of introducing the assumption of Gaussian velocity distributions, where velocities actually appear non Gaussian. For this reason,  the physical sense of the resulting kinematic quantities has been questioned.

 In spite of the uncertainties and concerns, different studies  suggest that the parametric analysis actually provides physically meaningful and coherent results.  This is especially true when the spatial continuity of the properties of the individual components can be tested.     
 
Applying the multi-Gaussian method different authors have isolated the outflowing gas in multiple optical emission lines in diverse samples of active and non-active  star forming galaxies. The trends  found must be physically real and are difficult to explain as artifacts of the fitting method. As an example, it is  found  that the outflowing gas in different objects has higher density and is more highly reddened than the ambient gas (e.g. Holt et al. \citeyear{holt11}, Villar Mart\'\i n et al. \citeyear{vm14}, Arribas et al. \citeyear{arr14}). It is also found that  the line ratios used on the standard BPT diagnostic diagrams (Baldwin, Philips \& Terlevich \citeyear{bal81}) place the different kinematic components isolated in the spectra of QSO2, including the outflowing gas, in    the AGN area of the diagrams. The same method places  the individual components in star forming systems  in the corresponding starburst region of the diagrams (Arribas et al. \citeyear{arr14}). In hybrid objects, such as ULIRGs with Seyfert nuclei a   clear gradient has been found from narrow to intermediate to broad components in terms of position on the BPT diagrams (Rodriguez Zaur\'\i n et al. \citeyear{rod13}).
 These and other studies   suggest that in spite of the uncertainties, the multi-Gaussian fitting procedure provides physically meaningful results which would be difficult to achieve with the non parametric analysis. 

Although the two approaches have their advantages and limitations, the parametric method allows one to isolate more efficiently the outflowing gas, trace its spatial distribution and the spatial variation of its kinematic properties, with the least uncertainties due to possible contamination by other gaseous components.  For these reasons, we adopt it here.

The spectral line profiles were fitted with the STARLINK package DIPSO following the same procedure described in \cite{vm11b} (see also Villar Mart\'\i n et al. \citeyear{vm14}). DIPSO is based on the optimization of fit coefficients, in the sense of minimizing the sum of the squares of the deviations of the fit from the spectrum data.
The minimum number of components required to produce an adequate fit was used. 
 The output from a complete fit consists of the optimized parameters (FWHM, central $\lambda$, peak and integrated fluxes) and their errors (calculated in the linear approximation, from the error matrix). All FWHM values were corrected for instrumental broadening by subtracting the instrumental profile in quadrature (\cite{hum15}).

\subsection{Methods to constrain the spatial extension of the outflows}
\label{Sec:methods4p2}

To determine whether the ionized outflows detected in our sample are spatially extended, the following three methods will be used.
They work at different spatial scales and are  complementary for our purposes.

 \begin{figure}
\includegraphics{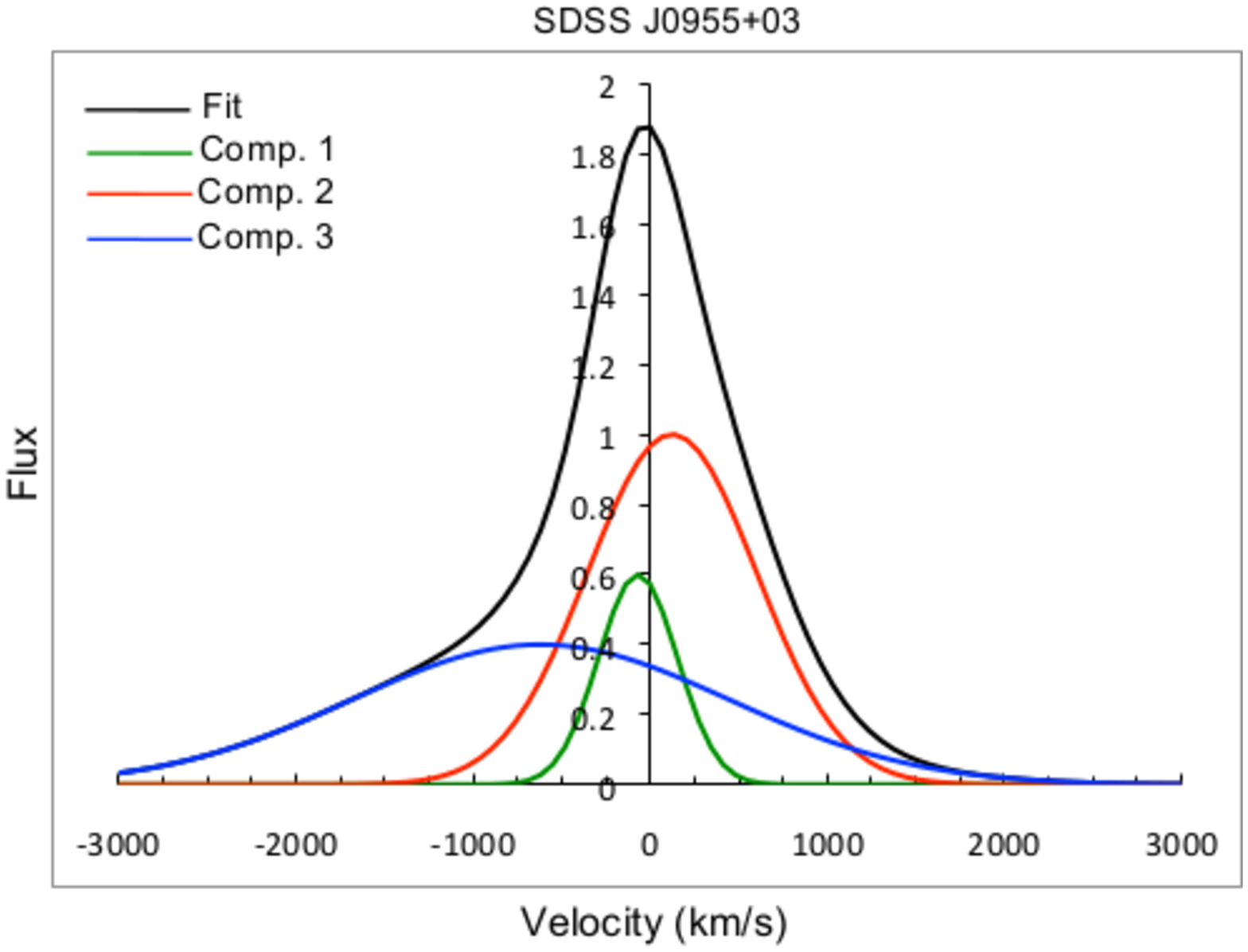}
\vspace{2.0in}
\includegraphics{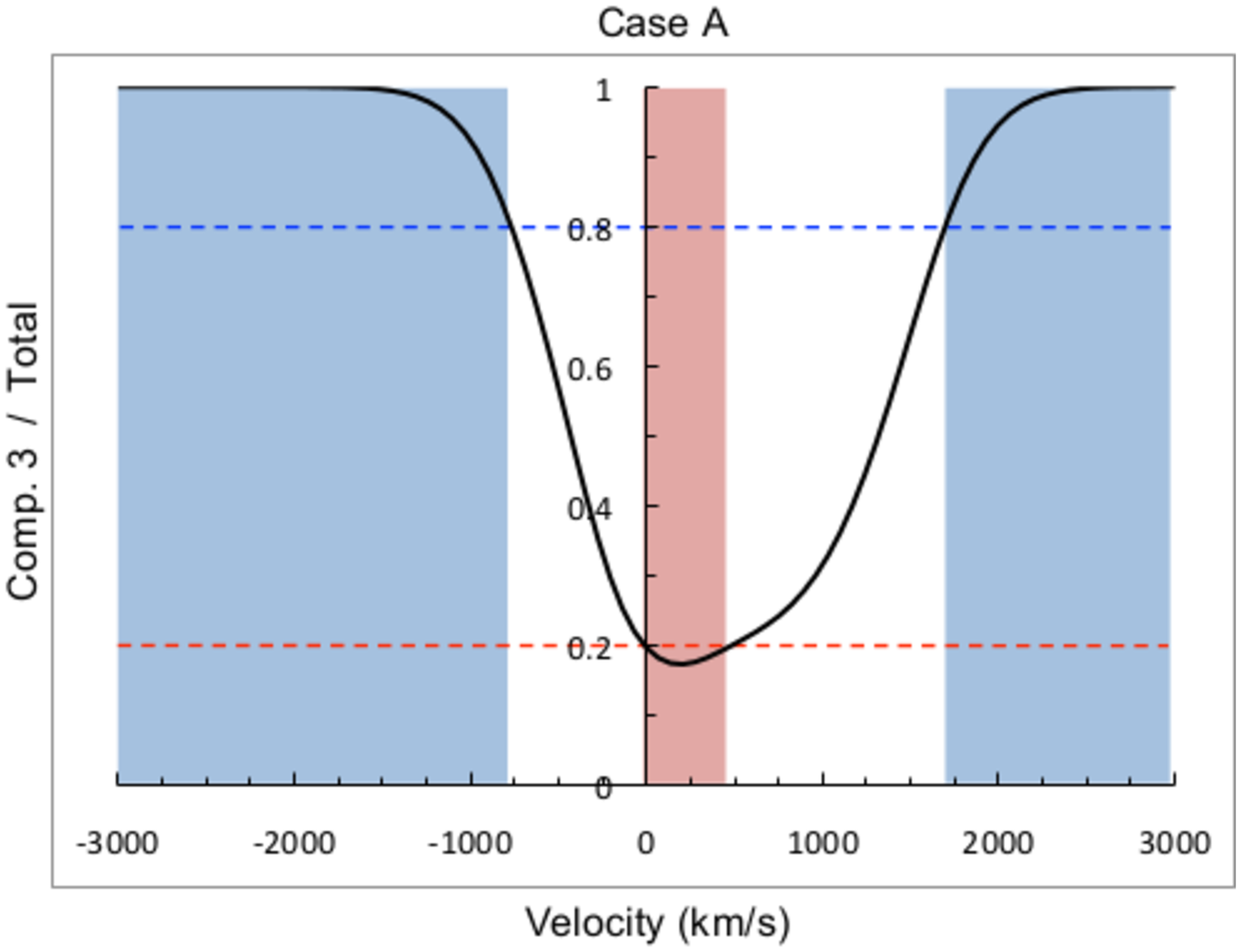}
\vspace{2.0in}
\includegraphics{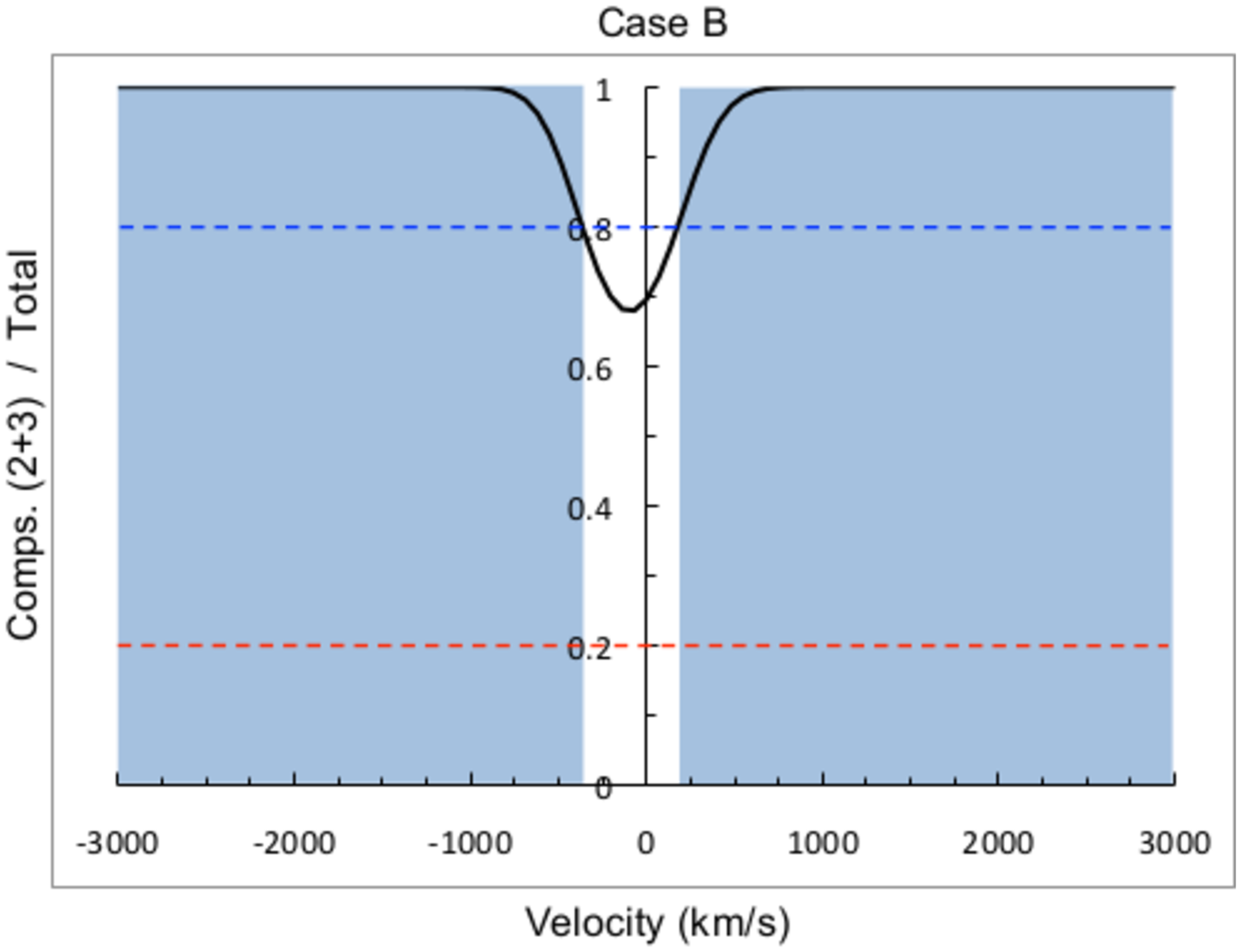}
\vspace{2.0in}
\caption{In order to constrain the sizes of the outflowing region applying the spectroastrometric 
technique it is first necessary to determine at which velocities the outflow emission dominates. The method is illustrated here for SDSS J0955+03.
The fit to the nuclear [OIII]$\lambda$5007 (black) and  the individual kinematic components (coloured
lines) are shown in the top panel (flux in arbitrary units). The middle and bottom panels show the relative contribution of the outflowing gas to the total line flux considering that the broadest (blue Gaussian) component traces the outflow (Case A) or if also the intermediate component (red Gaussian) participates (Case B). The colored areas mark  the velocity ranges at which this ratio is $\ge$0.8 (i.e. the outflow dominates the line flux, in blue) or at which the ambient gas dominates (ratio $<$0.2, in red). The inferred value of $R_{\rm o}$ might depend on the adopted scenario. Such uncertainty will be taken into account when relevant.}
\label{example0955}
\end{figure}

(i)  Comparison between the spatial profile of the outflowing gas emission and the seeing disk.  

For this  we perform a multi-Gaussian fit of the [OIII] lines  at every spatial pixel along the slit.  The FWHM, $V_{\rm s}$ and flux of the individual kinematic components are measured pixel by pixel.  Whenever this method could not be applied due to the complexity of the line profiles and/or the low S/N of the data, 1-dimensional spectra were extracted from several spatial apertures several pixels wide located at both sides of the emission lines spatial centroid.  

The spatial profile along the slit of the outflowing gas is thus generated and then compared with the seeing profile. This  was  built  specifically for each object as described in Sect. \ref{Sec:seeing}.

If the outflow is not resolved, method (i) provides an upper limit 
on its extension. This will be estimated as follows. Let us  consider a source of observed full width at half maximum 
 FWHM$_{\rm obs}\pm\Delta(\rm FWHM_{\rm obs})$. If the seeing has FWHM'$\pm\Delta$(FWHM'), we will assume that the source is spatially unresolved when:
 
  FWHM$_{\rm obs} <$ FWHM' + $\Delta$(FWHM') + $\Delta(\rm FWHM_{\rm obs}$) 

In such cases, an upper limit for the intrinsic  FWHM$_{\rm int}$ will be  set by estimating the value that will result in

 FWHM$_{\rm obs}$ = FWHM' + $\Delta$(FWHM') + $\Delta(\rm FWHM_{\rm obs}$).
 
 Given the $z$ of our targets  and the range of seeing sizes, method (i) is insensitive to FWHM$_{\rm int}\la$2-5 kpc depending on the seeing size. Similarly, for a barely resolved source we will  estimate FWHM$_{\rm int}$ by subtracting  the seeing FWHM from
FWHM$_{\rm obs}$  in quadrature.

 Notice that these calculations rely on the assumption that the intrinsic spatial profile of the outflowing gas is Gaussian. Results assuming an intrinsic S\'ersic or Voigt profile would be roughly equivalent, while a top hat profile would result in sizes $\sim$0.2 dex smaller (Hainline et al. \citeyear{hai14}).

(ii)  Search for kinematic substructure within the seeing disk.  Sometimes the spatial profiles of the seeing disk and the source will 
be found to be consistent within the errors. Discerning  whether the object is spatially resolved or not will be difficult, taking uncertainties on the object and
seeing sizes into account.

Method (ii) will allow  us to discern this by investigating the possible presence of kinematic spatial variations within the seeing disk, which  would not be expected for an unresolved source. If the pixel by pixel analyses described above
reveals spatial variations of the FWHM and/or $V_{\rm s}$ within the  seeing disk of a given kinematic component (in particular the outflowing gas), this will imply that it is spatially resolved. The outflow extension will be constrained in such cases as explained above for a barely resolved source.  If no kinematic variation is found, we will conclude that the source is not resolved in comparison with the seeing disk.

(iii) Spectroastrometry. This method allows one to study the spatial structure of the  outflowing gas  on scales well below 
the seeing size.   It involves measuring the relative position of the  centroid of the [OIII] spatial profile as a function of velocity.  The size of the outflow will be constrained from the spatial shifts measured for those velocities at which the outflow dominates (see below).

\begin{figure*} 
\includegraphics{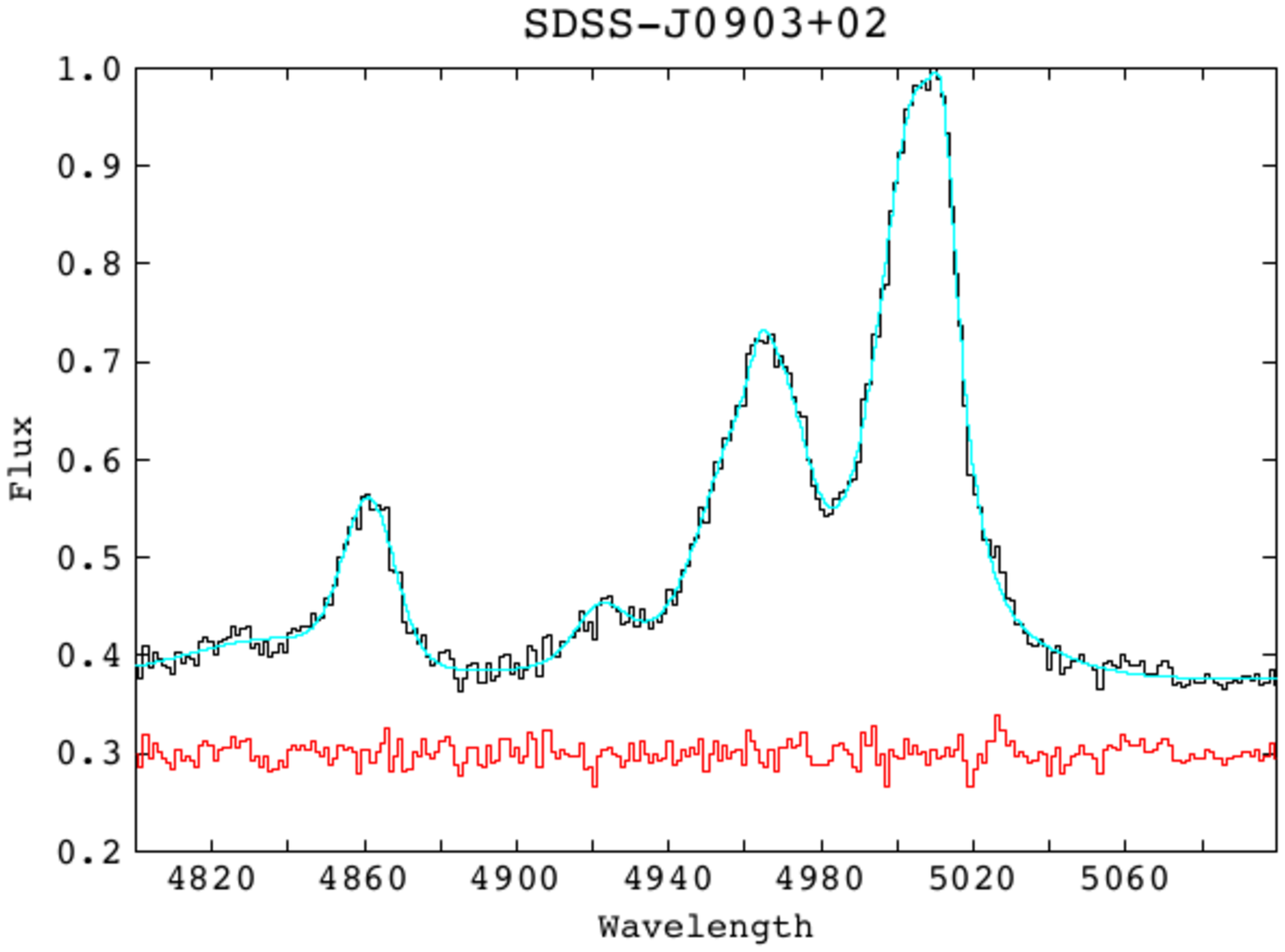}
\includegraphics{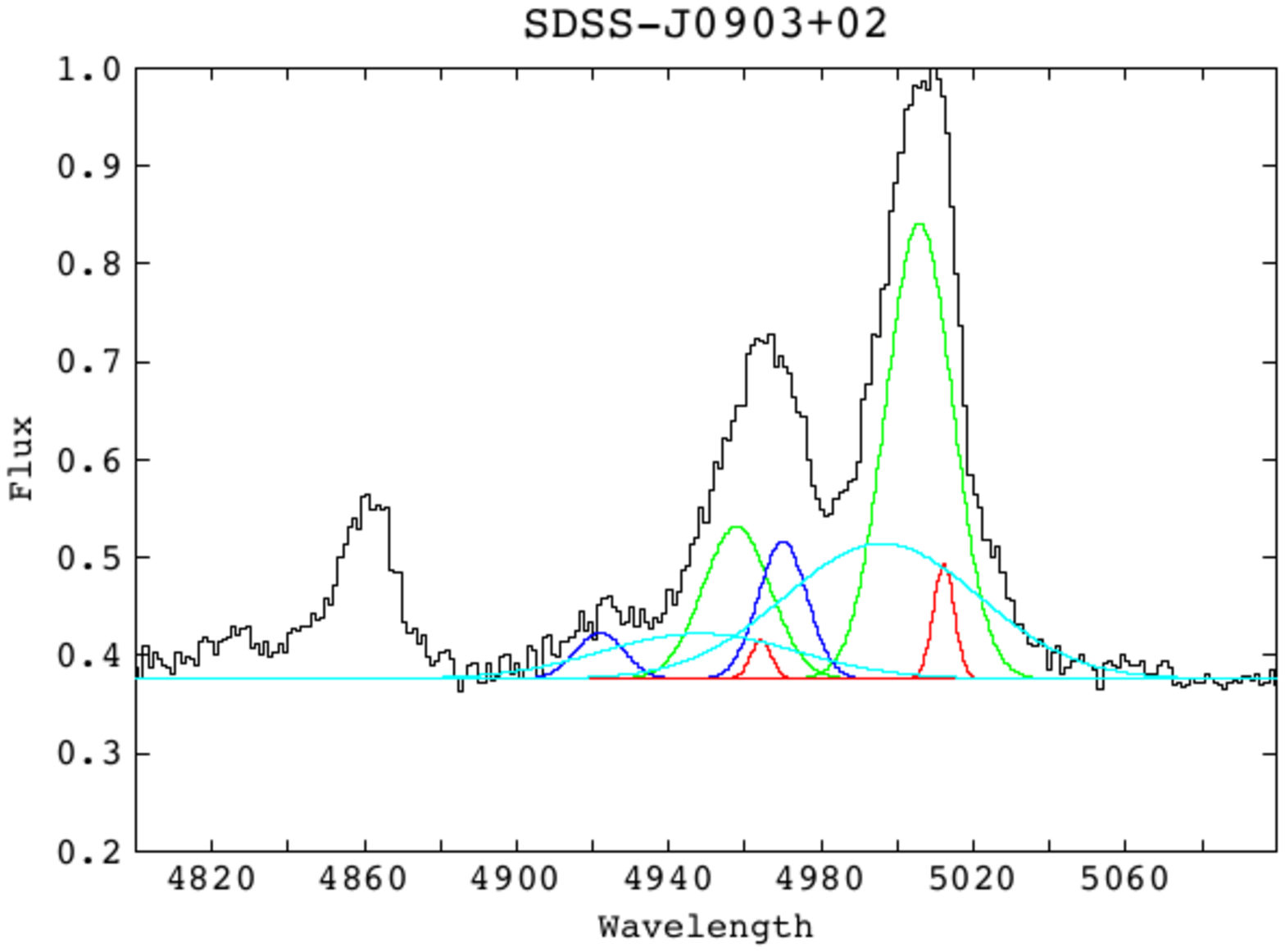}
\vspace{2.2in}
\includegraphics{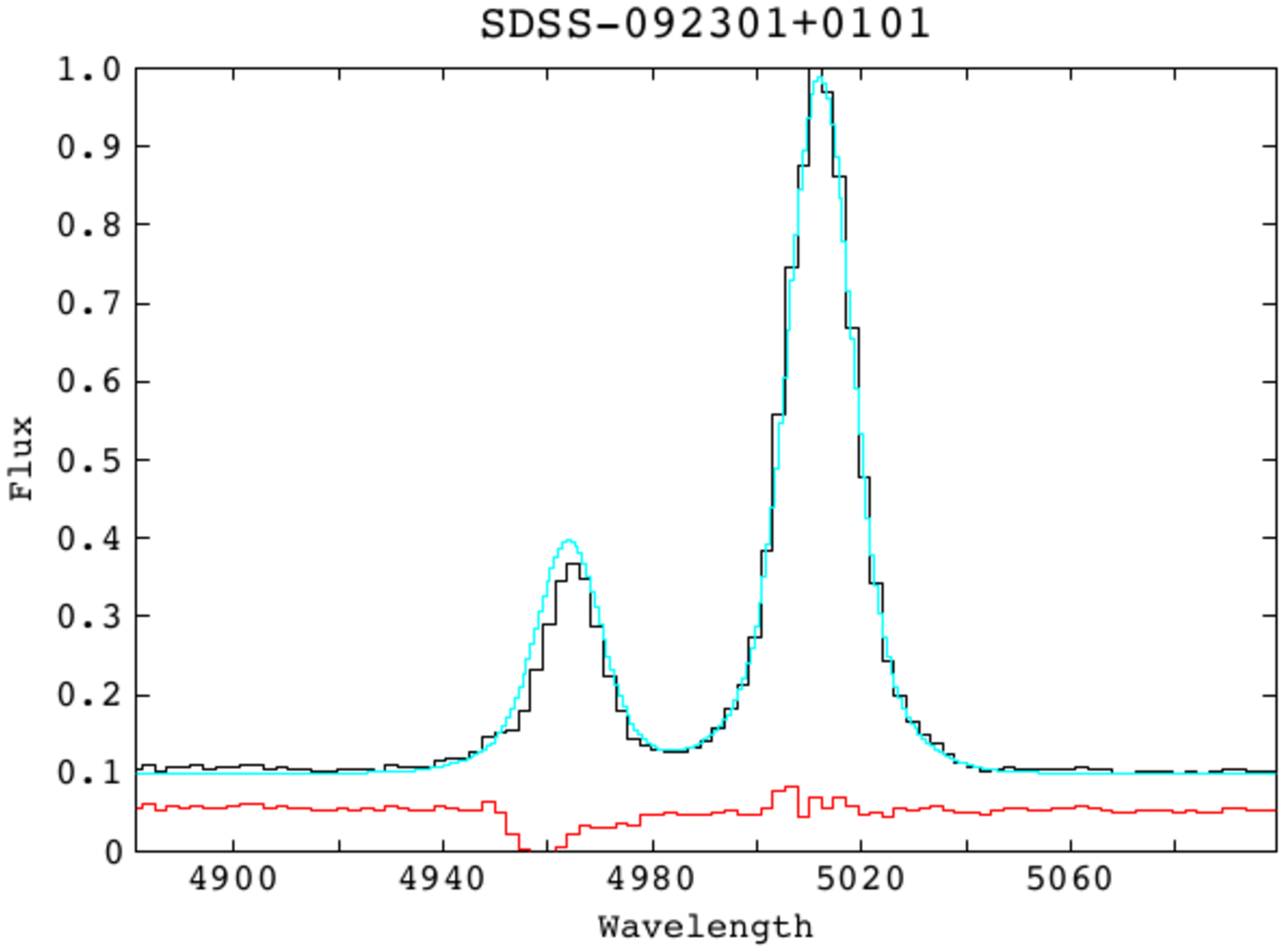}
\includegraphics{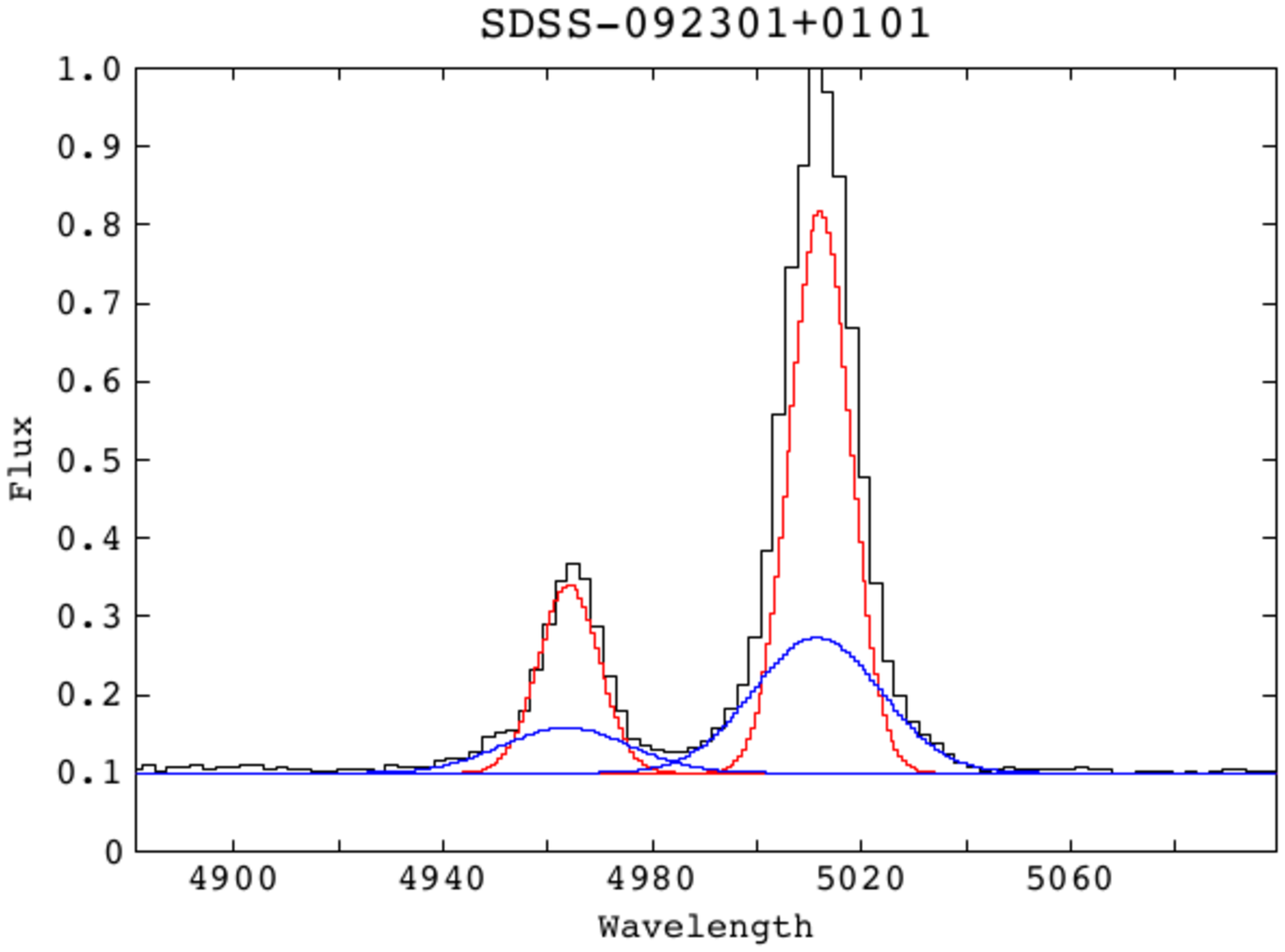}
\vspace{2.2in}
\includegraphics{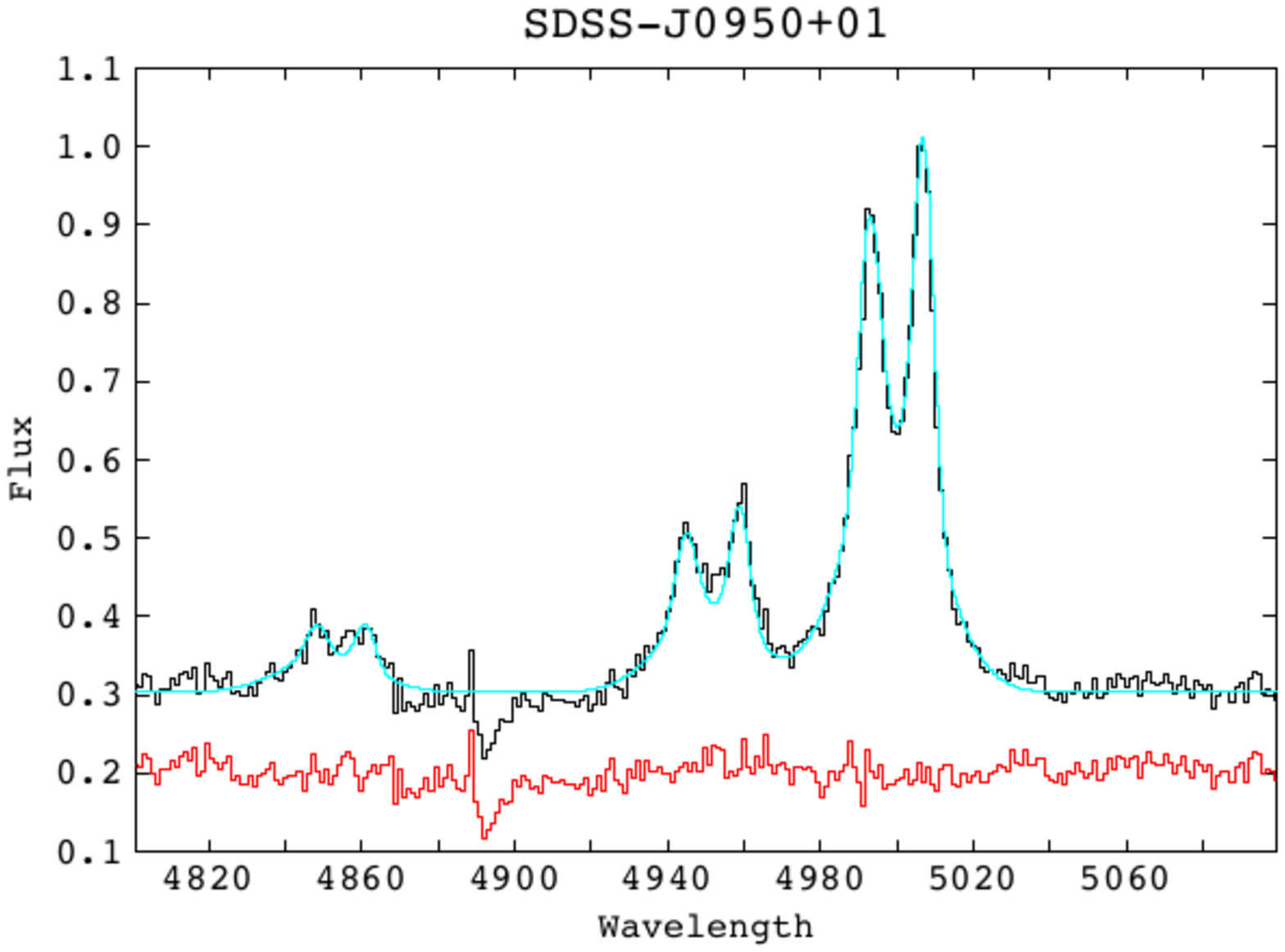}
\includegraphics{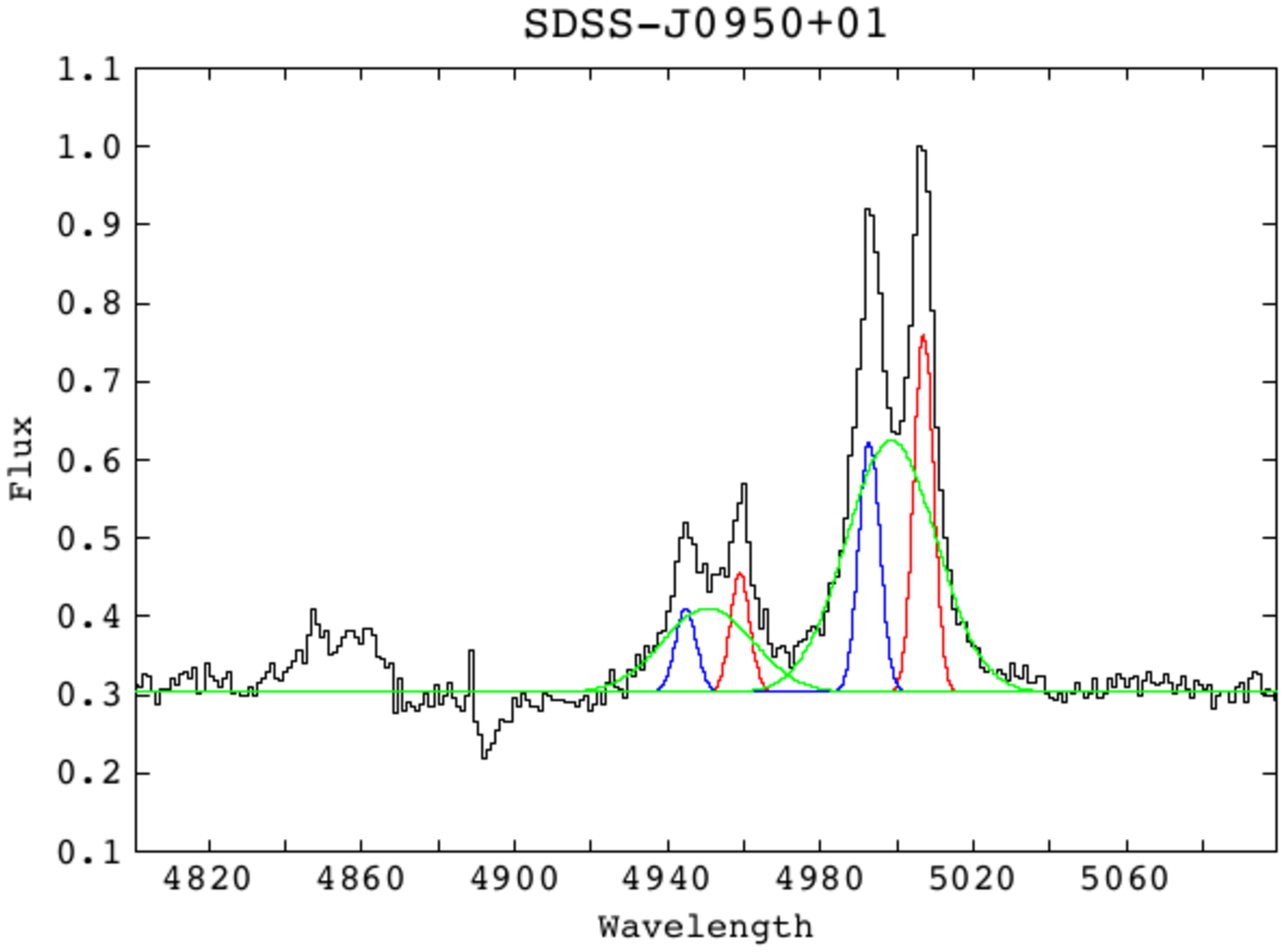}
\vspace{2.2in}
\includegraphics{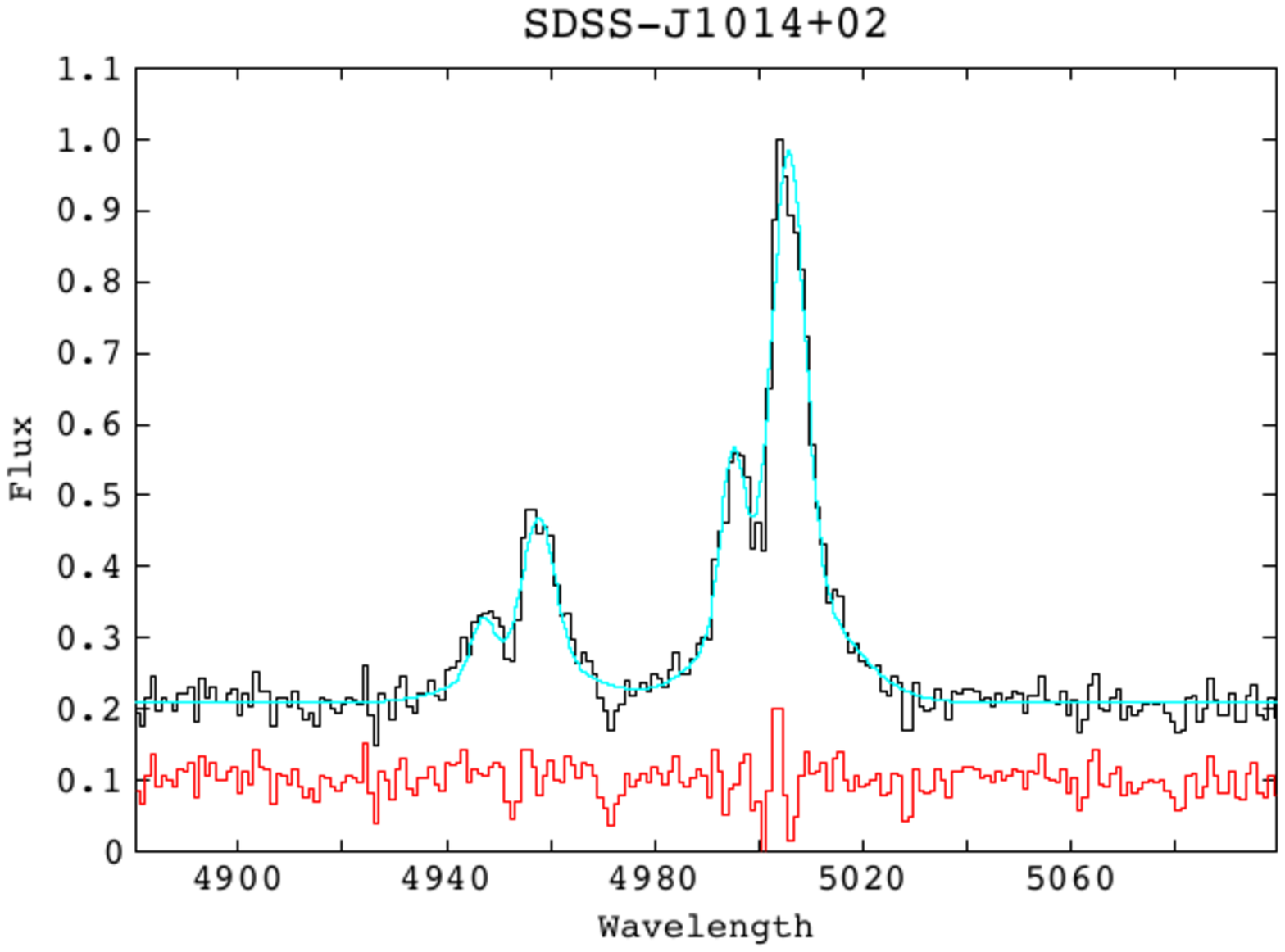}
\includegraphics{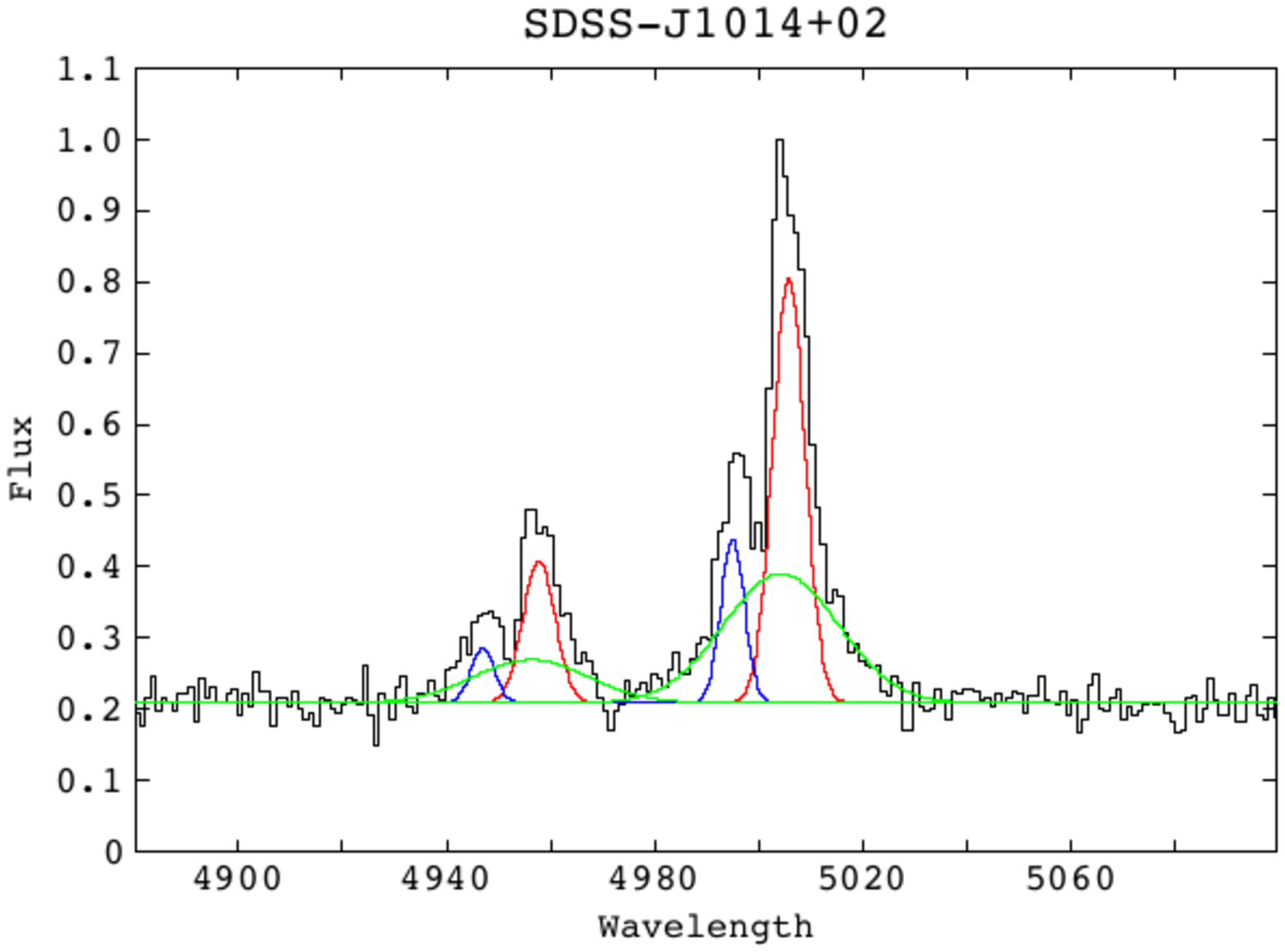}
\vspace{2.2in}
\caption{Fits of the nuclear spectra of the 2011 sample. The data, the fits and residuals are shown in black, cyan
and red respectively (left panels). The residuals have been scaled up for clarity.  The individual components are shown in the right panels with different colors. Cyan, blue, green and red are used  from the most blueshifted to the more redshifted components. Fluxes have been normalized to the peak of  [OIII]$\lambda$5007.}
\label{nuclei1}
\end{figure*}

The spectroastrometric method provides a lower limit on the maximum extent of the outflows  since it does not take into account the spread in radial distances and velocities of the outflowing clouds. For instance,  in the case that these clouds were fairly symmetrically distributed about the continuum centroid, the displacement measured could be small even if the outflow extended to several kpc. On the other hand, this method gives the displacement of the outflow from the continuum centroid at the most extreme velocities, for which its emission dominates the total line flux.  The spatial centroid at lower velocities possibly traces  the ambient gas, even if the outflow also contributes partially to the line flux.

One source of uncertainty is related to the exact location of the AGN, which should mark the spatial zero for the spatial shift measurements.
In QSO1 this is possible by determining the centroid of the continuum emission, since this is clearly dominated by  the  spatially unresolved accretion disk. In QSO2 this
is not necessarily the case because  the central engine is hidden from our line of sight and the optical continuum is likely to be dominated  by the stellar host. In spite of this uncertainty, we will use the continuum centroid (measured  on the red side  near   [OIII]$\lambda$5007) as the spatial zero. In most cases, this coincides with the centroid of the [OIII] line, which is reassuring.  For those few exceptions where both centroids show significant shifts, we will discuss the implications  on
 the estimated outflow sizes.

In order to constrain the outflow sizes, it is essential to identify at which velocities the outflow  emission
 is dominant and then measure the corresponding spatial shifts. This is not trivial, given the complexity of the line profiles which often involve
several kinematic components of different, sometimes uncertain nature. We will  achieve this using the results of the multi-Gaussian spectral decomposition of the nuclear spectra.

The method is illustrated  in Fig. \ref{example0955}.
The top panel shows an example of the fit to the nuclear [OIII] line (black) of one of our targets.
 The individual kinematic components isolated in the fit are shown with different colors. 
The middle and bottom panels show the ratio between the flux emitted by the outflow $\frac{F_{\rm outflow}}{F_{\rm total}}$  and the total line flux
at different velocities, in  two different scenarios described below.
We will assume that when this ratio is $\ge$0.8 (blue areas in the figure), the outflow dominates.  We will  then infer
the spatial centroid for these velocities and the shift relative to the spatial zero. This will give a lower limit on the outflow radial size $R_{\rm o}$. 

The sizes constrained in this way (also with methods (i) and (ii)) depend on our ability to isolate the outflow emission. This is confidently done when the line profiles are  dominated by a narrow core which is likely to trace the underlying gravitational motions and a much broader component with FWHM$\ga$1000 km s$^{-1}$ emitted by the outflow.
 In some cases, such as that shown in Fig. \ref{example0955},
 the line profiles are so complex  (a narrow, an intermediate and a broad component) that the identification of the outflow kinematic components is not  obvious.  The middle and bottom panels illustrate this uncertainty by showing the difference in the velocity ranges where the outflow dominates  if the broadest component is considered the only tracer of the outflow (middle panel) or if  the intermediate component also participates (bottom panel). 
 
This uncertainty has been taken into account by considering the different possibilities and will be mentioned when relevant.

\section{Analysis and Results}

\subsection{Nuclear kinematics}
\label{Sec:nuc}

\begin{figure} 
\includegraphics{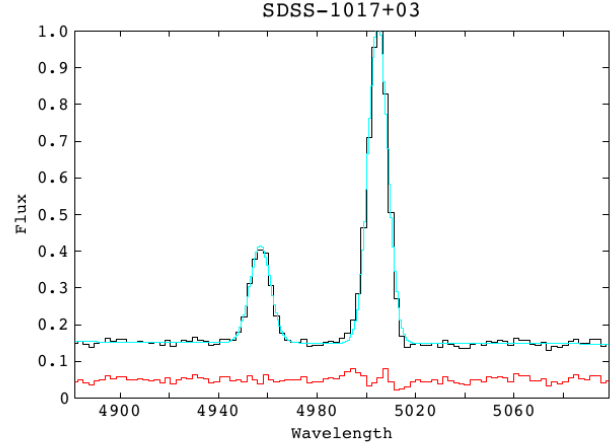}
\vspace{2.5in}
\caption{SDSS J1017+03. Fit (cyan) of the nuclear [OIII] lines (black). No clear evidence for an ionized outflow is found. Flux normalized to the peak of 
[OIII]$\lambda$5007}
\label{nuc1017}
\end{figure}

\begin{figure*} 
\includegraphics{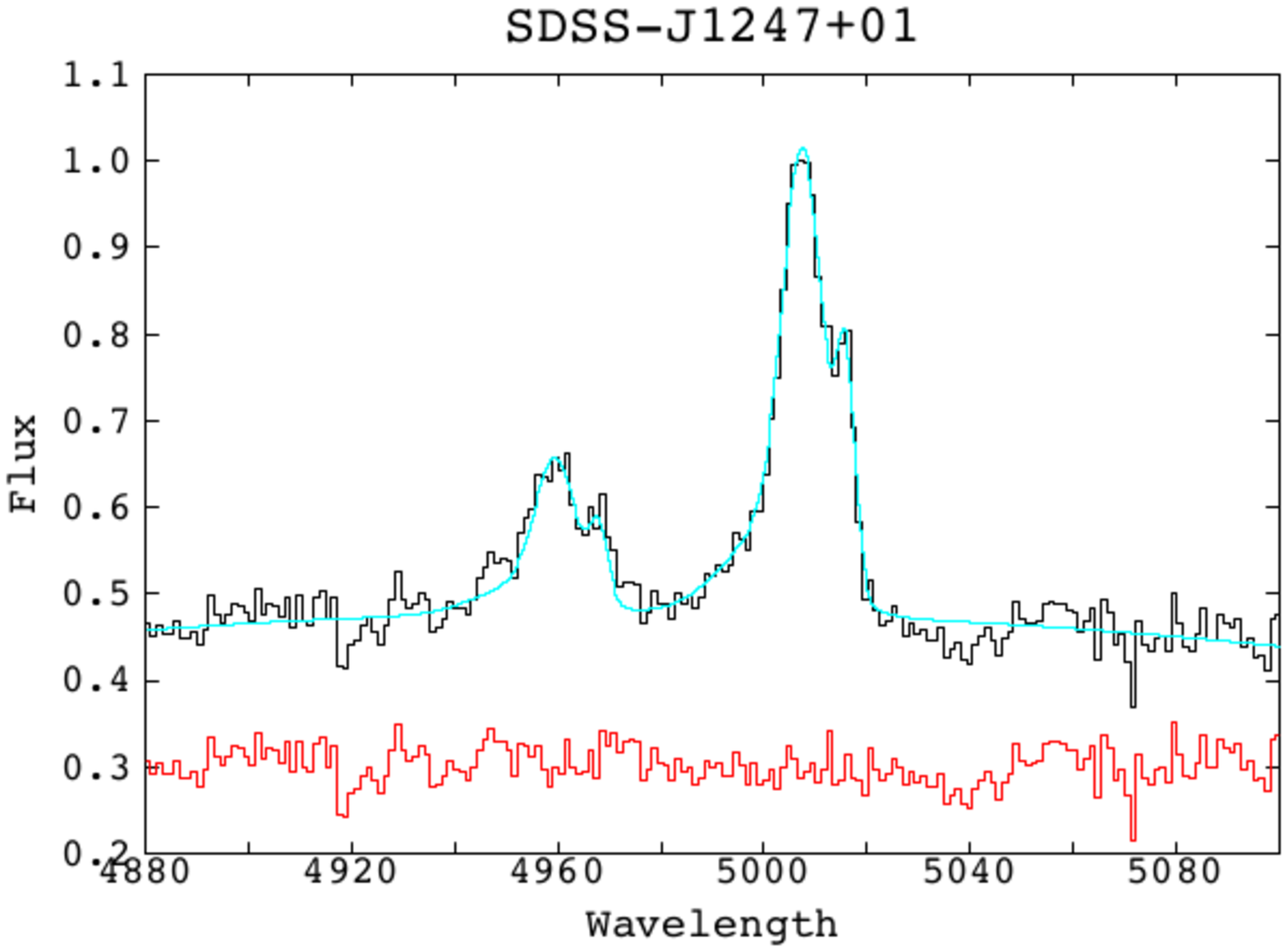}
\includegraphics{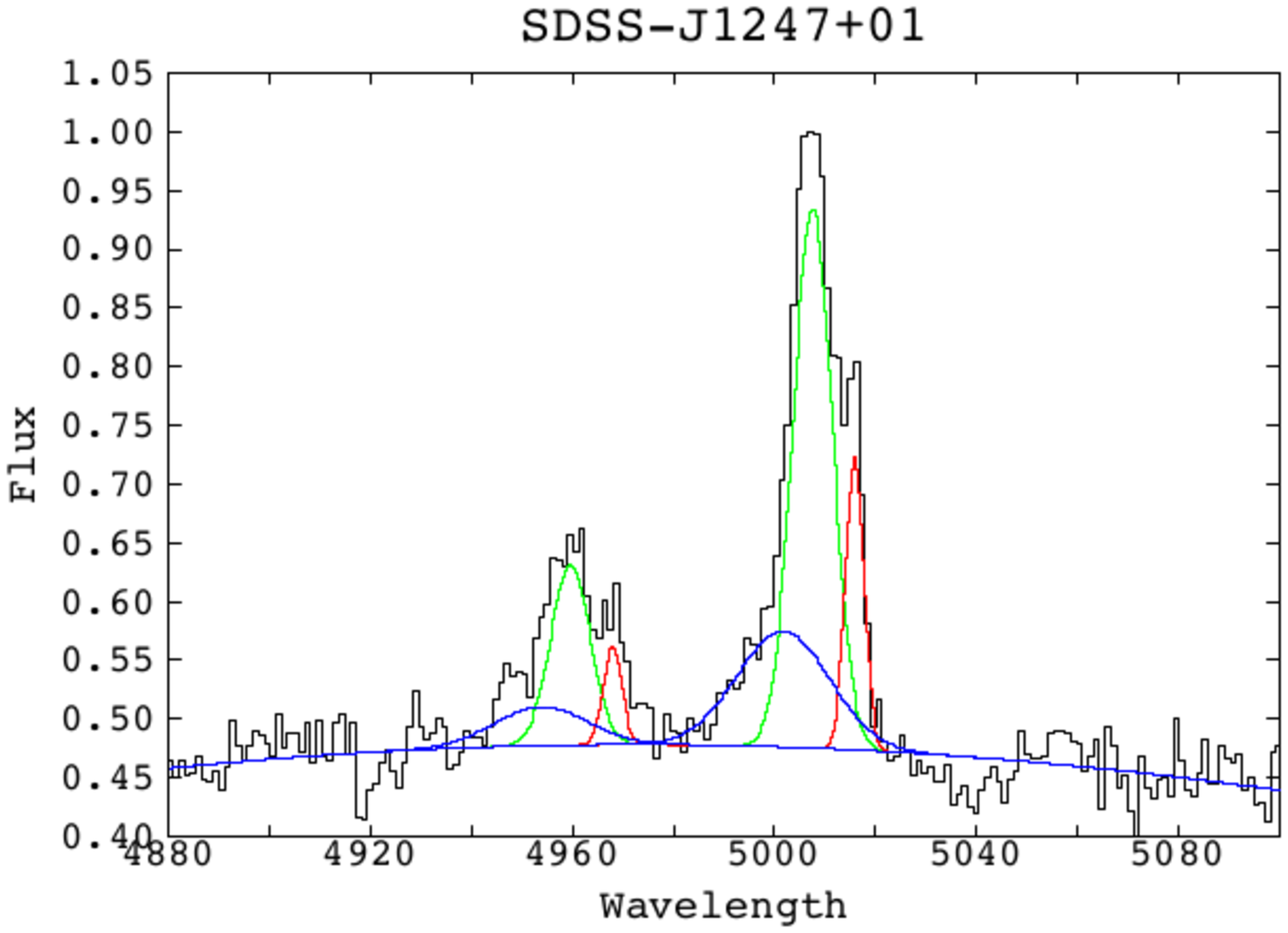}
\vspace{2.2in}
\includegraphics{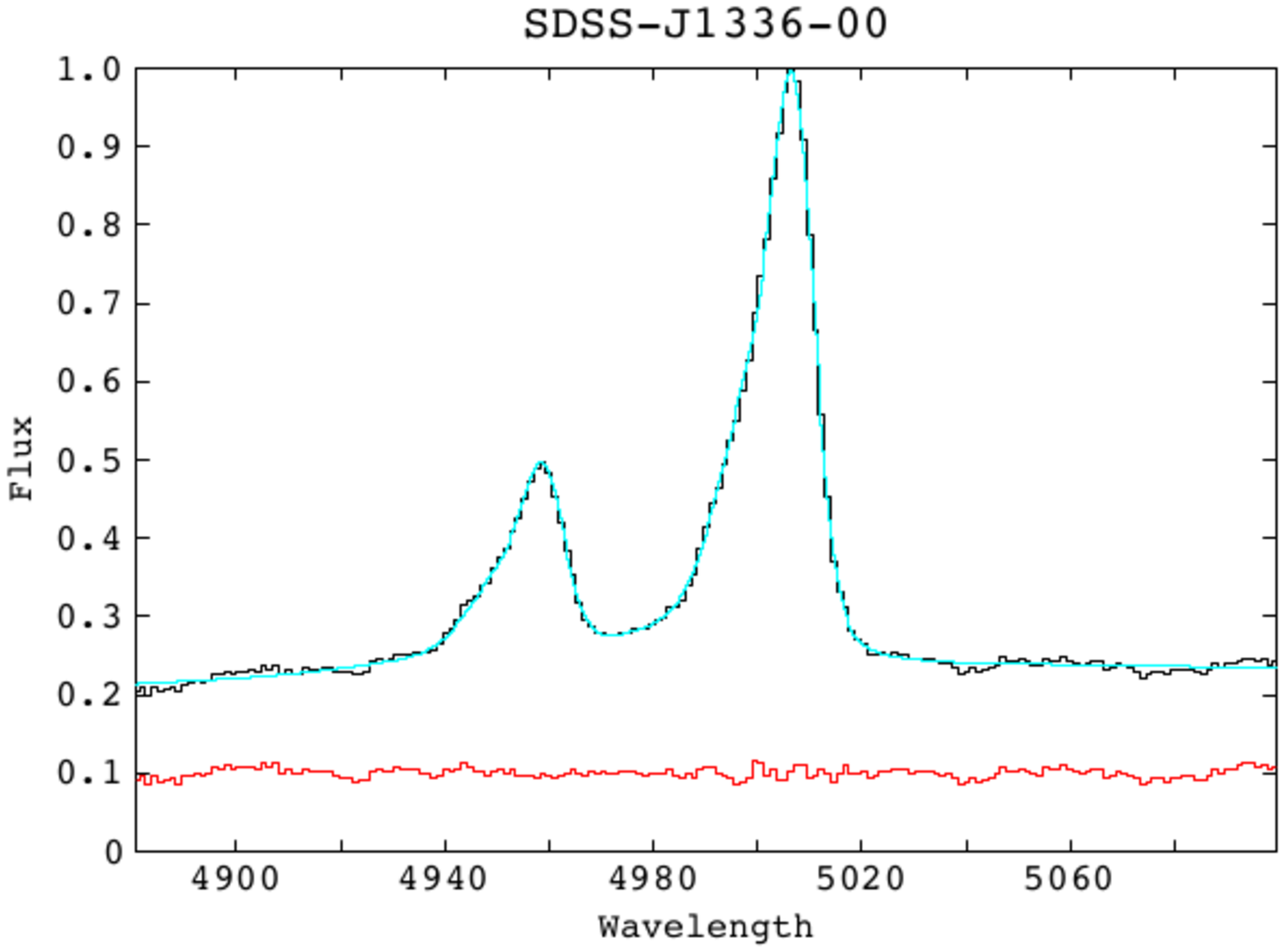}
\includegraphics{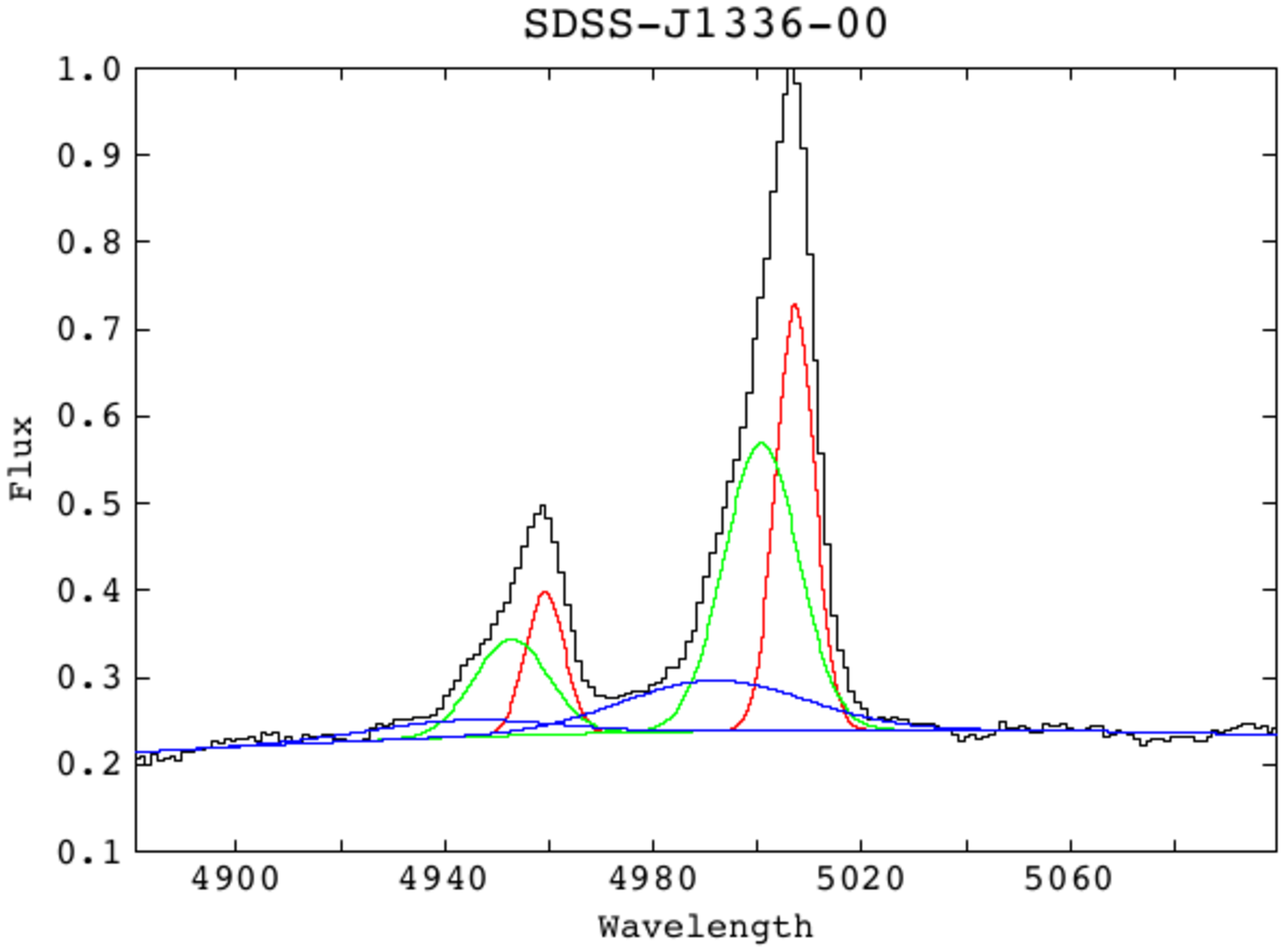}
\vspace{2.2in}
\includegraphics{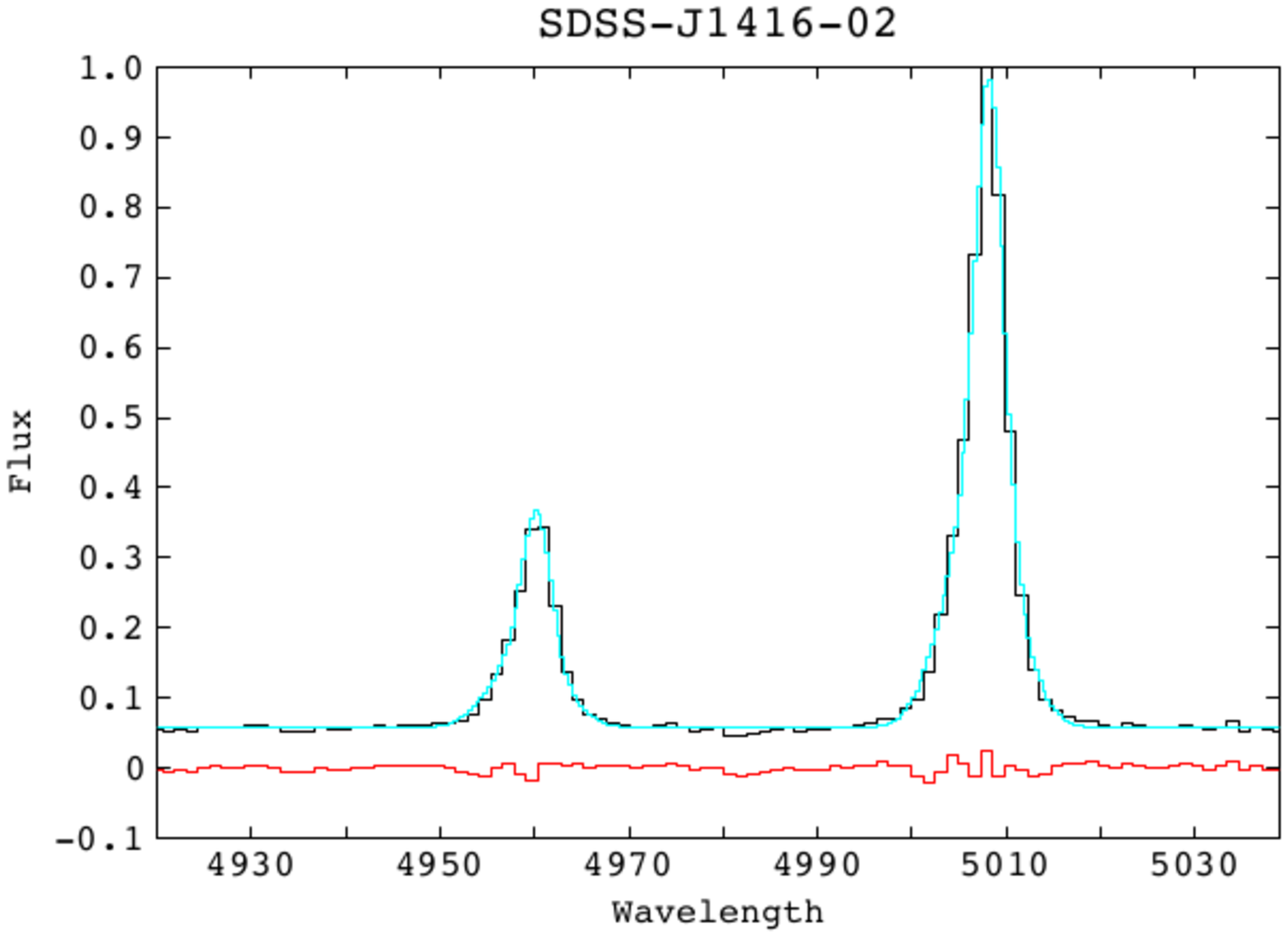}
\includegraphics{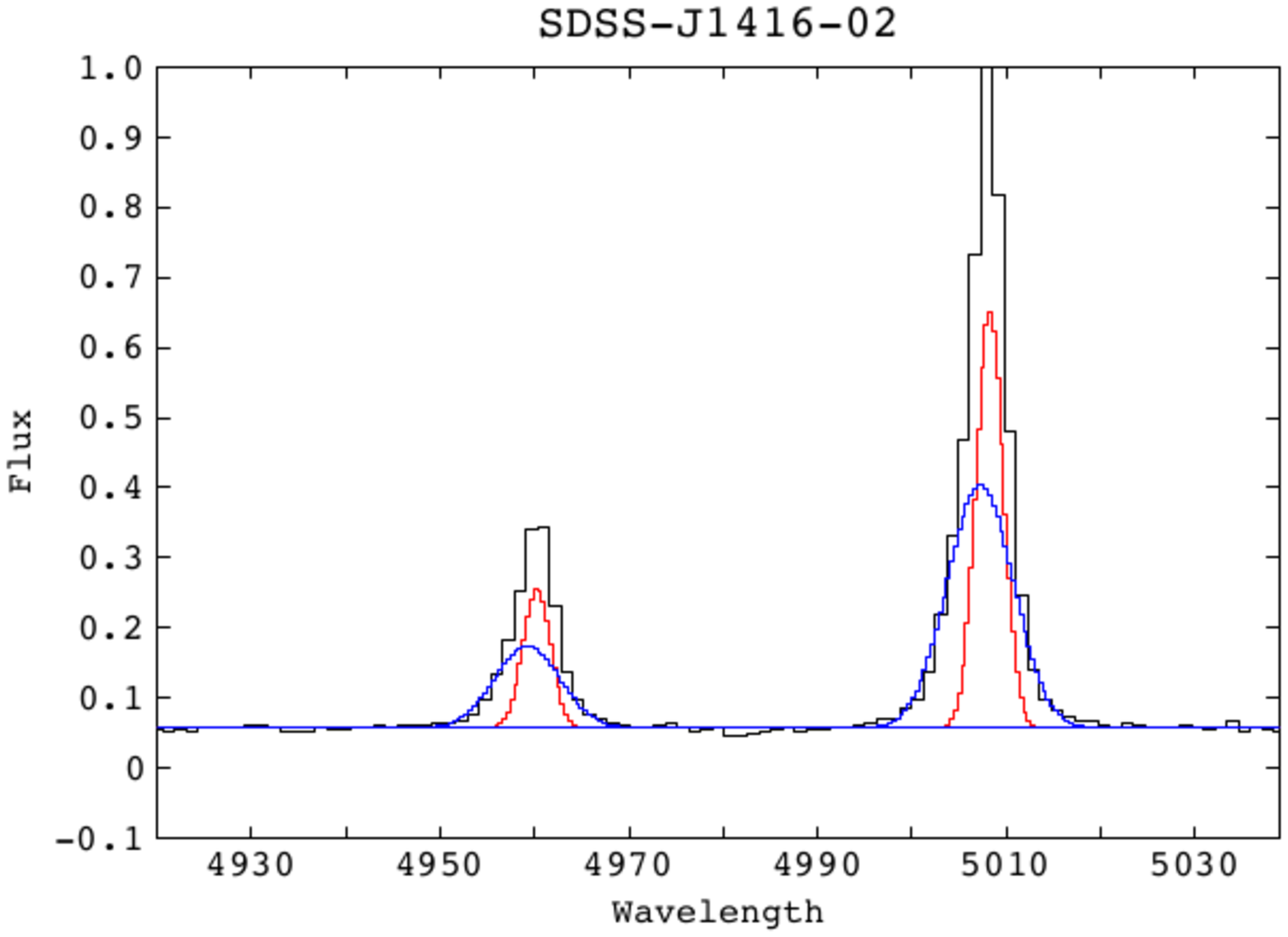}
\vspace{2.2in}
\includegraphics{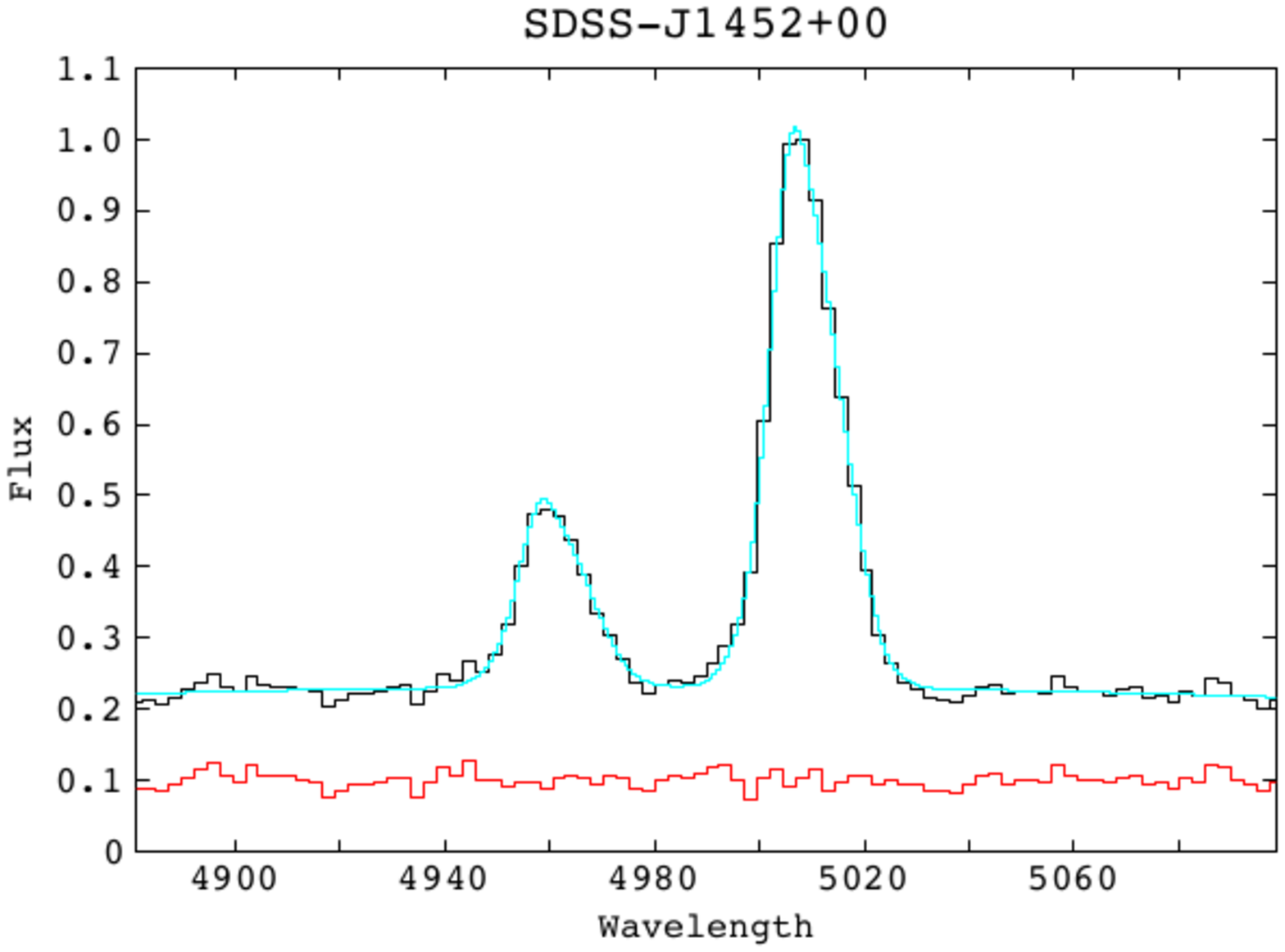}
\includegraphics{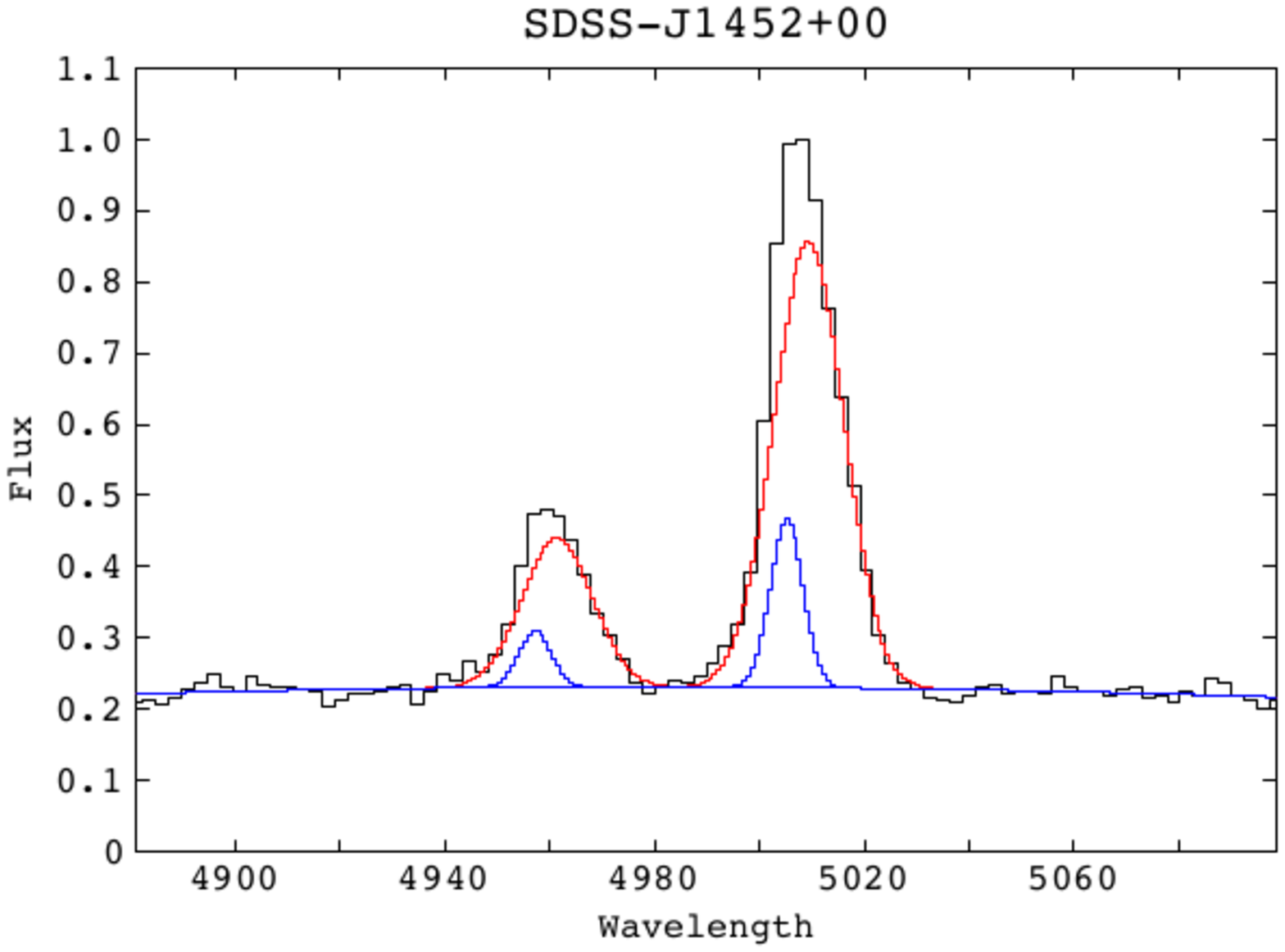}
\vspace{2.2in}
\caption{Fits of the nuclear spectra of the 2011 sample (cont. of Fig. 2).}
\label{nuclei2}
\end{figure*}

In \cite{vm11b} we analyzed the nuclear kinematics  in the 2009 sample. Signatures of ionized outflows were found in all objects.
We perform here a similar analysis  for the 2011 sample (except SDSS J1430-00, which was studied in Villar Mart\'\i n et al. \citeyear{vm12})  applying the same method and using the [OIII]$\lambda\lambda$4959,5007 
doublet. The nuclear spectra were extracted from
  4 to 6 pixel (1.0 to 1.5$\arcsec$) wide apertures centered at the spatial centroid of the line and 
selected to maximize the S/N of the  line wings, where potential outflow signatures are usually identified. 
The results of the fits are shown in Table \ref{tab:fitsnuc} and in   Fig. \ref{nuclei1} to 
 Fig. \ref{nuclei2}.  
Details on individual objects are given in Sect. \ref{Sec:ind} and   Appendix \ref{Sec:app}.  Here we focus on the global results.

In most objects the emission line profiles show kinematic substructure and a variety of  shapes: from highly asymmetric and broad profiles with more than one peak
to simple and symmetric shapes. Such diversity is common in QSO2 (\cite{vm11b}, Liu et al. \citeyear{liu13b}, Harrison et al. \citeyear{har14}, McElroy et al. \citeyear{mce15}).

\begin{table}
\centering
\caption{Results of the  nuclear fits  for the 2011 sample. The properties of the individual kinematic components are
shown. The velocity shifts $V_{\rm s}$ have
been calculated relative to the $\lambda$ of the narrow core of [OIII] (see the text). $F_{\rm rel}$ is the ratio of the flux of a given kinematic component relative to the total line flux.  The slit position angle PA  is quoted  for
each object and spectra (see Paper I for
detailes).} 
\begin{tabular}{lllllllll}
\hline
Comp. &  FWHM & $V_{\rm s}$ & $F_{\rm rel}$   \\
	&  km s$^{-1}$  &  km s$^{-1}$  &    \\   
\hline
&   SDSS J0903+02 & PA1-65.5$\degr$ & \\
Comp 1. & 202$\pm$80	&273$\pm$22	&	0.036$\pm$0.007&  \\ 
Comp 2. & 1194$\pm$40	& -114$\pm$21	& 0.47$\pm$0.03	 \\ 
Comp 3. &	855$\pm$78 & -2280$\pm$27	& 0.10$\pm$0.01	 \\ 
Comp 4. & 3546$\pm$174	& -522$\pm$83 	& 0.39$\pm$0.04	 \\ 
& &   PA2 63.4$\degr$ & \\
Comp 1. & $\la$154 	& -34$\pm$20	&	0.034$\pm$0.005 \\ 
Comp 2. & 1393$\pm$81	& 78$\pm$26	& 0.43$\pm$0.04	 \\ 
Comp 3. &	585$\pm$83 & -2135$\pm$30	& 0.08$\pm$0.01	 \\ 
Comp 4. & 3306$\pm$168	& -947$\pm$138 	& 0.42$\pm$0.05	 \\  \hline
&  SDSS J0923+01 & PA 40.9$\degr$ \\ 
Comp 1. & 708$\pm$34	& -3$\pm$9	&	0.66$\pm$0.11 \\ 
Comp 2. & 1667$\pm$177	& -36$\pm$40	& 0.34$\pm$0.12  \\ 
\hline
&   SDSS J0950+01 & PA 9.9$\degr$  \\
Comp 1. & 180$\pm$39	& 450$\pm$11	&	 0.20$\pm$0.01 \\ 
Comp 2. & 233$\pm$49	& -402$\pm$12	&  0.15$\pm$0.02 \\ 
Comp 3. &	1640$\pm$44 & -56$\pm$24	& 0.65$\pm$0.04	  \\  \hline
&  SDSS J1014+02  & PA1 -5.9$\degr$ &  \\
Comp 1. & 320$\pm$19	& 315$\pm$19	&	 0.42$\pm$0.04  \\ 
Comp 2. & $\la$312	& -316$\pm$21 & 0.12$\pm$0.02  \\ 
Comp 3. &	1537$\pm$72 & 189$\pm$34	& 0.46$\pm$0.06 	 \\ 
&    &  PA2 42.1$\degr$ & \\
Comp 1. & 388$\pm$21	& 334$\pm$19	&	 0.37$\pm$0.03 \\ 
Comp 2. & 254$\pm$32	& -320$\pm$21 & 0.06$\pm$0.02  \\ 
Comp 3. &	1580$\pm$115 & 262$\pm$53	& 0.57$\pm$0.05  \\ \hline
&   SDSS J1017+03 &  PA 37.7$\degr$ & No outflow? \\
Comp 1. & 398$\pm$15	& 0	&	 1.00  \\ 
\hline
&  SDSS J1247+01 & PA 60.7$\degr$ \\ 
Comp 1. & 422$\pm$50	& 23$\pm$21	& 0.53$\pm$0.11  \\ 
Comp 2. & $\la$170	& 525$\pm$22 &  0.16$\pm$0.04  \\ 
Comp 3. &	1173$\pm$198 & -278$\pm$173 &	 0.3$\pm$0.1 	 \\ \hline
&   SDSS J1336-00 & PA 57.4$\degr$ \\
Comp 1. & 479$\pm$14	& 48$\pm$9	& 0.39$\pm$0.03 \\ 
Comp 2. & 984$\pm$29	& -341$\pm$26	&  0.38$\pm$0.03  \\ 
Comp 3. &	2485$\pm$173 & -894$\pm$86 &	 0.23$\pm$0.02	 \\ \hline
&  SDSS J1416-02 &  PA 90.0$\degr$ & No outflow?\\
Comp 1. & Unres.	& 11$\pm$9	&	0.46$\pm$0.06 \\ 
Comp 2. & 370$\pm$37	& -72$\pm$15	& 0.54$\pm$0.07 \\ \hline
&  SDSS J1452+00 & PA -64.6$\degr$ &  No outflow? \\
Comp 1. & $\la$265	& -153$\pm$21	&	0.15$\pm$0.03 \\ 
Comp 2. & 846$\pm$21	& 97$\pm$23	& 0.85$\pm$0.05  \\  \hline
\end{tabular}
\label{tab:fitsnuc}
\end{table}

Given the lack of more adequate information to determine the systemic velocity of the host galaxies (e.g. from stellar features) the velocity shifts $V_{\rm s}$ have been calculated relative to the  $\lambda$ of  [OIII]$\lambda$5007 core. As discussed in  \cite{vm11b}, although the line  is complex and broad, it shows in most cases a narrow core whose $\lambda$ can be measured with high accuracy.  We  use its redshift as an approximation of  $z_{\rm sys}$, although it must be kept in mind that
the uncertainties are large regarding the validity of this assumption (Greene \& Ho   \citeyear{gree05}). For the double peaked sources with two prominent peaks  (SDS J0950+01, SDSS J1014+02) the middle $\lambda$ between the two  peaks has been used instead. 

Six out of nine objects (Table ~\ref{tab:fitsnuc}) show evidence for highly perturbed kinematics with the presence of at least one prominent broad component with FWHM$>$1000 km s$^{-1}$, which is in general blueshifted with respect to $z_{\rm sys}$. SDSS J1014+02 is an exception, but notice that the double peaked profile makes $z_{\rm sys}$ highly uncertain.
If the most prominent peak (Figs. ~\ref{nuclei1}) marks $z_{\rm sys}$, a blueshift of $\sim$-550 km s$^{-1}$ is inferred. If we applied this same criteria to the also double peaked SDSS J0950+01, the outflow would be blueshifted by $\sim$-500 km s$^{-1}$. Such broad components are signatures of ionized outflows (see also \cite{vm11b}).

Three objects show no clear  evidence for  outflows (see Appendix \ref{Sec:app}):
SDSS J1017+03, SDSS J1416-02 and SDSS 1452+00.   Interestingly, they have among the lowest $L_{\rm [OIII]}$ in the sample (Table \ref{tab:sample}).   We will discuss this in Sect. \ref{Sec:discussnuc}.

\subsection{Extended kinematics}
\label{Sec:ext}

\subsubsection{Individual objects}
\label{Sec:ind}

We present in this section the detailed analysis  of  four objects in the sample to highlight more clearly the main results of our study. SDSS J1307-02 and SDSS J1407+02 show  evidence for ionized outflows barely extended relative to the seeing disk;
  SDSS J1153+03  is also an   uncertain candidate to harbor a barely  resolved outflow; SDSS 0903+02  is the only object in our total sample of 18 objects for which the data are consistent with  a galaxy scale outflow.  They are, moreover, 
 good examples that illustrate well the difficulties inherent in this and related studies.

The study of the rest of the sample is  described in detail in  Appendix \ref{Sec:app}.

We summarize in Table \ref{tab:kinresults} the results of the  analysis of the 2009 and 2011 samples using the different methods described in Sect. \ref{Sec:methods}
The main results on the extent of the outflows in individual objects are presented in Table \ref{tab:sizes}.

\begin{table*}
\centering
\tiny
\caption{Main properties of the ionized outflows. W$_{max}$ and $V_{max}$ (columns 2 and 3)
 refer to the FWHM and $V_{\rm s}$ of the broadest component isolated in the nuclear spectra (Table 2). 
The values for the 2009 sample have been taken from VM11b. Column (4) indicates whether the ionized gas is spatially extended 
along the slit position angle. Column (5) specifies whether the objects shows merger/interactions signatures
(see VM11a and Paper I for 2009 and 2011 samples respectively).  Column (6) quotes whether the objects show unambiguous evidence for an ionized outflow.
Columns 7 to  9 quote whether the outflows are  spatially  extended as inferred from: the comparison between the  spatial profile of the outflowing gas and the seeing (method (i), column 7);  possible kinematic spatial variations of the outflowing gas accross the seeing disk (method (ii), column 8);    the spectroastrometric analysis (method (iii), column 9). N/A in these columns means that the method could not be applied. A question mark indicates that the conclusion (extended or not extended) is uncertain.  
The FWHM of the
lines in the  most  extended gas is quoted in column (10), when this can be clearly isolated from
the dominant central source. In general, the values are gross upper limits (see text). A ``?" in column (8) means that the necessary analysis could not be performed.} 
\begin{tabular}{llllllllllll}
Target  and slit PA &   W$_{max}$ & $V_{max}$ & Ionized gas & Interactions  & Outflow? & Method (i) & Method (ii) & Method (iii) & FWHM$_{\rm ext}$ \\
& km s$^{-1}$ &  km s$^{-1}$  &     extended? & /mergers? &  &  extended? & extended?  & extended? & km s$^{-1}$ \\
(1) & (2) & (3) & (4) & (5) & (6) & (7) & (8)  & (9) & (10)  \\ 
\hline
  &	&	& 2009	& sample	&	& \\ \hline
SDSS J0955+03 -65.0$\degr$ &     2500$\pm$200 & -620$\pm$140 &  Yes & No & Yes & N/A & No & ?  & $\la$267   \\
SDSS J1153+03 65.0$\degr$ & 1450$\pm$80 & 330$\pm$70  &  Yes & Yes & Yes & No & Yes? & Yes &$\la$270  \\
 SDSS J1228+00 -10.0$\degr$ &  2300$\pm$70 & 770$\pm$20  &  No & No & Yes & N/A & N/A & No &  N/A  \\
   SDSS J1307-02  76.0$\degr$&    1000$\pm$40 & -190$\pm$20 &  Yes & Yes  & Yes & Yes & Yes & Yes & 255$\pm$41  \\  
SDSS J1337-01 -21.0$\degr$ &    1280$\pm$60 & -290$\pm$10  & No &No & Yes  & No & No & Yes & N/A   \\
SDSS J1407+02 -30.0$\degr$&  1670$\pm$120 & -70$\pm$30  & No & No & Yes & Yes & Yes & Yes &  N/A   \\
 SDSS J1413-01 45.0$\degr$&     1190$\pm$50 & -145$\pm$30 &  No & No & Yes & No & No & Yes & N/A \\
 SDSS J1546-00 -71.0&    780$\pm$30 & -270$\pm$-35 & No & No & Yes & No  &  No & No & N/A  \\  \hline
  &	&	&	2011 &	sample &	& \\ \hline
SDSS J0903+02  PA1-65.5$\degr$   & 3546$\pm$174 & -522$\pm$83 & Yes &  Yes &  Yes&  Yes? & N/A  & Yes &  $\la$200  \\
PA2 63.4$\degr$ &      3306$\pm$168 & -947$\pm$138 &  Yes  &    &    Yes & Yes? & N/A  & Yes &  $\la$180,934$\pm$32   \\
SDSS J0923+01 40.9$\degr$  &  1667$\pm$177 & -36$\pm$40 &  Yes & Yes  & Yes &  No & No & Yes & 212$\pm$32 \\
 SDSS J0950+01 9.9$\degr$  &   1640$\pm$44 & -56$\pm$24  &  Yes & Yes & Yes &   No & No & Yes & $\la$235  \\
  SDSS J1014+02 PA1-5.9$\degr$ &    1537$\pm$72 & 189$\pm$34 & No &  Yes & Yes & No & No &  No &  N/A   \\
PA2 42.1$\degr$  &      1580$\pm$115 & 262$\pm$53 &  Yes  &   & Yes   & No  & No  & No & $\la$211 \\
 SDSS J1017+03  37.7$\degr$ &     N/A & N/A&  Yes & Yes &  No & N/A & N/A & N/A & $\la$140   \\
 SDSS J1247+01 60.7$\degr$ & 1173$\pm$198 & -278$\pm$173 & Yes& Yes & Yes & N/A & N/A & Yes & $\la$154  \\
 SDSS J1336-00 57.4$\degr$  &  2485$\pm$173 & -894$\pm$86  &  No & No & Yes & No & No & Yes & N/A   \\
 SDSS J1416-02   90.0$\degr$ &     N/A & N/A & Yes &Very likely   & No & N/A & N/A & N/A & $\la$200 \\
 SDSS J1430-00 9.8$\degr$ &   1600$\pm$200 &  -520$\pm$60 & Yes &  Yes &  Yes &  No & No & Yes & $\la$250  \\
  SDSS J1452+00  -64.4$\degr$ &  N/A & N/A &  Yes &  Yes & No &  N/A   & N/A & N/A & ?   \\
\hline
\end{tabular}
\label{tab:kinresults}
\end{table*}

\vspace{0.2cm}

{\it SDSS J1153+03}

\begin{figure*}
\includegraphics{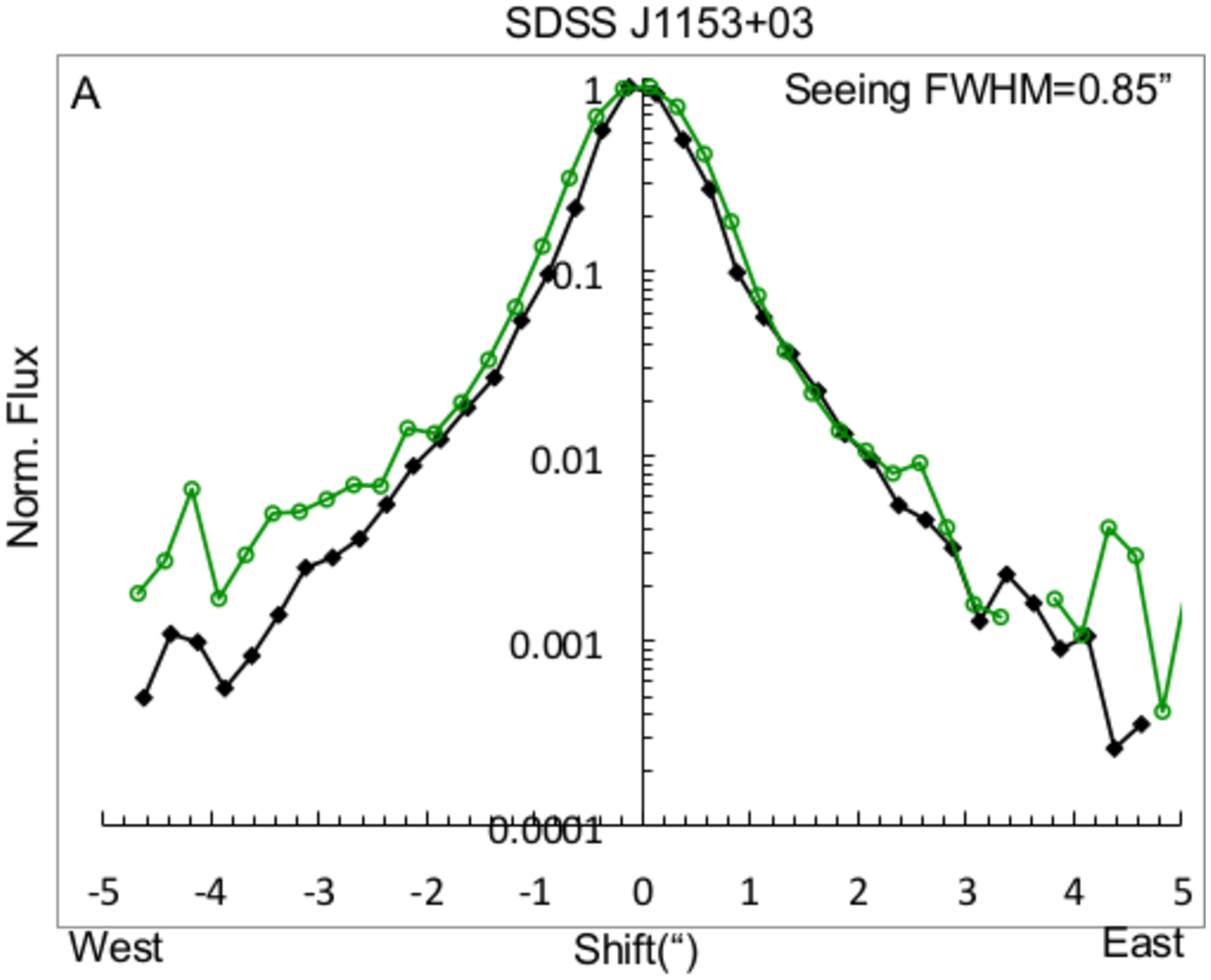}
\includegraphics{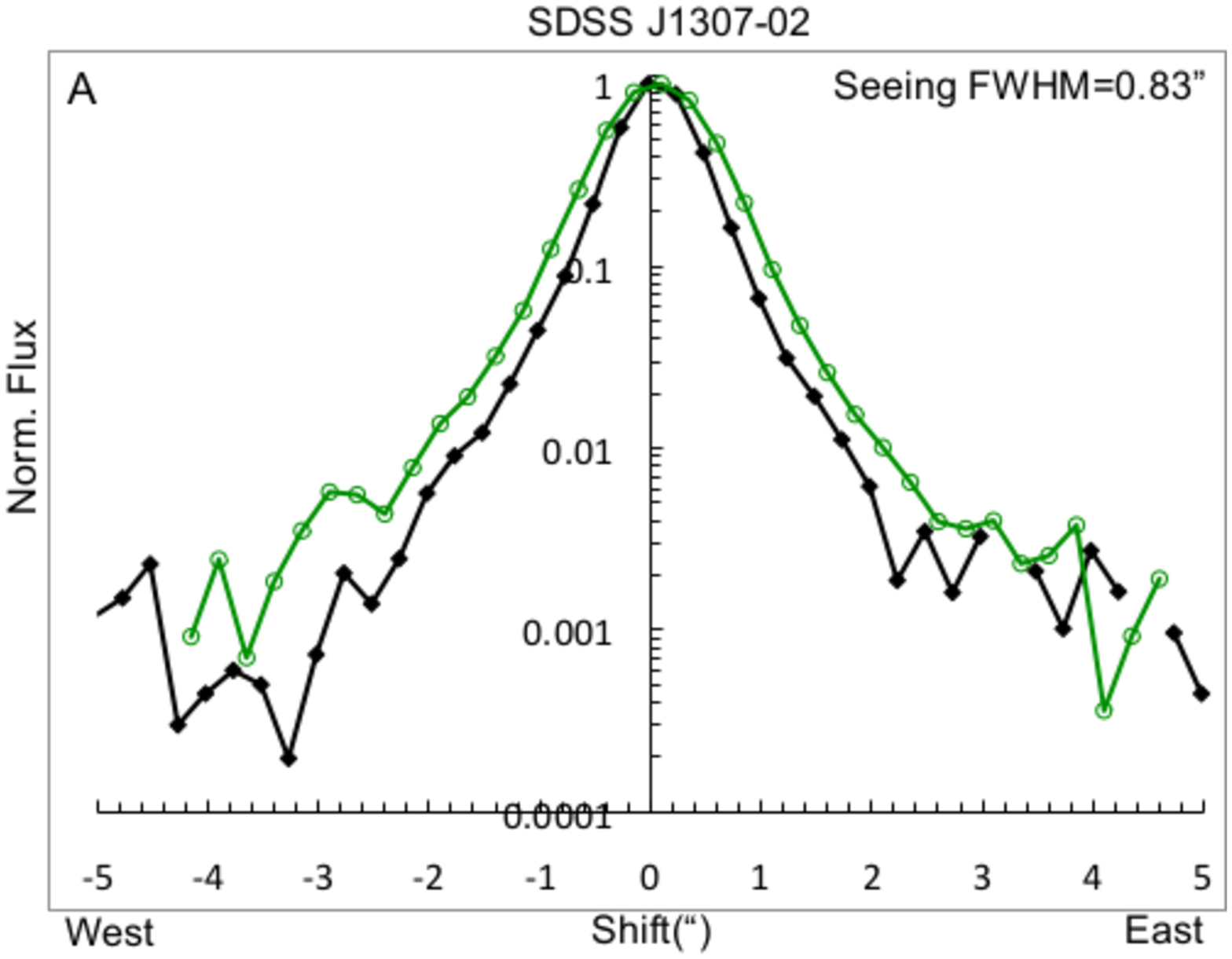}
\includegraphics{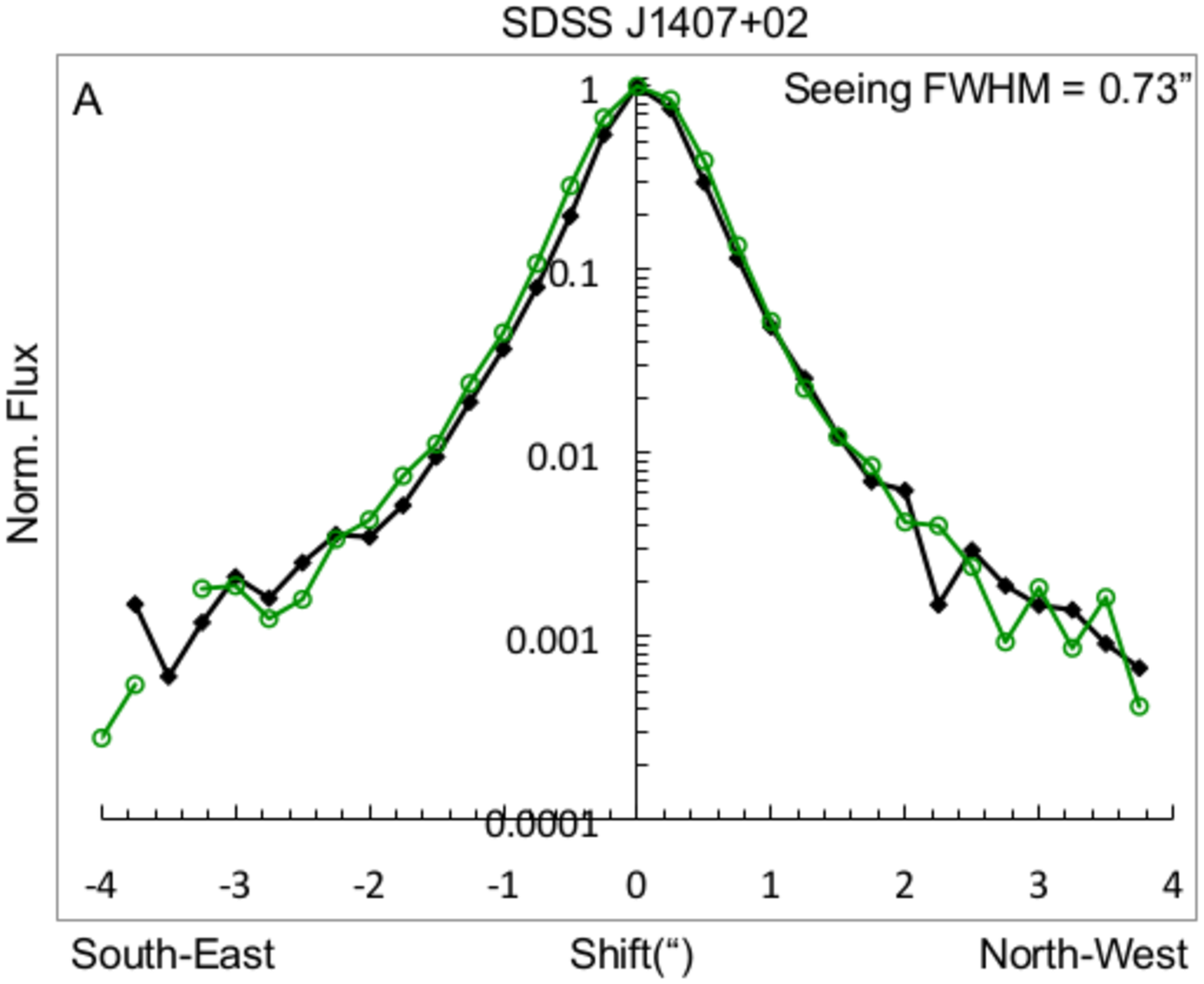}
\vspace{1.9in}
\includegraphics{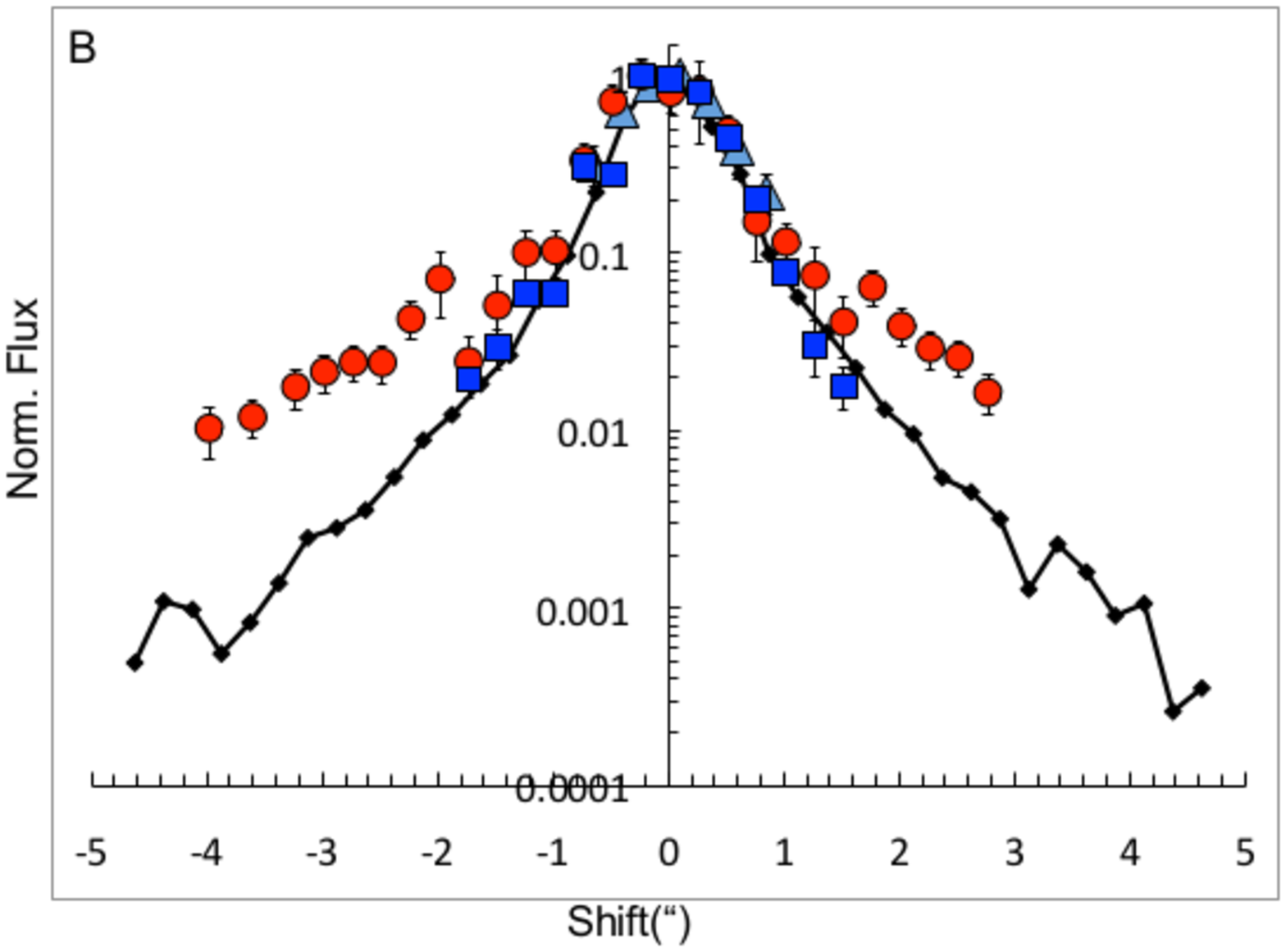}
\includegraphics{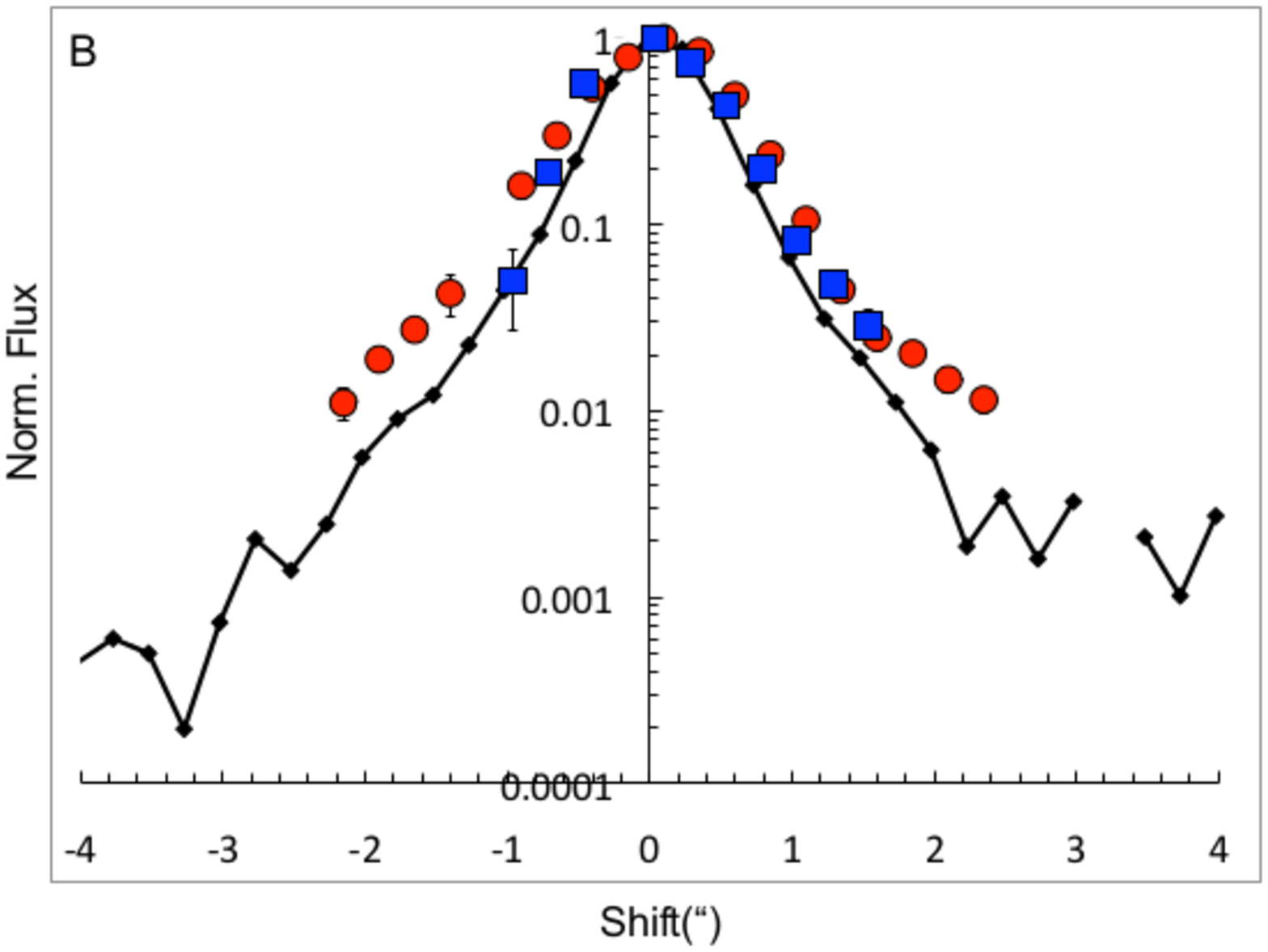}
\includegraphics{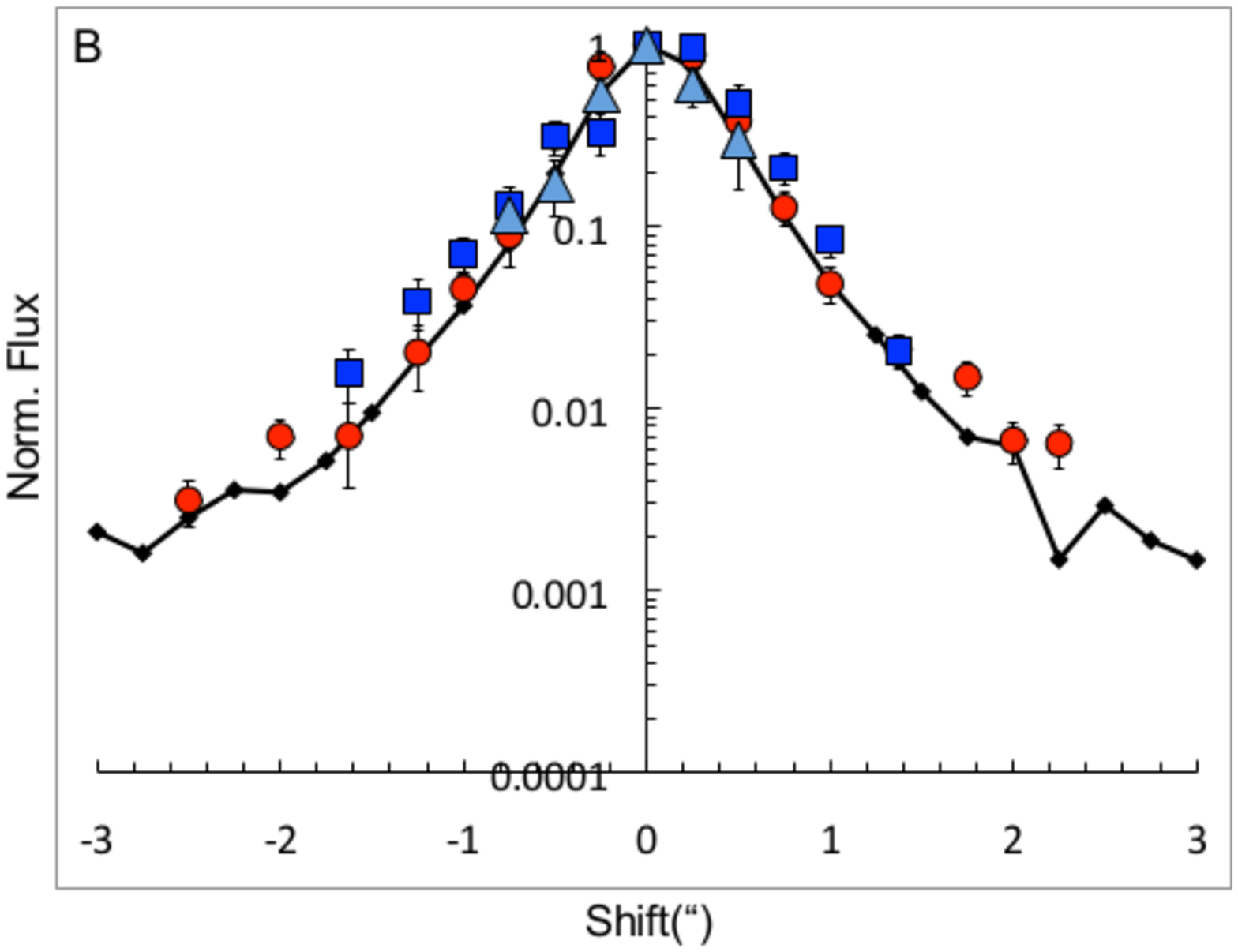}
\vspace{1.9in}
\includegraphics{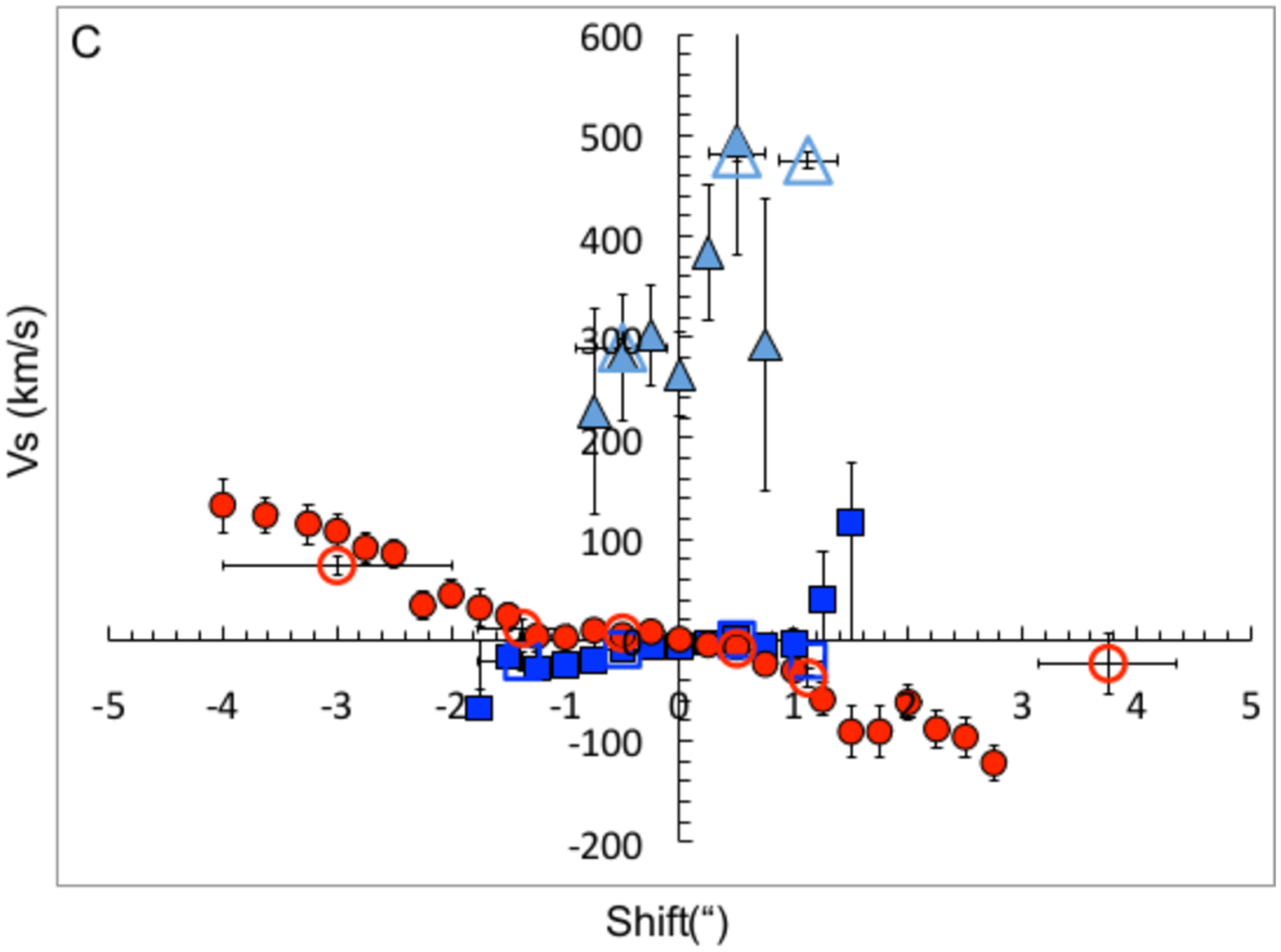}
\includegraphics{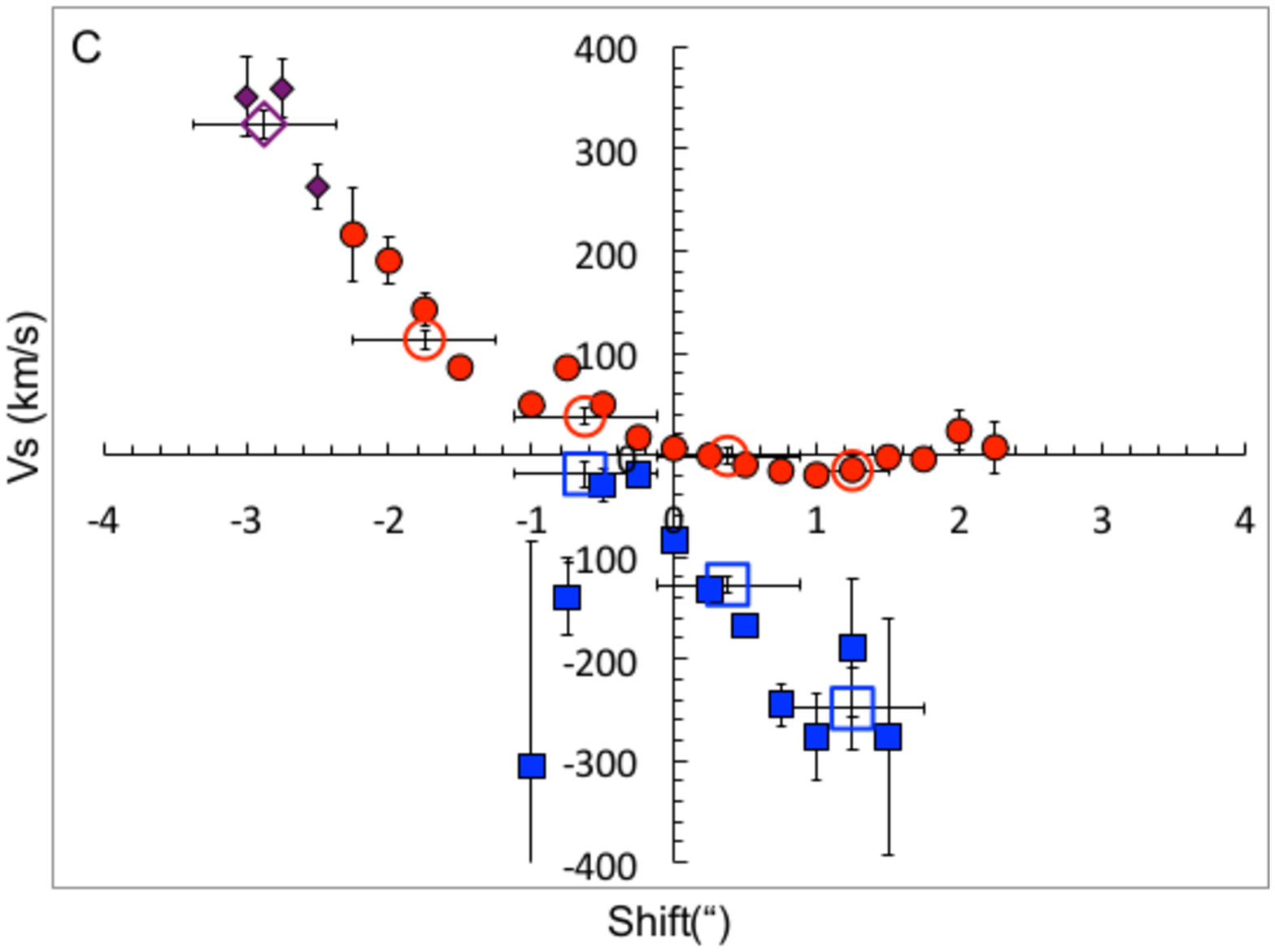}
\includegraphics{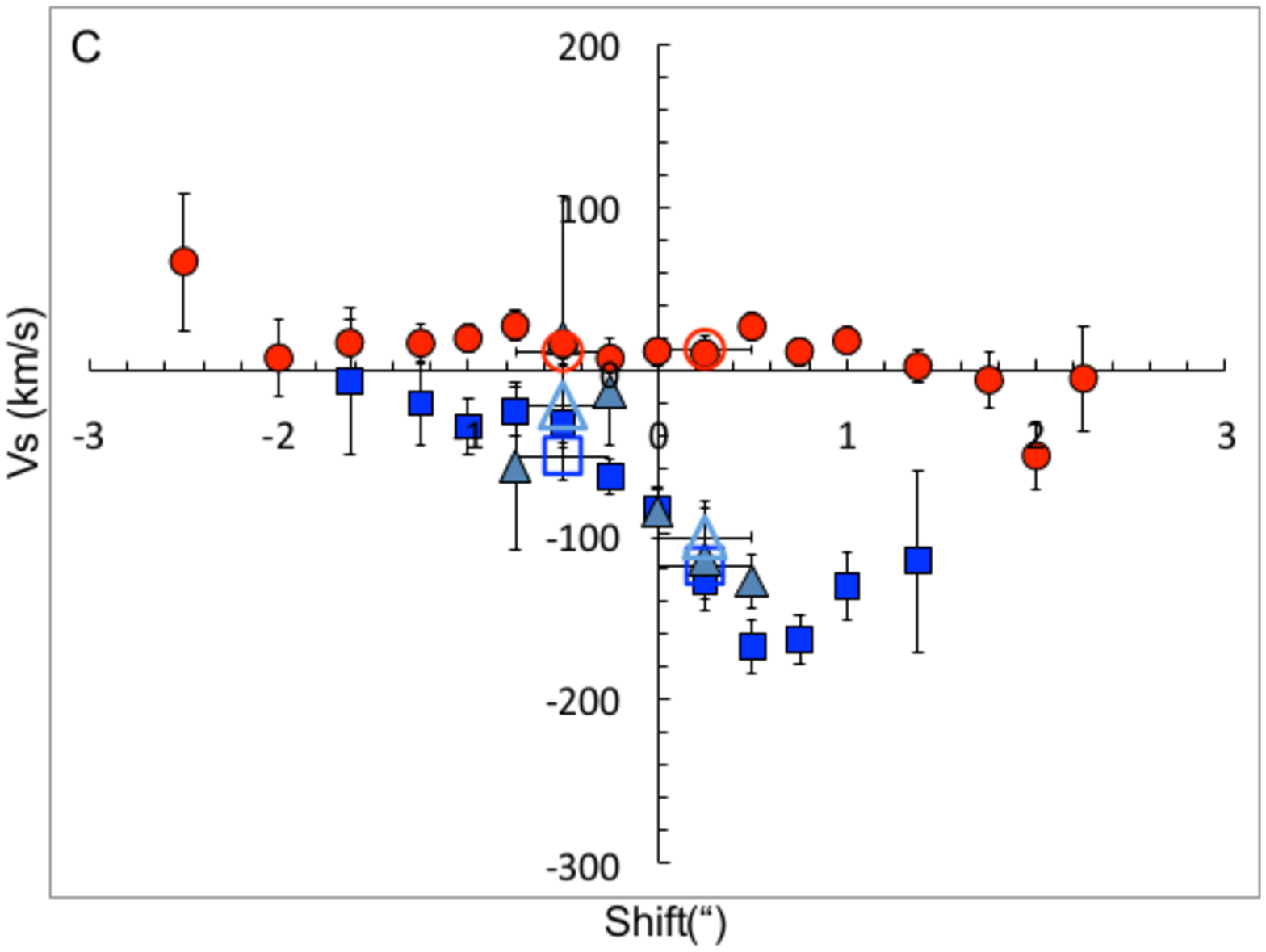}
\vspace{1.9in}
\includegraphics{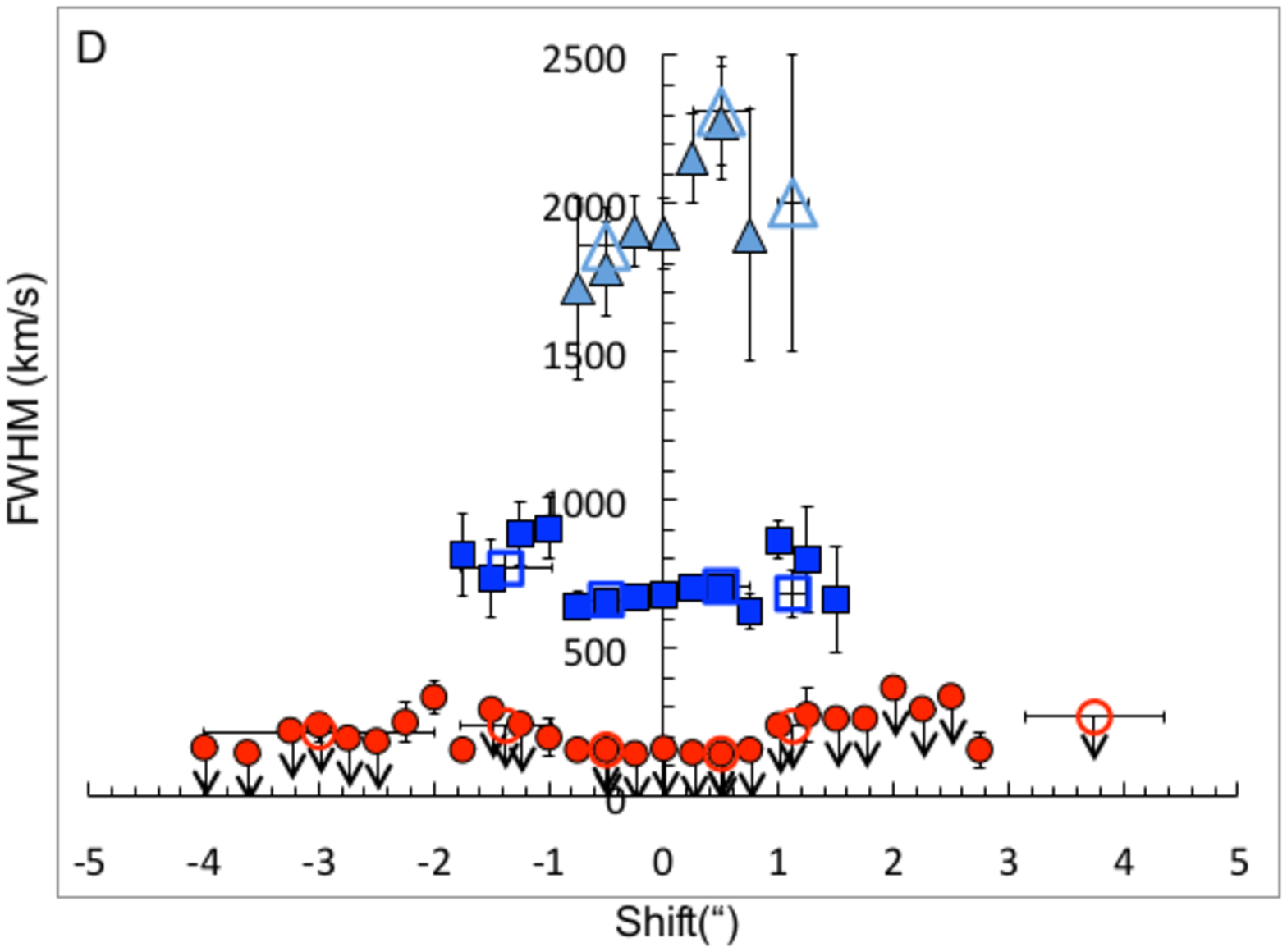}
\includegraphics{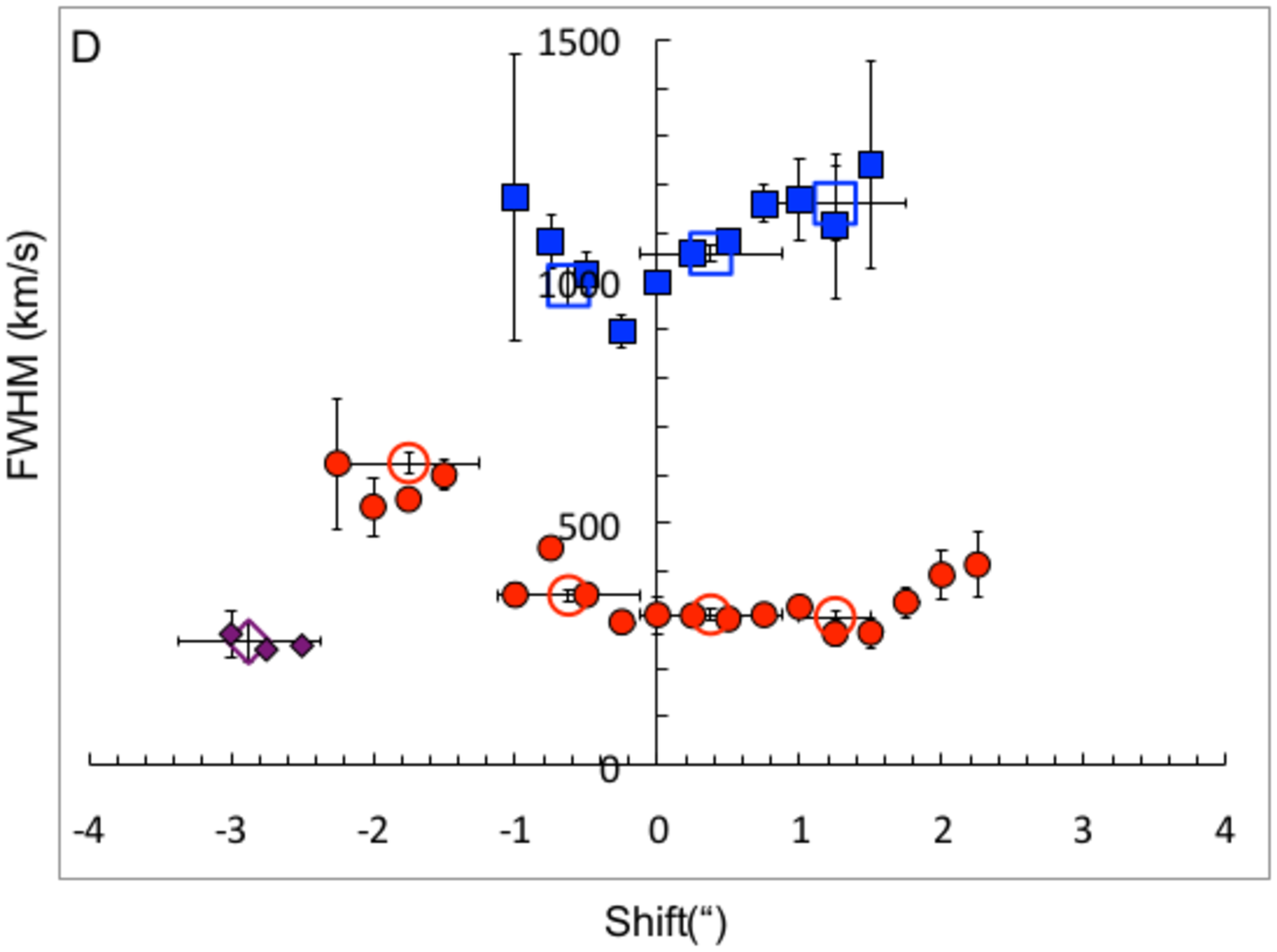}
\includegraphics{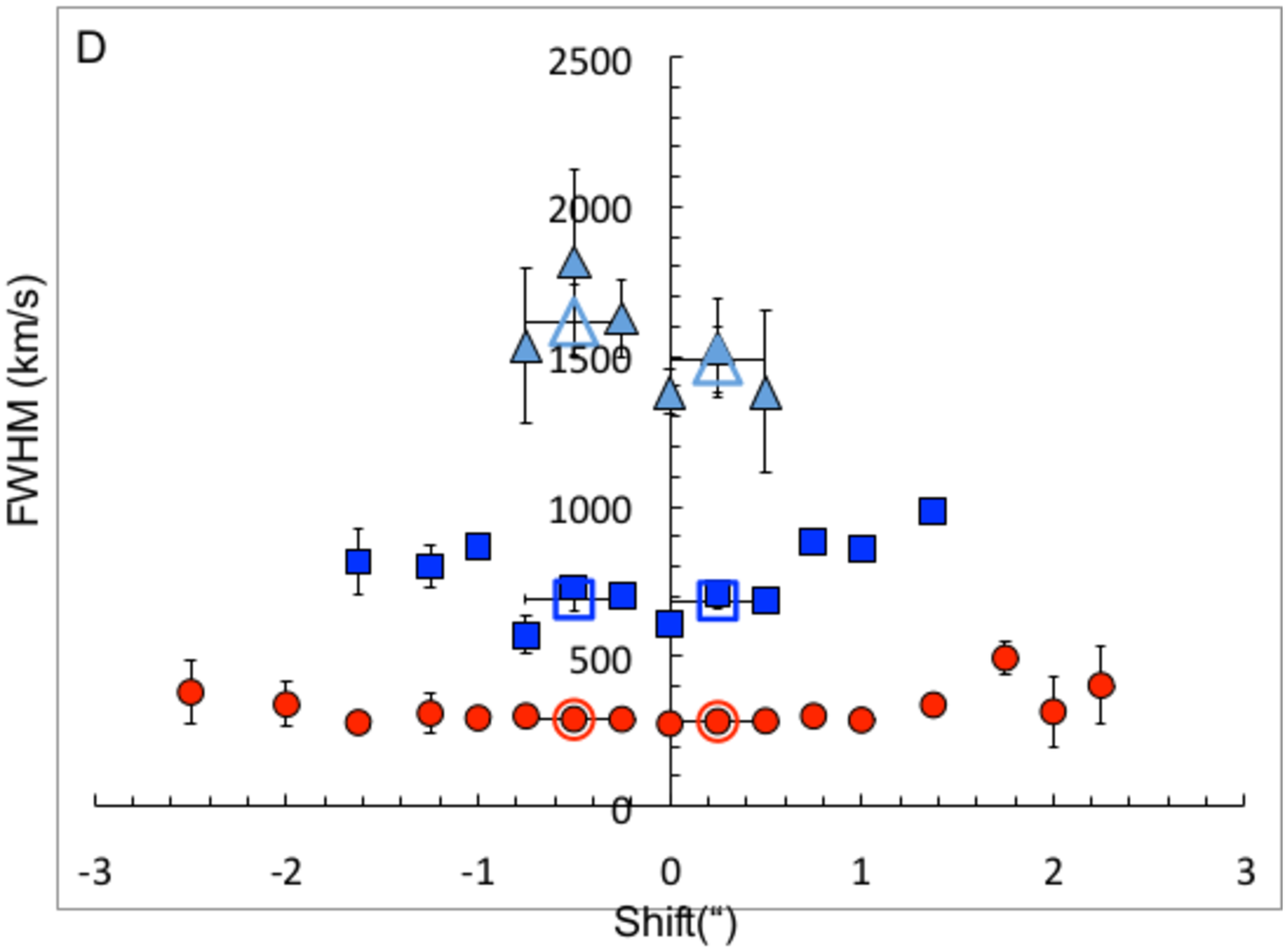}
\vspace{1.9in}
\caption{These  three QSO2 show  at least preliminary evidence for spatially extended outflows.
A) total [OIII] (continuum subtracted) spatial profile (green) compared with the seeing (black). B) spatial profile of the individual kinematic components isolated in the 
fits compared with the seeing. The narrow, intermediate and broad components are represented with red circles, dark blue squares and 
light blue triangles respectively. C) Velocity shift $V_{\rm s}$ and D) FWHM of the individual kinematic components.  Solid symbols correspond to pixel by pixel fits. Large hollow symbols correspond to large  (several pixels wide)  apertures,
selected to increased the S/N. The widths are indicated by the horizontal bars. For SDSS J1307-02 small purple diamonds are used in the outer region to the West  because these pixels show very distinct kinematics. Notice that for clarity sometimes we use a different scale for the the spatial axis of the A) panels compared with B), C) and D).}
\label{extended}
\end{figure*}

\begin{figure*}
\includegraphics{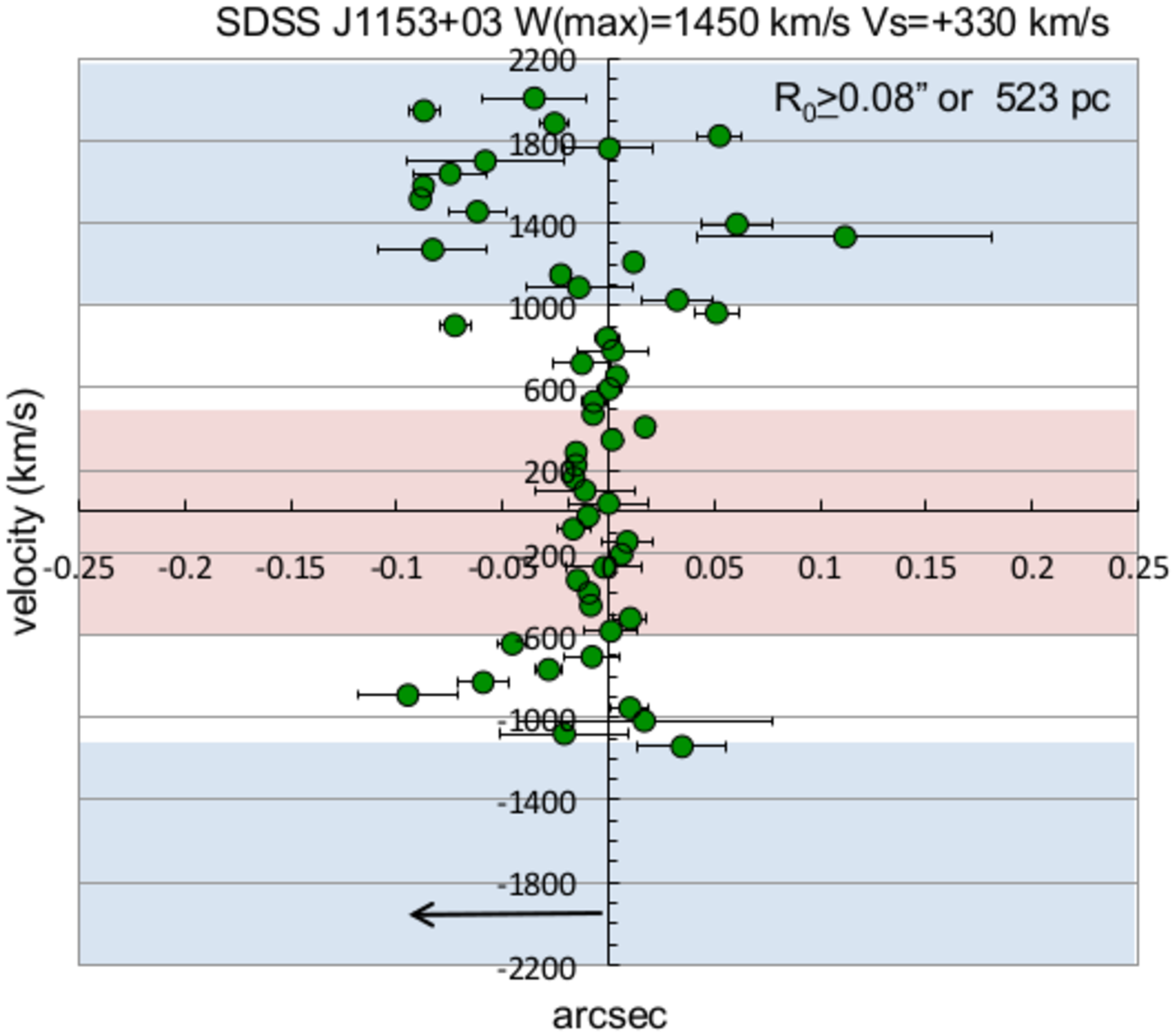}
\includegraphics{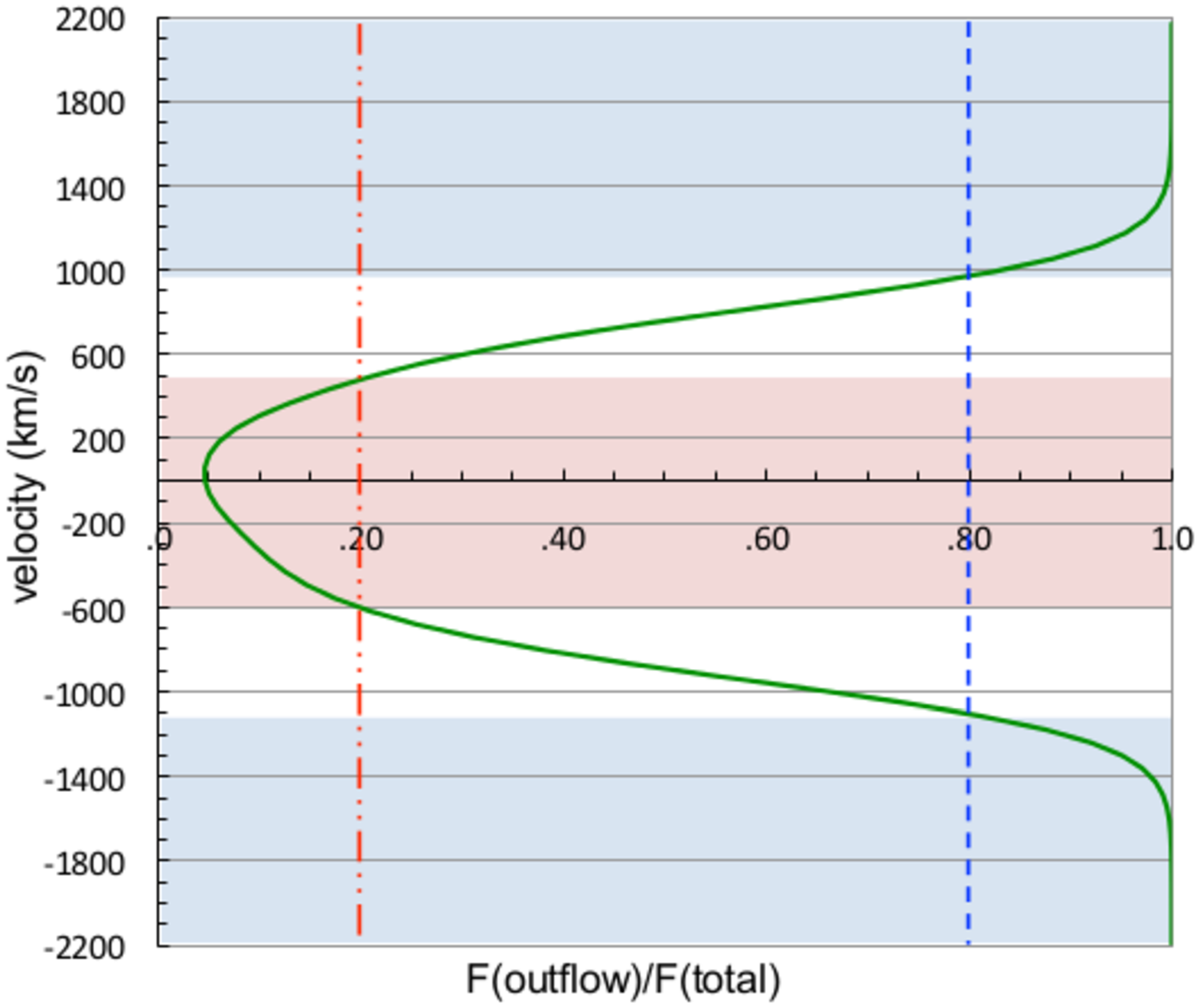}
\vspace{2.3in}
\caption{Spectroastrometric analysis for SDSS J1153+03. Left: Shift of the spatial centroid measured for [OIII] 
at different velocities relative to the continuum centroid. Right: Velocity shift versus the relative contribution of the outflowing gas to the total [OIII] line flux.  The blue areas mark the range of velocities for which the outflow dominates the line flux
($\frac{F_{\rm outflow}}{F_{\rm total}}\ge$0.8). The red areas mark the velocities for which its contribution is minimum ($\frac{F_{\rm outflow}}{F_{\rm total}}<$0.2). The emission at  velocities dominated by the outflow are
in general shifted relative to the continuum centroid. The arrow marks the assumed spatial shift representing the lower limit for the radial size of the outflow. In this case, $R_{\rm o}\ga$523 pc.}
\label{astrom1153}
\includegraphics{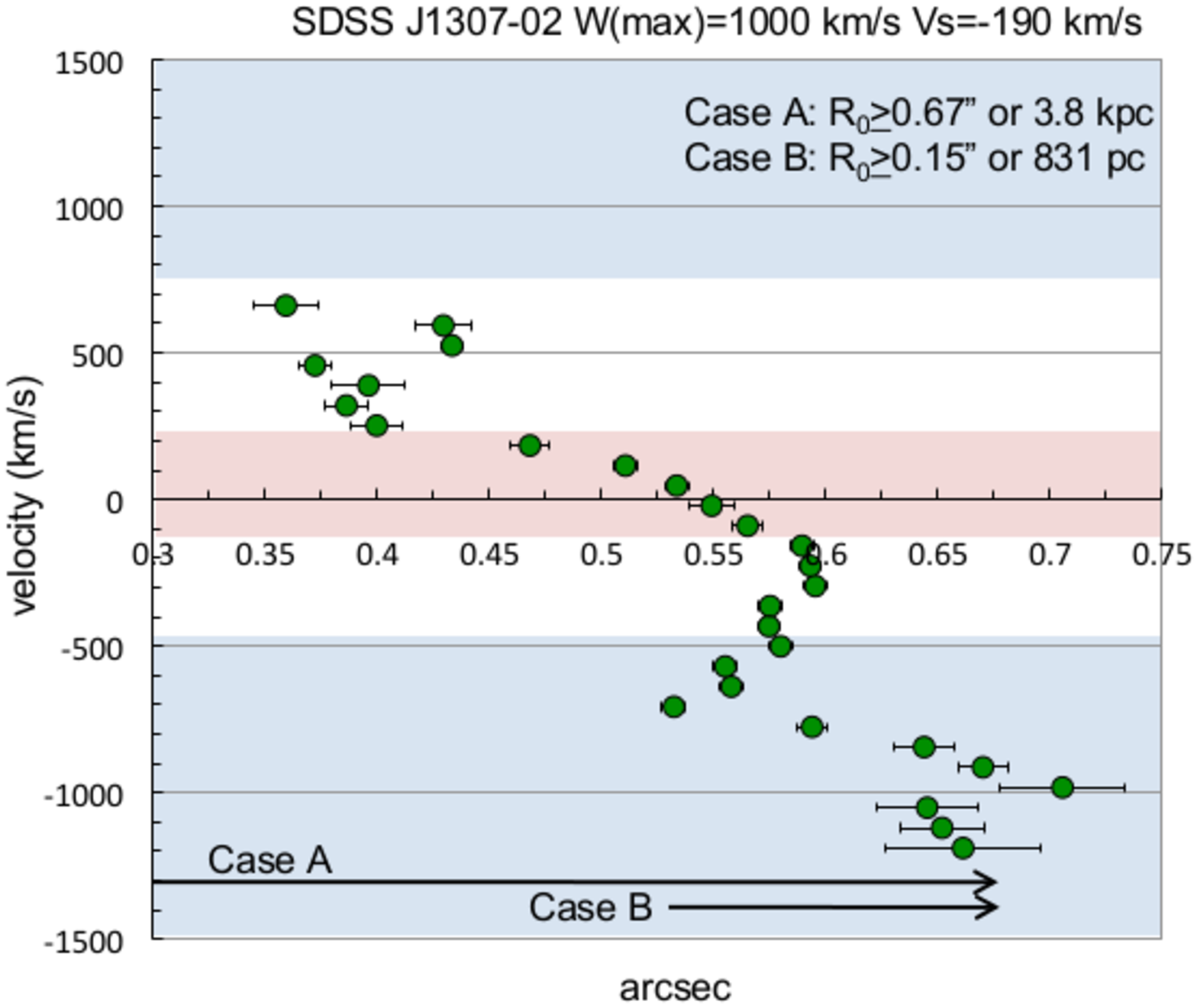}
\includegraphics{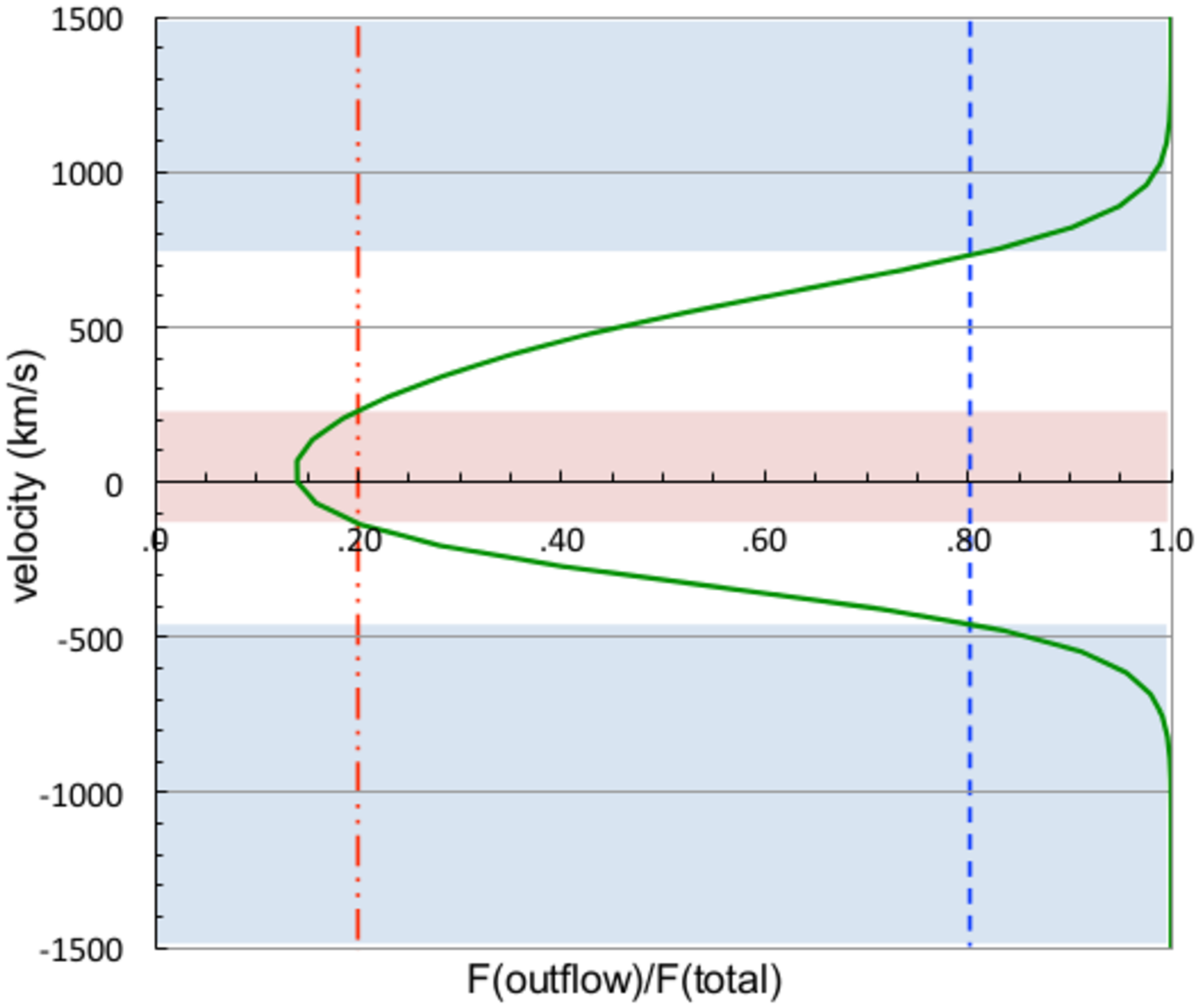}
\vspace{2.5in}
\caption{Spectroastrometry analysis for SDSS J1307-02.  The spatial
 centroids of the continuum and the emission lines location are significantly shifted at all velocities. This introduces uncertainties on the location of the AGN (i.e. the spatial zero). Two cases are considered to constrain $R_{\rm o}$: the continuum centroid marks the location of the AGN (Case A); the emission line centroid at zero velocity marks the location of the AGN
 (Case B). We will adopt $R_{\rm o}\ga$831 pc as a  conservative lower limit. Line, symbol and color codes as in Fig.  6.}
\label{astrom1307}
\includegraphics{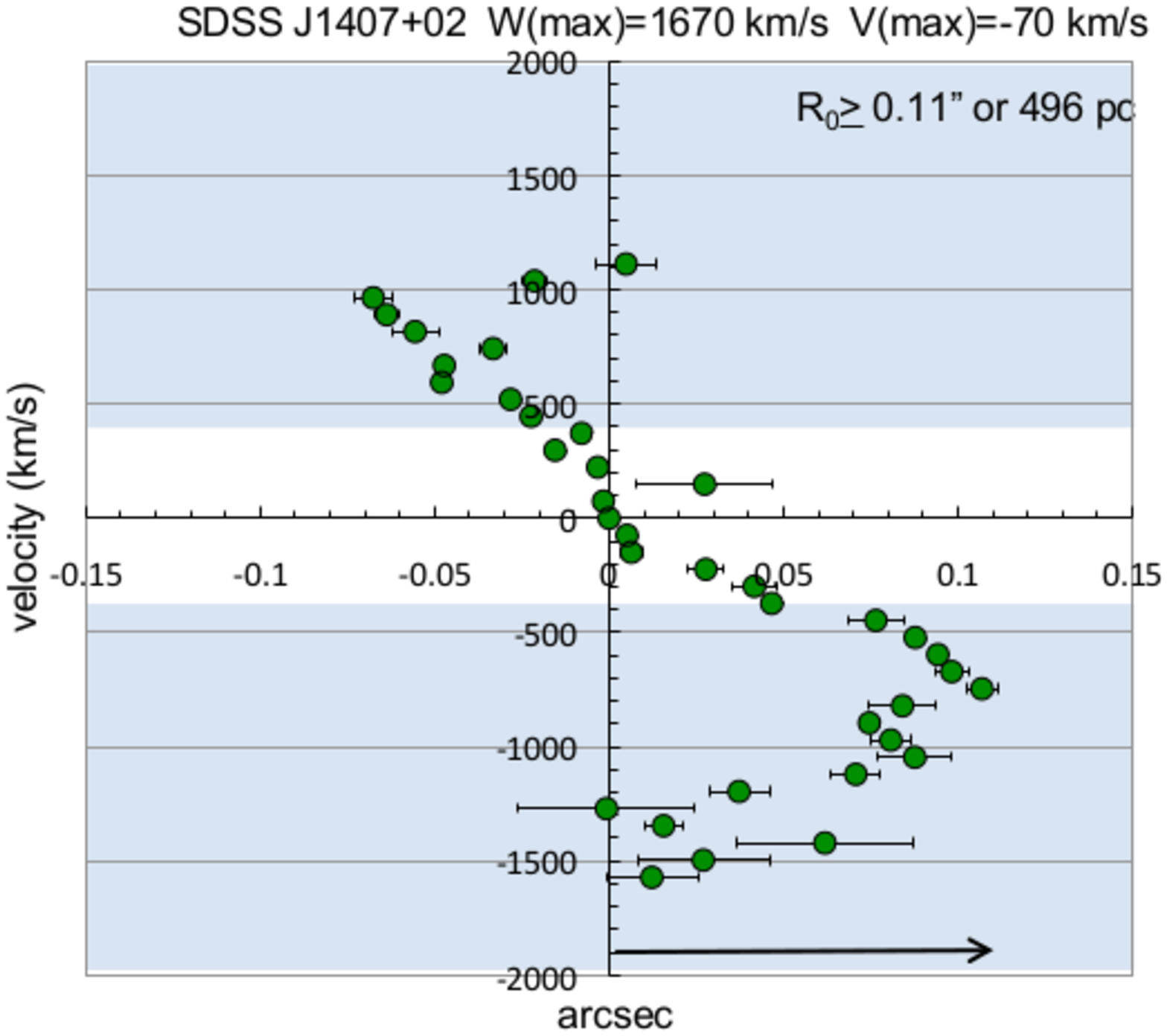}
\includegraphics{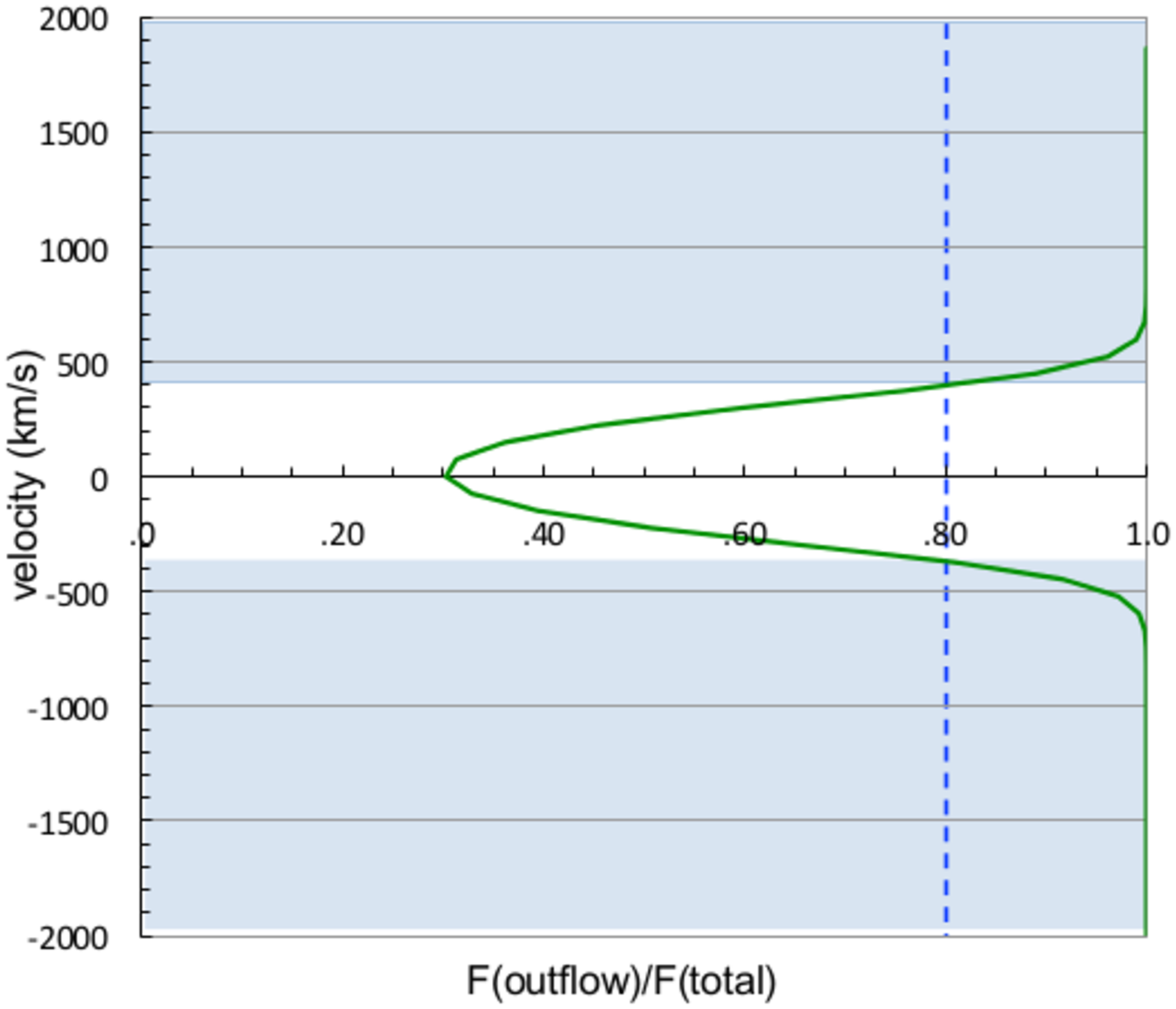}
\vspace{2.4in}
\caption{Spectroastrometry analysis for SDSS J1407+02. Line, symbol and color codes as in Fig.  6.}
\label{astrom1407}
\end{figure*}

% z=0.575, 6.50 kpc/arcsec

\vspace{0.2cm}
The nuclear spectrum of this radio-quiet QSO2 is best fitted with three components of FWHM=270$\pm$10, 730$\pm$10 and
1450$\pm$80 km s$^{-1}$ respectively.  The last one, which  is likely to trace the outflow, is redshifted by  $V_{\rm max}$=330$\pm$70 km s$^{-1}$ (\cite{vm11b}).  It is one of the exceptional  cases where the broadest component shows a clear redshift, rather than a blueshift (see also SDSS J1228+00).  Although rare, a  combination of reddening and  inclination angles of the outflow cone axis and a galactic disk    can be such  that  the  blueshifted   cone  
is entirely occulted by the disk and the redshifted cone is not (Crenshaw et al. \citeyear{cre10}).  In this situation, the observer's line of sight is approximately parallel to the cones edges, corresponding to an  orientation intermediate between type 1 and type 2. Such intermediate orientation was  proposed by \cite{vm11a} to explain the
detection of strong optical continuum and very broad underlying H$\beta$ emission
in the nuclear spectrum of this QSO2.

The [OIII] spatial profile is dominated by the central compact source (Fig. \ref{extended}, left panel A), which appears barely resolved 
in the central regions compared with the seeing  (FWHM=0.8$\pm$0.1). In addition, the excess of emission above the seeing wings at $>$2.5 arcsec to the West shows that the gas is extended (this is confirmed by  [OII]$\lambda$3727 or [OII] hereafter), which shows an even  clearer excess at both sides of the central source, although this is not plotted for simplicity).
\cite{vm11a} showed that at $\sim$+3.5$\arcsec$ to the East  the line has a clear contribution of a companion emission line galaxy, especially evident in the low ionization lines such as [OII].  
The maximum extension of the ionized gas towards the West is $\sim$4.7$\arcsec$  or $\sim$30 kpc.

The properties of the different kinematic components isolated in the [OIII] lines along the slit (on a pixel by pixel basis) are  shown in (Fig. \ref{extended}, right panels). The narrow component is clearly extended
and is relatively narrow, spectrally unresolved at most locations with FWHM$\la$270km s$^{-1}$ in general. On the contrary, the spatial distributions of the intermediate and broadest components are consistent with the seeing disk.  This fact sets an upper limit to the spatial extension of these components of FWHM$_{\rm int}\la$ 0.53 arcsec or 3.4 kpc (Sect. \ref{Sec:methods4p2}).

The kinematic plots (Fig. \ref{extended},  panels C and D) suggest kinematic spatial variations  of the broadest component (the outflow) within the seeing disk, which is also suggested by the analysis based on larger apertures (hollow symbols in  the figures).
 The FWHM varies from 1856$\pm$128 up to 2311$\pm$184 km s$^{-1}$. The velocity shift appears   also 
  different with $V_{\rm s}=$290$\pm$59 and 482$\pm$89 km s$^{-1}$ at both sides of the continuum centroid. 
  The changes in $V_{\rm s}$ and FWHM  are measured at less than $<$3$\sigma$ significance.  
  We have performed one more test.  We have forced the fits of the [OIII] lines in the two large apertures at both sides of the continuum centroid
to have exactly the same  FWHM and $V_{\rm s}$ for the broad component. This is what we would expect if the emission is smeared by the seeing. This was done in two ways: constraining the broad component of the West aperture using the free fits of the East aperture and vice versa.  We find that it is possible to find acceptable fits satisfying these premises.  From this, it cannot be discarded that the broadest component is spatially unresolved in comparison with the seeing. 
Thus, we consider it as a possible candidate to harbor a spatially resolved outflow, although follow up at higher spatial resolution is necessary.

\begin{table}
\centering
\caption{Constraints on the radial sizes of the outflows $R_ {\rm o}$ as derived from the spectroastrometric
analysis (lower limits) and the comparison with the seeing disk.
 $^a$This is a conservative lower limit. It has a large uncertainty due to the unknown   location of the AGN  (see the text). $^b$This is a gross upper limit, very uncertain due to the difficulty to apply methods (i) and (ii). $^c$See Sect. 5.2 for a detailed discussion of this object.} 
\begin{tabular}{llll}
\hline
Object &   $R_{\rm o}^{astrom}$ & $R_{\rm o}^{seeing}$\\ 
	&	 kpc 	& kpc  \\ \hline
	& 2009  sample & \\ \hline
SDSS J0955+03 & $>$0  & $\la$3.5$^b$   \\
SDSS J1153+03 &  $\ga$0.52&  $\la$1.7	 \\
SDSS J1228+00  & $>$0 & $\la$2.1 \\
SDSS J1307-02 & $\ga$0.83$^a$ &  1.3$\pm$0.4  \\
SDSS J1337-01 & $\ga$0.19 & $\la$1.7 \\
SDSS J1407+02 & $\ga$0.50    &  1.0$\pm$0.2 \\
SDSS J1413-01 & $\ga$0.21   &  $\la$1.5 \\
SDSS J1546-00 & $>$0   &  $\la$1.4\\  \hline
	& 2011  sample & \\
SDSS J1430-00 &  $\ga$0.12 & $\la$0.8  \\
SDSS J0903+02  & $\ga$1.1 & $\sim$10?$^c$ \\ 
SDSS J0923+01 & $\ga$0.16 & $\la$1.0 \\ 
SDSS J0950+01 & $\ga$0.38 & $\la$1.1 \\
SDSS J1014+02 & $>$0 & $\la$1.0  \\
SDSS J1217+01 & $\ga$0.28 & $\la$1.4 \\
SDSS J1336-00 & $\ga$0.19  &  $\la$1.0\\ 
\hline
\end{tabular}
\label{tab:sizes}
\end{table}

The results of the spectroastrometric analysis are shown in Fig. \ref{astrom1153}. We consider that the most natural identification of the outflowing gas is the broadest, redshifted component.
Notice that the results are not influenced by this choice or, instead, by assuming that the intermediate component is also part of the outflow. Fig. \ref{astrom1153} (left) shows that the highest velocities (which are  those where the outflow unambiguously 
dominates the line flux, as indicated by the blue areas) show a small spatial shift relative to the continuum centroid\footnote{This is  likely to be a reliable
indicator of the AGN is this object, due to its intermediate orientation.}. We infer  a lower limit $R_{\rm o}\ga$0.08 arcsec or 523 pc.

\vspace{0.2cm}

{\it SDSS J1307-02}
% z=0.425, 5.54 kpc/arcsec
\vspace{0.2cm}

The nuclear emission lines consist of 2 kinematic components of FWHM=237$\pm$22 and 1000$\pm$40 km s$^{-1}$
 respectively. The broadest  is blueshifted by $V_{\rm max}$=-190$\pm$20 km s$^{-1}$ and traces  the outflowing gas.

In addition, extended very low surface brightness emission is  detected at $\sim$3$\arcsec$ or $\sim$17 kpc to the West.
[OIII] is  narrow in these regions with FWHM=255$\pm$41 km s$^{-1}$.  This distant gas is ionized by stars 
and does not seem to be affected by the AGN (\cite{vm11a}).

The [OIII] spatial profile is dominated by a compact  but spatially resolved source with FWHM=1.04$\pm$0.03 arcsec,
compared with the  FWHM=0.83$\pm$0.05 arcsec seeing (Fig. \ref{extended}, middle, panel A). 
We infer FWHM$_{\rm int} \sim$0.63$\pm$0.09 arcsec or 3.5$\pm$0.5 kpc.

The spatially resolved kinematic analysis (Fig. \ref{extended} middle, panels B to D) reveals that  the narrow component is clearly spatially
extended. This seems to be the case as well for the outflowing gas, which shows an excess above the seeing disk especially obvious towards the
East.  We infer FWHM$_{\rm int}=$0.46$\pm$0.12 arcsec or  2.5$\pm$0.7  kpc for the outflow.

In addition,   the  nuclear ionized outflow  shows a significant
shift in velocity across its extension of 250$\pm$40 km s$^{-1}$. This kinematic substructure, which is
seen  as a result  of the pixel to pixel analysis and also using larger apertures (large hollow symbols in  Fig. \ref{extended}) 
reinforces that the outflow  is spatially resolved. We performed a similar test as described for SDSS J1153+03. It is found that no successful fits can be found forcing the broad component at both sides of the continuum centroid to have the same $V_{\rm s}$ and FWHM. This supports that the  spatial kinematic variation is real.

The spectroastrometric analysis  is  affected by the uncertainty in the AGN location (see Sect. \ref{Sec:methods}).   Fig. \ref{astrom1307} (left) shows that the  emission line  and continuum spatial centroids are significantly shifted  at all velocities ($\sim$0.55 arcsec on average).  
 We  consider two scenarios: the AGN is located at the continuum centroid (Case A) and at  the line centroid (Case B). We obtain  $R_{\rm o}\ga$0.67 arcsec (3.8 kpc) for Case A and $R_{\rm o}\ga$0.15 (831 pc) for case  B. We will adopt  the smallest value as a conservative lower limit for $R_{\rm o}$.

\vspace{0.2cm}

{\it SDSS J1407+02}
%z=0.309  4.51 kpc/arcsec

\vspace{0.2cm}

The nuclear spectrum is best fitted by three kinematic components of FWHM
  290$\pm$10, 720$\pm$20  and 1670$\pm$120 km s$^{-1}$ respectively, with the broadest  components  shifted by $V_{\rm max}$=-70$\pm$30 km s$^{-1}$ (\cite{vm11b}).
It is not clear whether the outflow is only traced by the broadest component or also the intermediate one.

The [OIII] spatial profile is dominated by the central compact source with no evidence for extended emission along the slit  (seeing FWHM=0.73$\pm$0.04; Fig. ~\ref{extended} right, panel A). 

The spatial profiles of the individual kinematic components  are consistent with the seeing disk (Fig. ~\ref{extended}, right  panel B),
except the intermediate component. This  shows a small excess above the seeing disk in most pixels.   We infer FWHM$_{\rm int}=$0.44$\pm$0.09  arcsec or 2.0$\pm$0.4 kpc for it.

The kinematic plots show that the narrow component (which traces  ambient not
outflowing gas) shows no spatial variation of its kinematics. On the contrary,  both the broad and the intermediate components show  similar shifts in velocity   across the spatial centroid (panel C). This suggests that they  
are spatially resolved. Both the pixel to pixel and the large aperture analysis lead to the same result. The shift
in velocity (81$\pm$32 km s$^{-1}$)  is measured at $\sim$3$\sigma$ level. 

If, as for the previous objects, we force the two large-aperture spectra  at both sides of  the spatial centroid to have identical broad components (V$_{\rm s}$ and FWHM), an acceptable fit is found. This means that there is a valid fit consistent with the spatial distribution of the broad component being spatially unresolved (the intermediate component still must present a significant velocity shift).  In spite of this uncertainty, the almost identical results found for the intermediate and the broad
components, the excess of the intermediate component above the seeing disk and the consistent trends found in the pixel to pixel kinematic analysis at both sides of the continuum centroid support that the spatial change of $V_{\rm s}$ is real.  We will thus consider this as a good candidate to harbor a spatially resolved ionized outflow.    From methods (i) and (ii) we constrain $R_{\rm o}$=1.0$\pm$02. kpc.

The  spectroastrometric analysis  also implies that the outflow is spatially extended (Figs. \ref{astrom1407}). 
We infer $R_{\rm o}\ga$496 pc if both the broad and intermediate components participate in the outflow. The similar behavior presented by the broad and intermediate components may support this. 
The  results vary only slightly   if only the broad component is considered, giving $R_{\rm o}\ga$338 pc.

We refer the reader to Appendix \ref{Sec:app} for a detailed description of the analysis and results on the rest of the sample. Table \ref{tab:kinresults} presents the main conclusions regarding whether the outflows are spatially
extended or not using the different methods described in Sect. \ref{Sec:methods}.
The  constraints (lower and upper limits) on the extent of the outflows in individual objects are presented in Table 
\ref{tab:sizes}.

  \vspace{0.2cm}

{\it SDSS J0903+02}

%z=0.329, 4.71 kpc/arcsec

\vspace{0.2cm}

\begin{figure*}
\includegraphics{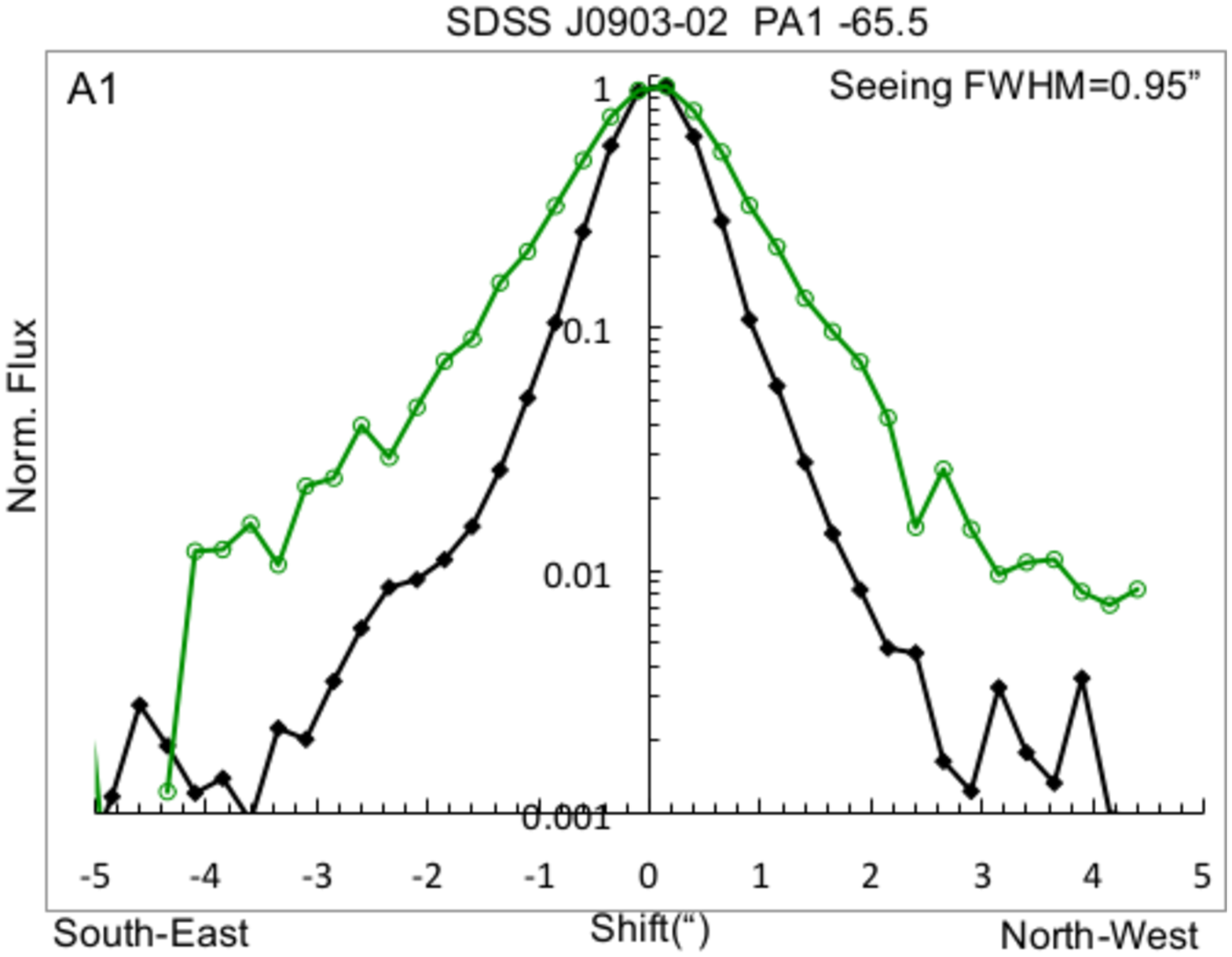}
\includegraphics{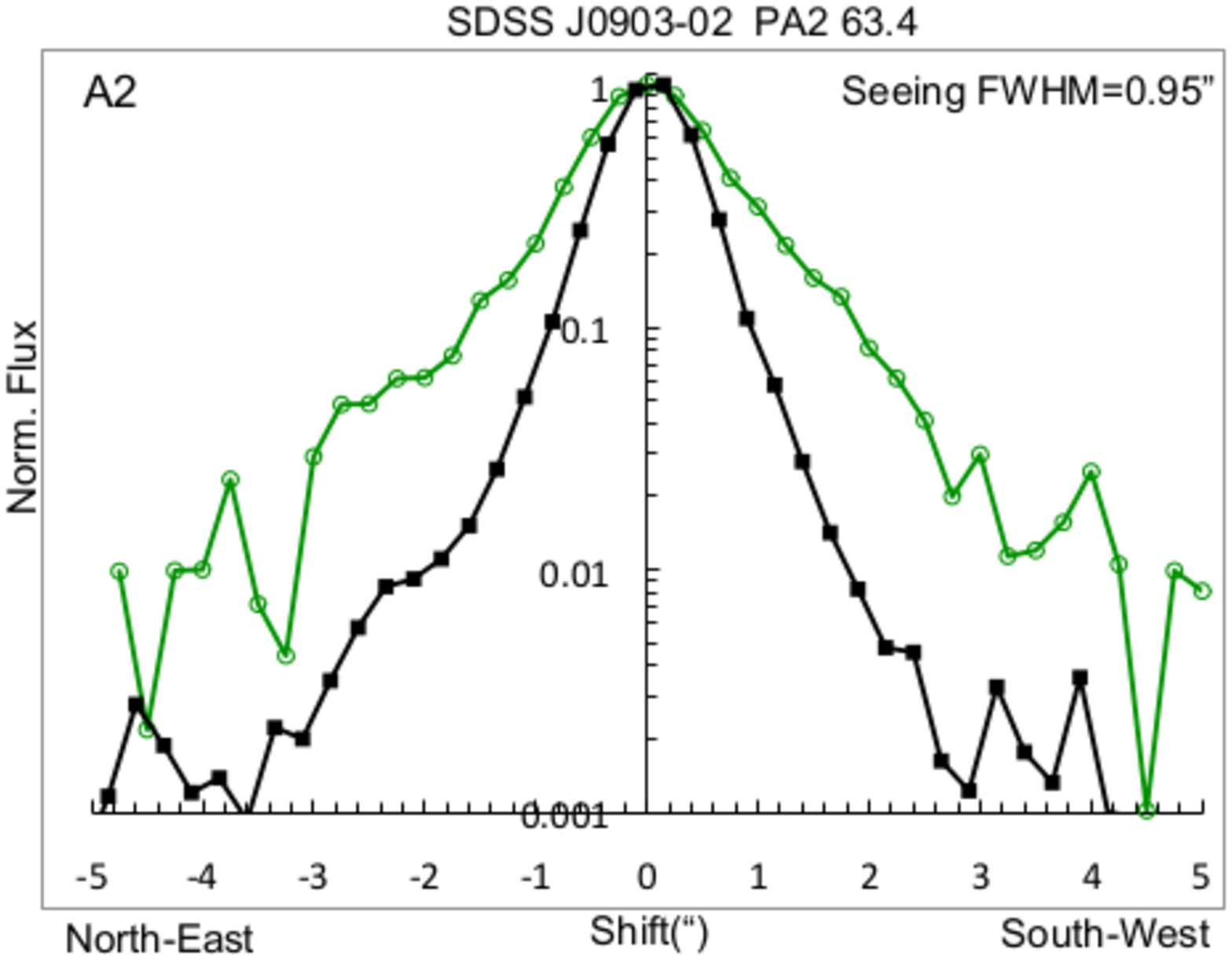}
\vspace{2.2in}
\includegraphics{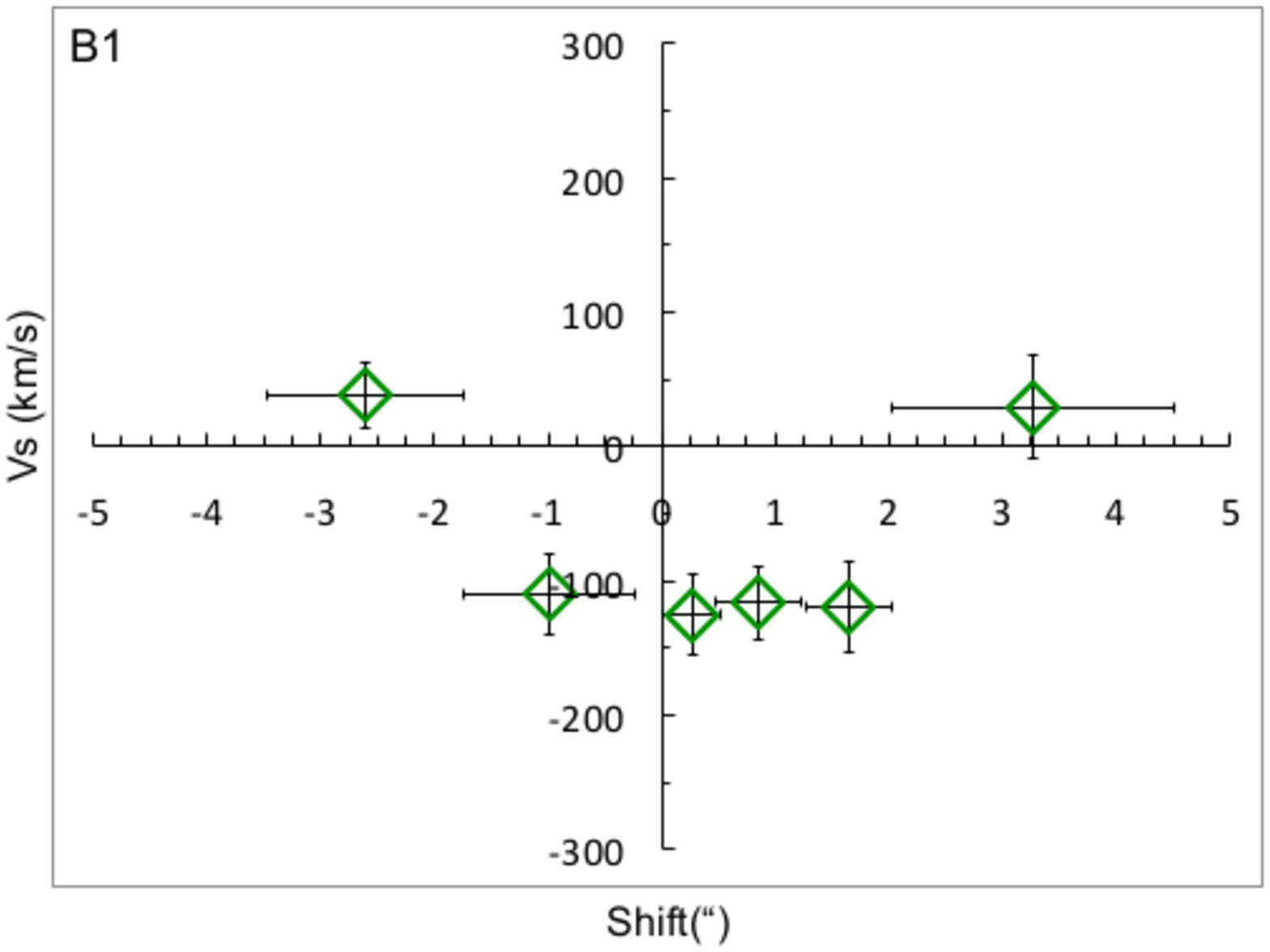}
\includegraphics{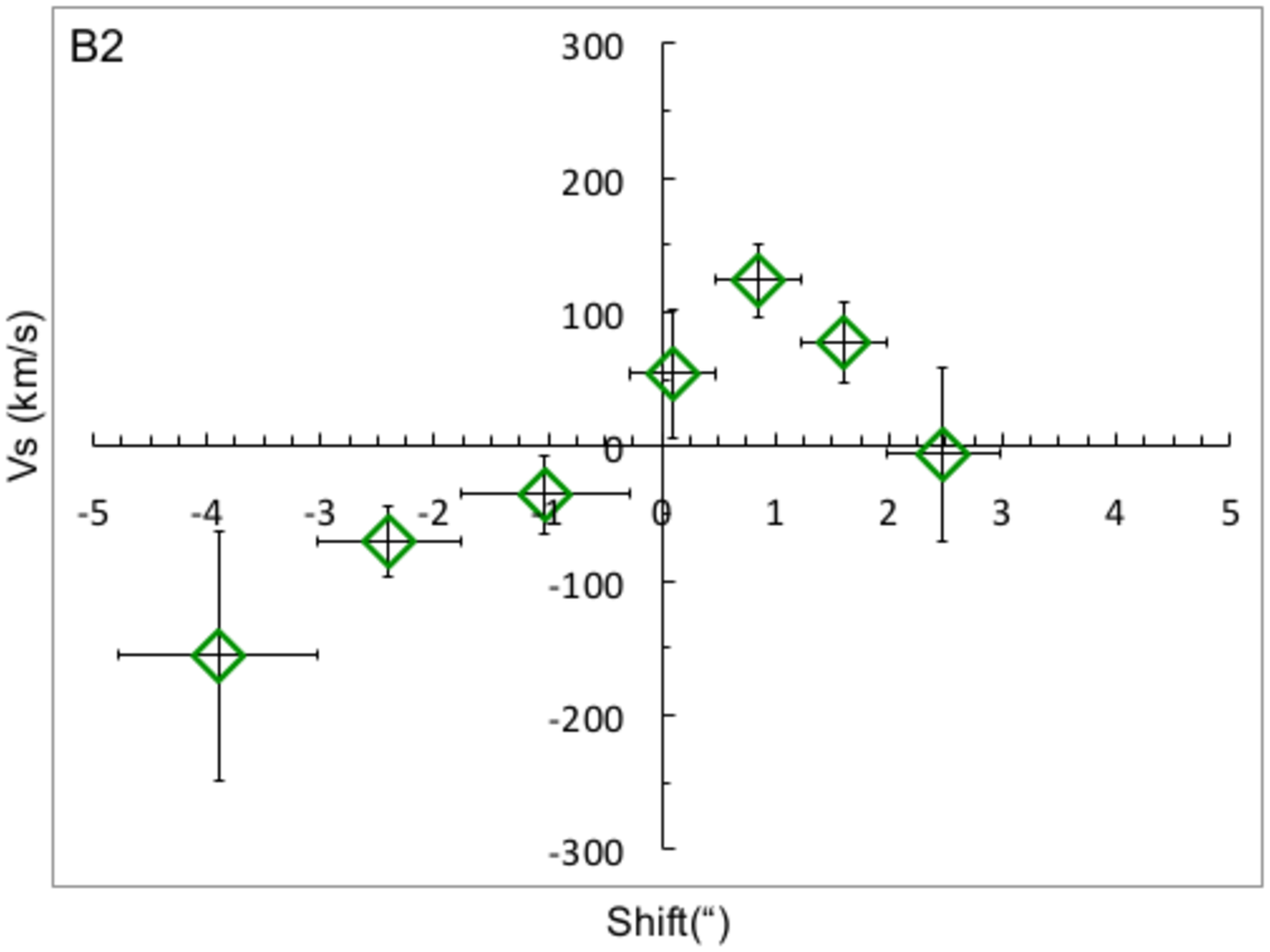}
\vspace{2.2in}
\includegraphics{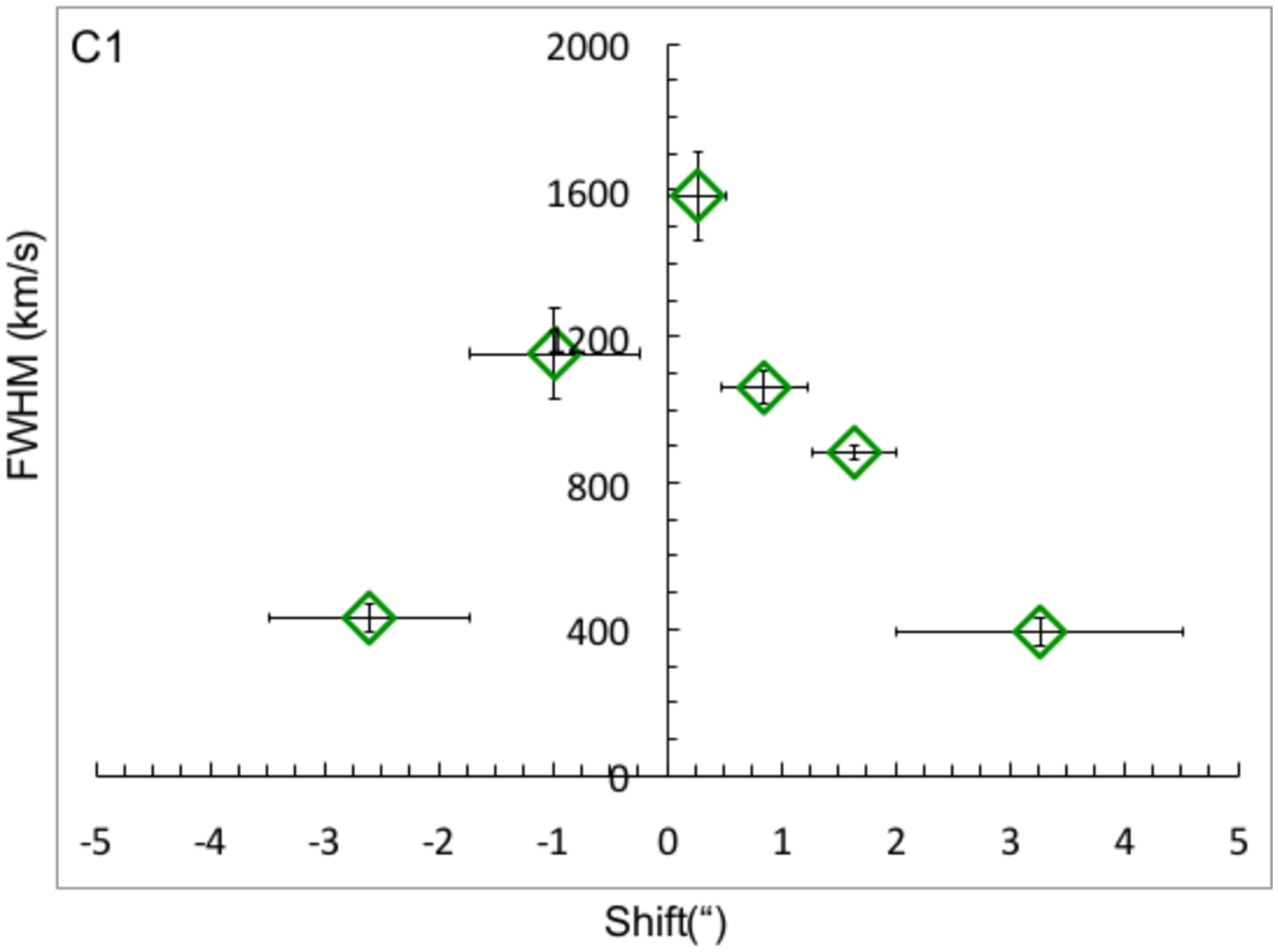}
\includegraphics{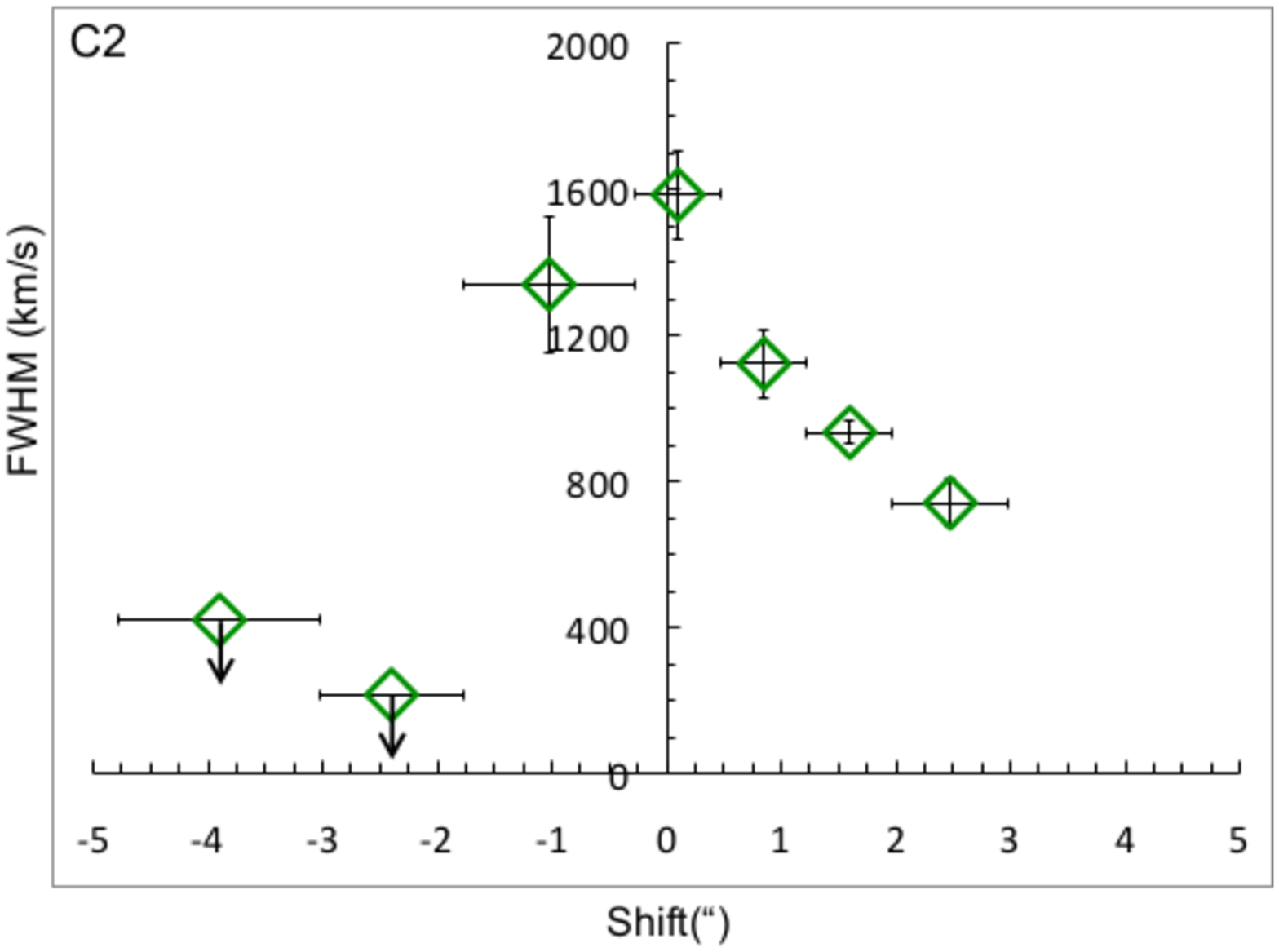}
\vspace{2.2in}
\caption{Spatially extended analysis of [OIII]$\lambda$5007 for SDSS 0903+02 along PA1 -65.5 (left) and PA2 +63.4 (right).
The spatially extended kinematic line decomposition could not be applied in this object. FWHM and $V_{\rm s}$  measurements could only be obtained from wide apertures. This is why large hollow symbols are used.}
\label{spat0903}
\end{figure*}

\begin{figure*}
\includegraphics{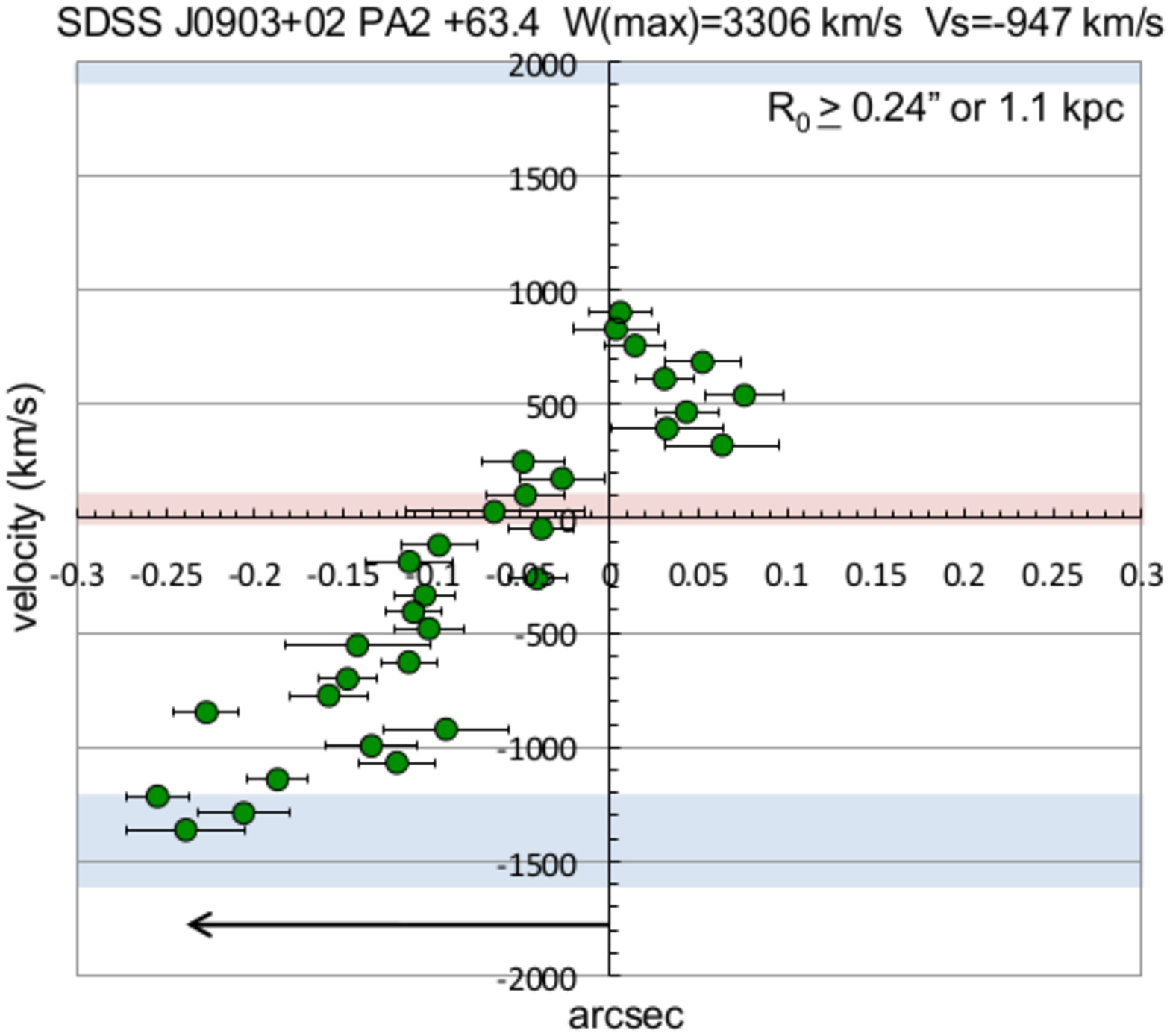}
\includegraphics{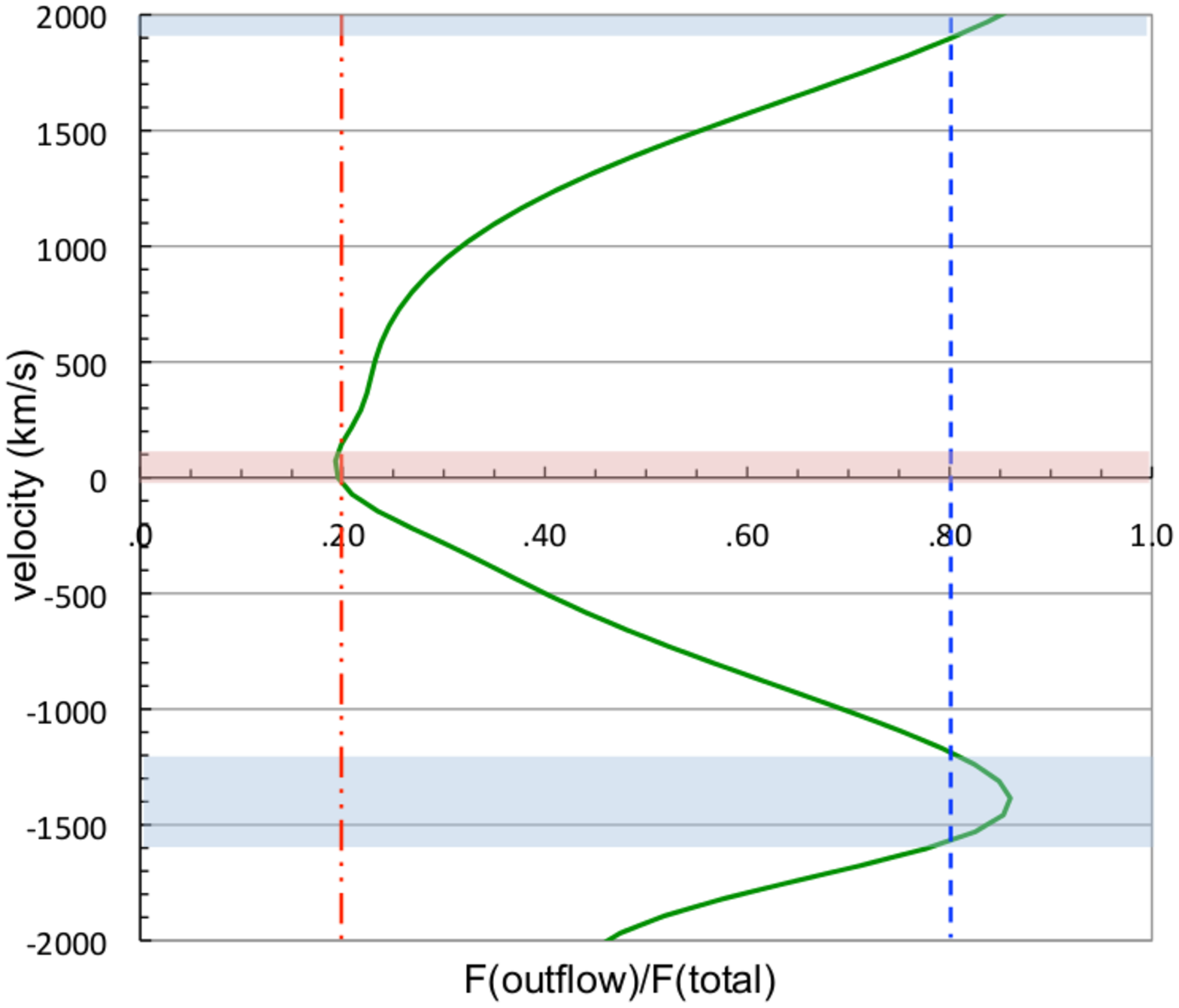}
\vspace{2.85in}
\caption{Spectroastrometry analysis for SDSS J0903+02 along PA2 (top panels). Line, symbol and color codes as in Fig. 6.}
\label{astrom0903}
\end{figure*}

This QSO2   shows extreme nuclear kinematics and  a highly disturbed optical  morphology 
(see \cite{hum15}). The [OIII] lines are broad  (Fig. \ref{nuclei1}) and the spectral decomposition requires 4 kinematic components to obtain a successful fit (Table \ref{tab:fitsnuc}). The broadest component has 
W$_{\rm max}$=3546$\pm$174 km s$^{-1}$ and $V_{\rm max}$=-522$\pm$83 km s$^{-1}$ as measured from the nuclear
spectra extracted from PA1 -65.5$\degr$ and W$_{\rm max}$=3306$\pm$168 km s$^{-1}$ and $V_{\rm max}$=-947$\pm$138 km s$^{-1}$ for PA2 64.4$\degr$.   
Based on \cite{vm14}, such extreme kinematics are suggestive of an outflow induced by the radio structures. SDSS J0903+02 is radio-intermediate according to our classification criteria (Sect. \ref{Sec:sample}). It has been proposed
by \cite{lal10} as a low radio power analogue of Compact Steep Spectrum sources (CSS), being  2-3 orders of magnitude less luminous. 

As discussed in \cite{hum15}, the emission  lines are spatially extended along both PA1 -65.5 and PA2 +63.4 (Fig. \ref{spat0903}, top panels)
compared with the seeing disk (FWHM=0.95$\pm$0.05 arcsec).  We also reported very narrow
lines  (FWHM$\la$180 km s$^{-1}$) at $\sim$5 arcsec  from the AGN.  Extended gas dominates the emission at this location. 

It is not possible to perform a spatially extended  analysis of the individual kinematic components due to the complexity of the line profiles and the low S/N of the individual
spectra.  We have extracted 
1-dim spectra from several apertures along both slit PA   and at both sides of the continuum centroid (Fig. \ref{spat0903},   panels B and C). The kinematic measurements correspond to single Gaussian fits. Although the kinematics are complex in some regions, the fits serve to characterize the global kinematics. It is clear, for instance, how the kinematics becomes more quiescent at increasing distances from the spatial centroid, with the FWHM varying from $\sim$1200-1600 km s$^{-1}$ at the center down to $\la$200-400 km s$^{-1}$ in the outer parts at $\ga$2.5 arcsec.  Within 2 arcsec from the AGN along both PA the lines are  broad  with FWHM$\sim$880-1600 km s$^{-1}$, also decreasing outwards.  

Most importantly, the lines are  also broad  at locations where the line emission is in clear excess relative to the seeing profile. For instance, FWHM$\sim$741-934 km s$^{-1}$   at $\sim$+1.6 and +2.5 arcsec  along PA2. If there happened to be a compact unresolved source at the center, it would contribute only $\sim$15\% of the total flux at these locations. 

This is the only object in the 2009+2011 sample (18 sources) for which broad lines are identified at spatial locations where 
the line emission is clearly  dominated by extended gas, which may be explained by large scale outflows\footnote{Another other object in our sample  with broad extended lines
is the double peaked emission line object SDSS J1247+01, but these are a consequence of the blending of the two narrow components. See Appendix A}.  
Even FWHM$\sim$400 km s$^{-1}$ measured at $\sim$3 arcsec is  broad compared with values one would expect at such large distances from the nucleus. The extended non outflowing ionized gas in mergers show typical FWHM$<$250 km s$^{-1}$, even in the most dynamically disturbed systems with signs of AGN activity (Bellochi et al. \citeyear{bel13}, Arribas et al. \citeyear{arr14}). Thus, the spatially resolved broad lines of SDSS J0903+02 are suggestive of an extended outflow. 

If the [OIII]  kinematics is dominated by an  outflow across its whole spatial extensions, we infer a radial size of $R_{\rm o}\sim$10 kpc.
SDSS J0903+02 is the only object in our sample for which the data are consistent with   a galactic scale outflow.

We do not discard an scenario where 2 or more emission line objects, plus an unresolved outflow, produce apparently extended highly disturbed kinematics thus
mimicking  an extended outflow (see for instance SDSS J1247+01 in Appendix \ref{Sec:app}). In fact, SDSS J0903+02
 shows   highly disturbed optical morphology indicative of a merger (\cite{hum15}) with
star forming knots (maybe also a double nucleus) and tidal features.    Higher S/N spectra would be required to fit the  line profiles at different locations, isolate the emission from the individual components and compare their
  spatial distribution  with the seeing disk.

For the spectroastrometric analysis we have used the PA2 spectrum, since it provides the most useful
constraints.  The results  are shown in  Fig.~\ref{astrom0903}.  We  have
considered that only
the broadest component contributes to the outflow. We emphasize that the same result would be obtained
if additional nuclear kinematic components also contributed.  The outflow is extended
along PA2 and has $R_{\rm o}\ga$0.24 arcsec or 1.1 kpc. This is the largest value of  the lower limits on $R_{\rm o}$ 
implied by the spectroastrometric
analysis of our sample (Table ~\ref{tab:sizes}).

\vspace{0.2cm}

\subsubsection{Global results}
\label{Sec:global}

 In this section we present the global results obtained after analyzing the whole sample.

 In 11 out of 18 objects  it was possible to isolate the spectra from extended regions far enough from the spatial centroid to ensure that seeing contamination is negligible. Except in one object, the lines  at these locations  are always relatively narrow (Table ~\ref{tab:kinresults}, column 10). They are spectrally unresolved in most cases with FWHM$<$270 km s$^{-1}$ or show FWHM$\sim$200-250 km s$^{-1}$
still consistent with a non-outflow scenario (e.g. Bellocchi et al. \citeyear{bel13}, Arribas et al. \citeyear{arr14}). 

The  exception  is the QSO2 SDSS J0903+02, for which broad FWHM$\sim$400-934 km s$^{-1}$ are measured at extended locations (see Sect. \ref{Sec:ind}). As explained there, this is a good candidate to harbor a large scale outflow ($R_{\rm o}\sim$10 kpc), although confirmation is needed to discard alternative explanations.

Our results show that three things are essential to assess whether the outflows are extended and constrain their sizes when dealing with seeing limited observations:

1) To characterize as accurately as possible the seeing spatial profile. In general, the emission line profiles are dominated by a compact unresolved or barely resolved source. Not only the seeing FWHM needs to be measured with the corresponding uncertainties, but also  the wings.  It is well known that the seeing  does not fit a simple Gaussian profile (e.g. Capaccioli \& de Vaucouleurs\citeyear{cap83}), so the FWHM does not characterize the wings emission. Underestimation of this contribution can lead to the inference of spatially extended outflows, since  the extended line emission is often identified as a faint excess above the seeing wings.

2)  To isolate the outflow emission by means of the spectral decomposition of the optical lines at different spatial locations. 
It is found that  the spatial distribution of the outflows is  in general
consistent with the seeing disk, while the total [OIII] flux profile  is often clearly extended.

When this method is applied, the outflows appear barely resolved in only  2 out of the 15 objects  with clear evidence for ionized outflows (SDSS J1307-02 and  SDSS J1407+02; maybe also SDSS J1153+03, although this object is dubious). In these two QSO2,
the outflows have radial sizes $R_{\rm o}=\frac{\rm FWHM_{\rm int}}{2}$=1.3$\pm$0.4 and 1.0$\pm$0.2 kpc respectively after correcting for seeing broadening.

For the rest of the sample,  we estimate upper limits  in the range $R_{\rm o}\la$0.8-3.5 kpc. In most cases $R_{\rm o}<$1-2 kpc (see Table \ref{tab:sizes}).

3)  To study the outflow spatial distribution on scales well below the seeing size. Based on the
spectroastrometric method we resolve the outflows in 11 objects. We constrain lower limits
 $R_{\rm o}\ga$0-0.5 kpc for all objects but 2, for which this lower limit is close to 1 kpc.
 For comparison, based on the spectroastrometry technique \cite{car15} have estimated radial sizes $R_{\rm o}\sim$0.4-1.9 kpc for the ionized outflows in  5 luminous quasars at $z\sim$2.4 with bolometric luminosities $L_{\rm bol}>$10$^{47}$ erg s$^{-1}$.

In general, therefore, the ionized outflows in our sample have   few$\times$100 pc $\la R_{\rm o} \la$ 2 kpc.
These  sizes are well below those quoted in the literature in related studies  of similar objects, which suggest   $R_ {\rm o}\ga$3-15 kpc (Liu et al. \citeyear{liu13a}, Harrison et al. \citeyear{har14}, McElroy et al. \citeyear{mce15}). We will discuss this apparent discrepancy in depth in Sect. \ref{Sec:noext}.

\section{Discussion}
\label{Sec:discuss}

\subsection{Nuclear ionized outflows}
\label{Sec:discussnuc}

In \cite{vm11b} we found ionized outflows in all objects in the  2009 sample. 
We have investigated this issue in the 2011 sample, which is a mixture of 4 QSO2 and 5 HSy2 (see also Villar Mart\'\i n et al. 
\citeyear{vm12}). Six out of nine objects show rather extreme kinematics due to nuclear ionized outflows, with the presence of at least one prominent broad component with FWHM$>$1000 km s$^{-1}$. This is in general blueshifted with respect to the assumed $z_{\rm sys}$. The FWHM and $V_{\rm s}$ of the broadest
components are in the range W$_{\rm max}\sim$[1173$\pm$198,3546$\pm$174] and $V_{\rm max}$ in the range [-894$\pm$86,
262$\pm$53] (Table ~\ref{tab:kinresults}).

There are three objects for which there is no convincing evidence for nuclear ionized outflows:  SDSS J1017+03, SDSS  J1416-01
and SDSS J1452+00. These happen to have among the lowest  $L_{\rm [OIII]}$. This is consistent with  \cite{zak14}
who found evidence for  quasar feedback  to operate above a threshold in bolometric luminosity of $L_{\rm bol}\sim$3$\times$10$^{45}$ 
erg s$^{-1}$.  The $L_{\rm [OIII]}$ for these four objects imply  $L_{\rm bol}\la$10$^{45}$ erg s$^{-1}$ (Stern \& Laor \citeyear{stern12}) in all cases below the threshold.

We indicate in Table ~\ref{tab:kinresults} whether the objects show signatures of mergers/interactions (see \cite{vm11a} and \cite{hum15}).  There is no obvious trend between the presence of ionized outflows and such signatures.
 As an example,  rather extreme outflows are found in SDSS J1336-00
(W$_{\rm max}\sim$2500 km s$^{-1}$,  $V_{\rm max}\sim$-900 km s$^{-1}$) and SDSS J0903+02 (W$_{\rm max}\sim$3500 km s$^{-1}$, $V_{\rm max}\sim$-500 km s$^{-1}$). The first object shows no traces of merger signatures, while the second does.  SDSS J1452+00   does not have  a detected  outflow, but presents morphological evidence for interactions.

Given the apparent diversity of merger status, or even the absence of merger/interaction signatures, among the host galaxies, there does not seem to be  a causal link between the
triggering of ionized outflows and merger status.  
In fact, this is not surprising. The ionized outflows are triggered by mechanisms related to the nuclear activity 
(Villar Mart\'\i n et al. 2011b, \citeyear{vm14}, Zakamska \& Greene \citeyear{zak14}) rather than stars.
Thus, outflows are expected basically any time a luminous AGN is active. Since this activity is
 not necessarily linked with a clear interaction/merger phase  (Bessiere et al. \citeyear{bes12}, Villar Mart\'\i n et al. \citeyear{vm12}, 
 Ramos Almeida et al. \citeyear{ram11}), nuclear outflows are not necessarily linked with this either.

 \subsection{No evidence for large scale, wide angle extended ionized outflows}
\label{Sec:noext}

Recent works  have reported that  wide angle  quasi spherical ionized outflows extended over large spatial scales 
are prevalent among QSO2 at $z\la$0.7. The observed projected total spatial extents are in the range $D_{\rm o}\ga$6 to $\ga$30 kpc.  Assuming  $R_{\rm o}= \frac{D_{\rm o}}{2}$, this corresponds to radial sizes $R_ {\rm}\ga$3-15 kpc. Some outflows are reported to   engulf  entire 
galaxies.  

 Our results are in apparent contradiction with these studies, while they agree  more with  \cite{kar16} who report outflow radial sizes 1.2-2.1 kpc in 5 QSO2 at $z<$0.1. The ionized outflows in our sample have   typical    $R_{\rm o}\la$1-2 kpc. The largest confirmed values are $R_{\rm o}\sim$1 kpc.  The only exception (although confirmation is needed) might be SDSS J0903+02, which
might harbor a $R_{\rm o}\sim$10 kpc outflow.

Previous studies  were  based on integral field spectroscopic data, while 
our analysis uses long-slit spectra along one or at most two position angles. During the observations  several objects showed   no peculiar features in the optical images  and 
the slit had to be placed blindly. In general, these objects show no evidence for extended ionized gas along the slit when compared with the seeing.
We are obviously limited by the lack of spatial coverage. 
However, if the outflows are wide angle (even quasi-spherical),  
we should have detected all or at least most  of them. 

We propose instead that  the interpretation of ubiquitous extended and wide angle outflows in QSO2 needs to be reconsidered. Based on our study, it is probable that seeing smearing was not adequately accounted for.  Our study shows that, even when extended nebula are confirmed, most of the reported  ionized outflows are  unresolved when compared with the seeing. We propose that the reported sizes in related works should also be  considered firm upper limits.

The fact that the line emission extends well beyond the seeing area enclosed within its FWHM is usually presented in the literature as an argument in favor of the large spatial extensions of the outflows. However, we have seen that  because of the huge surface brightness contrast,  the central emission is  detected at  large distances from its spatial centroid. It  can even dominate  $\sim$several times  beyond the seeing FWHM (even up to $\sim$10 times for our data and sample). This is particularly critical for distant and/or very luminous objects, for which the physical size of the  spatial resolution element is smaller and/or the emission from the central source can be detected and traced at large distances from the centroid due to its high brightness. We have also seen that whenever we have been able to safely isolate extended gas emission, not contaminated by the central source, it  shows in general quiescent kinematics.

Both our method of long-slit spectroscopy and our sample selection differ from earlier studies. We therefore also discuss other possible explanations for our discrepancy in the extent of the outflows

One issue to take into account is the sample selection. \cite{liu13b}  studied   the most luminous radio-quiet quasars in the SDSS catalogue, with $l_{\rm O3}>9.2$ at $z\sim$0.45-0.65. 
The authors claim sizes of up 15 kpc   from 
the active  nucleus\footnote{\cite{hai14} warn that these sizes are overestimated by $\sim$0.1-0.2 dex, as compared with values inferred after  accounting for seeing broadening.}. 
 High luminosity QSO2 are  associated with larger 
(Hainline et al. \citeyear{hai14}) and more luminous nebulae, which may favour the detection of the extended outflows.   If AGN induced outflows are preferentially triggered above certain threshold in luminosity ($L_{\rm bol}\ga$3$\times$10$^{45}$ erg s$^{-1}$, Greene \& Zakamska \citeyear{zak14}), it is more likely to detect them in more luminous samples.

Interestingly,  three out of the four candidates to harbor spatially resolved outflows in comparison with the seeing (SDSS J1307-02, SDSS J1407+02 and tentatively SDSS J1153+03) have among the highest [OIII] luminosities with $l_{\rm O3}\ge9.0$,
 but  in all three cases the spatially-resolved outflows are relatively compact: $R_{\rm o}<$2 kpc.

\cite{har14}  studied more nearby sources (z$\sim$0.09-0.20) covering a wider range of luminosities ($l_{\rm O3}$=8.3-9.6). They selected objects with a 
luminous broad [O III]$\lambda$5007 emission-line component seen in their SDSS spectrum,  that contributes at least 30\% of the total flux and has FWHM$>$700 km s$^{-1}$. Thus, their sample is
focussed on systems with highly perturbed nuclear kinematics, a selection criterion that was not applied by us and may   favor the detection of extended outflows.  The fact that these are at significantly lower redshifts   may  also favor the detection,  because of their higher apparent brightness. The
authors infer  $D_{\rm o}\ga$6-16 kpc  ($R_{\rm o}  \ga$3-8 kpc)  in all targets, all of which would have been spatially resolved in our observations, 
if detected (see also Harrison et al. \citeyear{har12}). 

 \cite{mce15}  studied nearby ($z\sim$0.06-0.11)   SDSS QSO2 with $l_{\rm O3}$=8.7-9.8 with no prior
 requirements regarding nuclear kinematic properties. They  prioritized  objects
 with extinction correction $A_{\rm [OIII]}<4$ to restrict the impact of uncertainty in the extinction correction
 of the [OIII] luminosities.
 They  find   that extended outflows are present throughout the whole sample. They infer $R_{\rm o}>$3 kpc in general
 with maximum size $\sim$8 kpc.  

The combination of an incomplete spatial coverage by our long-slit data and a sample which is less biased towards high luminosity and highly kinematically perturbed objects  might run against the effectiveness of detecting extended outflows. However, this is not likely to be the explanation.  Not even the objects in our sample with high luminosities and/or highly perturbed nuclear kinematics show  such large scale outflows.

All three studies  mentioned above, showed that the region containing  the ionized outflows 
has in general a round morphology. Such appearance already suggests that the emission  is dominated by the central, unresolved nuclear emission  smeared by the seeing disk. 
Another potential source of concern is that the spatial profiles of the $W_{\rm 80}$ parameter measured  for most objects 
are remarkably flat across the  outflow regions (just like the FWHM 
profiles shown for the broad components of most of our sources). This is difficult to explain if they are truly spatially extended outflows, but  is entirely consistent with what would be expected from  fits done at different locations across the seeing disk  for a spatially unresolved source.  

Unfortunately, even for studies were a detailed comparison with the radial seeing profile was done, the method generally differs from our approach, which makes a direct comparison difficult. For example,  \cite{liu13a} used observations of the standard star broadened to the width of the seeing size for each object. We note, however, that while this method is certainly adequate to reproduce the core of the seeing profile, it may significantly underestimate the contribution of the wings. 

We also note that the round morphologies are actually 
difficult to reconcile with the nature of theses systems, since the  outflowing gas is 
ionized by AGN related processes (e.g. Villar Mart\'\i n et al. \citeyear{vm14}). The continuum source is hidden from the line of sight and
thus  the  ionized gas (including the outflow) must be preferentially distributed  within two opposite ionization cones whose opening angle is determined by the obscuring torus. Circular morphologies could be expected in objects with type 1 orientation (the projection of a cone with the axis close to the line of sight would result in a circular morphology), but not type 2s. Interestingly, those very few objects for which  perturbed gas seems to be clearly extended in comparison with the seeing show elongated morphologies and are radio-loud or radio-intermediate (Liu et al. \citeyear{liu13b}). This is naturally explained if a large ($>$1 kpc) scale radio jet triggers the outflows (Tadhunter et al.  \citeyear{tad94},  Sol\'orzano-I\~narrea et al. \citeyear{sol01}, Humphrey et al. \citeyear{hum06}).

%It is clear that in order to asses the true extension of the outflows two things are essential: 1)  to isolate the outflow emission by means of the spectral decomposition of the optical lines at different locations in order to account for possible blends with the emission from other gaseous  components not associated with the outflows, and  2)  to study the spatial distribution of the outflows on scales well below the seeing size. Not considering these issues (as occurs the   studies quoted above) can lead to misinterpretations of the true spatial extensions of the outflows. 

Typical radial sizes of the NLR in QSO2 as measured from the [OIII] line increase with AGN luminosity from $\sim$3 kpc up to a limiting maximum size of $\sim$6-8 kpc (Hainline et al. \citeyear{hai14}). 
We have seen that the outflowing gas has typically few$\times$100 pc $\la R_{\rm o}\la$1-2  kpc (also constrained using [OIII]),
 well below the total NLR size. These are  in good agreement with the reported sizes of typical molecular outflows in Seyfert galaxies and quasars (Cicone et al. \citeyear{cic14}, Alatalo et al. \citeyear{ala15}; although see also Cicone et al. \citeyear{cic15} for an exception). We find no evidence for the outflows to transport mass and energy to large distances across  the galaxy interstellar medium (ISM).

\subsection{Implications about the outflow impact on the host}

In the previous sections we have found that, though ionized outflows exist in most of the luminous AGN of our sample, they are in general extended over scales of $R_{\rm o}\la$1-2 kpc. SDSS J0903+02 might be the only exception and it will be  discussed  separately in Sect. \ref{Sec:j0903}.

 We analyze the implications regarding:

a) the outflow density;

b) the outflow mass $M_{\rm o}$, mass injection rate $\dot M_{\rm o}$ and   energy injection rate $\dot E_{\rm o}$;

c) the potential impact on the galaxy host;
    
\subsubsection{$M_{\rm o}$, $\dot M_{\rm o}$ and $\dot E_{\rm o}$}

For the purpose of comparison with other studies , we will calculate the  outflow mass, mass injection rates and 
energetics as:

$$M_{\rm o} = 2.8 \times 10^9~ {\rm M_{\odot}} \bigg(\frac{L_{\rm H\beta}^{\rm o}}{10^{43}~ \rm erg ~s^{-1}}\bigg)  \bigg(\frac{n}{100 \rm ~ cm^{-3}}\bigg)^{-1} \rm ~~~~ [eq. 1] $$ 

$$\dot M_{\rm o} = \frac{M_{\rm o} ~V_{\rm o}}{R_{\rm o}}  \rm ~~~~ [eq. 2] $$

$$\dot E_{\rm o} = \frac{1}{2}{\dot M_{\rm o} ~V_{\rm o}^2} = \frac{1}{2} \frac{M_{\rm o}}{R_{\rm o}} ~V_{\rm o}^3 \rm ~~~~ [eq. 3] $$

where  $L_{\rm H\beta}^{\rm o}$ is  the H$\beta$ luminosity emitted by the outflowing gas and $V_{\rm o}$ is the 
the outflow velocity  (e.g. Nesvadba et al. \citeyear{nes11},  Liu et al. \citeyear{liu13b},  Arribas et al. \citeyear{arr14}).

It is clear that  $R_{\rm o}$, $V_{\rm o}$ and $n$ play a critical role in the determination of these quantities.

\vspace{0.2cm}

{\it The outflow density: $n$}

\vspace{0.2cm}

Values $n\sim$100 cm$^{-3}$ are commonly used in the literature for the outflowing gas. We argue next  that $n\ga$1000 cm$^{-3}$ are likely to be more realistic.

The fact that $R_{\rm o}\la$1-2 kpc
implies that the outflows entrain the inner region of the NLR or at least the bulk of the outflow  emission is originated there. They possibly expand from the accretion disk 
and propagate through the inner NLR. In some cases, a weak radio jet might drive the outflow. 

Thus, the expected outflow densities are  those characteristic of the inner NLR (or higher, if compression is produced by shocks). Average densities typically assumed for the NLR in AGN are $\sim$several$\times$100-10$^4$ cm$^{-3}$. Indeed,  $n\sim$10$^3$-10$^4$ cm$^{-3}$ for the outflowing gas are supported by different studies     (e.g. Holt et al. \citeyear{holt11}, Villar Mart\'\i n et al. \citeyear{vm14}, Villar Mart\'\i n et al. \citeyear{vm15}).

 Because such densities are usually inferred from optical lines of low critical densities (most frequently the [SII]$\lambda\lambda$6717,6731 doublet), they most likely trace moderate density gas, since they are not sensitive to high densities. Indeed, both observational and theoretical arguments suggest the presence of a density gradient in the NLR with $n\la$100 cm$^{-3}$ (low density limit) in the outer parts  (e.g. Bennert et al. \citeyear{ben06a}, \citeyear{ben06b}) and possibly  $n\sim$10$^{7-8}$ cm$^{-3}$ for the inner gas (e.g. De Robertis \& Osterbrock \citeyear{derob84}, Baskin \& Laor \citeyear{bas05}, Stern \& Laor \citeyear{stern14}). According to different authors (e.g. Stern \& Laor \citeyear{stern14})  for a broad distribution of densities, the emission of a certain forbidden line is expected to peak at gas with $n$
 similar to the critical density of the line $n_{\rm crit}$. If this is the case for the outflowing gas, given that it  emits, for instance, very strong [OIII]$\lambda\lambda$4959,5007 
 $n\sim n_{\rm crit}=$8.5$\times$10$^{5}$ cm$^{-3}$ are expected.
 
 The possibility that lower density ionized gas is involved in the outflow is  not discarded (Hopkins et al. \citeyear{hop12}), but its contribution to the outflow  optical line fluxes  would be comparatively very faint.  The outflowing gas we can measure  traces higher density gas.

\vspace{0.2cm}

{\it The outflow velocity: $V_{\rm o}$}

\vspace{0.2cm}

The NLR clouds are known to span a range in velocities revealed for instance by  the different FWHM shown by
the optical emission lines  (De Robertis \& Osterbrock \citeyear{derob84}).
A range of velocities is expected for the outflowing gas as well (Liu et al. \citeyear{liu13b}).
Eq. 2 and 3 above require the approximation of considering a single velocity  for a given object. Different
approaches are taken in the literature, such as (a) $V_{\rm o}=\frac{1}{2} \rm {\rm W}_{\rm max}$ + $V_{\rm max}$,
(b) $V_{\rm o}= V_{\max}$ or (c) $V_{\rm o}=\frac{1}{3}$ W$_{\rm 80}$. 

This is especially critical for the estimated energies.  $V_{\rm o}$ differs in cases (a) and (b) by a factor of typically $\sim$3-5,
an uncertainty that affects $\dot M_{\rm o}$. More importantly, this introduces a factor of
$\sim$27-125 uncertainty in $\dot E_{\rm o}$.

We will consider the two extreme cases (a) and (b) to obtain a range
of possible mass outflow rates and kinetic energies of the outflows.

 \vspace{0.2cm}

{\it The H$\beta$ luminosity: $L_{\rm H\beta}^{\rm o}$}

\vspace{0.2cm}

Only a fraction of the total $L_{\rm H\beta}$ is emitted by the outflow.  To estimate $L_{\rm H\beta}^{\rm o}$   we  used   $\frac{L_{\rm [OIII]}^{\rm o}}{L_{\rm [OIII]}}$   (Table \ref{tab:fitsnuc} and \cite{vm11b}) and  a typical $\big(\frac{[OIII]}{H\beta}\big)^{\rm o}=$12.6 for the outflowing gas. This is the median value (with standard deviation 4.0) measured in QSO2
at $z\la$0.7 (\cite{vm11b}, Villar Mart\'\i n et al. \citeyear{vm14}).  The mass uncertainty associated with the assumed ratio is
comparatively small (a factor $\la$1.5) and we will ignore it. Uncertainties in the identification of the nuclear kinematic components contributing to the outflow are also expected to be $\la$2 and will also be ignored. We will assume in all cases that the broadest kinematic component traces the outflow emission.

 We will ignore reddening also. Based on measurements of the Balmer decrement of the outflowing gas (Villar Mart\'\i n et al. \citeyear{vm14}), this can result in underestimations of the outflow mass by a factor of $\la$3.  This uncertainty is negligible compared with that derived from the density.

 \subsubsection{Outflow mass  $M_{\rm o}$, mass injection rate $\dot M_{\rm o}$ and energy injection rate
$\dot E_{\rm o}$}

We show in  Table \ref{tab:calc-outflows} the values estimated for  $M_{\rm o}$,  $\dot M_{\rm o}$
and  $\dot E_{\rm o}$ for $n=$1000 cm$^{-3}$. As discussed above, significantly higher densities
are not discarded. Thus, all the quoted values  are upper limits. As an example, a density 10 times higher would decrease all
calculated quantities by the same factor. 

%As mentioned above, lower densities are not discarded either, but this gas has a minor contribution of the integrated flux of the outflowing gas.

The outflow mass is quoted in column (5). Excluding SDSS J0903+02, the range of masses spanned by our sample is $\sim$(0.4-12)$\times$10$^6$
M$_{\odot}$ with a median value $M_{\rm o}=$1.1$\times$10$^6$ M$_{\odot}$. These values are well below those estimated for the large scale outflows
identified in other studies :  $M_{\rm o}\sim$(0.2-1)$\times$10$^9$ 
M$_{\odot}$     (Liu et al. \citeyear{liu13b});  $\sim$(1.0-20)$\times$10$^9$ 
M$_{\odot}$ (McElroy et al. \citeyear{mce15});  $\sim$(2-40)$\times$10$^7$ M$_{\odot}$  (Harrison et al. \citeyear{har14}). The discrepancy is a consequence of the  $n=$100 cm$^{-3}$  value used in those studies   and 
  the assumption that  all the observed ionized gas is involved in the  outflow.

As explained above, $\dot M_{\rm o}$ and  $\dot E_{\rm o}$ are even more uncertain. Not only the uncertainty in $n$
is involved, but also in $R_{\rm o}$ and $V_{\rm o}$.
We have calculated  the minimum and maximum values  allowed by the range
of $R_{\rm o}$ and $V_{\rm o}$ values constrained from the observations. This means, $R_{\rm o}^{astrom} \le R_{\rm o} \le R_{\rm o}^{seeing}$ (Table \ref{tab:sizes}) and $V_{\rm max} \le V_{\rm o} \le V_{\rm max} + \frac{FWHM_{\rm max}}{2} $ (see above).
For those objects for which $R_{\rm o}^{astrom}$ could not be constrained (i.e. $R_{\rm o}^{astrom}>$0) we adopt an
illustrative 200 pc   value. 

With all this information, the maximum and minimum $\dot M_{\rm o}$ and  $\dot E_{\rm o}$  are thus calculated
(Table \ref{tab:calc-outflows}, columns 6 and 7).   The first obvious conclusion is that even when $n$ is fixed, 
when all other uncertainties are considered, the quantities characterizing the outflows are  poorly constrained.
For a given object, the allowed $\dot M_{\rm o}$  and $\dot E_{\rm o}$  values differ by typically $\sim$several$\times$10s 
and $\sim$several$\times$100s respectively.  Introducing the uncertainty in the density will make the situation much worse, since  these 
uncertainty factors will be multiplied by the same $n$ factor.

The second conclusion is that the mass and energy outflow rates are in general   modest or low compared with values
estimated in related works. Even in the most optimistic cases (i.e. the most favorable assumptions that produce the highest values),
the typical estimated mass outflow rates 
are  $\dot M_{\rm o}\la$10 M$_{\odot}$ yr$^{-1}$.  For comparison, other studies have found 
 $\dot M_{\rm o}\sim$(0.2-2)$\times$10$^4$ 
 M$_{\odot}$ yr$^{-1}$  (Liu et al. \citeyear{liu13b}); $\sim$370-2700
 M$_{\odot}$ yr$^{-1}$ (McElroy et al. \citeyear{mce15}) and ú$\sim$3-70 M$_{\odot}$ yr$^{-1}$ (Harrison et al. \citeyear{har14}). The discrepancy would be even larger if higher densities were considered. 
 
The $\dot M_{\rm o}$ values can be compared with the star forming rates SFR, although this   must be considered cautiously since the stars do not form  out of ionized, but molecular gas. 
While  $\dot M_{\rm o} \ga SFR$  for the molecular gas suggests  that  star forming quenching can occur, this is not necessarily the case for the ionized gas.
 \cite{zak16} have estimated upper limits of SFR
  in obscured quasars in the range $\sim$few to $\sim$100 M$_{\odot}$ yr$^{-1}$ 
 with a median value of $\sim$18 M$_{\odot}$ yr$^{-1}$.  All uncertainties considered, it is not possible to conclude whether the  mass outflow rate  of ionized gas  surpasses the star formation rates although, as explained above,
 the mass outflow rates are likely to be significantly lower.   

\begin{table*}
\centering
\caption{Outflows masses, mass outflow rates and kinetic energies. Column (2) gives  the [OIII] luminosity emitted
by the outflow relative to the total line luminosity, estimated from the nuclear spectra. $V_{\rm o}$ in column (3) is the outflow
velocity. The minimum and maximum possible values are shown. The minimum and maximum $R_{\rm o}$ values
are shown in column (5) (see Table 4). For those objects for which $R_{\rm o}^{astrom}$ could not be constrained (i.e. $R_{\rm o}^{astrom}>$0) we adopt an
illustrative 0.2 kpc   value. 
$M_{\rm o}$ in column (5) is the outflow mass assuming $n$=1000 cm$^{-3}$. Values for $n$=100 cm$^{-1}$ are also quoted for
SDSS J0923+02, the only candidate in our sample to harbor a galaxy scale outflow.
The mass rate and energy rate of the outflows are given in columns (6) and (7). The minimum and maximum values allowed
by the calculations are quoted taking into account uncertainties on $R_{\rm o}$ and $V_{\rm o}$. (8) gives the bolometric luminosity $L_{\rm bol}$ estimated from L$_{\rm [OIII]}$ (Stern \& Laor 2012). } 
\begin{tabular}{llllllll}
\hline
 &  &  &    n=1000 cm$^{-3}$&  &    \\ 
Object  & $\frac{L_ {\rm [OIII]}^{\rm o}}{L_ {\rm [OIII]}^{\rm tot}}$ &   $V_{\rm o}$ & $R_{\rm o}$ & $M_{\rm o}$ & $\dot M_{\rm o}$   & $\dot E_{\rm o}$  & $L_{\rm bol}$ \\
 	& &  min/max 	& 	min/max  &  &  min/max & min/max \\
 & &   km s$^{-1}$ & kpc  & $\times$10$^6$ M$_ {\odot}$ & M$_ {\odot}$ yr$^{-1}$ & $\times$10$^{42}$ erg s$^{-1}$ & $\times$10$^{42}$ erg s$^{-1}$\\ 	
(1) & (2) & (3) & (4) & (5)   & (6) & (7) & (8)  \\ 
\hline
  &		& 2009	& sample	&	& \\ \hline
SDSS J0955+03 &  0.40 & 620/1870 & {\it 0.20}/3.5 & 1.1 & 0.2/10.3 & 0.02/11.2 & 2.1$\times$10$^3$\\
SDSS J1153+03 &  0.10 &  330/1055 & 0.52/1.7 & 3.9 & 0.8/8.1 & 0.03/2.8 & 8.7$\times$10$^4$ \\
 SDSS J1228+00 & 0.67& 770/1920& {\it 0.20}/2.1 & 11.6 & 4.4/114 & 0.8/130 & 2.8$\times$10$^4$ \\
   SDSS J1307-02  &    0.32 & 190/690 &  1.3$\pm$0.4 &  2.8 & 0.3/1.5 & 0.004/0.2 & 1.1$\times$10$^4$\\  
SDSS J1337-01 &   0.70 &  290/930 & 0.19/1.7 & 3.2 & 0.6/16.0  & 0.01/4.3 &  4.4$\times$10$^3$\\
SDSS J1407+02 &  0.13& 70/905 &  1.0$\pm$0.2 & 1.1 & 0.4/1.0 & 0.001/0.3 & 1.1$\times$10$^4$ \\
 SDSS J1413-01 &   0.19 & 145/740 & 0.21/1.5 & 2.5  & 0.2/8.8 & 0.002/1.5 & 1.9$\times$10$^4$\\
 SDSS J1546-00 &  0.42 &  270/660 & {\it 0.20}/1.4 & 0.5 & 0.1/1.8 & 0.002/0.3 & 7.6$\times$10$^2$ \\ \hline
  &	&		2011 & sample	&	& \\ \hline
 SDSS J0903+02  n=1000 & 0.42  &  947/2600 &  1.1  &  2.2 & 1.9/5.3 &  0.5/11.4 &  5.4$\times$10$^3$ \\
SDSS J0903+02  n=100 &  0.42  &  947/2600 &   10.0  &  22 &  2.1/5.9 &  0.6/12.5 &  5.4$\times$10$^3$ \\
SDSS J0923+01  &   0.34 & 36/870 & 0.16/1.0  & 1.7 & 0.1/9.9 & 3$\times$10$^{-5}$/2.3 &  5.2$\times$10$^3$\\
 SDSS J0950+01 &    0.65 &  56/876 & 0.38/1.1 & 0.9 & 0.1/2.2  &  5$\times$10$^{-5}$/0.5 & 8.7$\times$10$^2$ \\
  SDSS J1014+02 &   0.46 & 189/958 & {\it 0.20}/1.0 & 0.8 & 0.2/3.7 & 0.002/1.1 & 1.1$\times$10$^3$ \\ 
SDSS J1247+01 &  0.30&   278/865 & 0.28/1.4 & 0.4 & 0.1/1.4 & 0.002/0.3 & 9.0$\times$10$^2$ \\
 SDSS J1336-00 &  0.23 & 894/2137 & 0.19/1.0 & 0.9 & 0.8/9.8 & 0.2/13.9 & 3.3$\times$10$^3$\\
 SDSS J1430-00 & 0.24 & 520/1320 &  0.12/0.80  & 0.6 & 0.4/6.3 & 0.03/3.4 & 1.8$\times$10$^3$ \\
\hline
\end{tabular}
\label{tab:calc-outflows}
\end{table*}

 For the energy injection rates, the maximum  $\dot E_{\rm o}$ our calculations allow are in general $\la$10$^{43}$  erg s$^{-1}$.  Values  $\sim$(0.4-3)$\times$10$^{45}$,  $\sim$(5-19)$\times$10$^{42}$  erg s$^{-1}$ and $\sim$(0.3-30)$\times$10$^{42}$  were obtained by \cite{liu13b}, \cite{mce15} and  \cite{har14}
 respectively. The values are not far from those obtained by \cite{har14} and \cite{mce15}. The reasons are that the authors assume a larger radius $R_{\rm o}$ and
 $V_{\rm o}=W_{80}$/1.3. This implies significant lower velocities (e.g. $V_{\rm o}$=510-1100 km s$^{-1}$ for Harrison et al. \citeyear{har14}), compared with our range of maximum possible velocities $\sim$690-2600
km s$^{-1}$. 
 
The maximum kinetic energy of the outflows   allowed by the calculations  is $<$1\% of  the bolometric luminosity of the objects (Table \ref{tab:calc-outflows}, column 8).
This percentage could be much lower taking into account the uncertainties in  the outflow energy rates.
Thus, low conversion rates of $L_{\rm bol}$ to kinetic energy of the ionized gas are enough to explain the estimated outflow energy rates. From the energetic point of view, the outflows can thus be easily triggered by the active nuclei.

In conclusion, our results suggest that the sizes, masses, mass and energy injection rates of the typcial ionized 
 outflows are low to moderate compared with related studies. No evidence is found supporting that the typical outflows expand into  and affect the  large scale ($>$several kpc) interstellar medium. The uncertainties in  the $\dot M_ {\rm o}$ and on the SFR prevent a meaningful comparison, although $\dot M_ {\rm o}>$SFR seems unlikely.   The outflows might have a much more dramatic impact on other  phases of the interestellar ISM, such as the molecular phase (e.g. Cicone et al. \citeyear{cic14}, Alatalo et al. \citeyear{ala15}).
 
 In this context, it is interesting that at high redshifts seeing limited observations (e.g. Alexander et al. \citeyear{ale10}; Harrison et al. \citeyear{har12}) tend to report larger extensions for the outflows than those that use AO systems to improve spatial sampling (Forster-Schreiber et al. \citeyear{for14}). The samples so far are relatively small and the intrinsic properties of the targets may well explain these results. However,  these results may also suggest that the true sizes of seeing limited observations might be affected by
 seeing smearing.
 
There is evidence that the radio jets/lobes can trigger ionized outflows on scales of $\ga$10 kpc in high $z$ powerful radio galaxies (Humphrey et al. \citeyear{hum06}, Nesvadba et al. \citeyear{nes08}).   Whether such extended outflows exist in the general population of AGNs from low to high redshifts and in non-radio-loud systems
remains open. 
On this respect, it is interesting to mention the  nearby radio quiet QSO2 SDSS J1430+1339 at $z=$0.085 (Harrison et al. \citeyear{har14}).
It shows clear evidence for the impact of AGN induced feedback in the form of ``bubbles"  of radio emission that are extended $\sim$10-12 kpc at both sides of the nucleus and correlate tightly with the  morphology of the [OIII] extended ionized gas. It is possible that extended radio structures are necessary for AGN feedback to work across large spatial scales. Alternatively, it might be the mechanism whose imprints are easier to identify across large distances.

\subsection{SDSS J0903+02: a galaxy scale outflow candidate}
\label{Sec:j0903}

 The radio intermediate QSO2 SDSS J0903+02 is the only object in our sample for which the data are consistent with   a galactic scale outflow ($R_{\rm o}\sim$10 kpc), 
although alternative explanations cannot be discarded (see Sect. \ref{Sec:ind}).   It has been proposed  as a low radio power analogue of Compact Steep Spectrum sources (CSS) (Lal \& Ho \citeyear{lal10}).  Extreme nuclear kinematics are  often observed in  CSS  and it is likely to be due to young radio jet induced outflows (Holt et al. \citeyear{holt11}), whose effects can be seen in the extended gas
as well (Shih et al. \citeyear{shih13}).

 If an outflow is confirmed as responsible for the highly extended
apparently perturbed kinematics in SDSS J0903+02, low densities $n\la$100 cm$^{-3}$ are more realistic in the extended gas. We show in Table \ref{tab:calc-outflows} the predicted outflow parameters for the two extreme outflow radial sizes we have inferred 1.1 and 10 kpc and densities $n=$1000 and 100 cm$^{-3}$ respectively, following the same simple scenario where  where all the outflowing gas has  similar density,
In the low density case, M$_ {\rm o}$=2.2$\times$10$^7$  M$_ {\odot}$, $\sim$10 times higher than for the rest of the sample.
On the other hand, this cancels out with the same factor 10 on the radial size, resulting on similar $\dot M_{\rm o}\sim$6 M$_{\odot}$ 
yr$^{-1}$ and $\dot E_{\rm o}\sim$0.002 L$_ {\rm bol}$ at most. 

If this outflow extends for $\sim$10 kpc, it can clearly reach regions very distant from the galactic nucleus, although the mass, mass injection
and energy injection rates are also low compared with related studies. 

\section{Summary and Conclusions}

We have analyzed the spatially extended kinematic properties of the ionized gas in a sample of 18 luminous type 2 AGN: 11  QSO2
and 7 high luminosity Seyfert 2 (HLSy2). This sample includes those analysed in Villar-Mart\'\i n et al. (2011a,b, 2012)
and nine new objects with no previous kinematic studies. The main results and conclusions can be summarized as follow:

(i) The nuclear spectra for the new sample of 9 objects provide convincing evidence for nuclear ionized outflows in 6 of them. The remaining three objects, which are HLSy2,  have among the
lowest  [OIII] luminosities in our sample. This is coherent with other works which suggest   that  quasar feedback  operates above the threshold in  bolometric luminosity of $L_{\rm bol}\sim$3$\times$10$^{45}$ 
erg s$^{-1}$.   This study further reinforces previous results which demonstrate that nuclear ionized outflows  are ubiquitous in QSO2.

(ii)  Ionized outflows are found in luminous type 2 AGN with host galaxies showing a broad diversity of merger status (or  absence of merger/interaction signatures). There does not seem to be  a causal link between the
triggering of ionized outflows and merger status.   Outflows are expected basically any time a quasar is active
and are  not necessarily linked with a clear interaction/merger phase.

(iii)  In order to reliably asses whether the outflows are spatially extended and constrain their sizes based on seeing limited observations three things are essential:  1) to characterize as accurately as possible the seeing size and shape, especially the wings for each individual object/observation; 
2)  to isolate the outflow emission  at different spatial locations to avoid contamination by other gaseous  components not  related to the outflow and compare it with the seeing profile;  3)  to study the outflow spatial distribution on scales well below the seeing size.

 (iv)  When this is done, it is found that only for the radio intermediate QSO2 SDSS J0903+02  the extended gas kinematics is consistent with
a giant outflow of radial size $R_{\rm o}\sim$10 kpc, although alternative scenarios cannot be discarded.

(v) In most objects  (12/15) the outflows are found to be unresolved as compared with the seeing disk.  This sets upper limits on the radial sizes
typically $R_{\rm o}\la$1-2 kpc.  In only  2/15 objects (possibly 3/15)  we find  convincing evidence for seeing-resolved outflows.
Both have  $R_{\rm o}\sim$1 kpc. Based on the spectroastrometric method, the outflows are found to be extended in 11/15 objects. Lower limits on their radial sizes are  typically $R_ {\rm o}\sim$several$\times$100 pc.  Thus, our results constrain the location of the outflows within the inner 1-2 kpc of the Narrow Line Region. Except
maybe for SDSS J0903+02, we  find no evidence for the ionized outflows to expand well outside the central, dense galactic nucleus and out into the  extragalactic medium.

This is  inconsistent with related studies  which claim that large scale wide angle ionized outflows 
are ubiquitous in QSO2. Such outflows are proposed to extend $\ga$several kpc from the AGN and can sometimes  engulf entire galaxies.  We argue that  the apparently very extended turbulent kinematics identified in other studies  are   in general a consequence of seeing smearing. A careful account of this effect would probably result in spatially unresolved (or barely resolved) outflows. 

(vi) We argue that the expected outflow  gas  density is $n\ga$1000 cm$^{-3}$. In fact, a wide range of densities (up to possibly $n\sim$10$^7$ cm$^{-3}$), similar to those characteristic of the inner Narrow Line Region, may be plausible. Lower $n$ are not discarded, but the contribution of this  gas to the outflow optical line emission is comparatively negligible. 

(vii)   The implications  in our understanding of the feedback phenomenon are important since all parameters characterizing the masses and energetics of the outflows  become highly uncertain.  One conclusions seems unavoidable: 
  the  outflow sizes, masses, mass injection and energy injection rates are in general  modest or low  compared with related studies. 
We obtain outflows masses
 $M_{\rm o}\sim$(0.4-12)$\times$10$^{6}$ M$_{\odot}$ assuming $n=$1000 cm$^{-3}$. These are $\sim$10$^2$-10$^4$ times lower than values reported in the literature.
Even under the most favorable asumptions,   we obtain  $\dot M_{\rm o}\la$10 M$_{\odot}$ yr$^{-1}$ in general,
100-1000 times lower than claimed in other studies.  The discrepancy would be even larger if higher densities  and/or the less favorable assumptions on the possible velocities and radial sizes were considered. Taking into account typical star forming rates in these systems ($\sim$few-100 M$_{\odot}$ yr$^{-1}$), given the large uncertainties involved, it is not possible to conclude whether  $\dot M_{\rm o}\ga$SFR. However, since $\dot M_{\rm o}$ are likely to be significantly smaller, this seems unlikely.

In conclusion,  we find no evidence supporting that the  {\it ionized} outflows have a significant impact on their host galaxies, not by injecting energy and mass beyond the inner $\sim$1-2 kpc of the Narrow Line Region, well into   the interstellar medium, neither by triggering large mass outflow rates. The outflows might have a much more dramatic impact on other  phases of the ISM, such as the molecular phase.
 
 The loose constraints on  $n$ (probably $\ga$1000 cm$^{-3}$), $V_ {\rm o}$ (uncertain by  factors of $\sim$ a few) and $R_{\rm o}$ (typically $\la$1-2 kpc) imply that for a given object the outflow mass and energy injection rates can be very uncertain by factors $>$10$^2$. High spatial resolution and high S/N spectroscopic studies  from the ground with adaptive optics  or from space  will help overcome these limitations. Facilities such as MUSE with AO on VLT or
NIRSpec on the JWST  will allow  to measure the size, geometry, density and velocity distribution of the ionized outflows
with significantly improved accuracy.  

\section*{Acknowledgments}

We  thank the staff at Paranal Observatory for their support during the
observations. MVM, SA  and BE acklowledge  support  from the Spanish Ministerio de
Econom\'\i a y Competitividad through the grants AYA2012-32295 and AYA2012-39408-C02-01. AH acknowledges Funda\c{c}\~{a}o para a Ci\^{e}ncia e a
Tecnologia (FCT) support through UID/FIS/04434/2013, and through
project FCOMP-01-0124-FEDER-029170 (Reference FCT
PTDC/FIS-AST/3214/2012) funded by FCT-MEC (PIDDAC) and FEDER
(COMPETE), in addition to FP7 project PIRSES-GA-2013-612701. AH also
acknowledges a Marie Curie Fellowship co-funded by the FP7 and the FCT
(DFRH/WIIA/57/2011) and FP7 / FCT Complementary Support grant
SFRH/BI/52155/2013. BE is grateful that the research leading to these results has been funded
by the European Union 7th Framework Programme (FP7-PEOPLE-2013-IEF)
under grant 624351.  CRA is supported by a Marie Curie
Intra European Fellowship within the 7th European Community
Framework Programme (PIEF-GA-2012-327934).

 This research has made use of the VizieR catalogue access tool, CDS,
 Strasbourg, France. The original description of the VizieR service was
 published in A\&AS 143, 23.

\newcommand{\noopsort}[1]{}

\appendix

\section{Notes on individual objects.}
\label{Sec:app}

We present in this Appendix the detailed spatially extended analysis of all objects, but those three discussed in the main body of the paper.

\subsection{2009 sample}
\vspace{0.2cm}

{\it SDSS J0955+03}
% Z=0.422, 5.52 kpc/arcsec
\vspace{0.2cm}

This radio-intermediate  HLSy2 shows extreme nuclear kinematics, with three kinematic components having FWHM
  470$\pm$80, 1110$\pm$80  and 2500$\pm$200 km s$^{-1}$ respectively, with the broadest  blueshifted by $V_{\rm max}$=-620$\pm$140 km s$^{-3}$ (\cite{vm11b}). 
 
  \begin{figure*}
\includegraphics{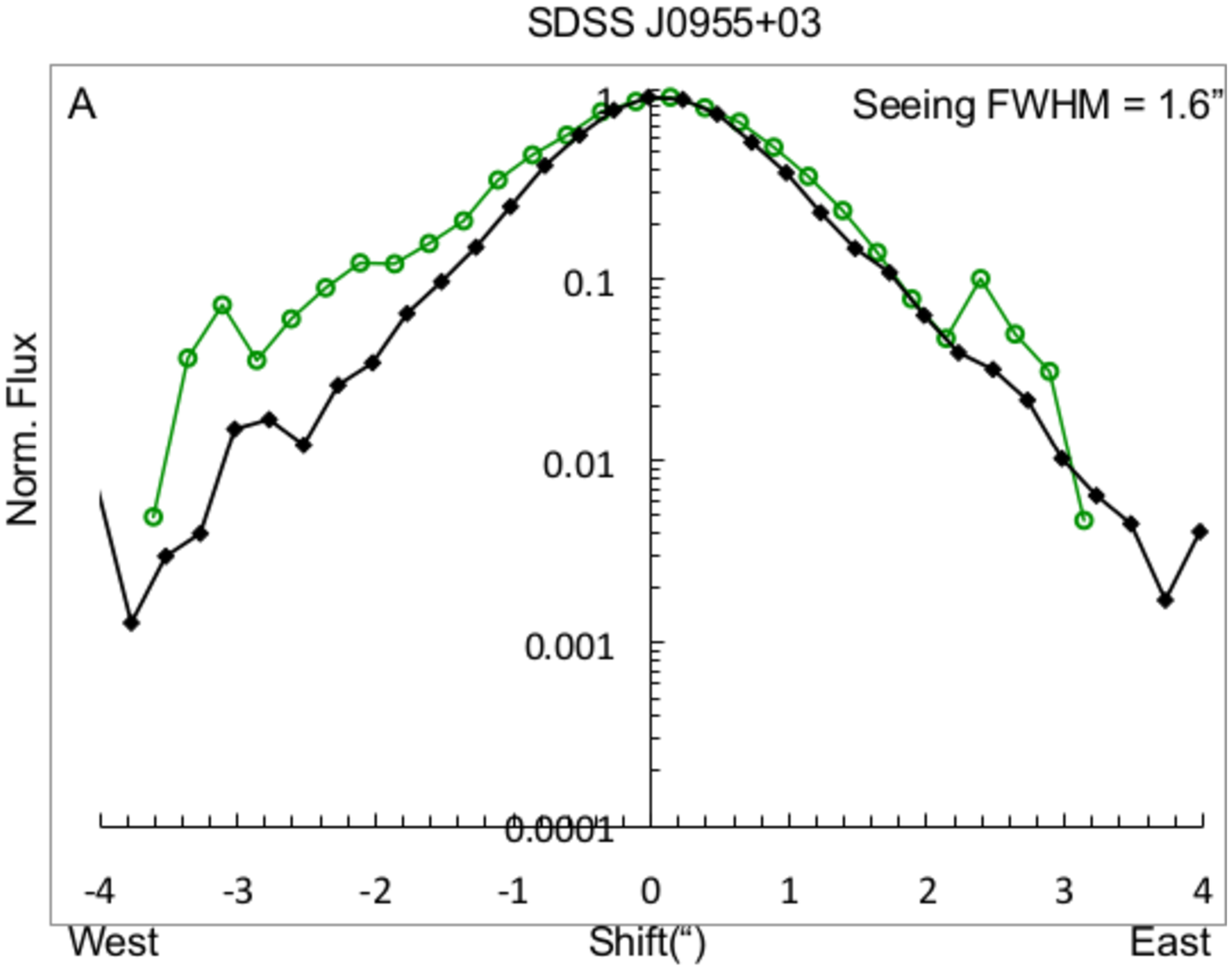}
\includegraphics{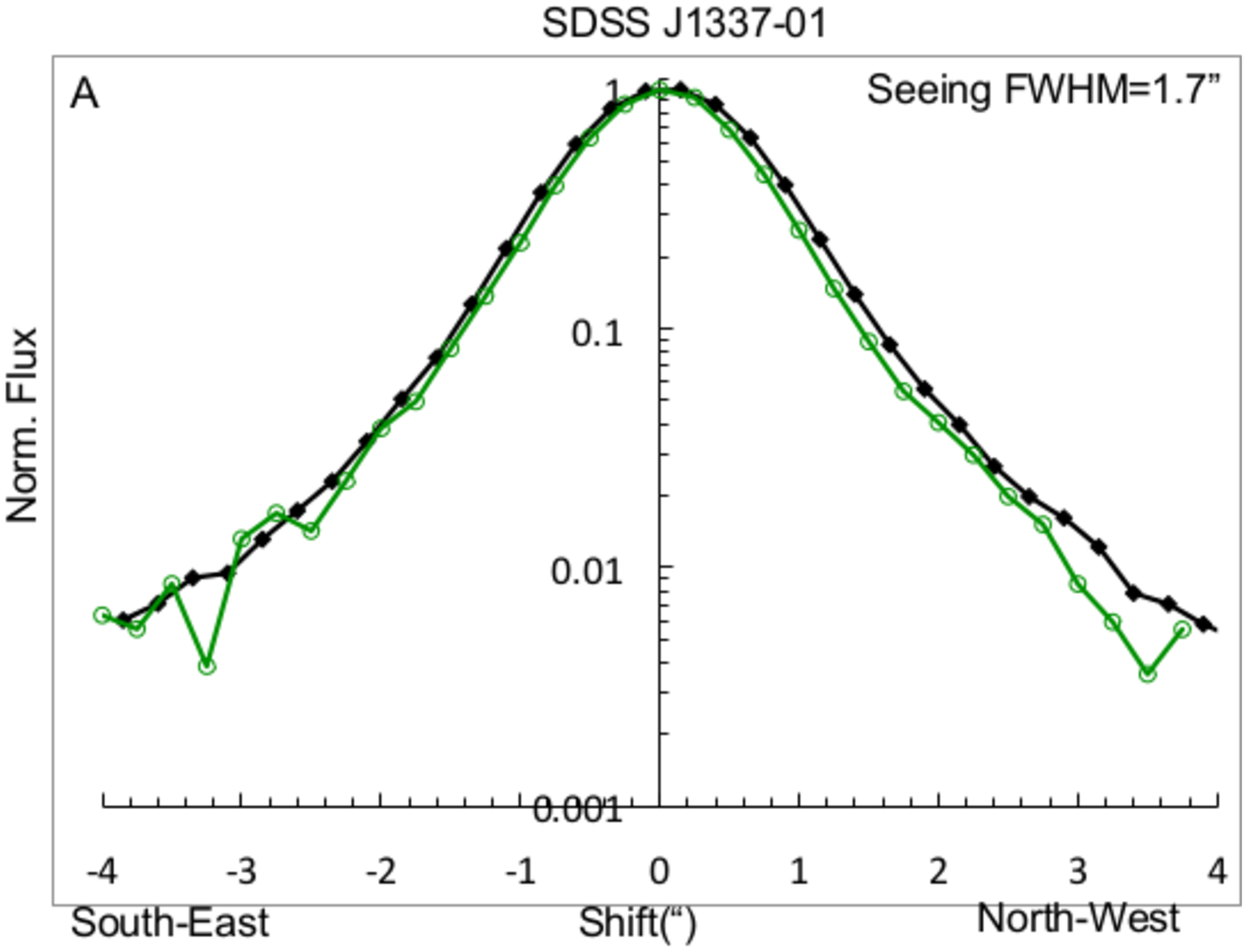}
\vspace{2.2in}
\includegraphics{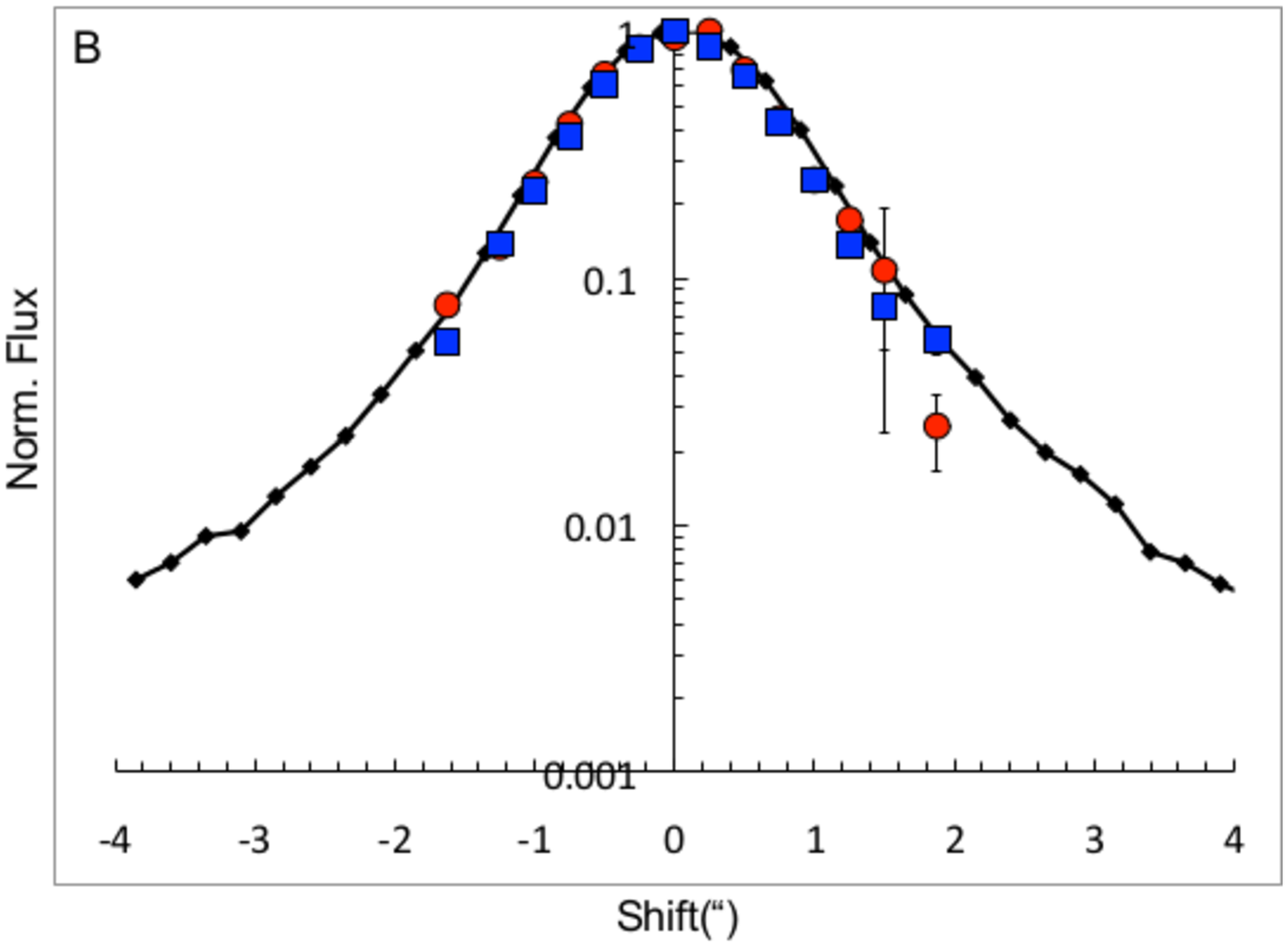}
\vspace{2.2in}
\includegraphics{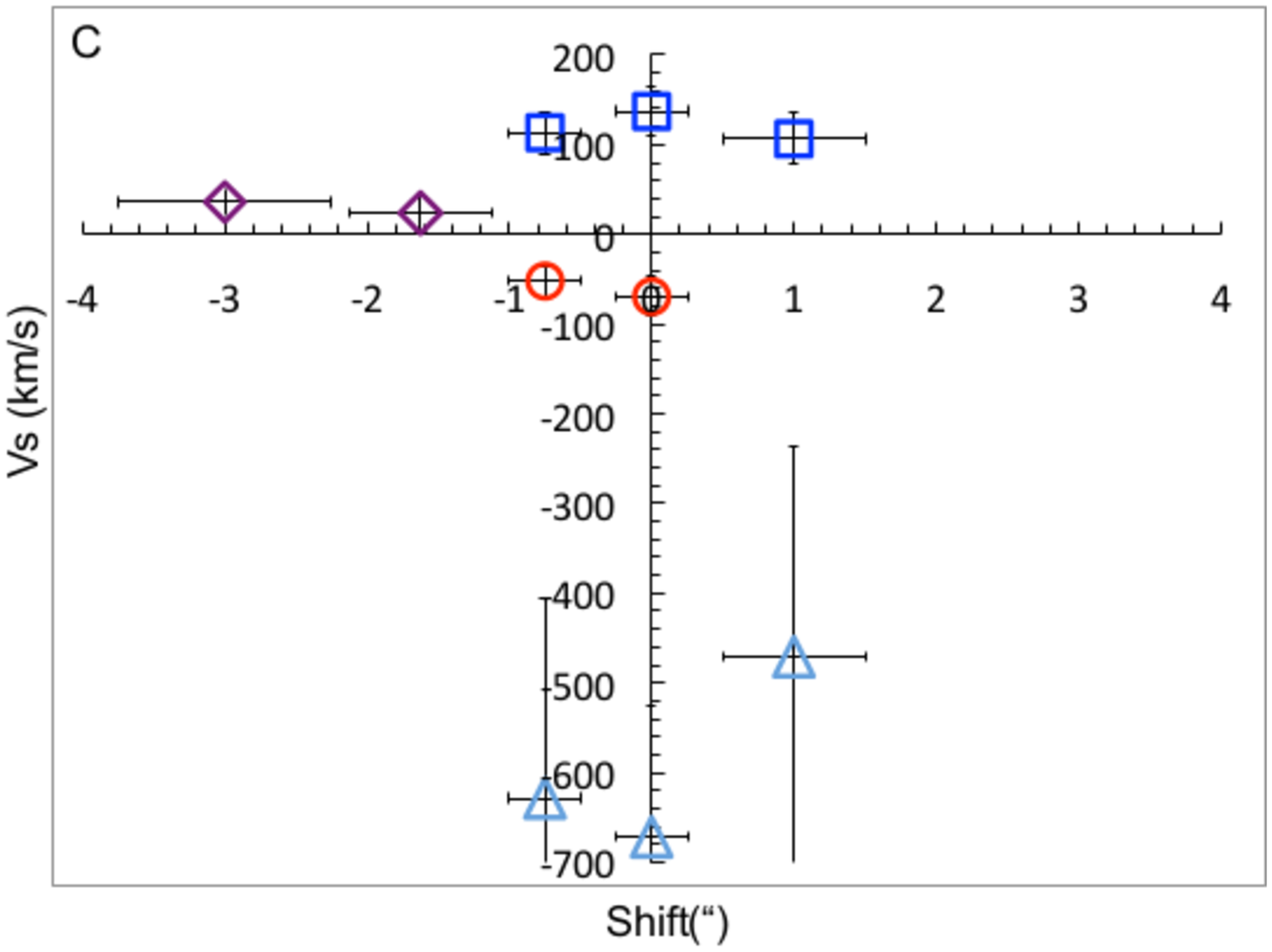}
\includegraphics{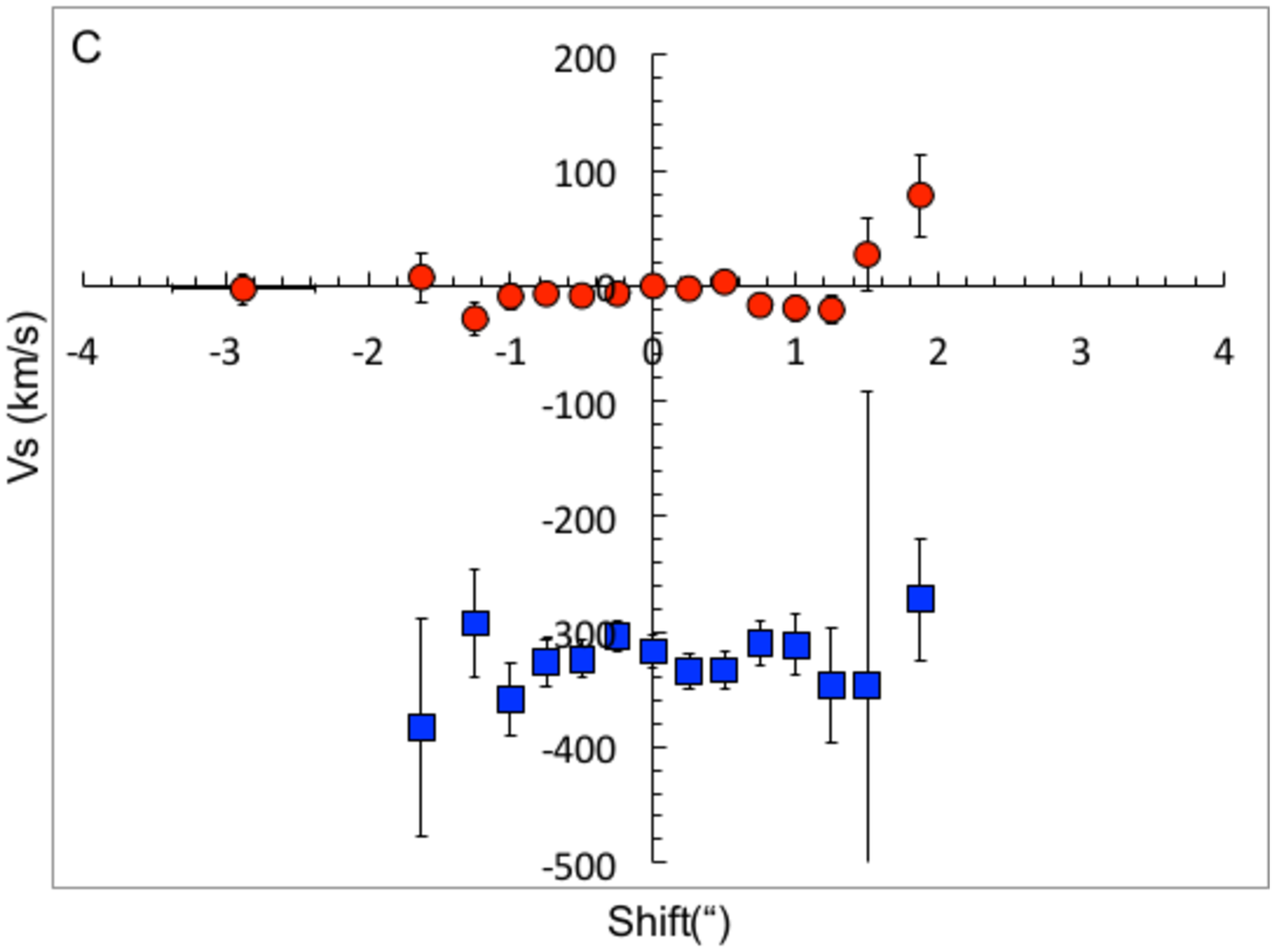}
\vspace{2.2in}
\includegraphics{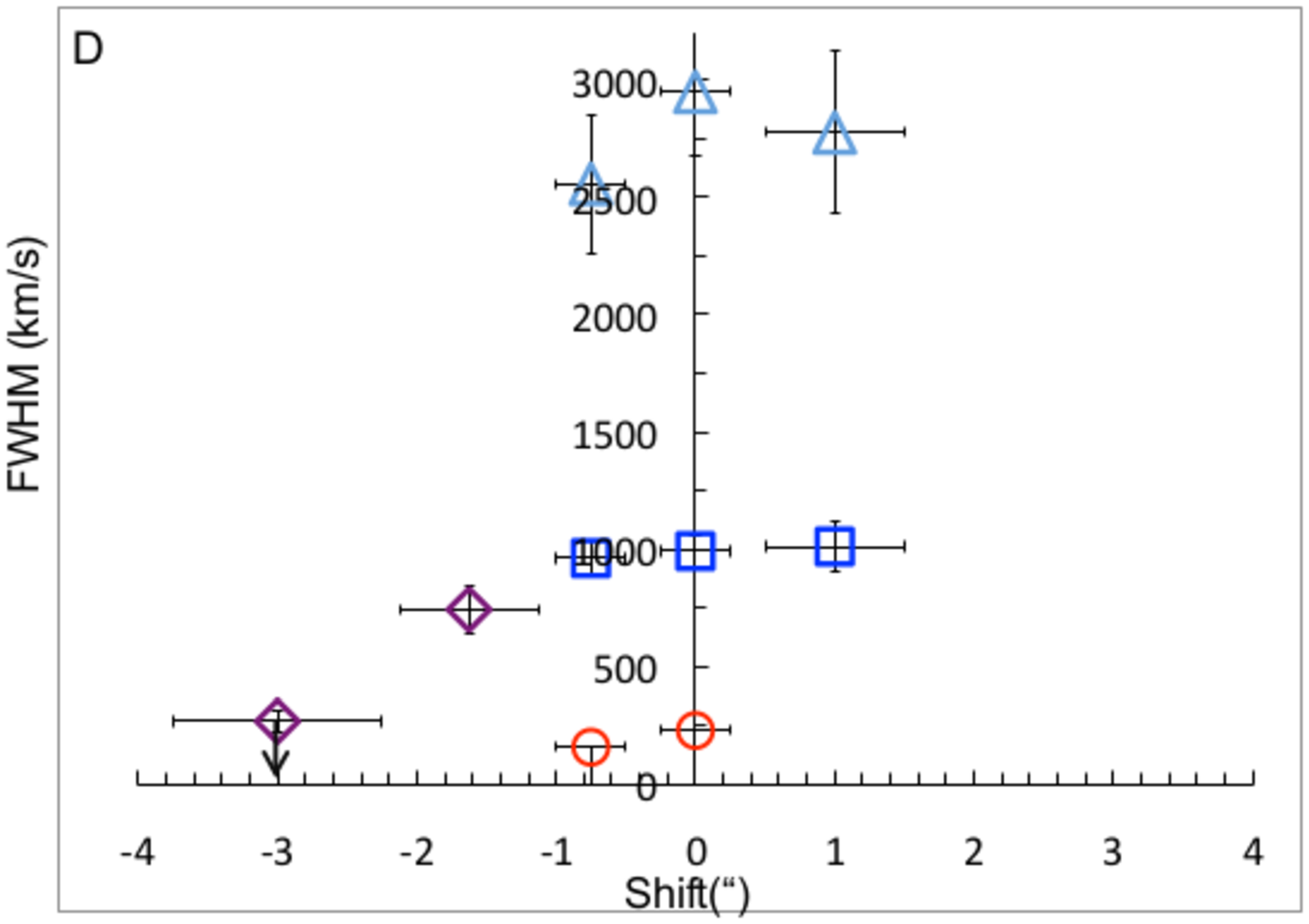}
\includegraphics{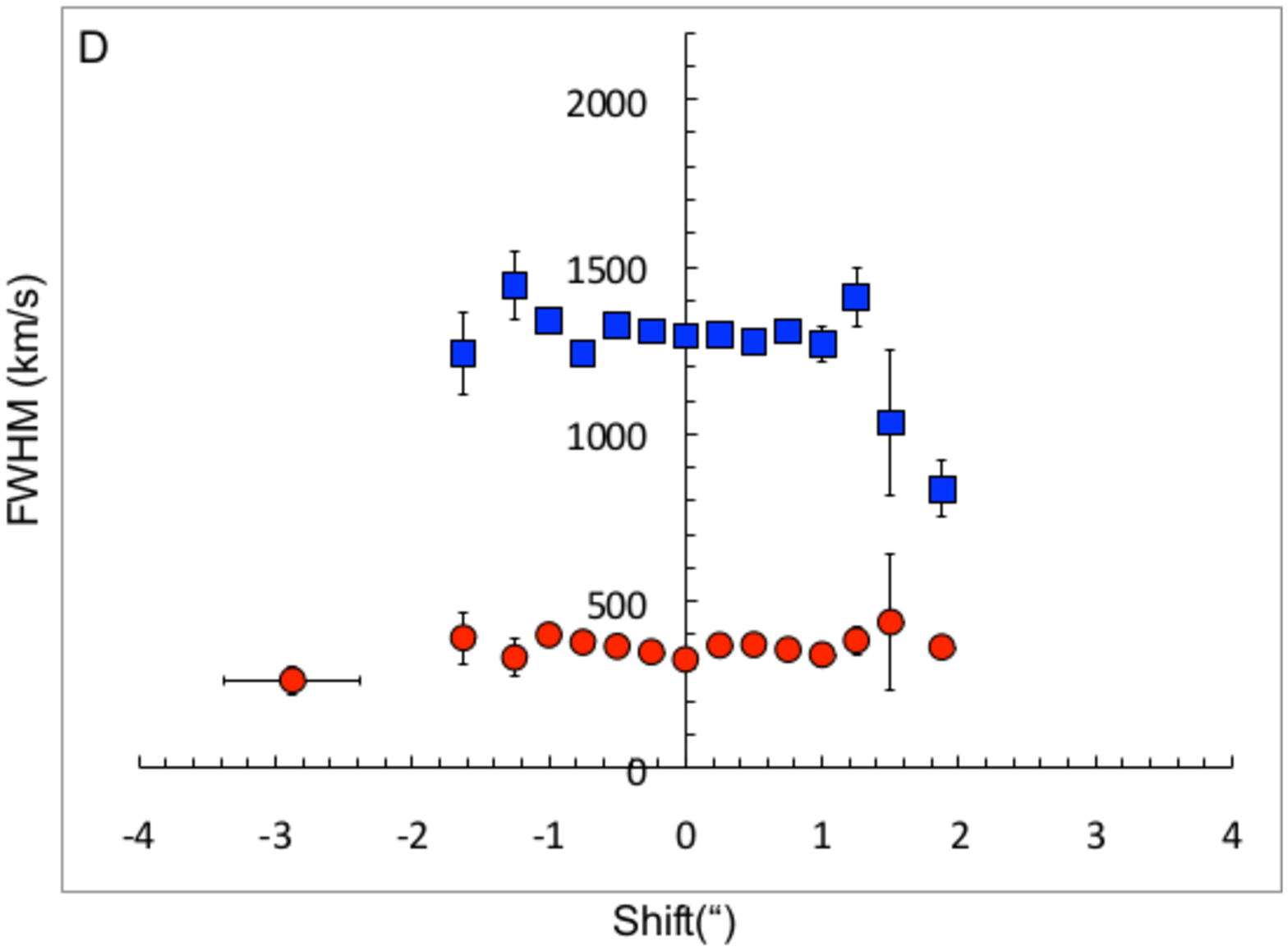}
\vspace{2.2in}
\caption{Spatially extended analysis of [OIII]$\lambda$5007  for SDSS J0955+03 (left) and SDSS J1337-01 (right). Colour and line style codes as in Fig. 5.}
\label{spat0955v1337}
\end{figure*}

\begin{figure*}
\includegraphics{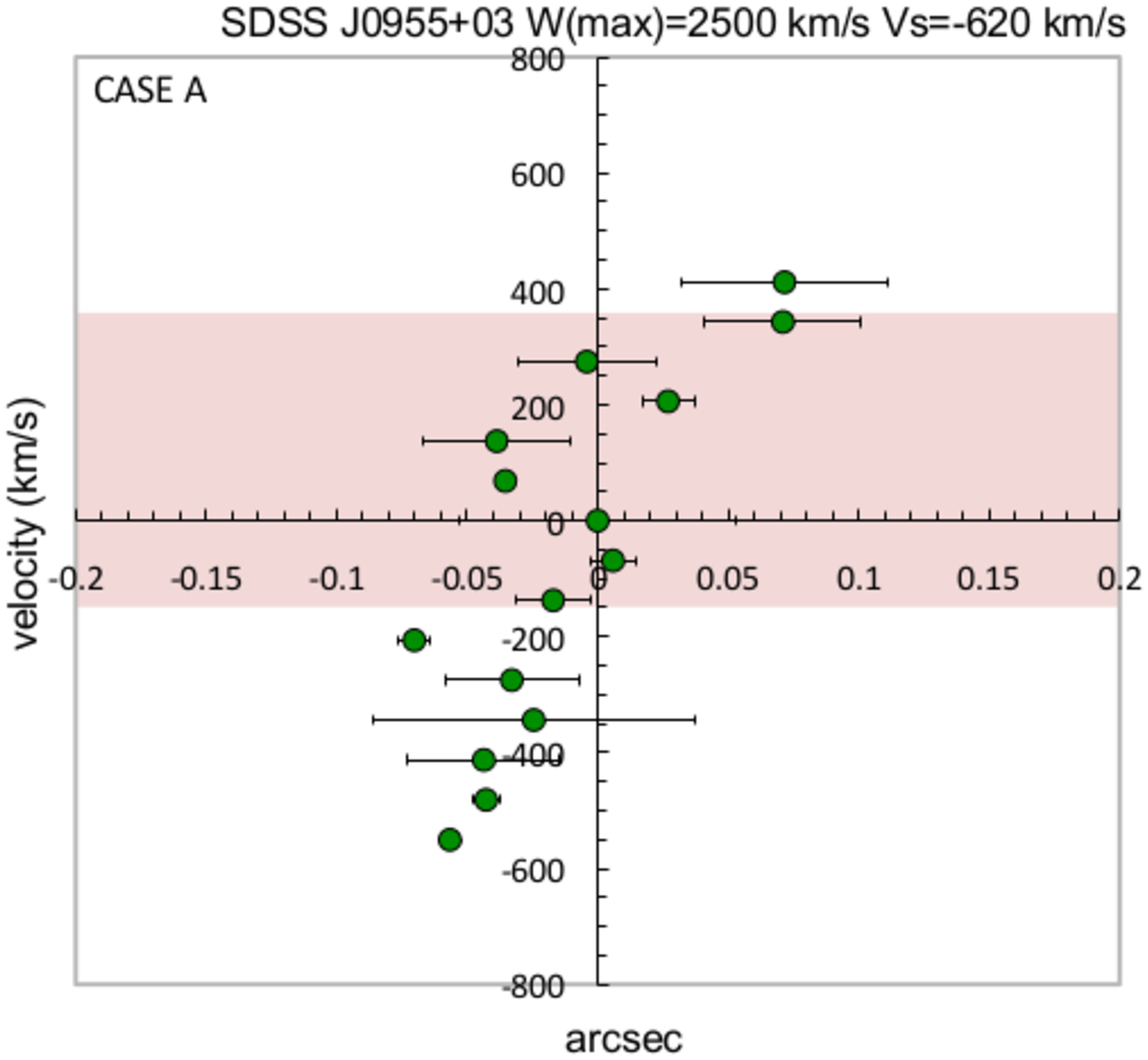}
\includegraphics{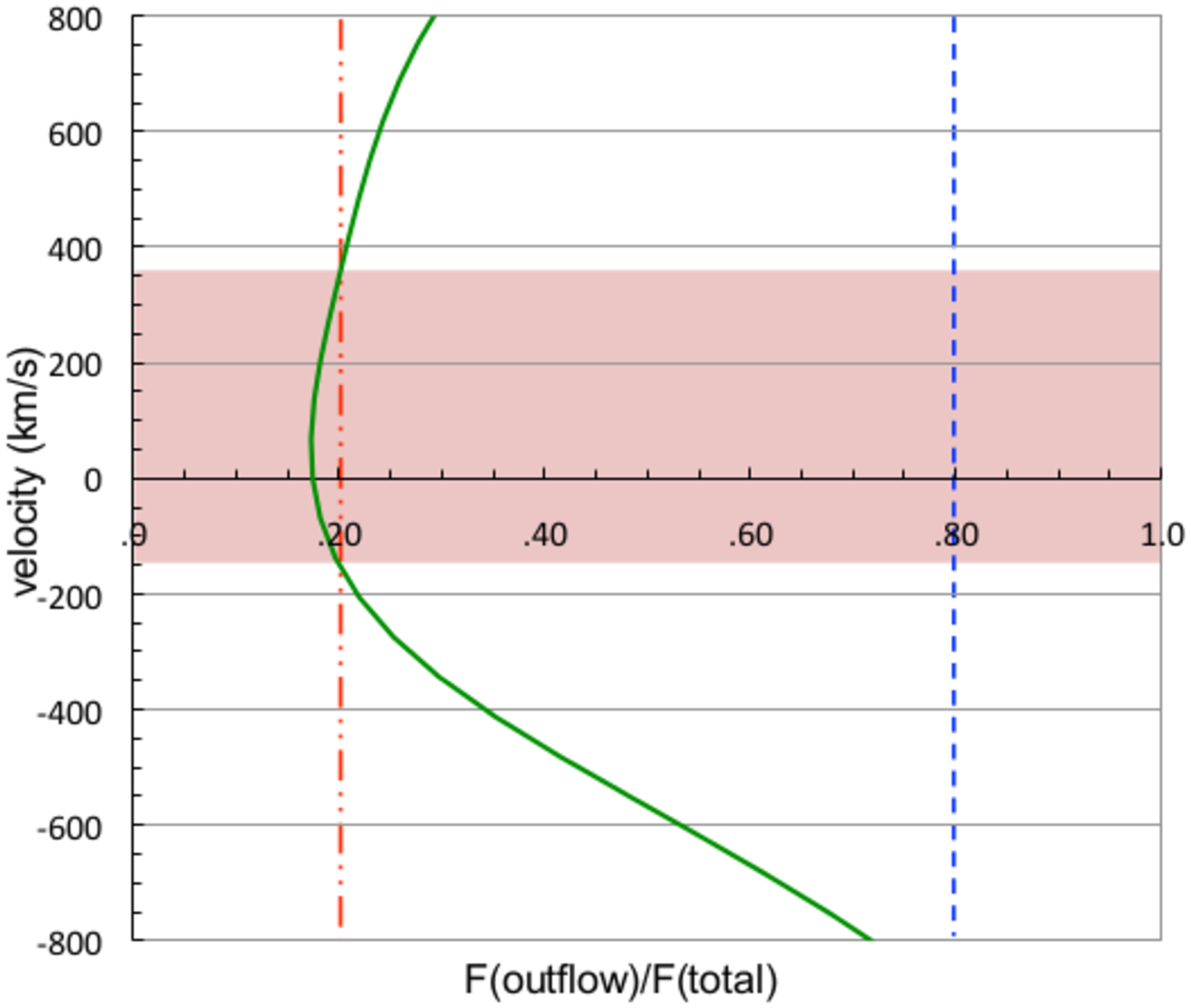}
\vspace{2.8in}
\includegraphics{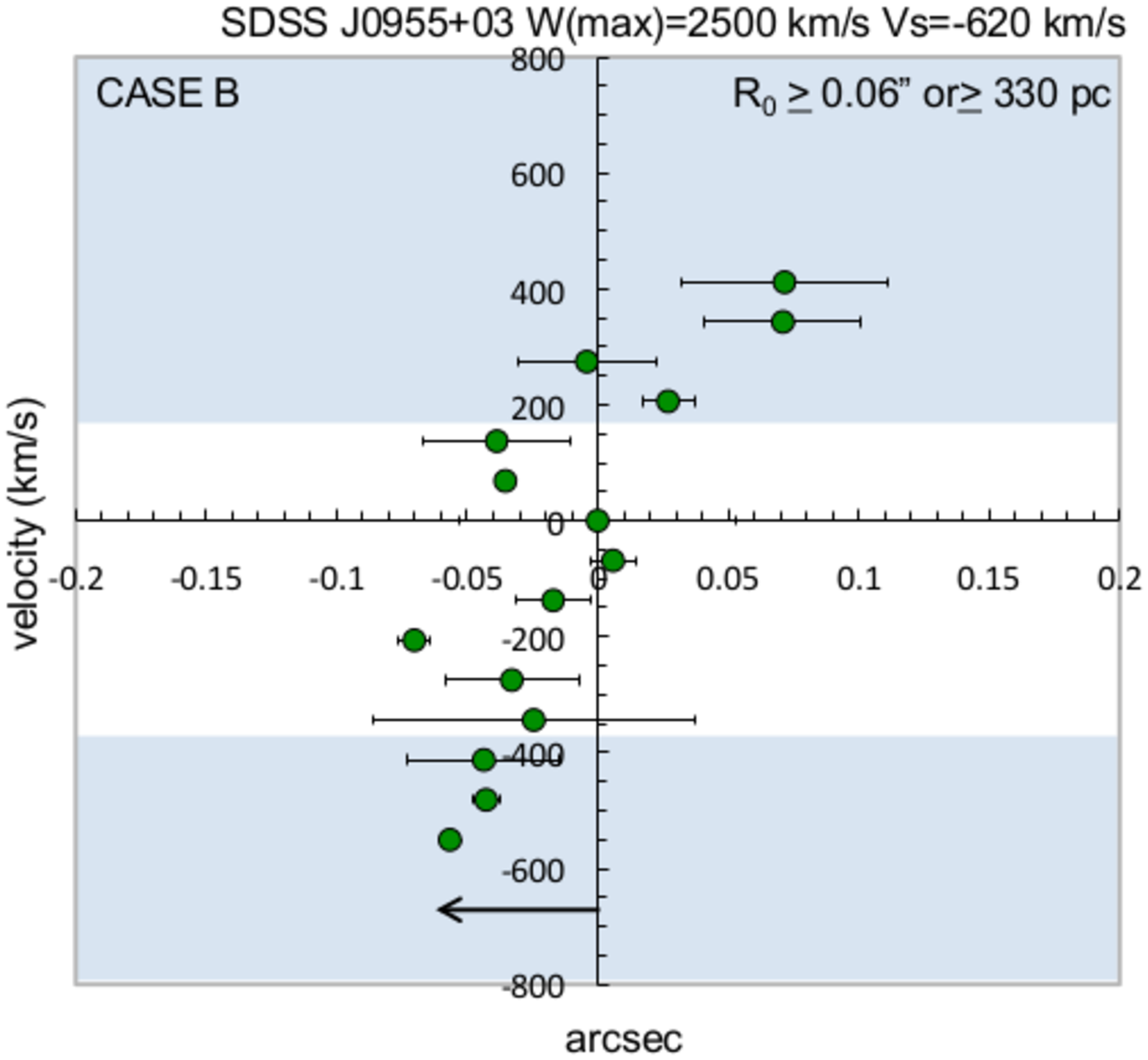}
\includegraphics{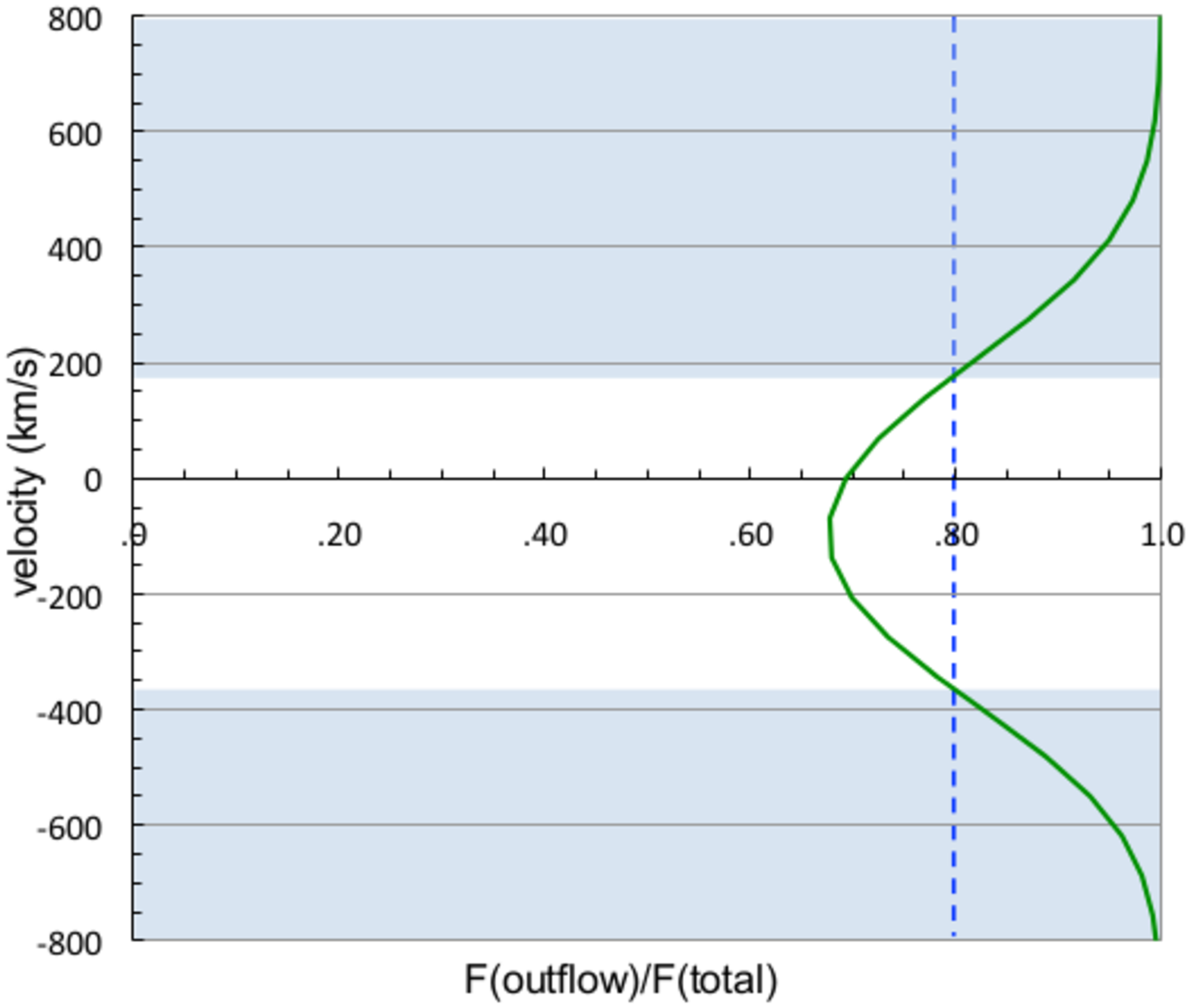}
\vspace{2.8in}
\caption{Spectroastrometry analysis for SDSS J0955+03. Line, symbol and color codes as in Fig. 6. Two scenarios are considered. Case A (top panels): the broadest component traces the outflow. Case B (bottom panels): the intermediate and broad components participate in the outflow (see the text).}
\label{astrom0955}
\end{figure*}

   The results of the spatially extended analysis are shown in  Fig. ~\ref{spat0955v1337} (left panels). 
The  [OIII]  spatial distribution (panel A)  is  dominated   by a central rather compact source  (FWHM=1.92$\pm$0.08 arcsec), but resolved compared
with the seeing FWHM=1.6$\pm$0.1 arcsec.  Thus, 
FWHM$_{\rm int}$=1.1$\pm$0.2 arcsec or 5.9$\pm$1.2 kpc.  Very faint emission might be detected at $\sim$3$\sigma$ level up to $\sim$7 arcsec
 or 39 kpc according to \cite{vm11a}. 

\vspace{0.2cm}
\begin{figure}
\includegraphics{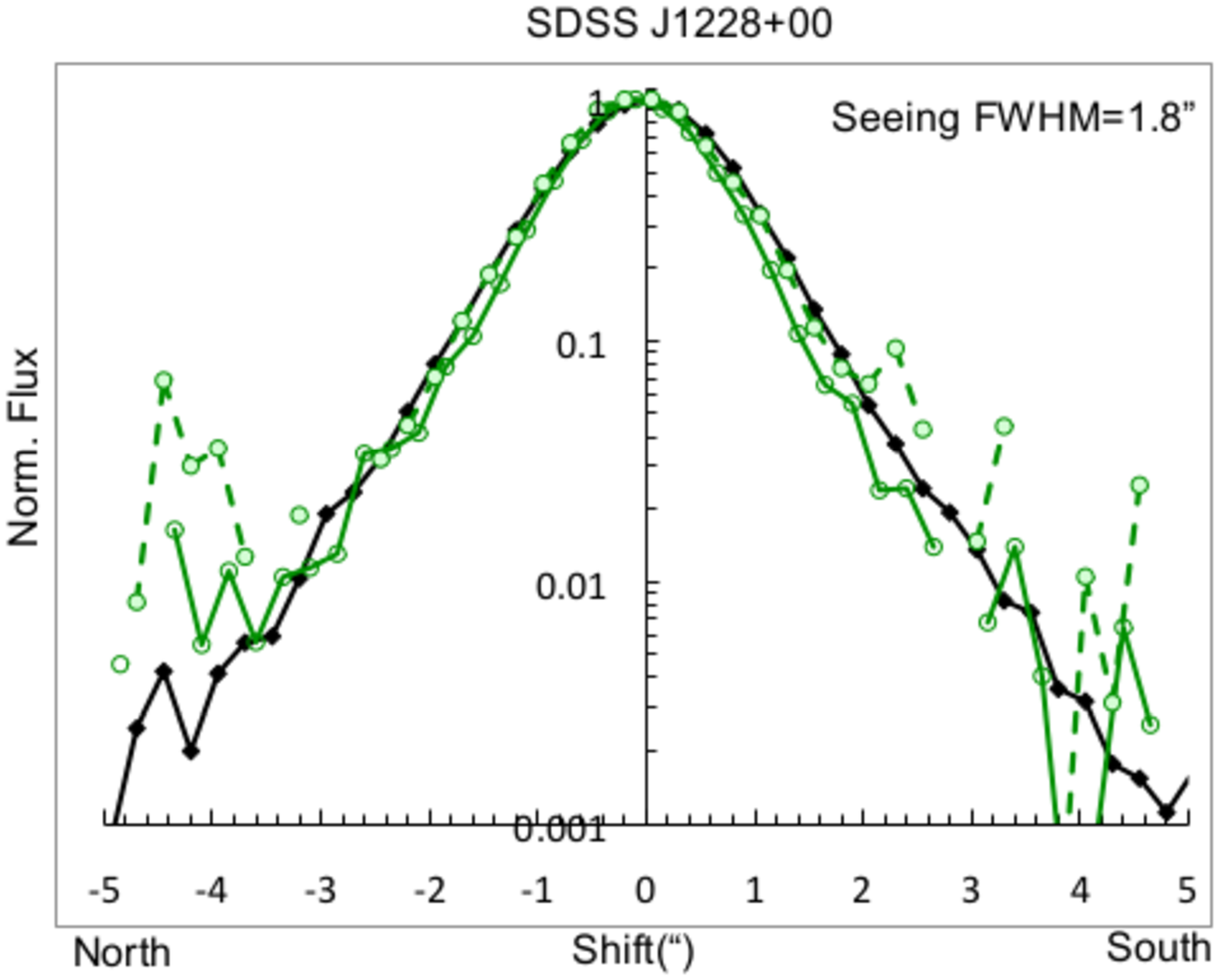}
\vspace{2.4in}
\caption{SDSS J1228+00. Neither [OII] (dashed green line) or [OIII] (solid green line) are spatially resolved along the slit.}
\label{spat1228}
\end{figure}

 The spatial distribution of the three nuclear components   cannot be traced pixel by pixel
due to the complexity of the lines and the relatively low signal to noise of the spectra. Instead, 
  1-dim spectra were extracted from     apertures several pixels wide  (Fig. ~\ref{spat0955v1337} left, panels C and D). 
 It is found that in the outer region (purple diamond, -3 arcsec West), where the excess above the seeing disk is clearer, the lines are rather narrow with FWHM$\la$267 km s$^{-1}$. 

The complex, turbulent kinematics is concentrated in the central region, well within the seeing FWHM. None of the three kinematic components 
 show  obvious spatial  variations  within the seeing FWHM, in consistency with being unresolved. 
At $\sim$-1.6 arcsec West  the line emission is  clearly extended, although the excess above the seeing disk is lower. The lines are broad (FWHM$\sim$740 km s$^{-1}$). However, because a mult-Gaussian fit cannot be applied  due to the low S/N of the spectrum, it cannot be concluded that the central nuclear outflow is spatially extended. 
Due to the uncertainty in the true spatial distributions, we will adopt the conservative assumption  
FWHM$_{\rm int}\la$7.1 kpc or $R_{\rm o}\la$3.5 kpc for the bulk of the outflow emission.

The results based on the spectroastrometric analysis  depend on what components are considered as
 tracers of the outflow  which is uncertain for this object  (Fig. \ref{example0955}). While it seems clear that
  the broadest component is emitted by the outflow,  the contribution of the intermediate component   is doubtful.   FWHM$\sim$1000 km s$^{-1}$ is broad, but not necessarily inconsistent with gravitational motions. As en example, according to the correlation between black hole mass and stellar velocity dispersion (e.g. Tremaine et al. \citeyear{tre02})  a bulge containing a black hole of mass M(BH)$\sim$3$\times$10$^9$ M$_{\odot}$ would have a FWHM$_{\rm stars}\sim$1000 km s$^{-1}$.
  
   If only the broadest component contributes to the outflow (Case A), this gas does not dominate the line emission 
   at any velocity with measured spatial centroid available (Figs. \ref{astrom0955}, top panels). In this case $R_{\rm o}$ cannot be constrained. If both the broad and intermediate components  are taken into account (Case B in the figure), the outflow dominates ($\ge$80\%) the line flux at a broad range of velocities. A lower limit on its radial size  is set to $R_{\rm o}\ge$0.06 arcsec
or 330 pc.

 Thus, the most conservative interpretation
of the complete analysis of this HLSy2 implies that the outflow has 0$< R_{\rm o} \la$3.5 kpc.

\begin{figure*}
\includegraphics{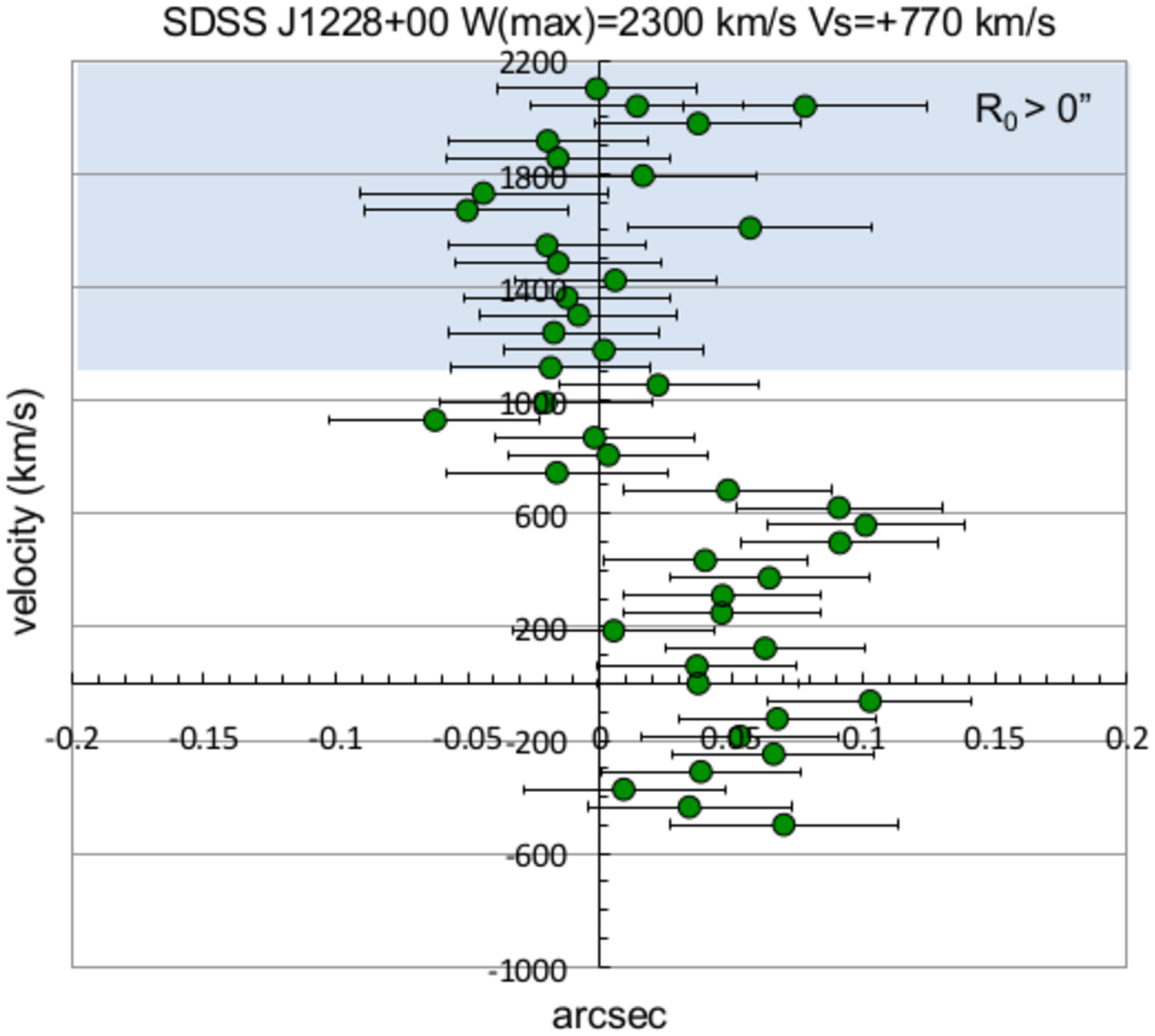}
\includegraphics{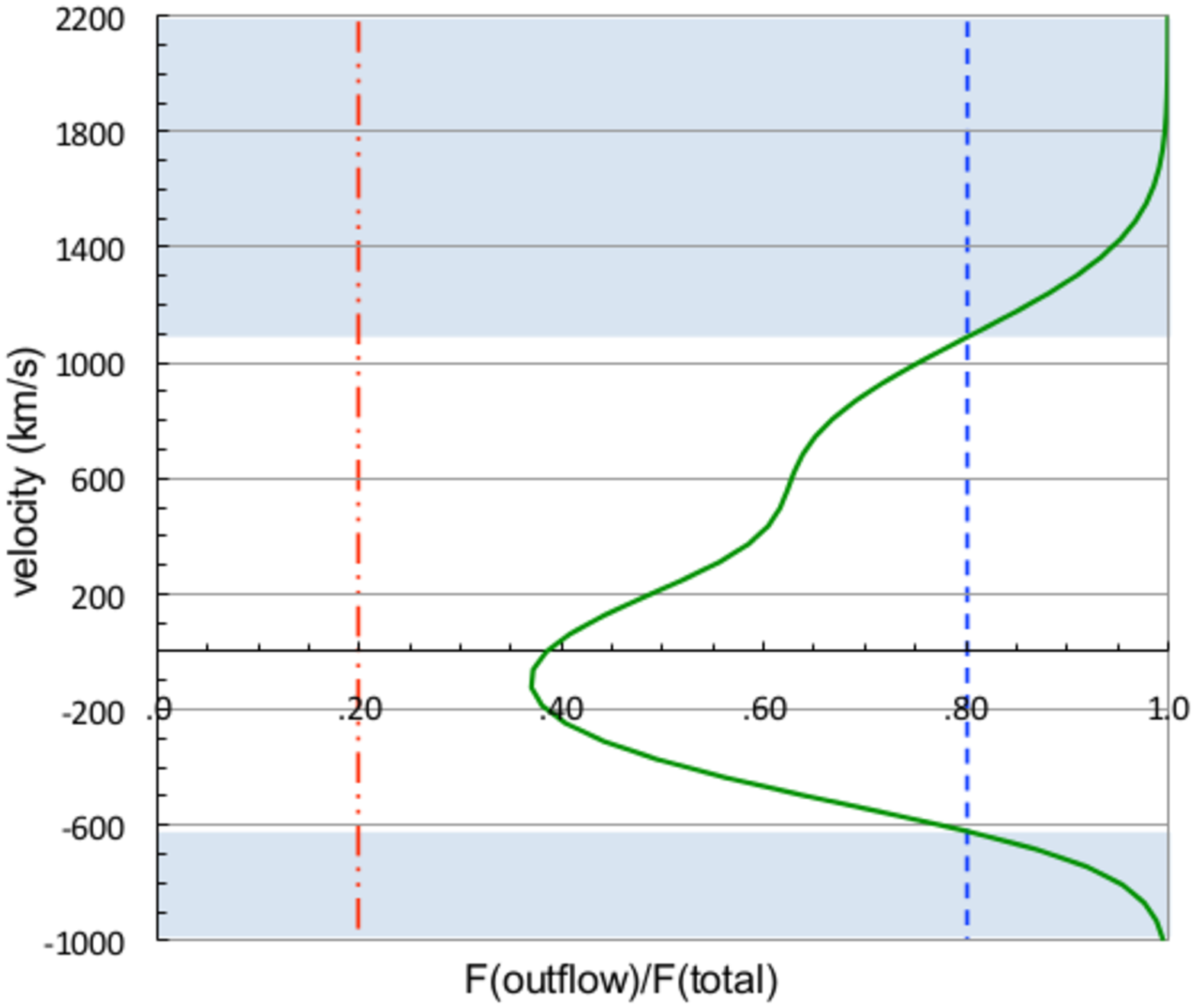}
\vspace{2.85in}
\caption{Spectroastrometry analysis for SDSS J1228+00.  Line, symbol and color codes as in Fig. 6.}
\label{astrom1228}
\end{figure*}

\vspace{0.2cm}

{\it SDSS J1228+00}
\vspace{0.2cm}

%z=0.575, 6.54 kpc/arcsec

This  radio-intermediate QSO2 has a
rather extreme nuclear outflow with
 W$_{\rm max}$=2300$\pm$70 km s$^{-1}$ and  redshifted by $V_{\rm max}$=770$\pm$20 km s$^{-1}$ (\cite{vm11b}).  
 As  explained for SDSS J1153+03,  a redshifted outflow component  may be  favored   at an intermediate type 1-type 2 orientation.  Such scenario   was proposed by \cite{vm11a}  based on the optical nuclear emission line properties.
Two additional components of FWHM=550$\pm$20 and 730$\pm$40 km s$^{-1}$ are also present. 

Neither  [OII] or [OIII]   are spatially extended  along the slit (Fig. ~\ref{spat1228}).  The slightly broader profile
of [OII] might be a consequence of the seeing  dependence with wavelength.
The seeing was rather poor (FWHM=1.75$\pm$0.09 arcsec) during the observations. We estimate FWHM$_{\rm int}\la$0.66 arcsec
or $\la$4.3 kpc.

The complexity of the line profiles (see \cite{vm11b}) and the fact that sky residuals affect the [OIII] profile prevents a pixel by pixel (or wide aperture) kinematic analysis. I.e. methods (i) and (ii) cannot be applied.

\begin{figure*}
\includegraphics{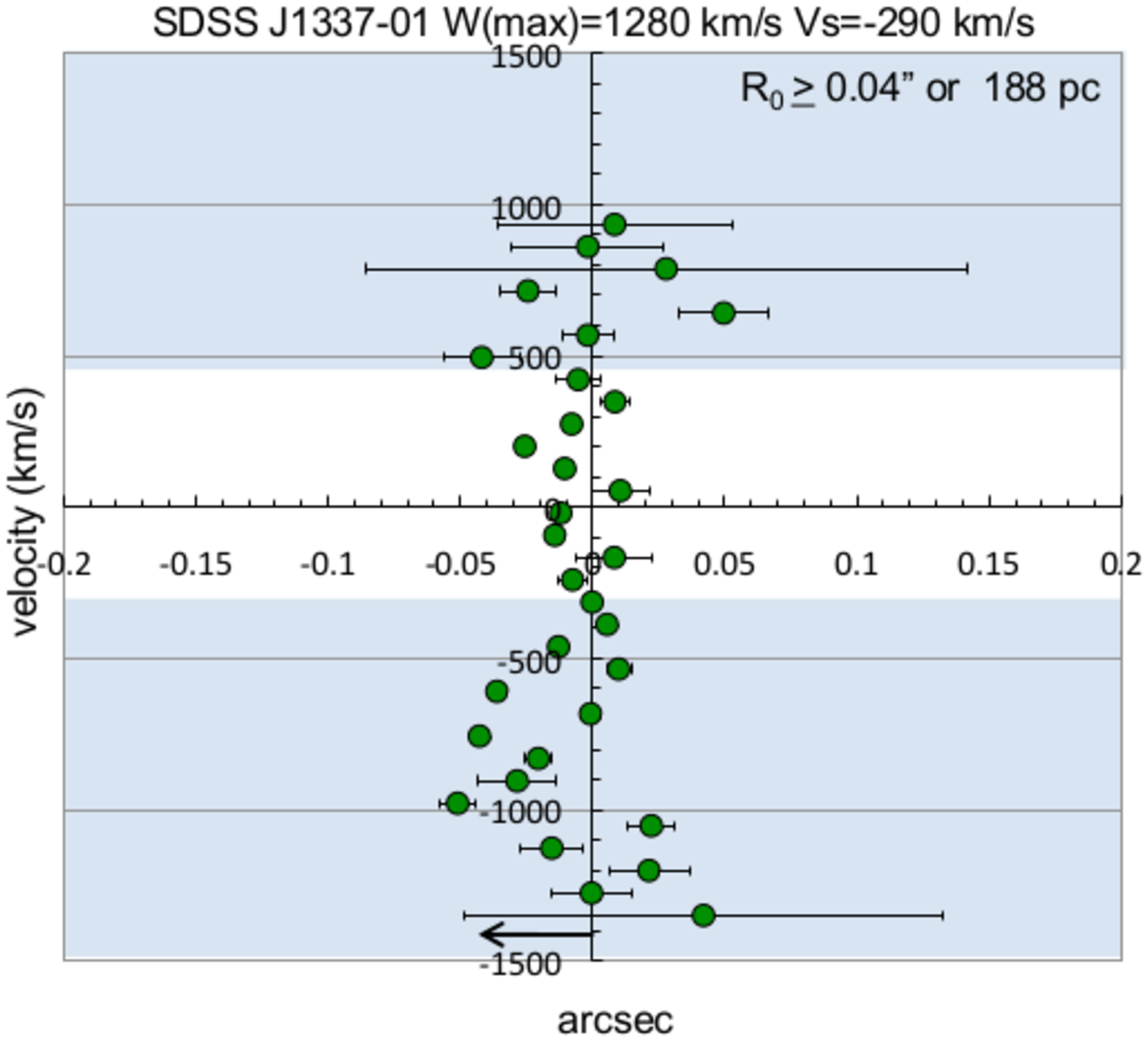}
\includegraphics{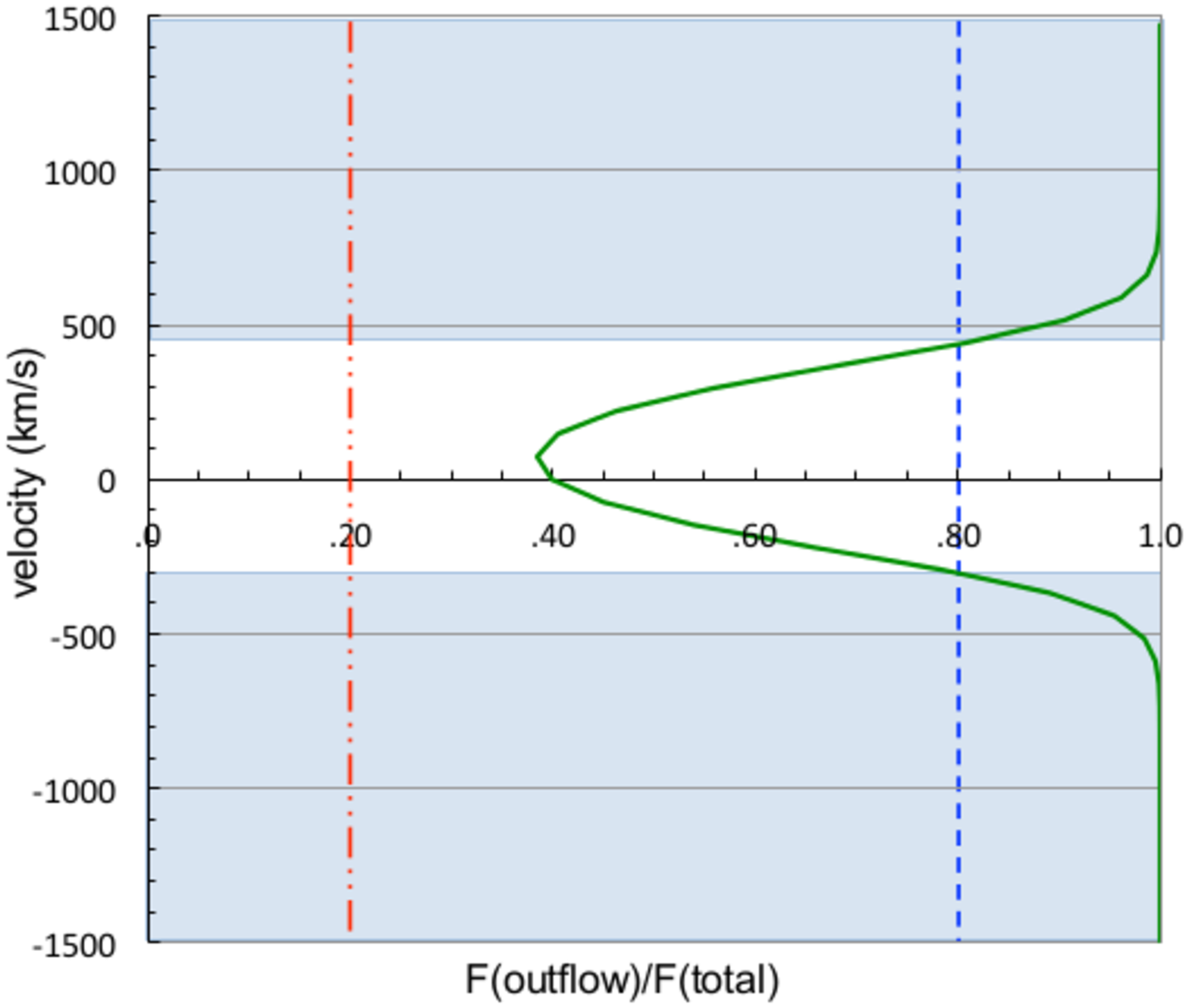}
\vspace{2.85in}
\includegraphics{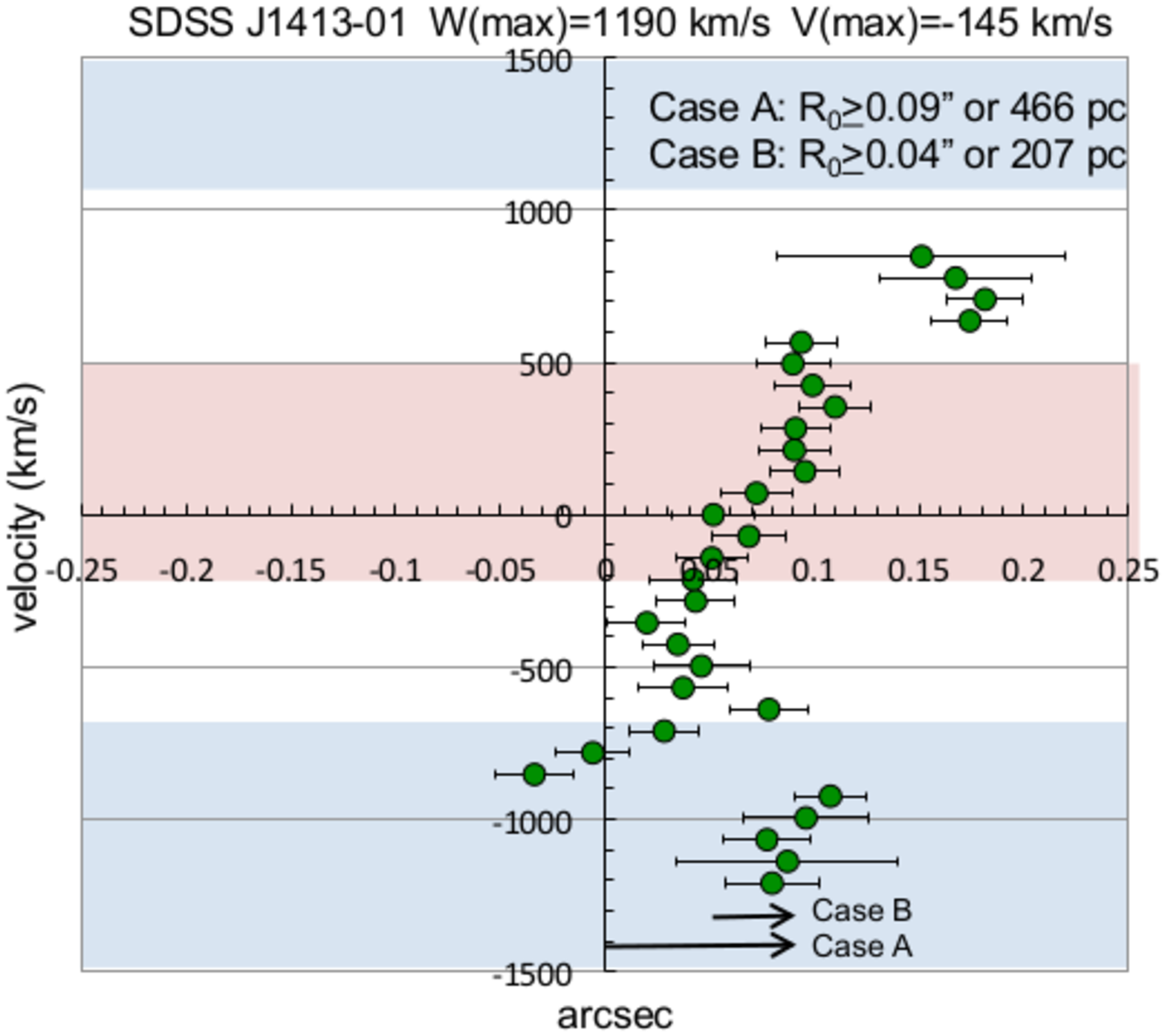}
\includegraphics{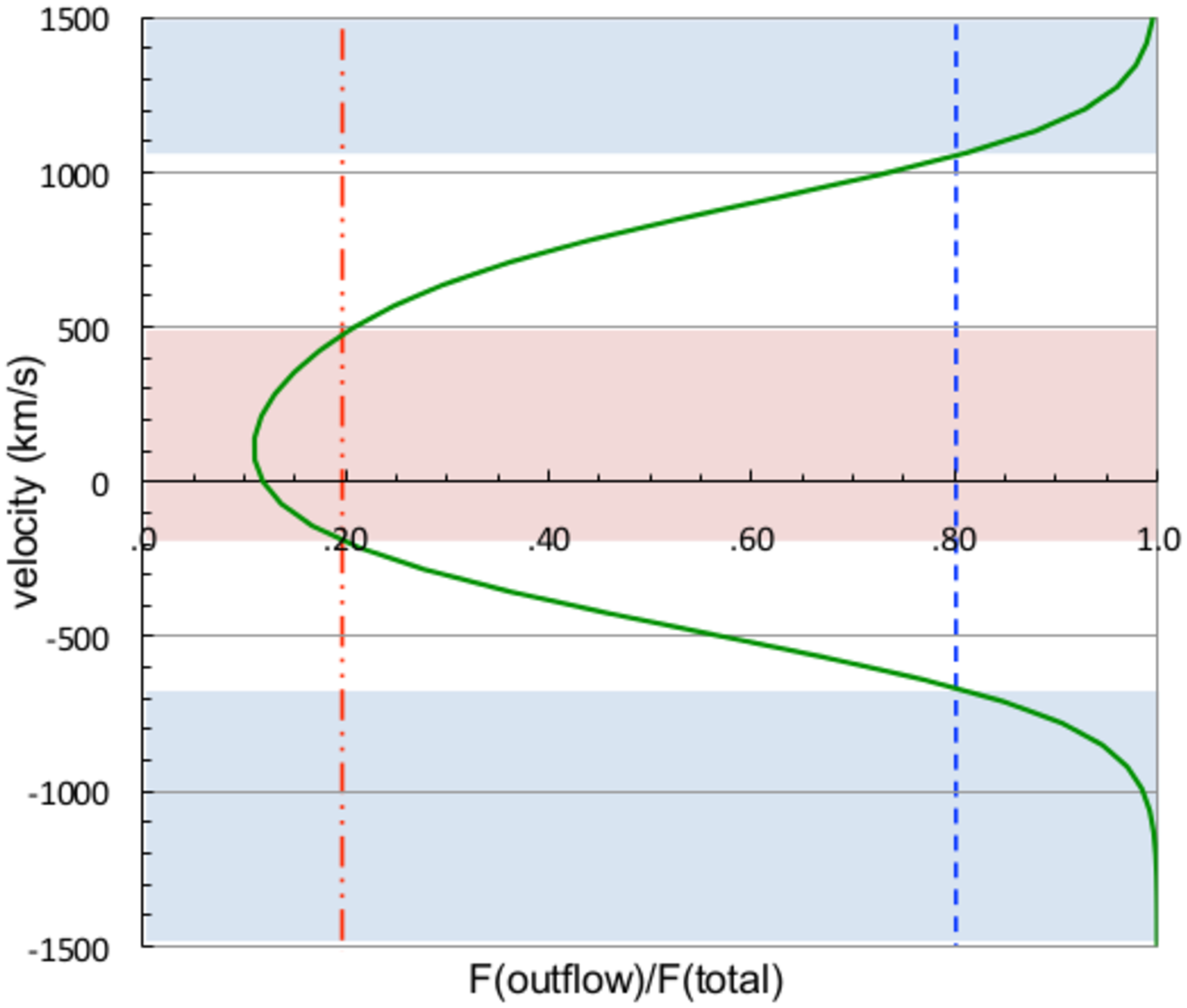}
\vspace{2.85in}
\caption{Spectroastrometry analysis for SDSS J1337-01 (top panels) and SDSS J1413-01 (bottom). Line, symbol and color codes as in Fig. 6.}
\label{astrom1337v1413}
\end{figure*}

The spectroastrometric analysis does not resolve the outflow (Fig. ~\ref{astrom1228}). Those velocities for which the line flux is dominated
by the outflowing gas appear co-spatial with the continuum centroid, while lower velocities, not dominated
by the outflow emission show  a clear shift of   $\sim$0.05 arcsec on average.  
If this QSO2 has an intermediate viewing angle, the  continuum centroid is expected to be a reliable indicator of
 the AGN location. Thus, while the ambient gas appears
spatially extended, the outflowing gas does not.

Based on the limited  amount of information,  there is no evidence for an spatially extended nuclear ionized outflow in this system along the slit. We conclude 0$< R_{\rm o}\la$2.1 kpc.

\vspace{0.2cm}

{\it SDSS J1337-01}

% z=0.239. 4.71 kpc per arcsec

\vspace{0.2cm}

\begin{figure*}
\includegraphics{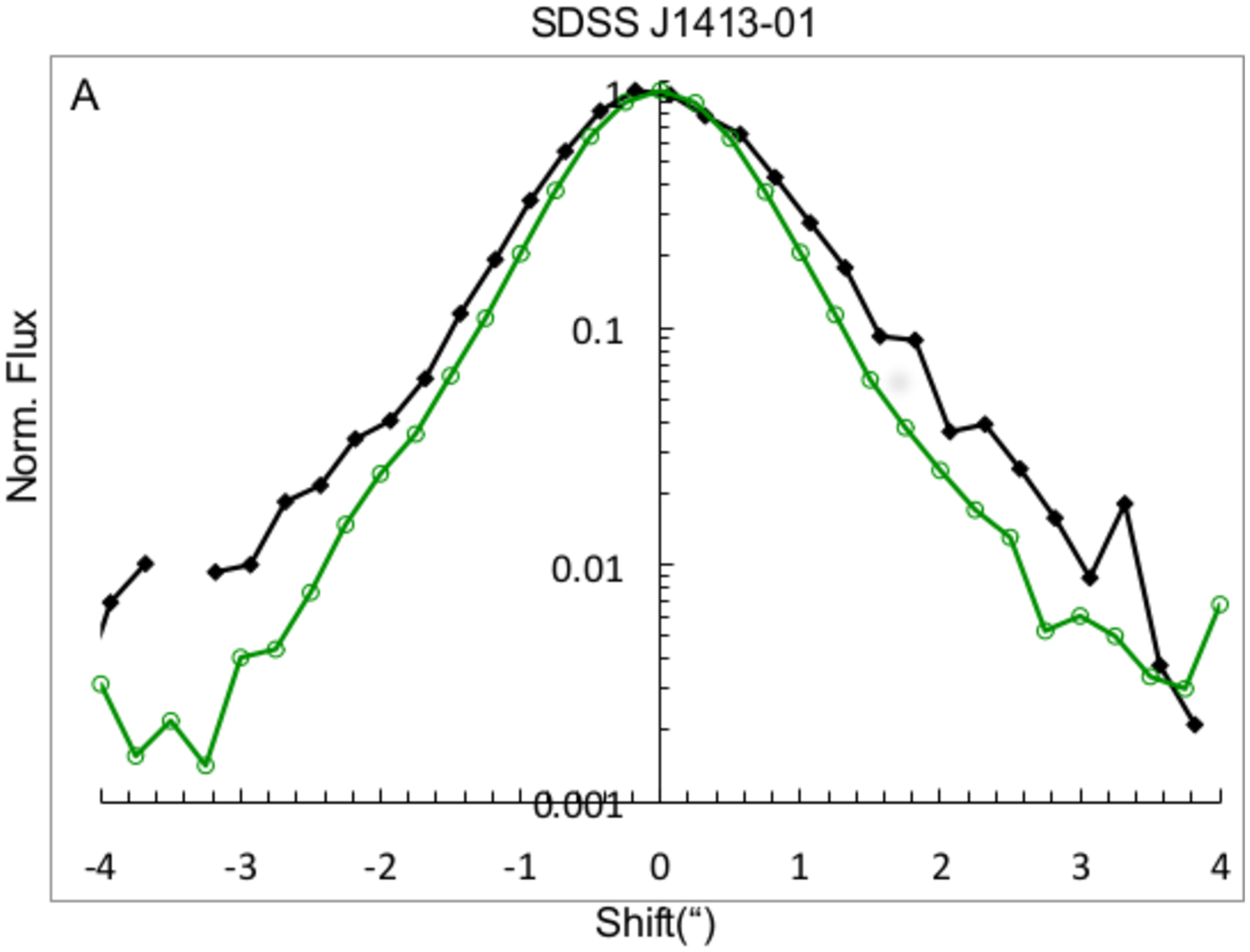}
\includegraphics{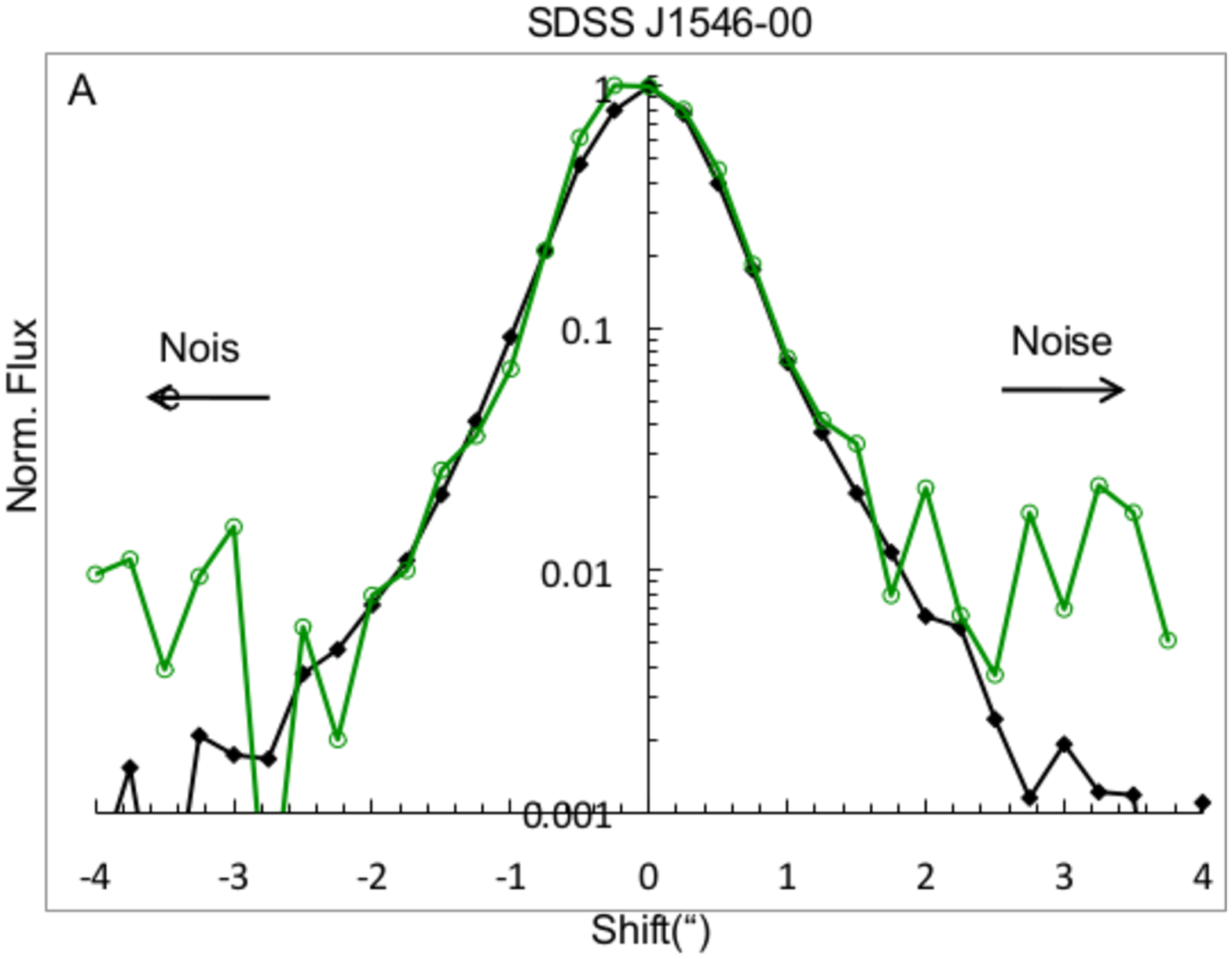}
\includegraphics{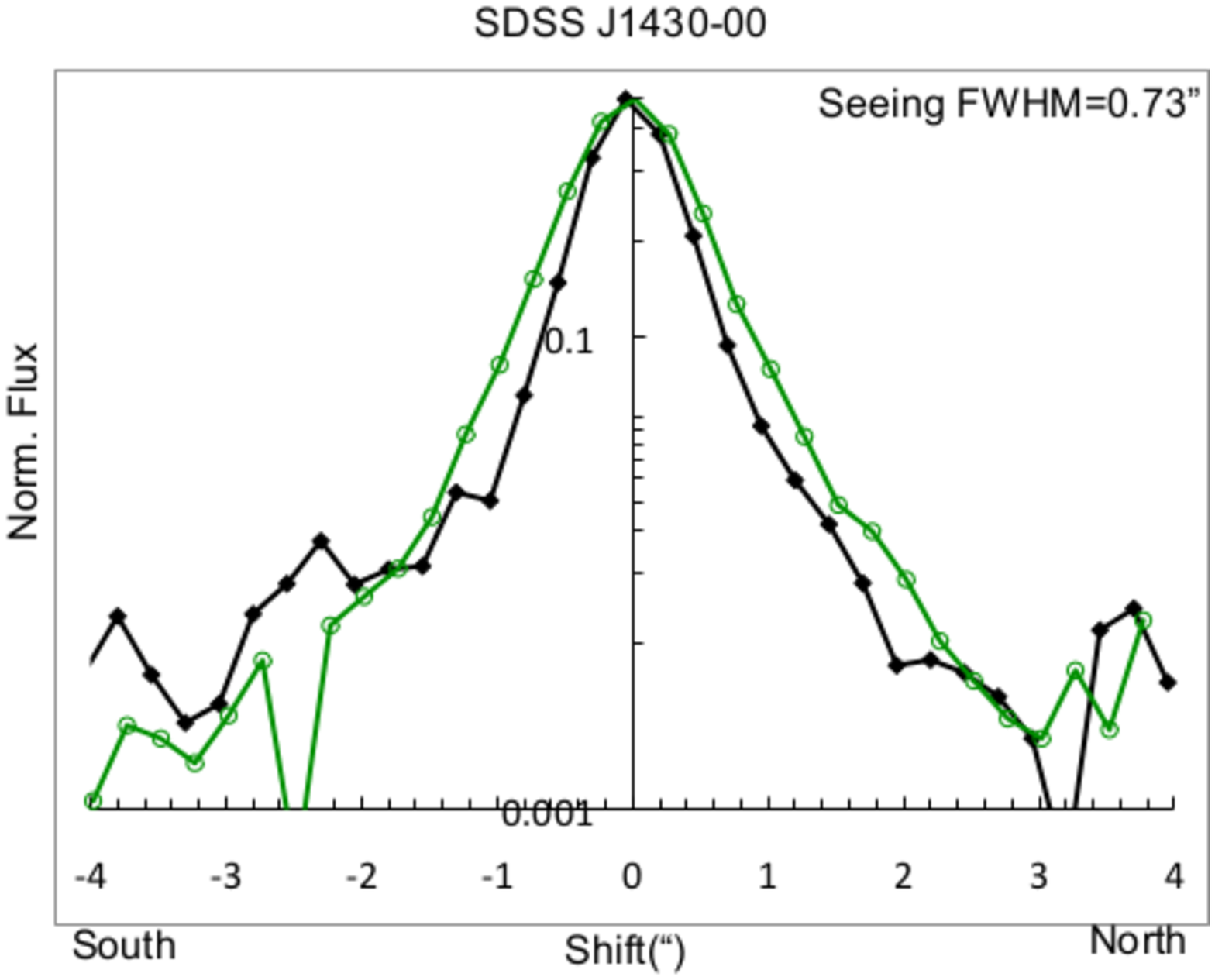}
\vspace{1.9in}
\includegraphics{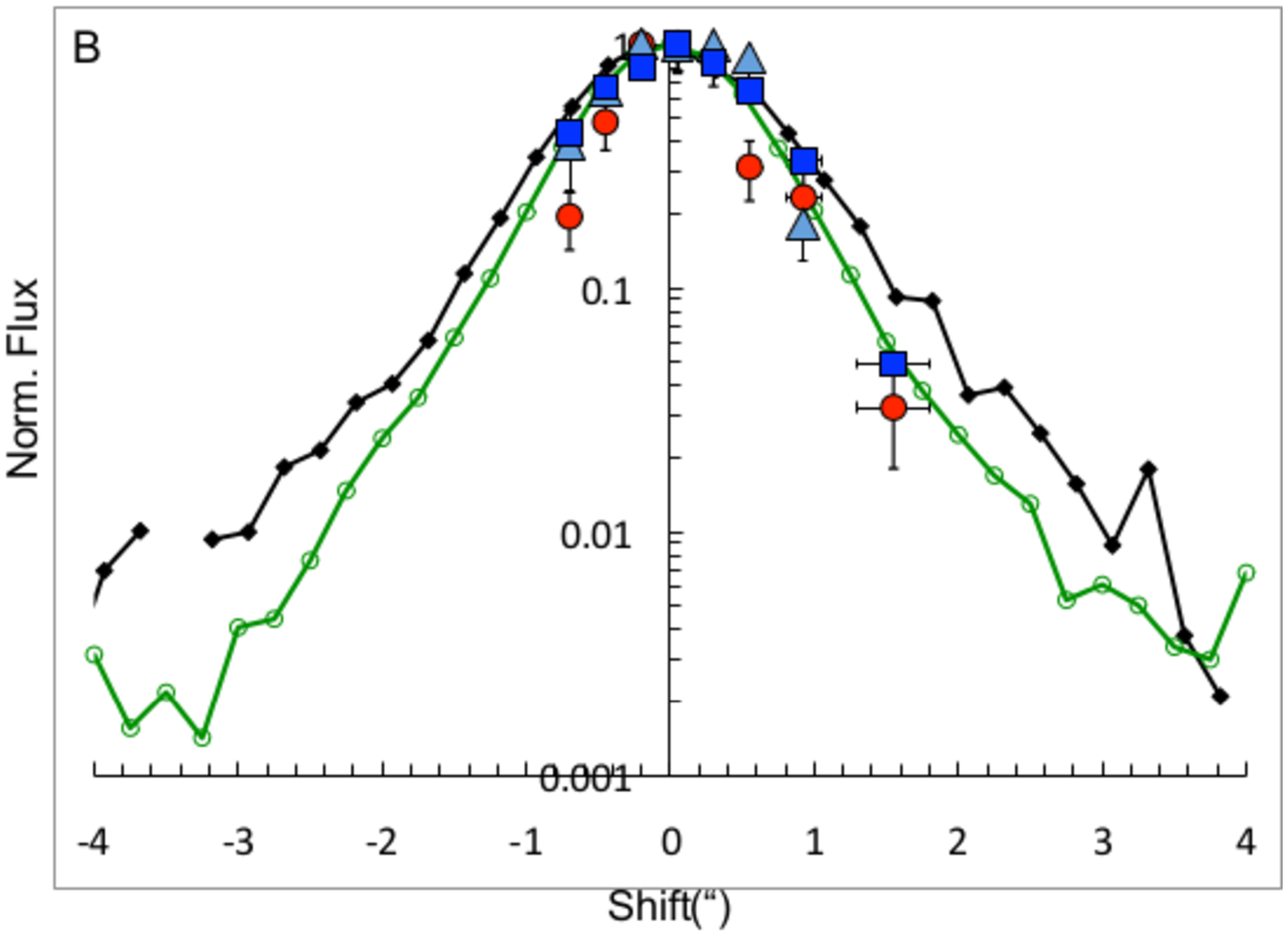}
\includegraphics{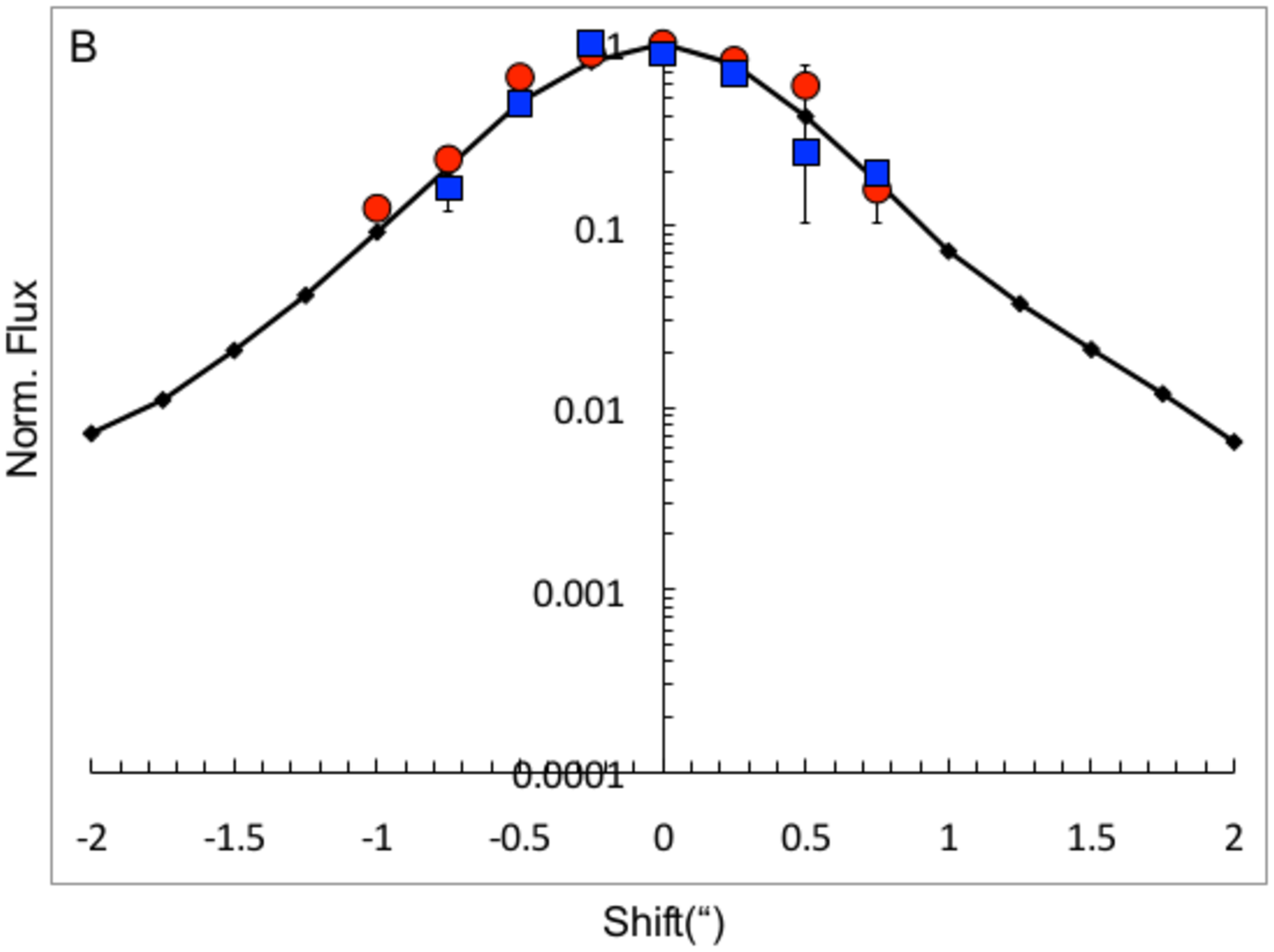}
\includegraphics{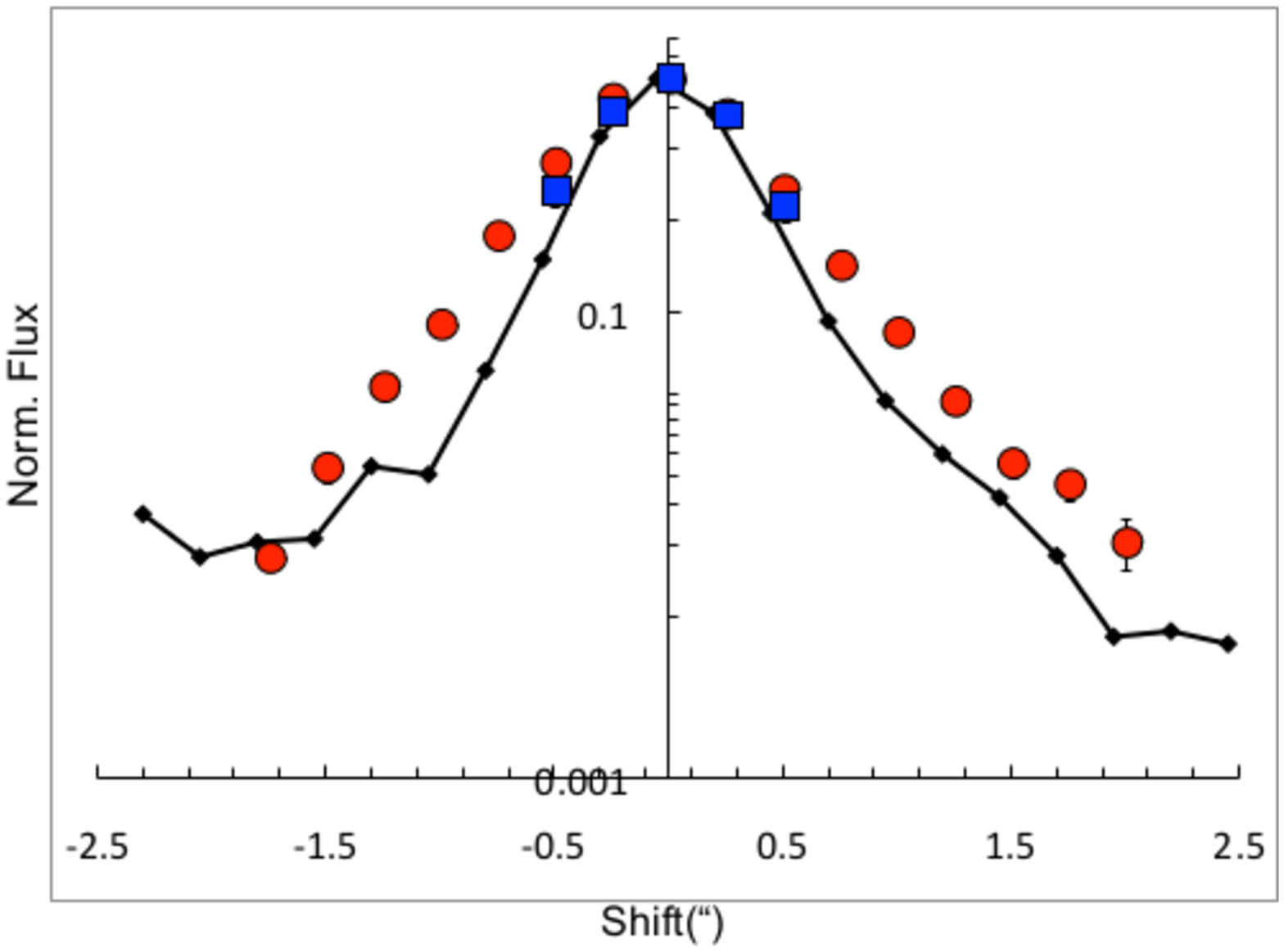}
\vspace{1.9in}
\includegraphics{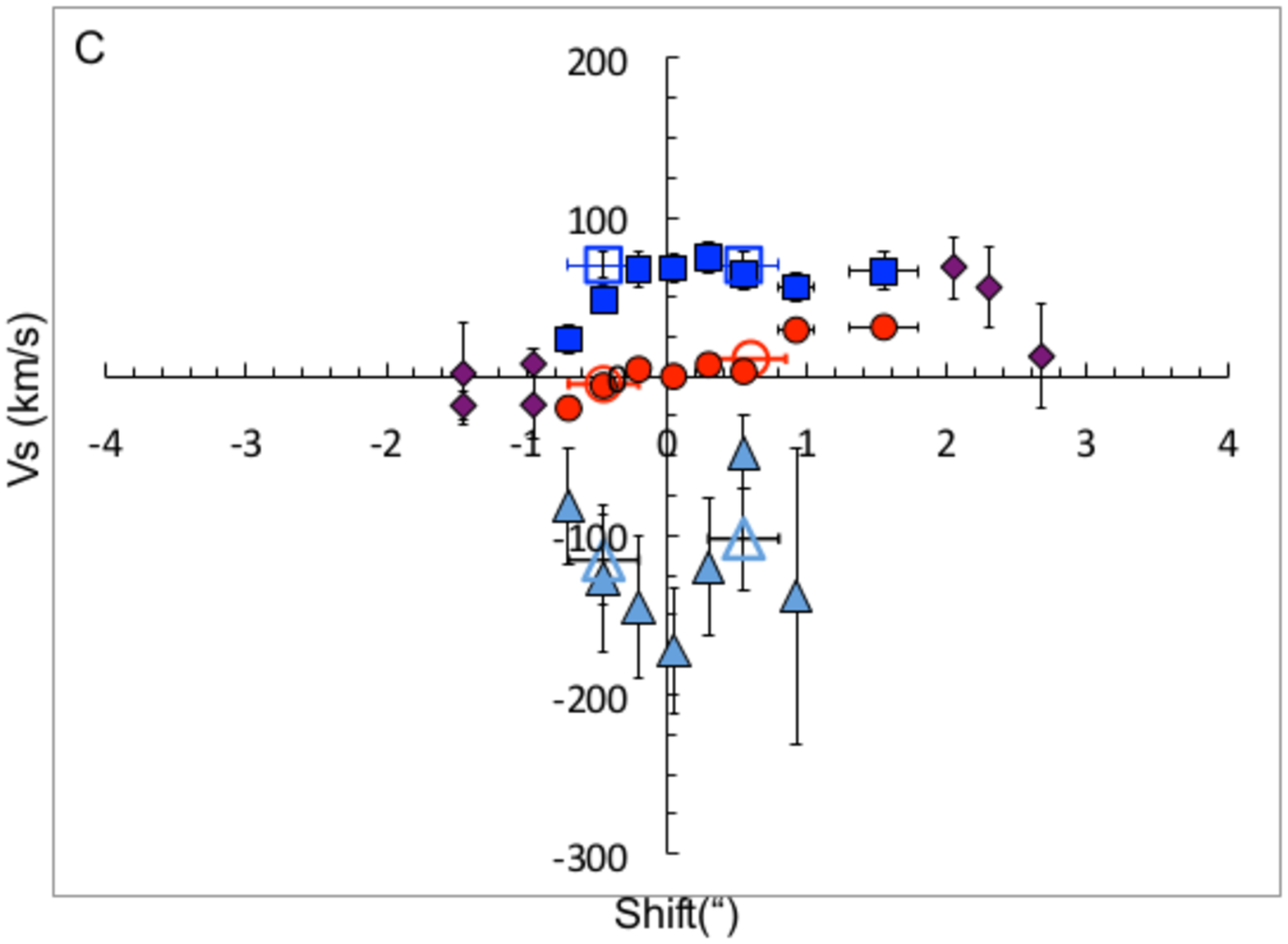}
\includegraphics{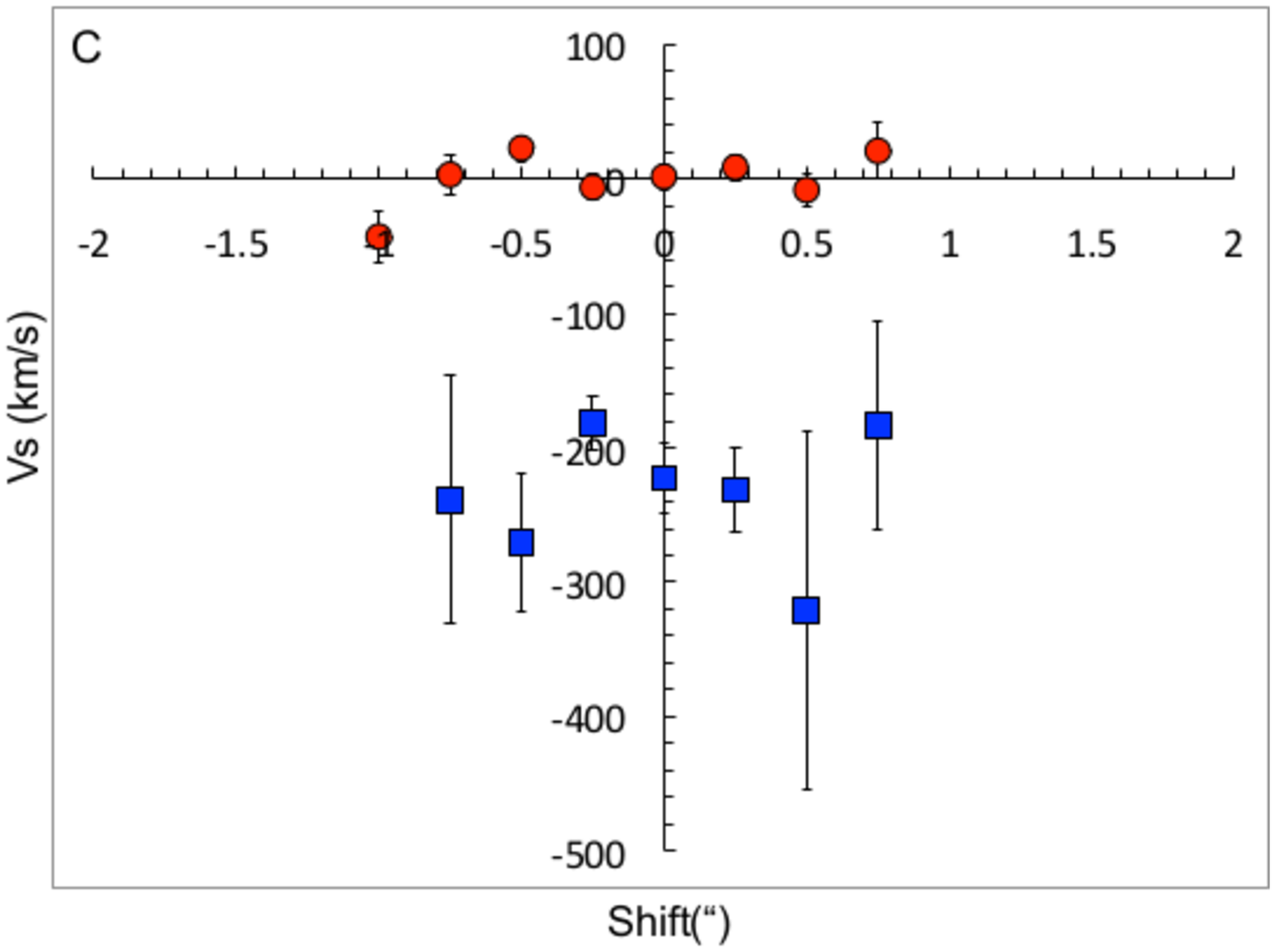}
\includegraphics{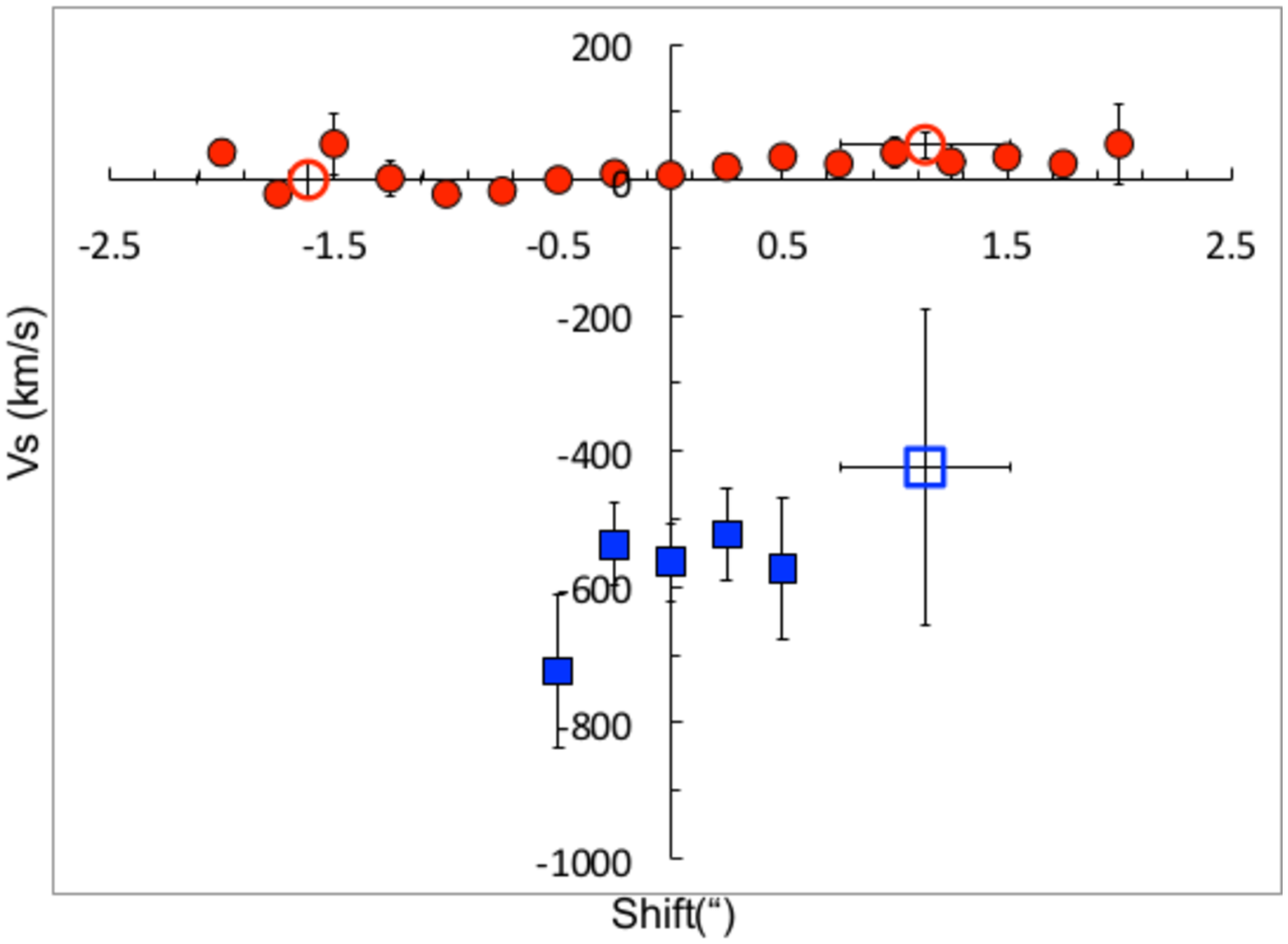}
\vspace{1.9in}
\includegraphics{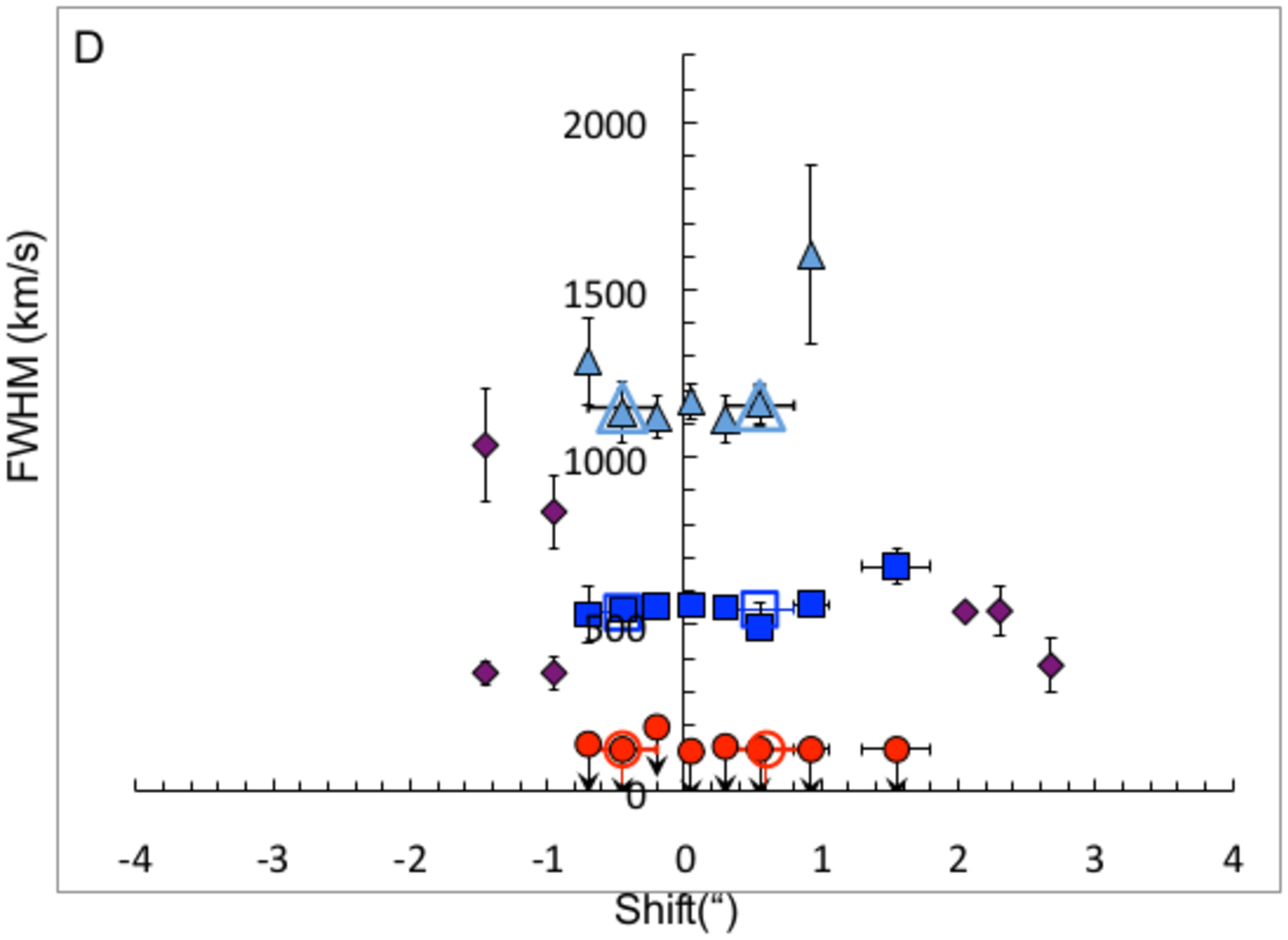}
\includegraphics{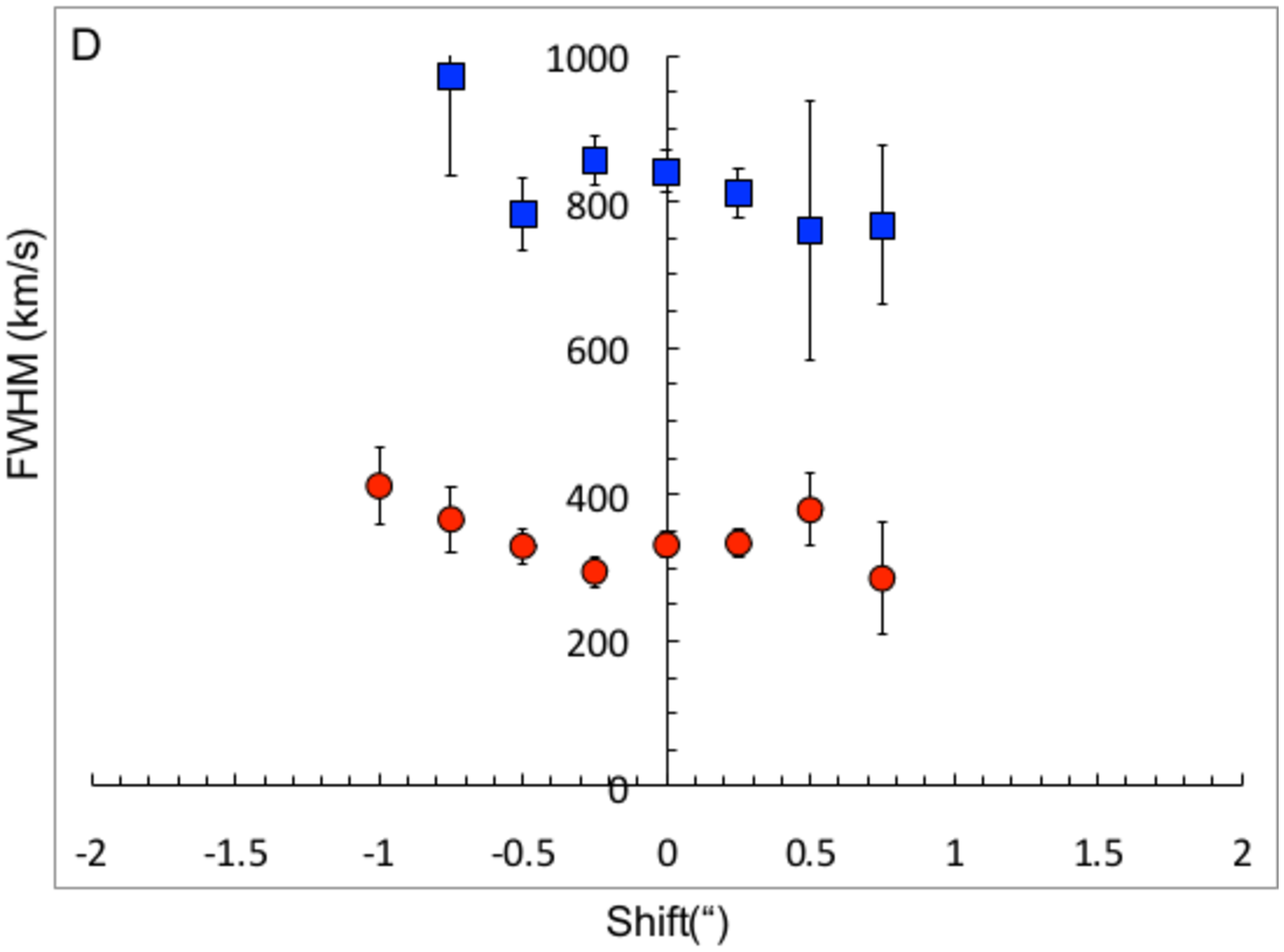}
\includegraphics{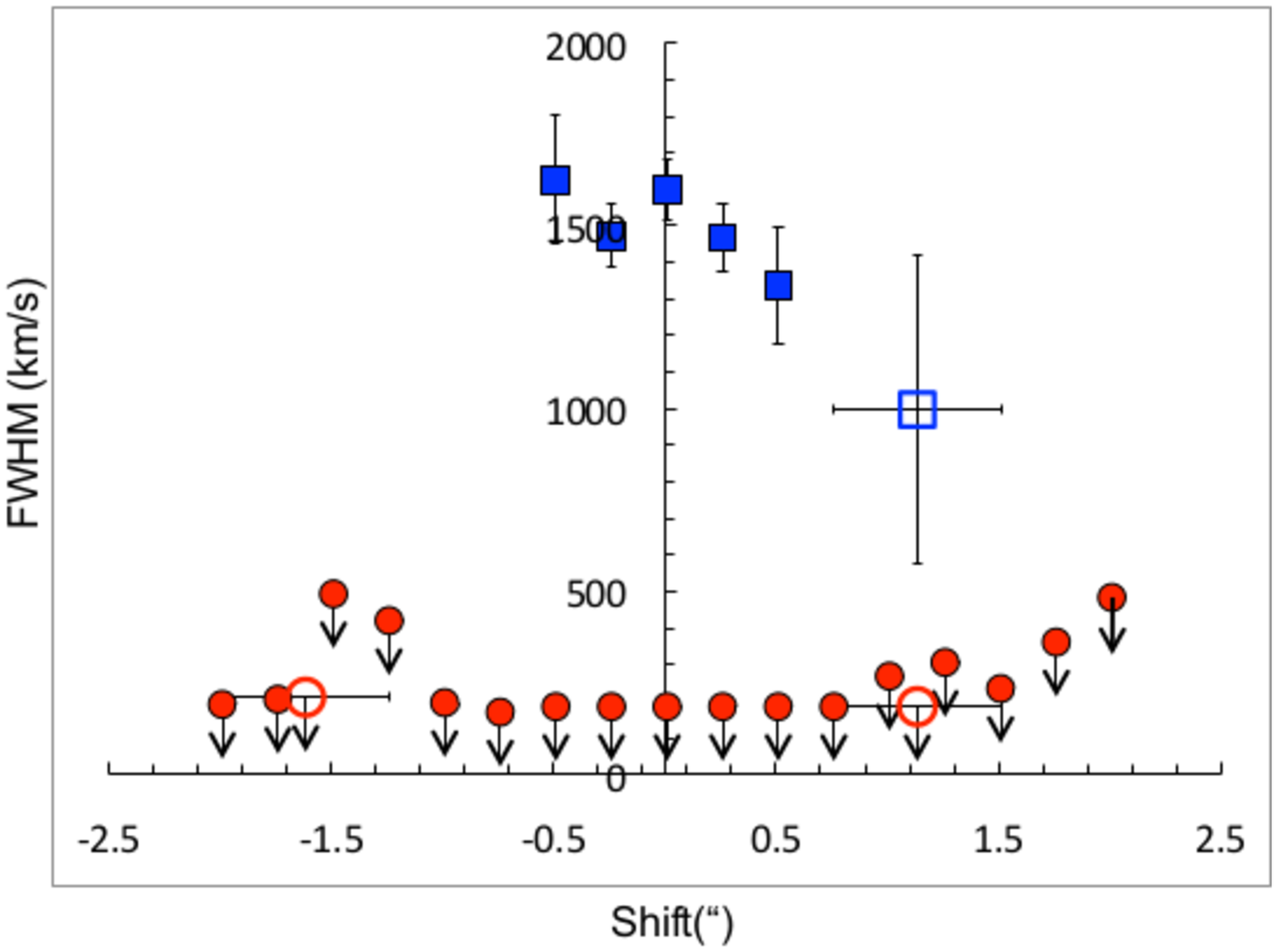}
\vspace{1.9in}
\caption{Spatially extended analysis of [OIII]$\lambda$5007   
for SDSS J1413-01 (left), SDSS J1546-00 (middle) and SDSS J1430-00 (right).  Symbol, line and color codes as in Fig. 5.  Because the [OIII] spatial profile (green solid line) is narrower than the seeing (black solid line) for SDSS J1413-01, the spatial profiles of the individual kinematic components are compared
with both in panel B (left). The results of the kinematic fits in the outer pixels for this QSO2 are shown with different symbols (purple diamods in panels C and D) because due to the low S/N of the spectra multiple Gaussian fits were uncertain. The lines  at these locations seem to be  a mixture of the three kinematic components identified in the central pixels.}
\label{spat1413}
\end{figure*}

This radio-quiet QSO2 harbors a nuclear outflow with W$_{\rm max}$=1280$\pm$60 km s$^{-1}$  bluesfhited by  
$V_{\rm max}$=-320$\pm$12 km s$^{-1}$ relative to the second, narrower component of FWHM=305$\pm$20 km s$^{-1}$ (\cite{vm11b}).

  The [OIII] spatial profile is
dominated by the central unresolved source (seeing FWHM=1.6$\pm$0.1 arcsec). We cannot confirm 
 extended emission  up to $\sim$3.5$\arcsec$ from the central source as reported in our previous analysis (\cite{vm11a}).
The careful comparison with the seeing disk demonstrates that  the line flux at this location is still dominated by the wings
of the central, unresolved source (Fig. ~\ref{spat0955v1337} right, panel A). 

Both kinematic components  isolated in the fits shows a spatial distribution
 consistent with the seeing disk (Fig. ~\ref{spat0955v1337} right, panel B) and show no clear kinematic substructure (panels C and D).  
 Thus, methods (i) and (ii) do not resolve the outflow.  We infer FWHM$_{\rm int}\la$0.71 arcsec or 3.3 kpc.  
 
The spectroastrometric analysis (Fig. ~\ref{astrom1337v1413}, top panels) shows that the spatial centroid of the line emission is approximately
cospatial with the continuum centroid for most velocities.   A  small spatial shift of $\sim$0.04 arcsec  is hinted for some blueshifted  velocities
dominated by the outflow. We will thus assume $R_{\rm o}\ga$0.04 arcsec or 188 pc.

\vspace{0.2cm}

{\it SDSS J1413-01}

% 5.18 kpc per arcec z=0.380

\vspace{0.2cm}

Three kinematic  components are isolated in the nuclear spectrum of this radio-quiet QSO2. Two of them have
 FWHM=260$\pm$20 and 600$\pm$20  km s$^{-1}$. The broadest, which is associated with the outflow,  has W$_{\rm max}$=1190$\pm$50 and $V_{\rm max}$=-145$\pm$30 km s$^{-1}$ (\cite{vm11b}). 

The [OIII] spatial profile is dominated by the central compact source (Fig. ~\ref{spat1413} left, panel A), which is actually narrower (FWHM$\sim$1.3 arcsec) than the seeing disk (FWHM=1.54$\pm$0.09$\arcsec$) measured from stars in the broad band image. No extended emission is detected (see also \cite{vm11a}).

The spatially extended analysis reveals that the  3 kinematics components isolated in the fits 
 are mostly concentrated in the central  part of the seeing disk and are consistent with being spatially unresolved (Fig. ~\ref{spat1413} left,  panel B).   We infer FWHM$_{\rm int}\la$0.60 arcsec or 3.1 kpc.

Although there is some hint of kinematic substructure for the outflowing gas within the seeing disk (panel C),
 the error bars are large and the apparent shifts are  
within the uncertainties.
In fact, there is no sign of such sub-structure when extracting larger apertures (large hollow symbols in the figures).

Only the spectroastrometric analysis reveals extension for the ionized outflow Fig. ~\ref{astrom1337v1413} (bottom). 
The  spatial line centroid  is shifted relative to the continuum centroid at most velocities. This shift is $\sim$0.05 arcsec or 259 pc 
for gas moving at the assumed systemic velocity ($V_{\rm s}$= 0 km s$^{-1}$). As discussed for SDSS J1307-02 the location of
the AGN is unclear, although the uncertainty was much larger in that object ($\sim$0.55 arscsec). If the AGN is at the continuum centroid (Case A), we infer  a lower limit $R_{\rm o}\ga$0.09 arcsec or  466 pc. If it is  located at the emission line centroid (Case B), then $R_{\rm o}\ga$0.04 arcsec or 207 pc.

 In the most conservative case, 0.21$\la R_{\rm o}\la$1.5 kpc.
\vspace{0.2cm}

{\it SDSS J1546-00}

% z=0.383  5.20 kpc per arcsec

\vspace{0.2cm}

 The nuclear ionized outflow of this radio-intermediate HLSy2 has W$_{\rm max}$=780$\pm$30  km s$^{-1}$. It is blueshifted by $V_{\rm max}$=-270$\pm$35  km s$^{-1}$ relative to the dominant  component which has FWHM=320$\pm$20 km s$^{-1}$ (\cite{vm11b}).

The  [OIII] spatial profile   is consistent
with the seeing disk (FWHM=0.90$\pm$0.05$\arcsec$; Fig. \ref{spat1413} middle,  panel A).There is no evidence for the individual
kinematic components  to be spatially resolved and  they show
no  signs of  kinematic spatial substructure (panels B, C and D).
We infer FWHM$_{\rm int}\la$0.54 arcsec or $\la$2.8 kpc.

\begin{figure*}
\includegraphics{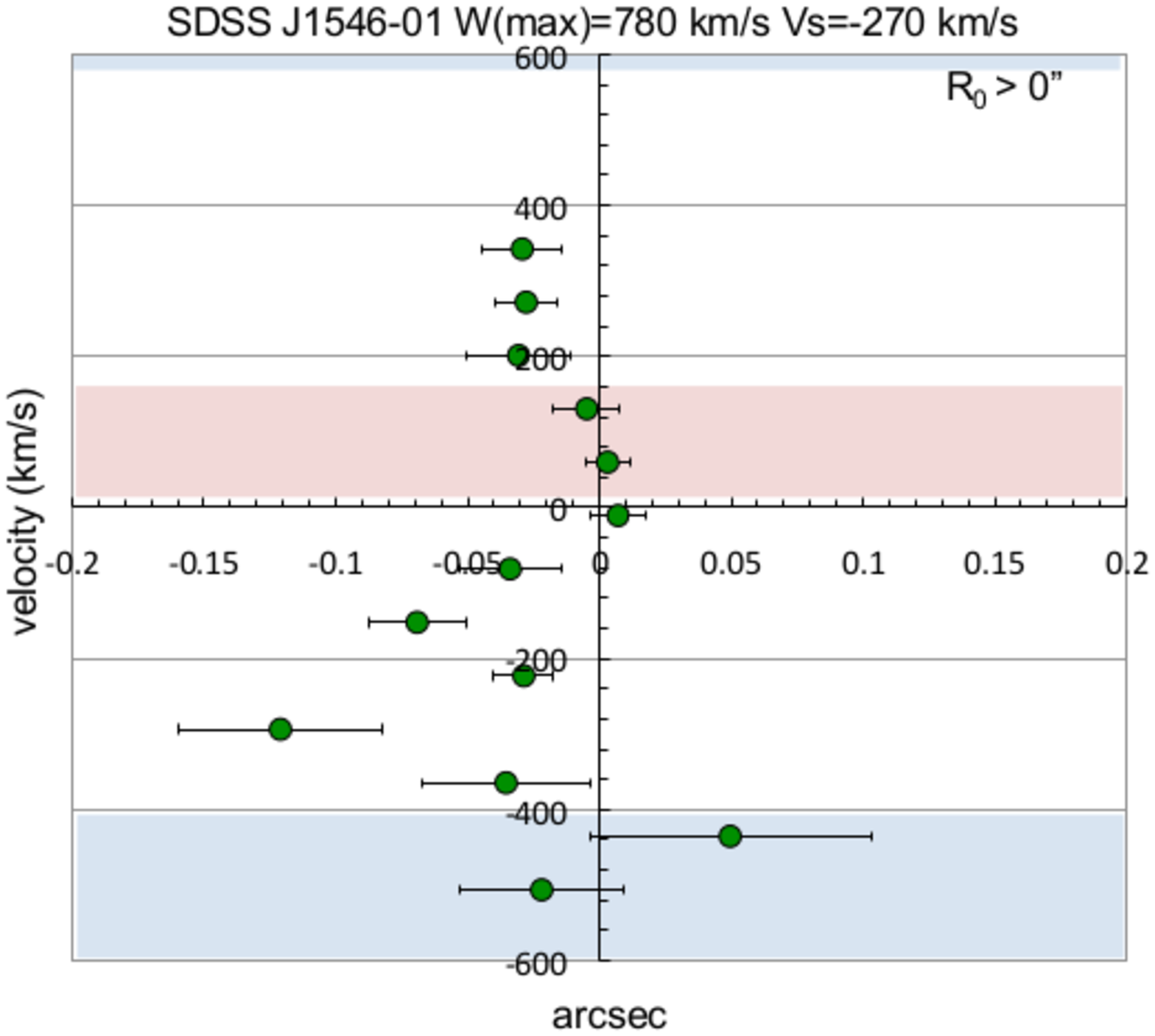}
\includegraphics{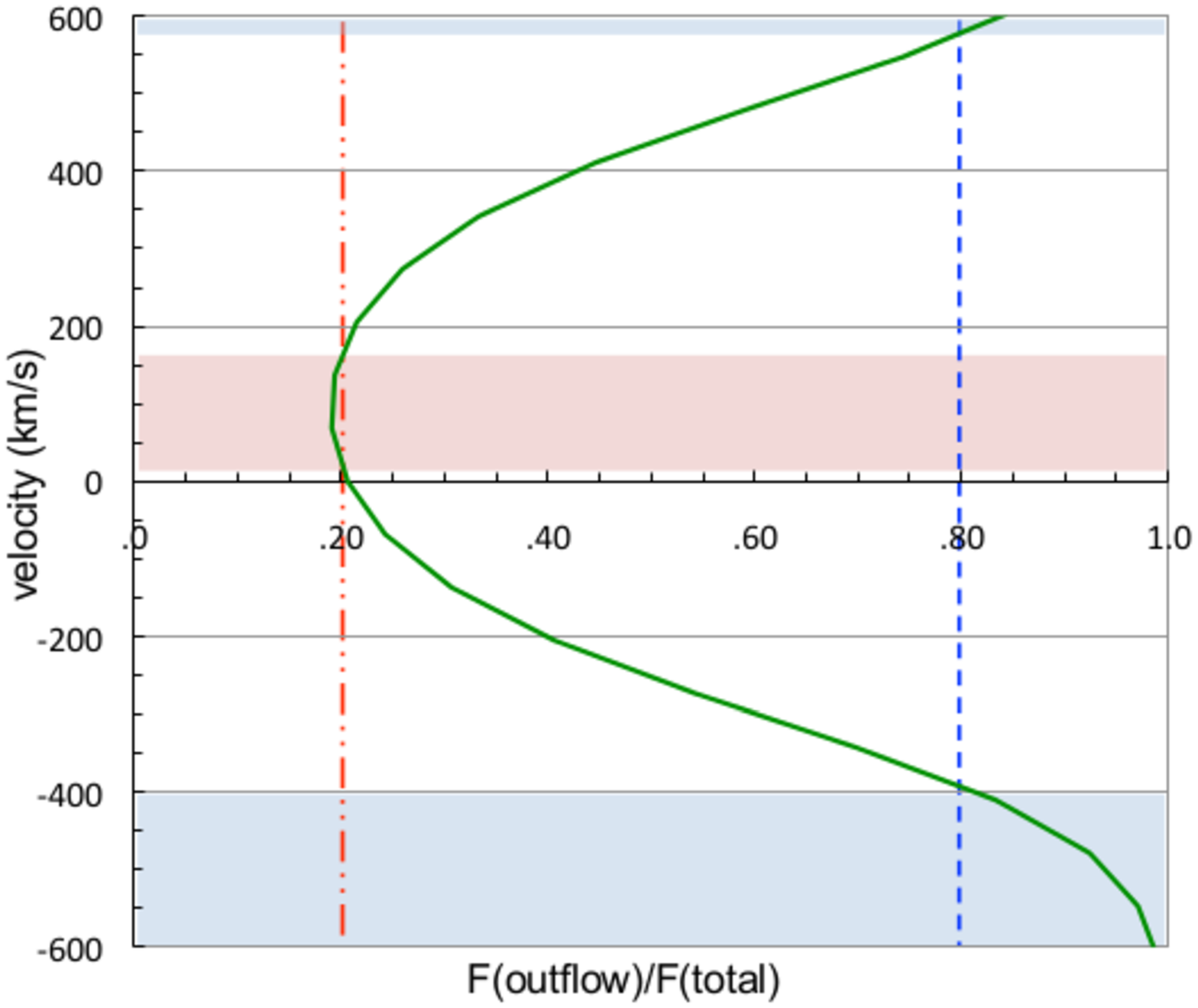}
\vspace{2.85in}
\includegraphics{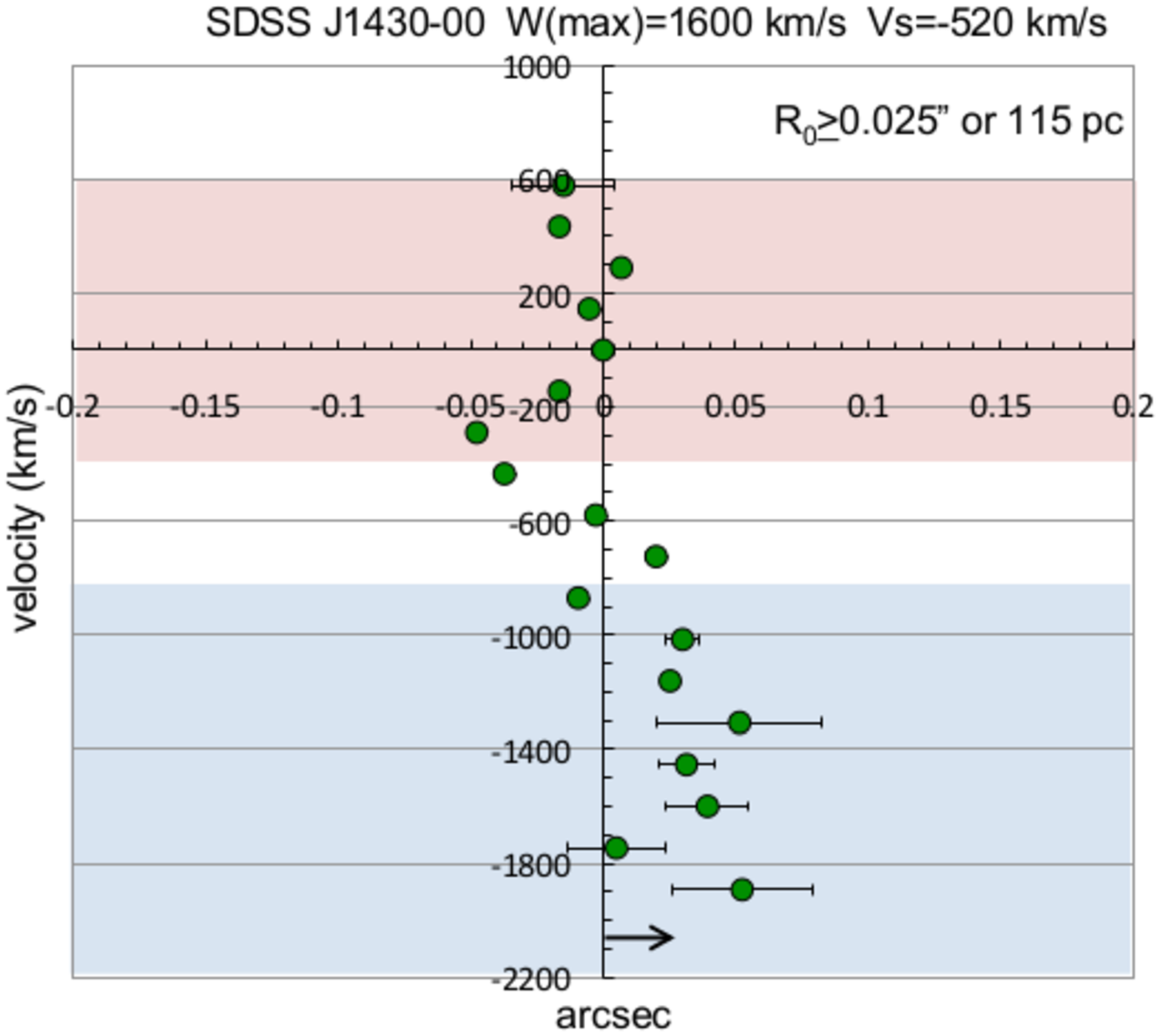}
\includegraphics{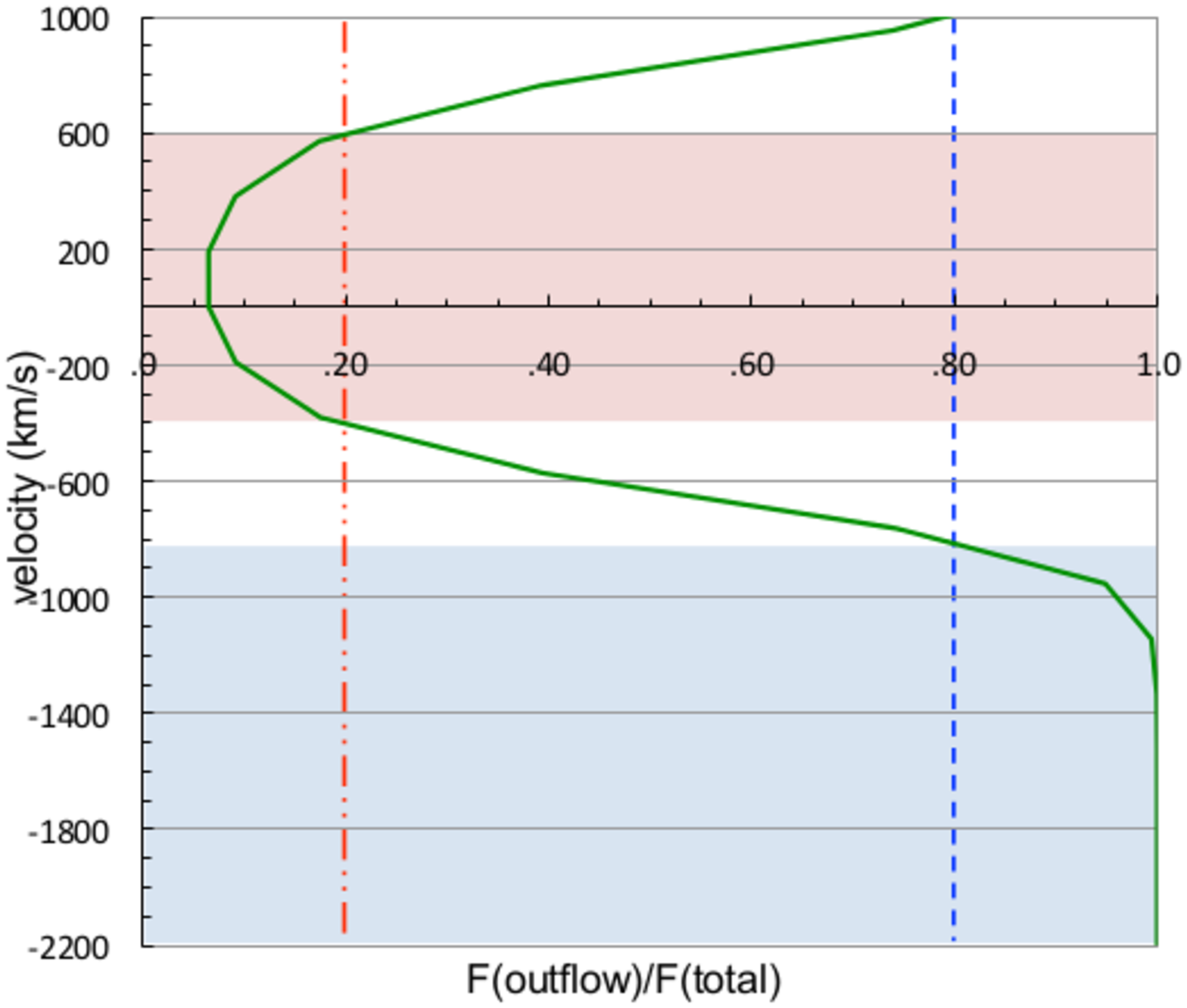}
\vspace{2.85in}
\caption{Spectroastrometry analysis for SDSS J1546-00 (top panels) and SDSS J1430-00 (bottom). Line, symbol and color codes as in Fig. 6.}
\label{astrom1546v1430}
\end{figure*}

Based on the spectroastrometric analysis we cannot confirm whether the outflow is extended
(Fig. ~\ref{astrom1546v1430}, top panels).  

\subsection{2011 sample}
~

{\it SDSS J1430-00}

% z=0.318, 4.60 kpc per arcsec

\vspace{0.2cm}

The nuclear [OIII] lines of this radio-quiet QSO2  consist of a narrow
(FWHM=280$\pm$20 km s$^{-1}$) and a broad component (FWHM$_{\rm max}$=1600$\pm$200 km s$^{-1}$). This
is blueshifted by $V_{\rm max}$=-520$\pm$60 km s$^{-1}$ and traces the ionized outflow (Villar Mart\'\i n et al. \citeyear{vm12}).  

The  [OIII] spatial profile   is clearly resolved  compared with the seeing disk (FWHM=0.73$\pm$0.05 arcsec; Fig. \ref{spat1413} right,  panel A).
It has FWHM$_{\rm int}$=0.58$\pm$0.07 arcsec or 2.7$\pm$0.3 kpc. VM12 showed the existence of ionized gas over a maximum total extension of 23 kpc (r$\sim$13 kpc from
the quasar).  The FWHM of the lines in the extended gas
 is  typically FWHM$\la$250 km s$^{-1}$ and thus rather quiescent.

 The pixel by pixel analysis  of the individual kinematic components    shows that while the narrow component is resolved (Fig. \ref{spat1413} right,  panel B), the 
  broadest component can only be isolated in the inner pixels and its spatial distribution is consistent with the seeing disk. It would be necessary to trace its emission further out from the spatial centroid to confirm  whether there is an excess above the seeing wings.
There is no clear evidence for kinematic spatial substructure (panels C and D).    We infer FWHM$_{\rm int}\la$0.35 arcsec or 1.6 kpc.

The spectroastrometric analysis (Fig. ~\ref{astrom1546v1430}) shows that the velocities for which the outflow emission dominates
present a small shift relative to the continuum centroid of $\sim$0.025 arcsec on average. They are also spatially shifted relative to those velocities where the ambient gas
clearly dominates the emission.  We infer $R_{\rm o}\ga$0.025 arcsec or 115 pc.

{\it SDSS J0923+01}

% z=0.386 5.23 kpc/arcsec

\vspace{0.2cm}

\begin{figure*}
\includegraphics{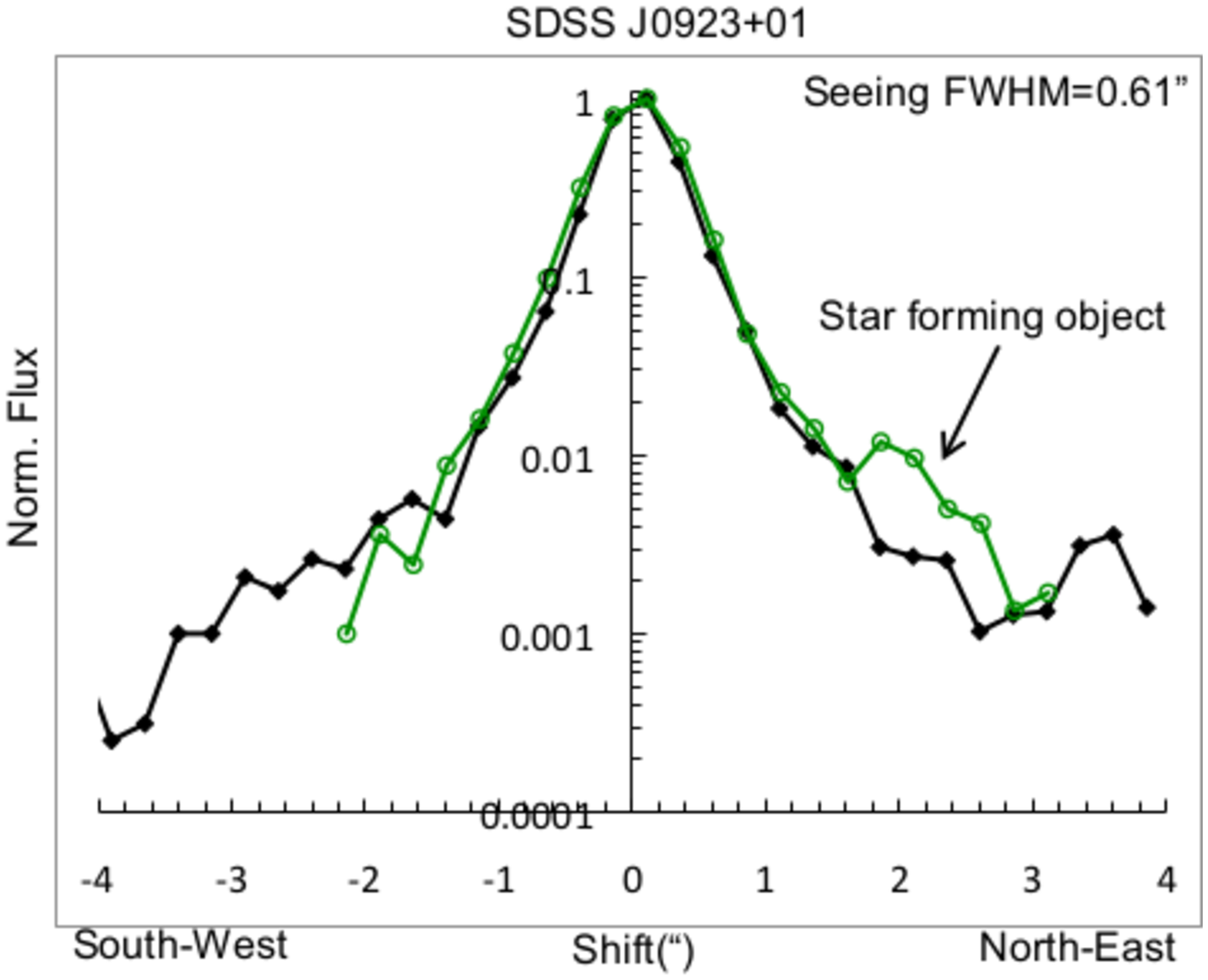} 
\includegraphics{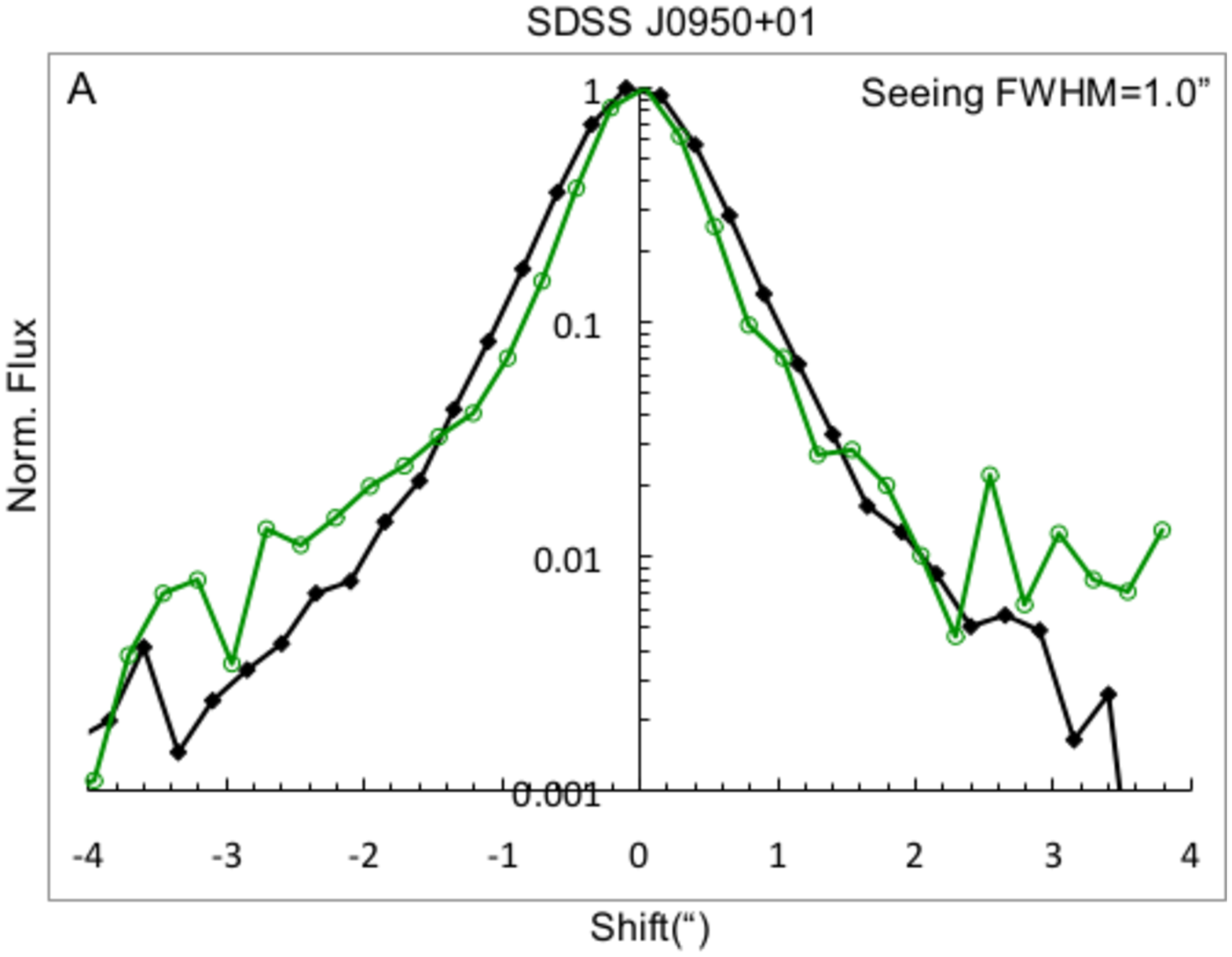}
\vspace{2.2in}
\includegraphics{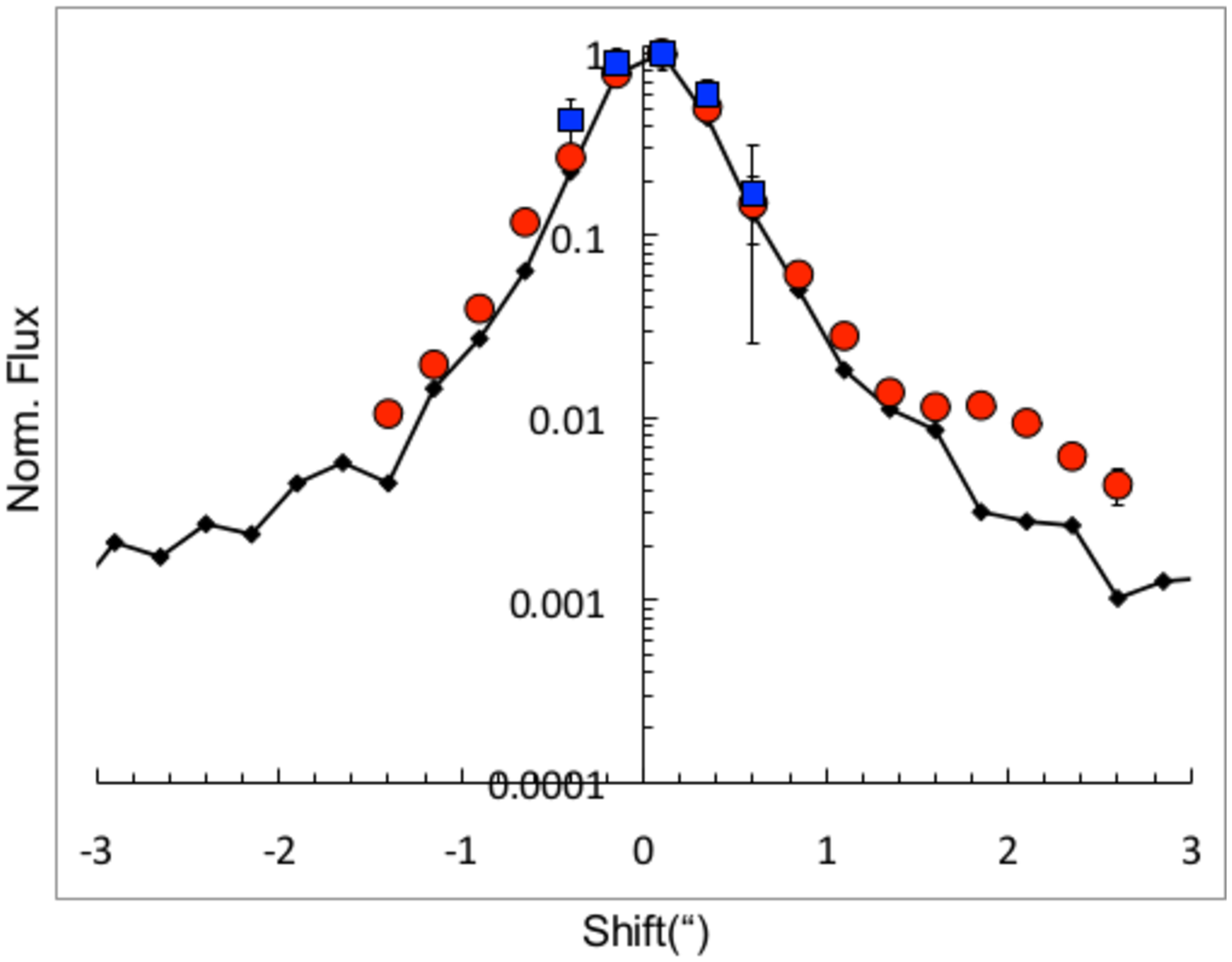}
\includegraphics{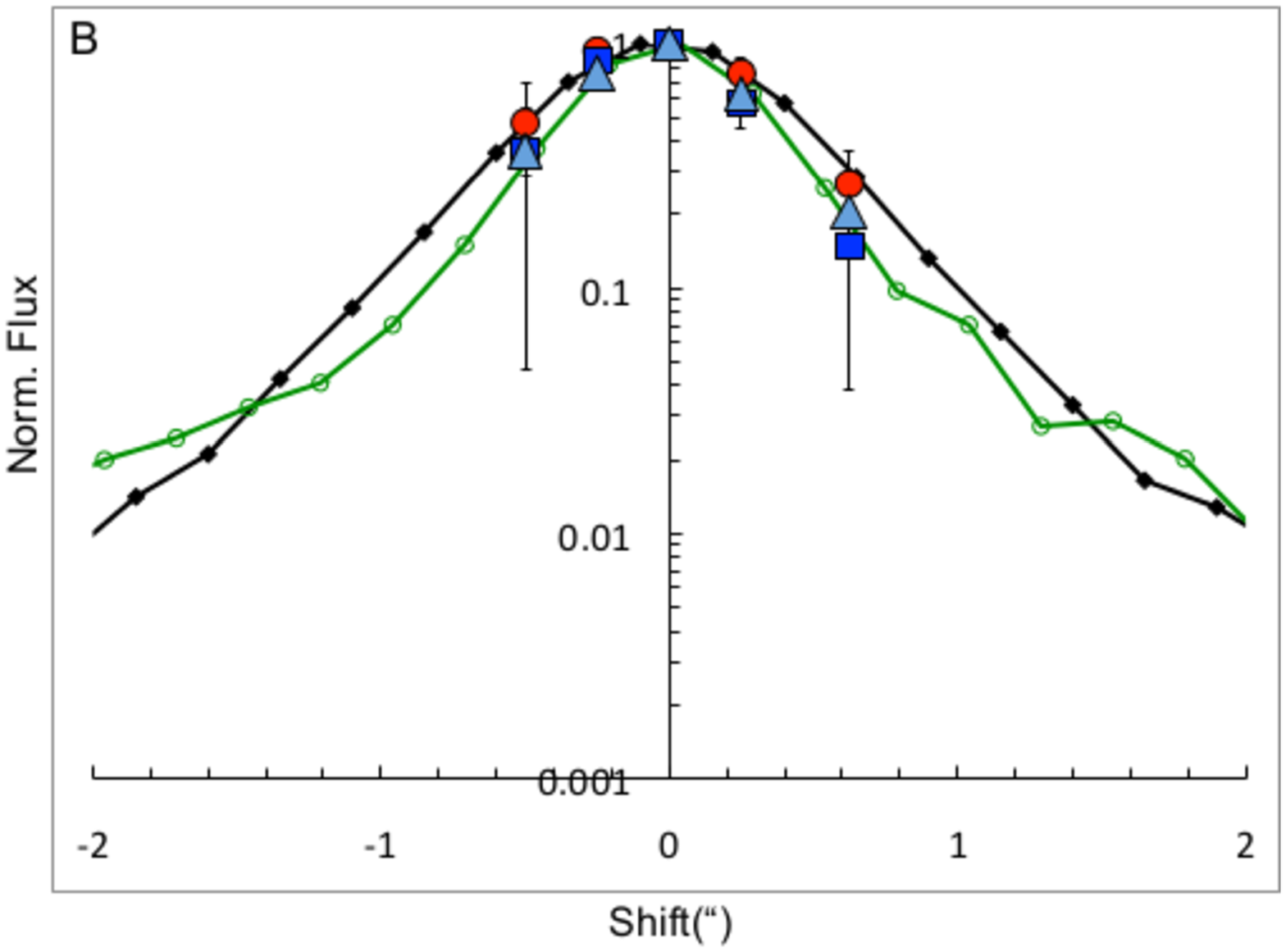}
\vspace{2.2in}
\includegraphics{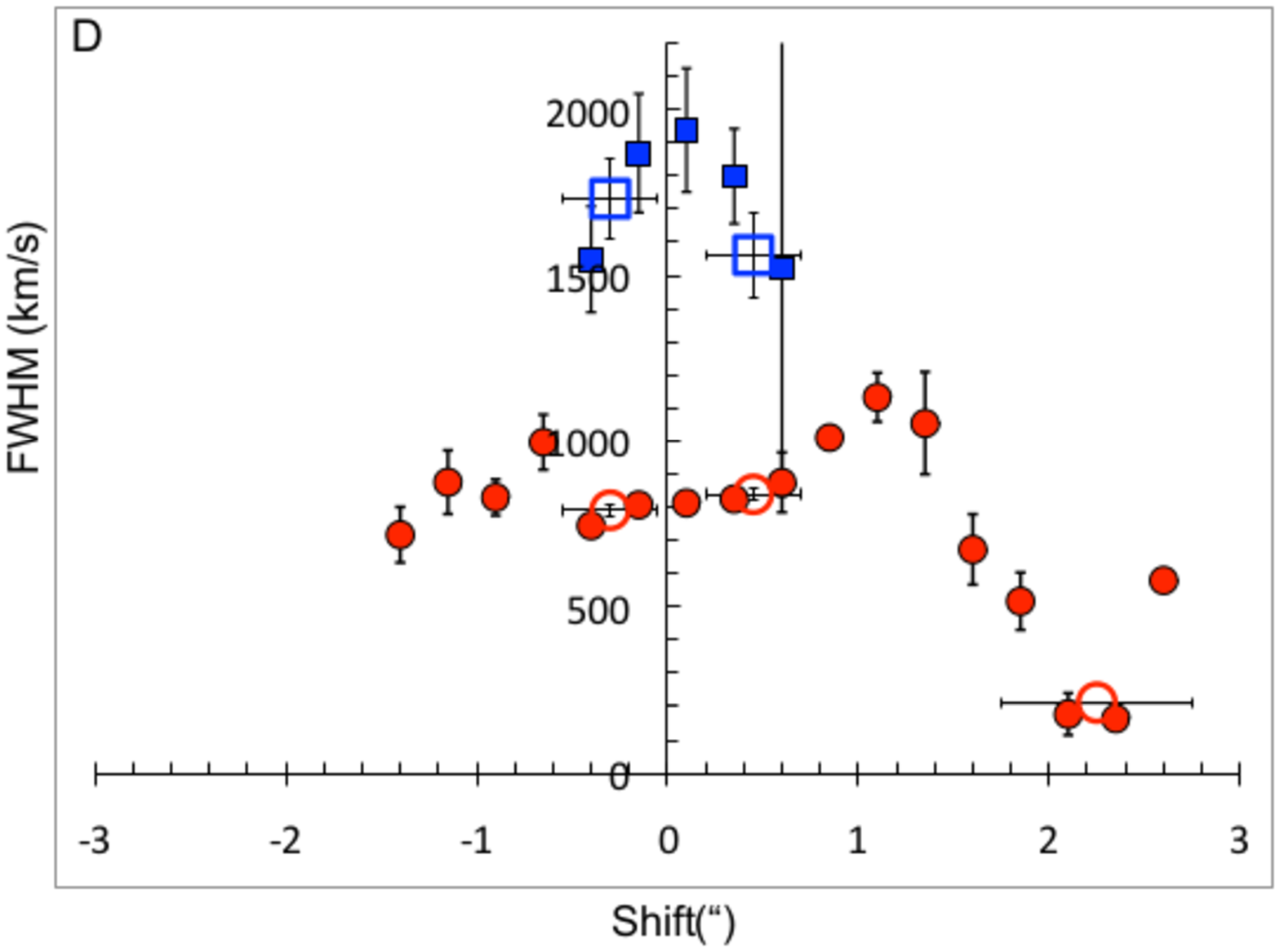}
\includegraphics{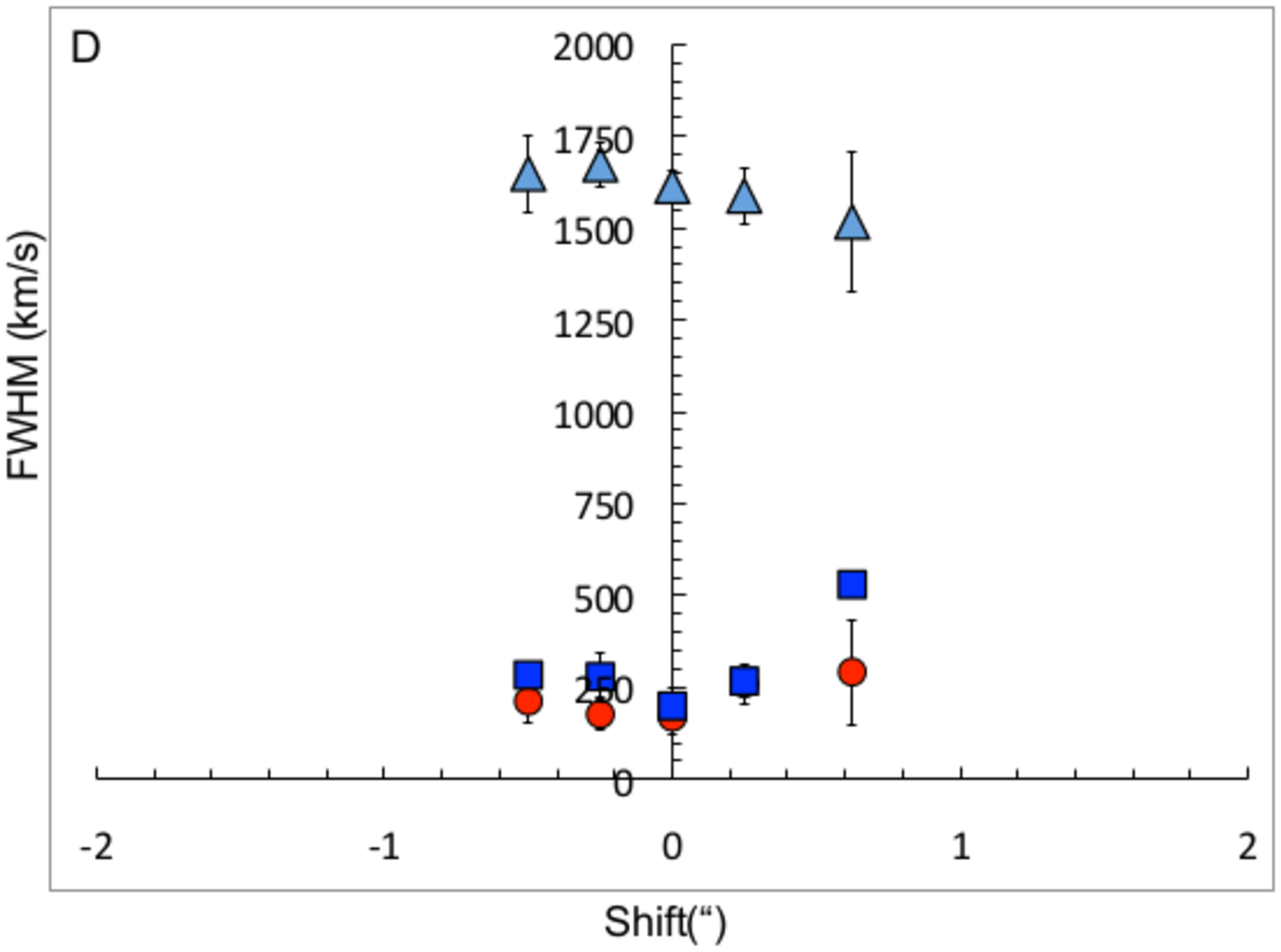}
\vspace{2.2in}
\includegraphics{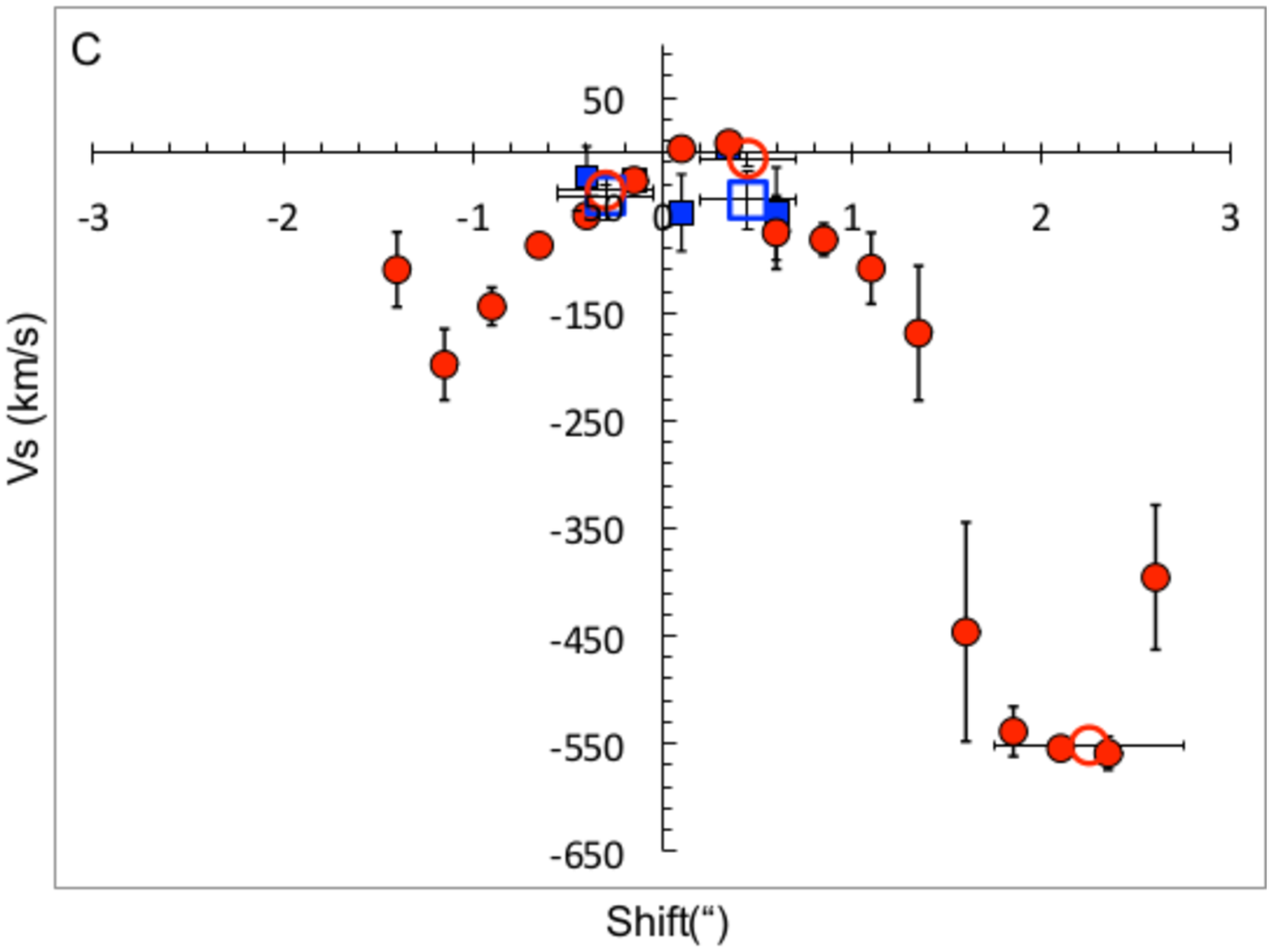}
\includegraphics{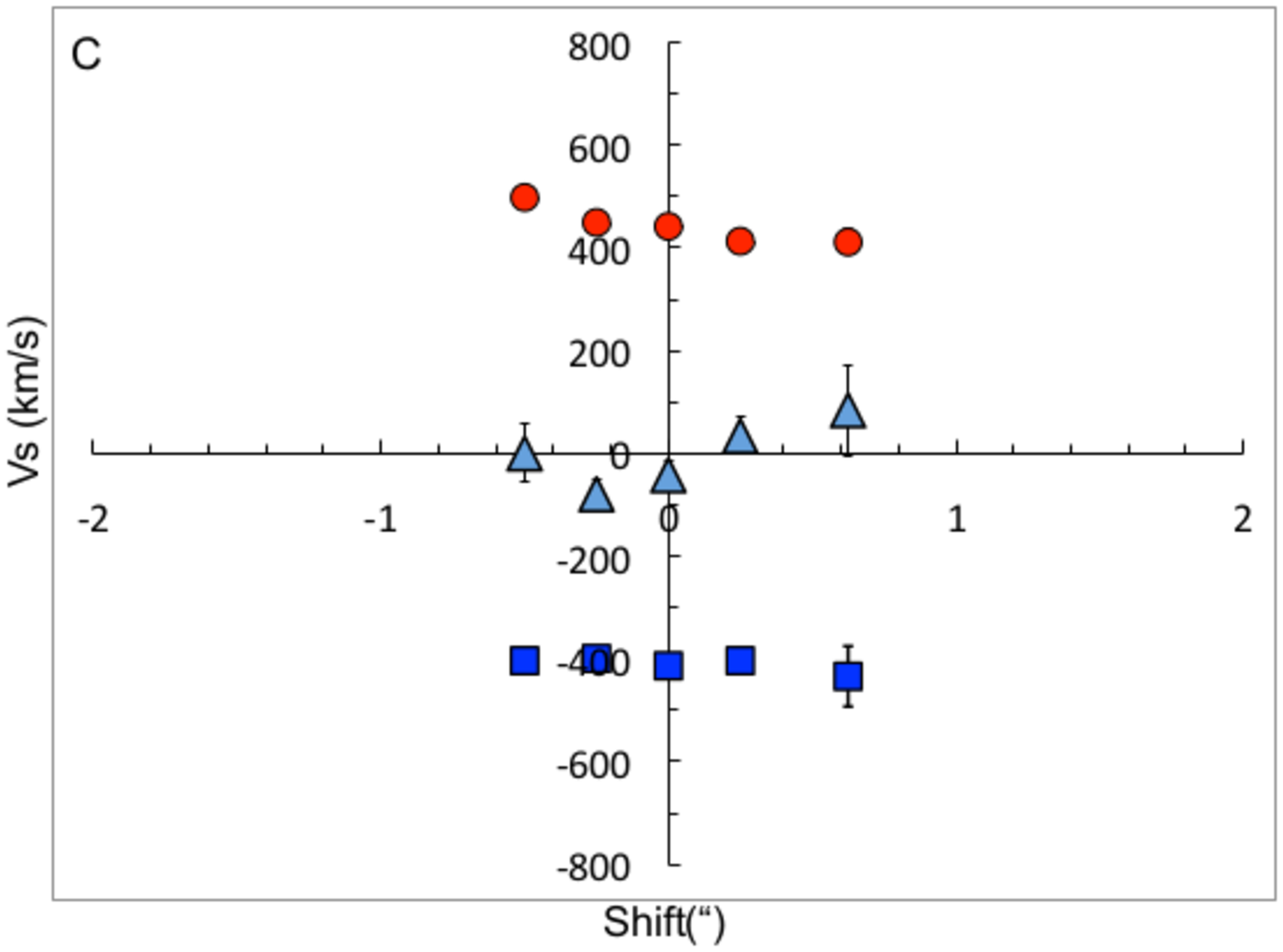}
\vspace{2.2in}
\caption{Spatially extended analysis of [OIII]$\lambda$5007 for  SDSS J0923+01 (left)  and SDSS J0950+01 (right). 
 Because the [OIII] spatial profile (green solid line) is narrower than the seeing (black solid line) for SDSS J0950+01, the spatial profiles of the individual kinematic components are compared
with both in panel B (right). Line, symbol and color codes as in Fig. 5. The spatial scale of the B, C and D panels is zoomed relative to panel A for clarity.}
\label{spat0923v0950}
\end{figure*}

\begin{figure*}
\includegraphics{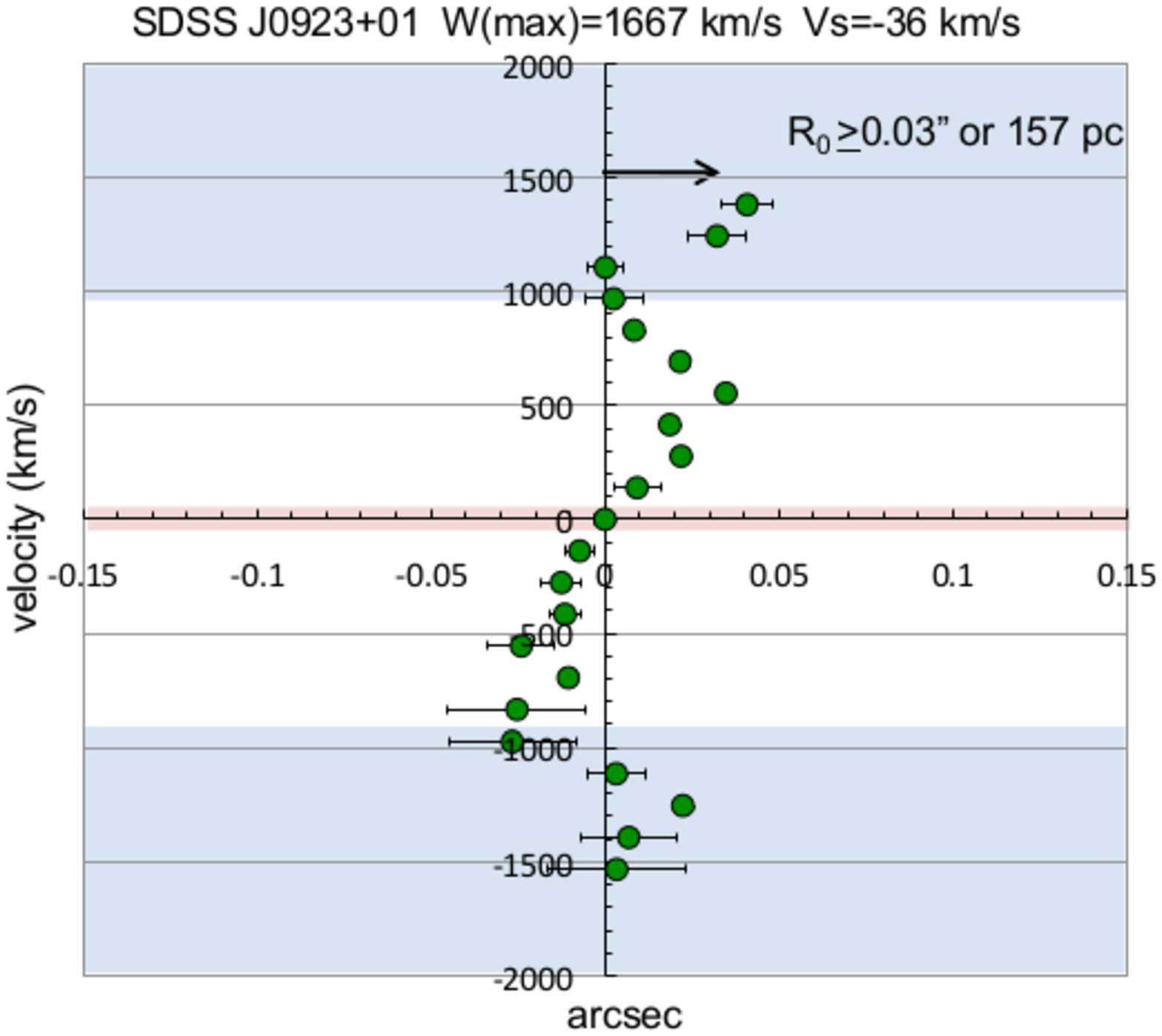}
\includegraphics{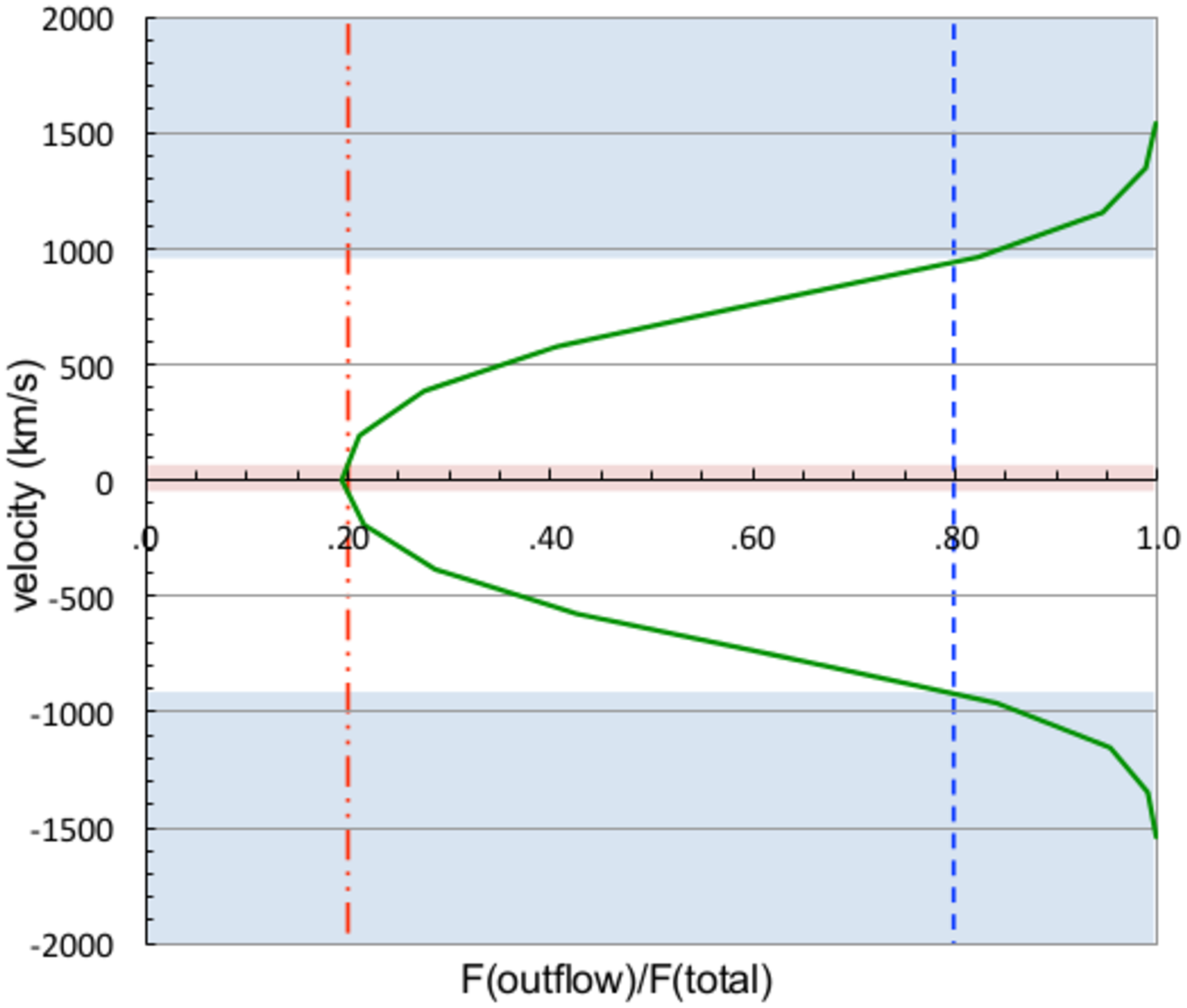}
\vspace{2.85in}
\includegraphics{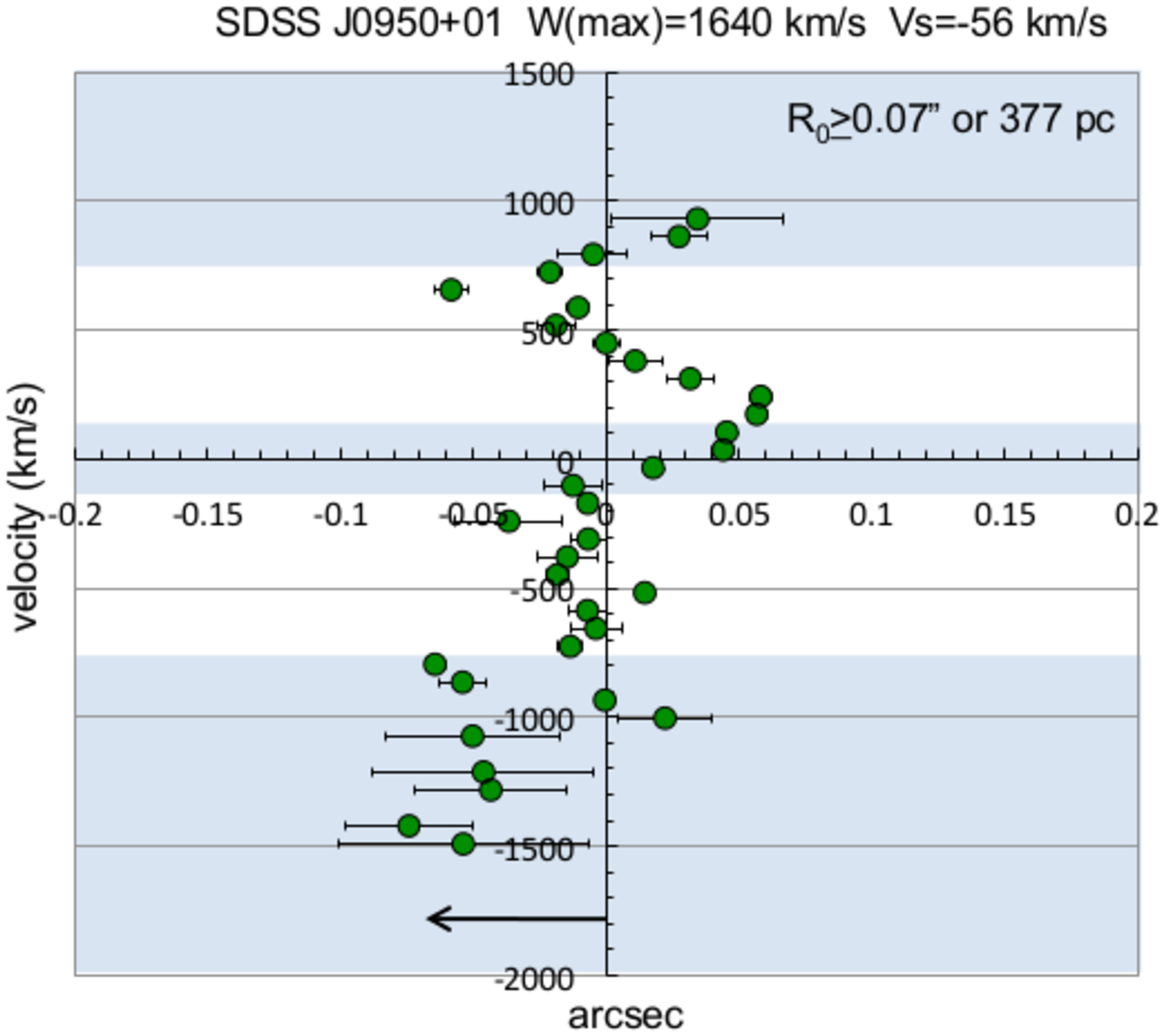}
\includegraphics{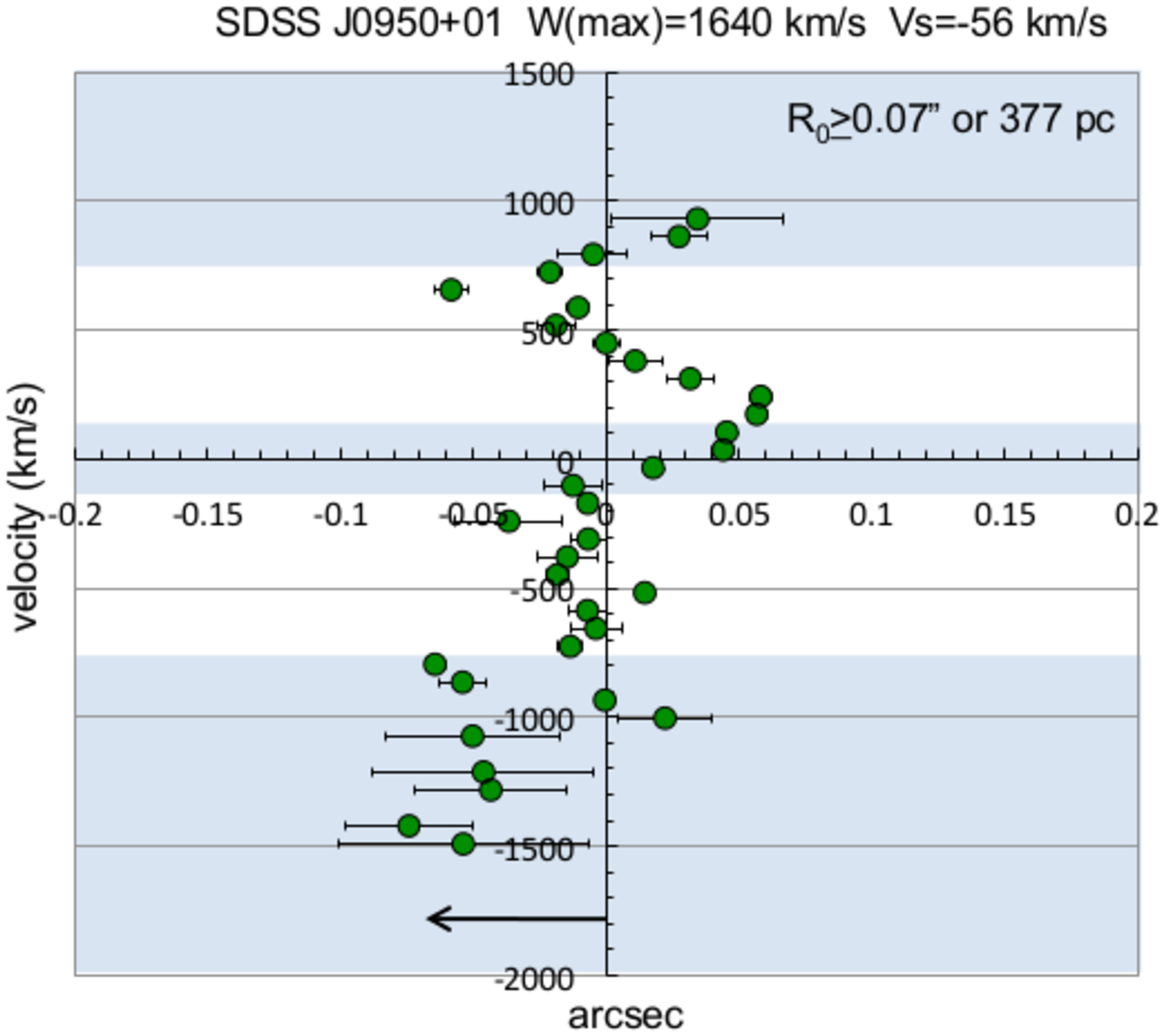}
\vspace{2.85in}
\caption{Spectroastrometry analysis for SDSS J0923+01 (top) and SDSS J0950+01 (bottom). Line, symbol and color codes as in Fig.  6.}
\label{astrom0923v0950}
\end{figure*}

The nuclear [OIII] lines of this radio-quiet QSO2 are best fitted with two kinematic components of FWHM=708$\pm$34 and 1667$\pm$177 km
s$^{-1}$ respectively. The broadest one, which traces the outflow,  is shifted by $V_{\rm max}\sim$-36$\pm$40 km s$^{-1}$ (Table \ref{tab:fitsnuc}, Fig.\ref{nuclei1}).

As reported in \cite{hum15}, the spatial profile of the emission lines along the slit is dominated by a spatially unresolved central component (seeing FWHM=0.66$\pm$0.06 arcsec; see also  Fig. \ref{spat0923v0950}, left panels).  In addition,  a faint and compact  star forming object is detected in excess above the seeing disk at 2 arcsec (10 kpc) north east of the AGN. It shows   very distinct kinematics to the nuclear region with 
narrow lines (FWHM=212$\pm$32 km s$^{-1}$) and $V_{\rm s}\sim$-550 km s$^{-1}$. 

The spatially extended kinematic analysis shows no evidence for spatial extension of the broadest component (Fig. \ref{spat0923v0950}, panel B left) neither convincing evidence for kinematic spatial substructure within the seeing disk (panels B and C). We estimate FWHM$_{\rm int}\la$0.40 arcsec or 2.1 kpc.

Beyond the inner few pixels where the broad component can be confidently isolated, broadening of the narrow component
is appreciated with FWHM$\sim$1000 km s$^{-1}$. This is evidence that the broad component also contributes to the line flux
at these locations. The S/N is not enough for a proper kinematic decomposition of the lines.  The
flux in these regions is also dominated by the central, spatially unresolved source.

According to the spectroastrometric analysis, the outflow is spatially extended. We infer  $R_{\rm o}\ga$0.03 arcsec or 157 pc (Fig. ~\ref{astrom0923v0950}).

\vspace{0.2cm}

{\it SDSS J0950+01}

% z=0.404 5.38 kpc/arcsec
\vspace{0.2cm}

This is a double peaked radio-intermediate HLSy2 (Table \ref{tab:fitsnuc}, Fig.\ref{nuclei1}), with two narrow components of
similar flux. They have FWHM=180$\pm$39
and 233$\pm$49 km s$^{-1}$ shifted relative to each other by 852$\pm$16 km s$^{-1}$. In addition, a broad component
(the ionized outflow) is isolated in the fit,
 with intermediate velocity shift between both narrow components and W$_{\rm max}$=1640$\pm$44 km s$^{-1}$. 
 
The [OIII] profile  is dominated by a compact core which appears  narrower than the seeing disk derived from stars in the image (FWHM=1.00$\pm$0.05 arcsec; Fig.\ref{spat0923v0950} right panels). An excess  of emission is detected $\ga$1.8 arcsec.  In \cite{hum15}  we showed that [OII] is also clearly extended at both sides of the continuum centroid, up to $\sim$2.3 arcsec from the AGN. We reported  narrow lines $\la$235 km s$^{-1}$ in the extended gas.

The kinematic decomposition of [OIII] in the central pixels  
shows that the spatial distributions of the individual components are consistent with the seeing disk.  
None of the three components show  clear evidence for kinematic spatial substructure.
We estimate FWHM$_{\rm int}\la$0.41 arcsec or 2.2 kpc for the outflow.

According to the spectroastrometric analysis, the ionized outflow is extended. We estimate  $R_{\rm o}\ga$0.07 arcsec or 377 pc (Fig. ~\ref{astrom0923v0950}).

\vspace{0.2cm}

{\it SDSS J1014+02}
%z=0.573, 6.52 kpc/arcsec
\vspace{0.2cm}

Two spectra were obtained for this radio-intermediate HLSy2  along PA1 -5.9 and PA2 +42.1 (\cite{hum15}).
It is is a double peaked system (Table \ref{tab:fitsnuc}, Fig. \ref{nuclei1}). The two narrow components 
 have FWHM$\sim$388$\pm$21 and $\sim$254$\pm$32 km s$^{-1}$ respectively  and and are shifted in velocity by $\sim$654$\pm$28 km s$^{-1}$ as measured from the PA1 spectrum.
A broad component  (the ionized outflow) is  moreover required by the fit. The two spectra  imply 
 FWHM$_{\rm max}$=1537$\pm$72 km s$^{-1}$ and  $V_{\rm max}$=139$\pm$34 km s$^{-1}$ for PA1 and  W$_{\rm max}$=1580 $\pm$115 for  $V_{\rm max}$=262$\pm$53 km s$^{-1}$  for PA2, which are in good agreement.

\begin{figure*}
\includegraphics{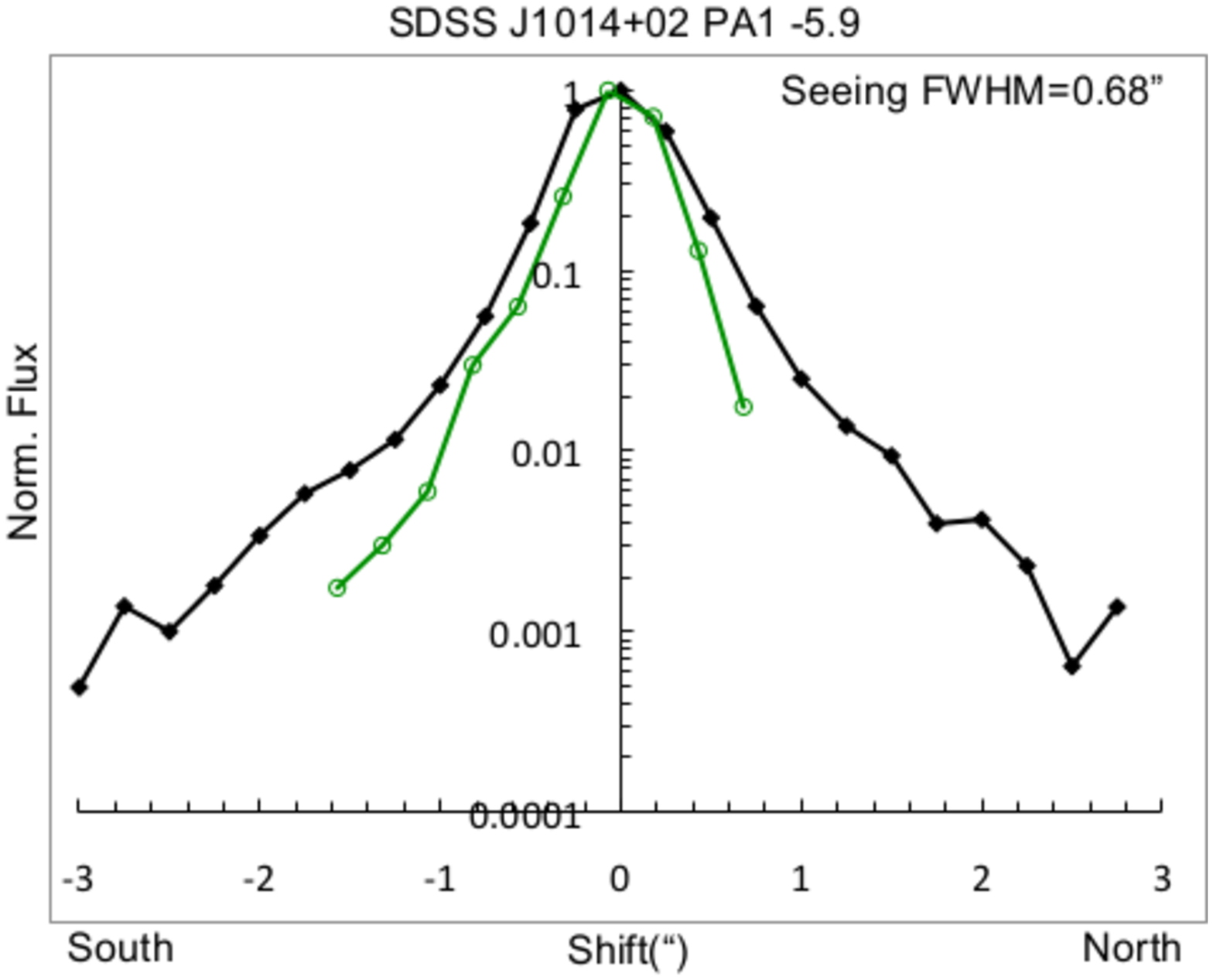}
\includegraphics{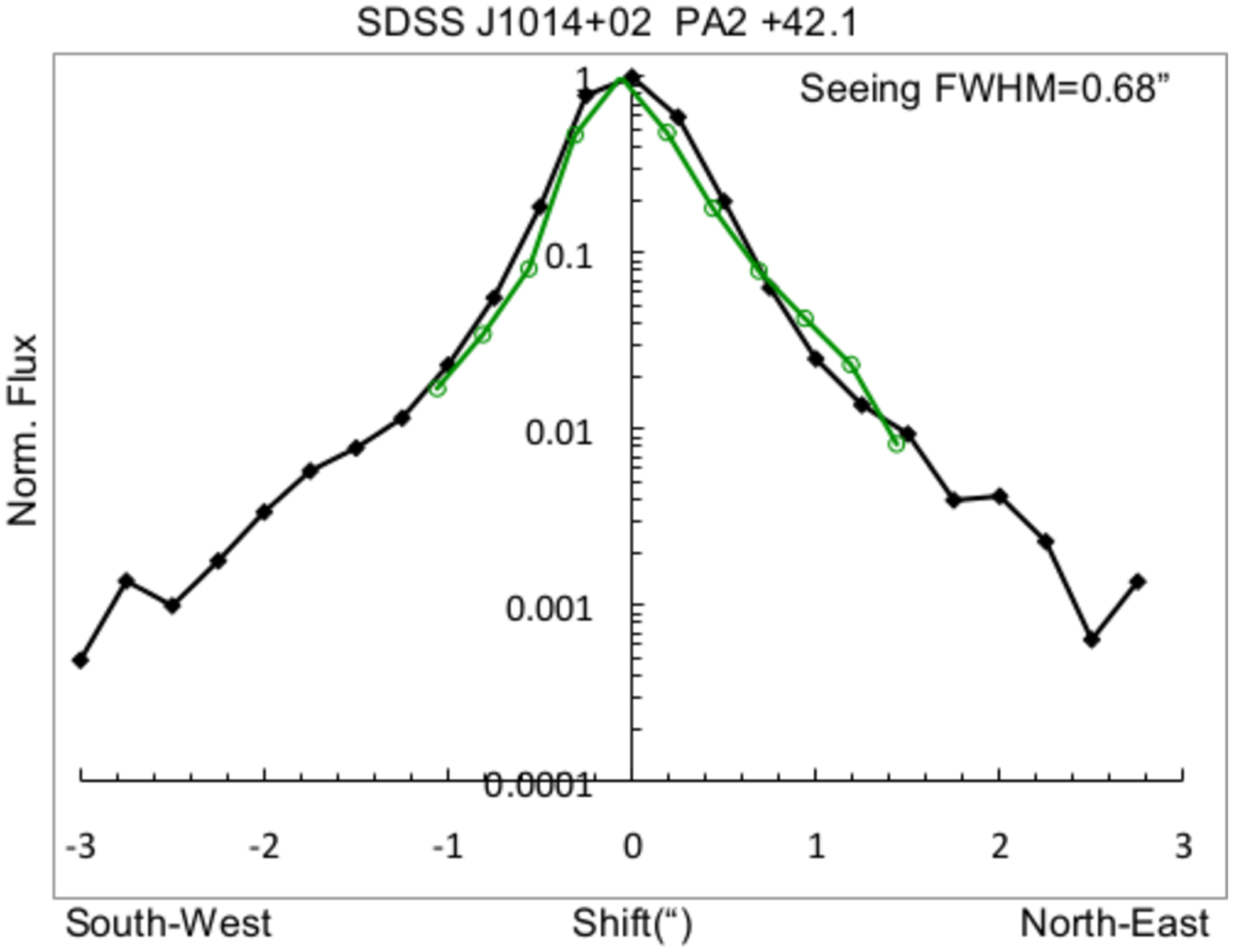}
\vspace{2.2in}
\includegraphics{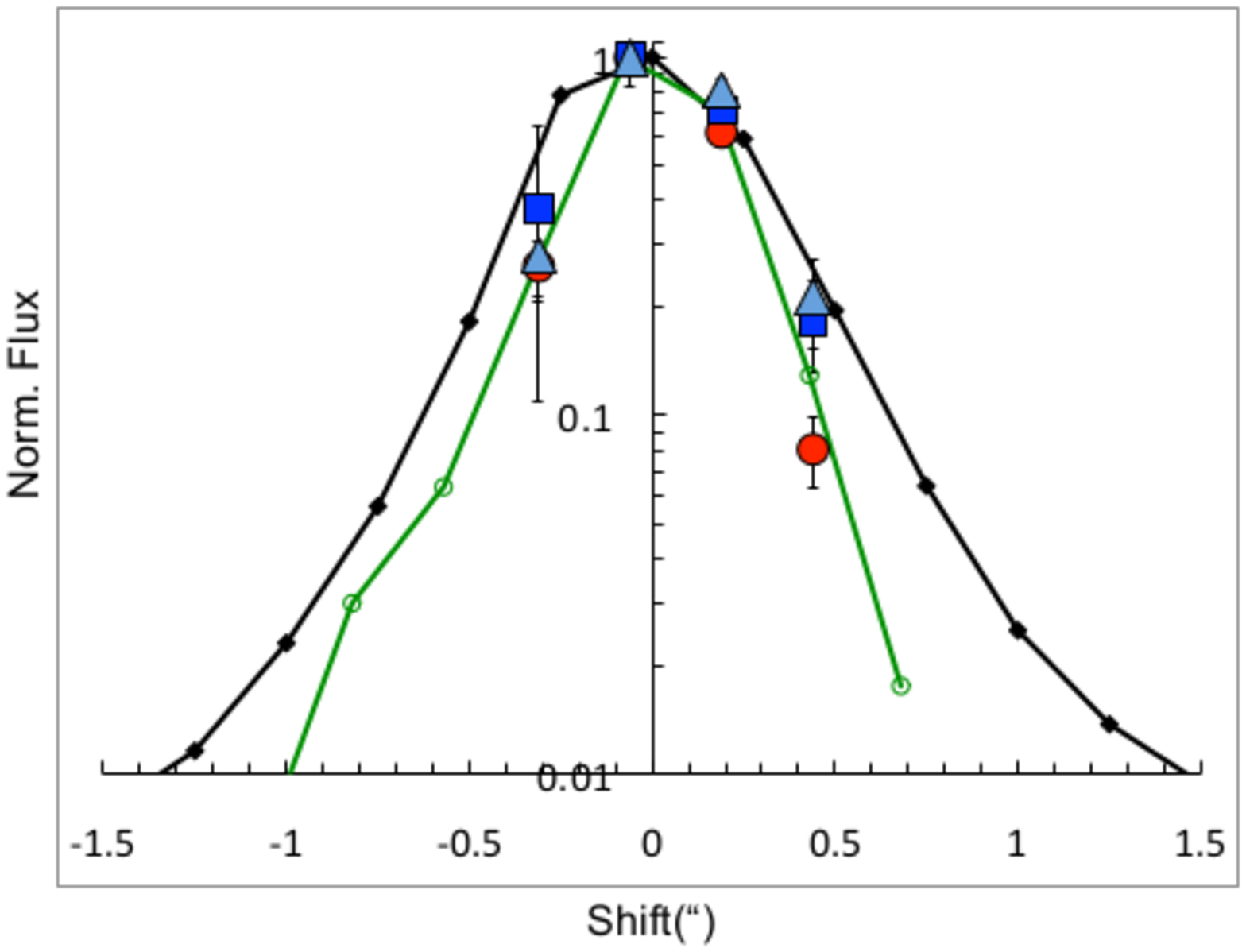}
\includegraphics{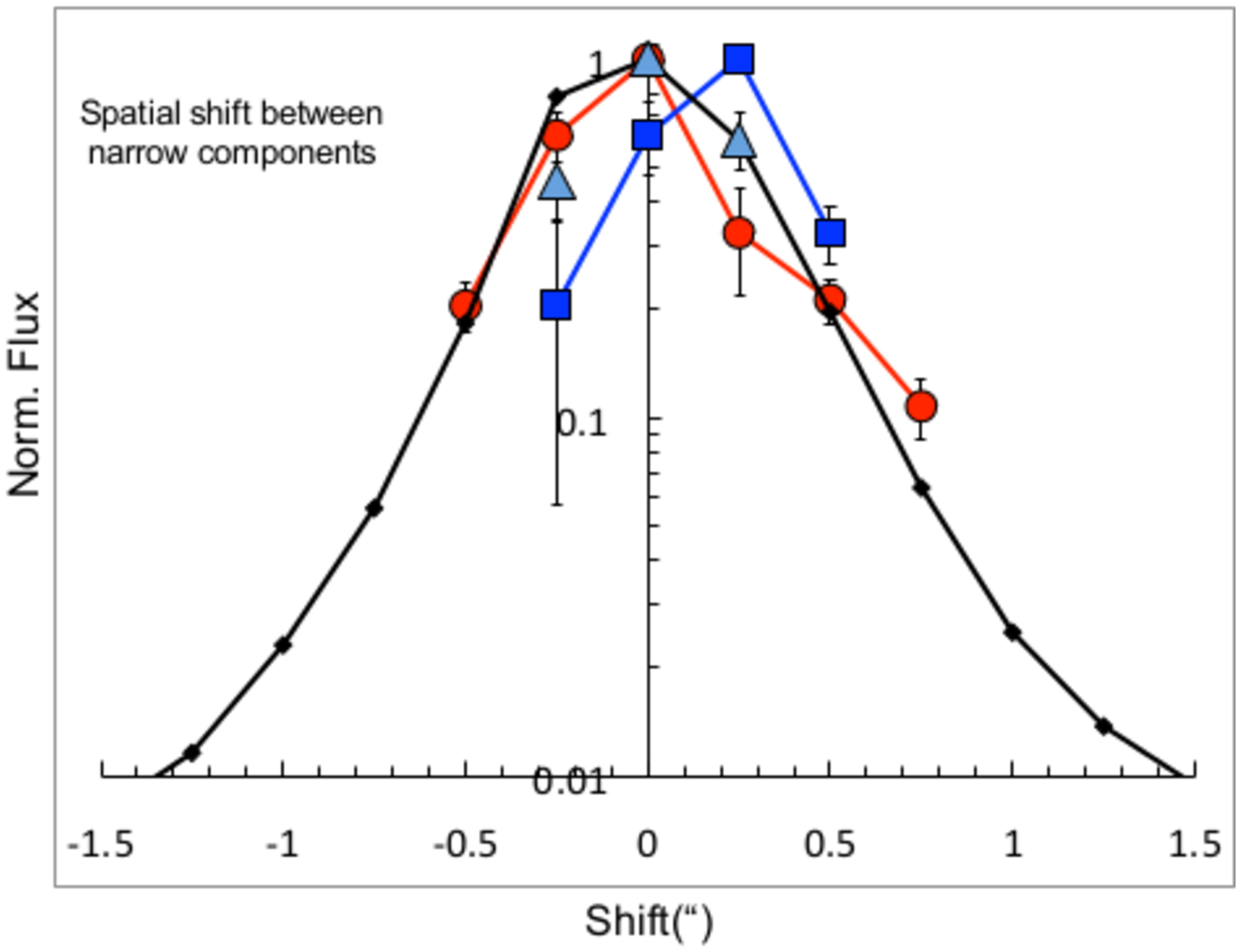}
\vspace{2.2in}
\includegraphics{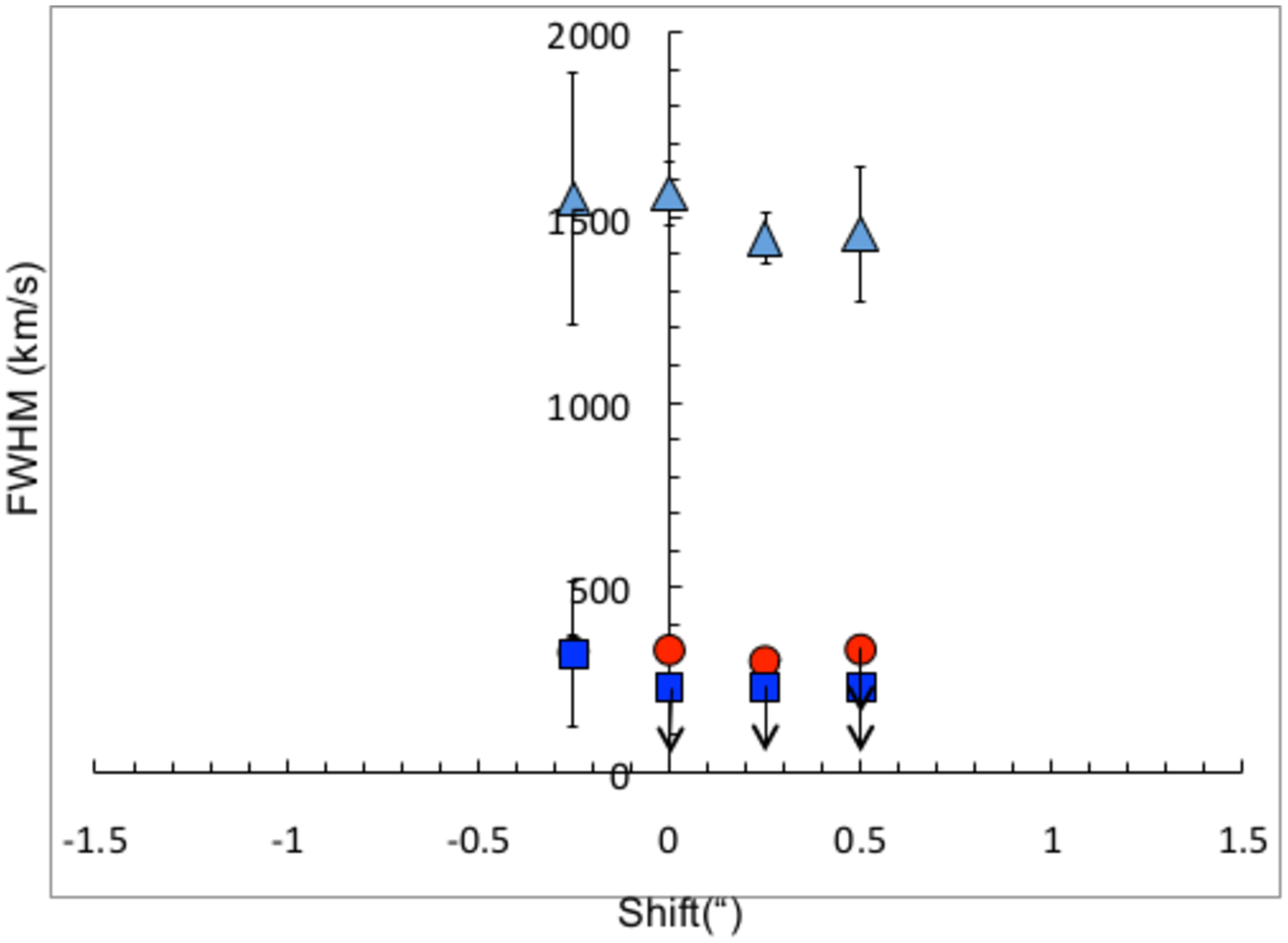}
\includegraphics{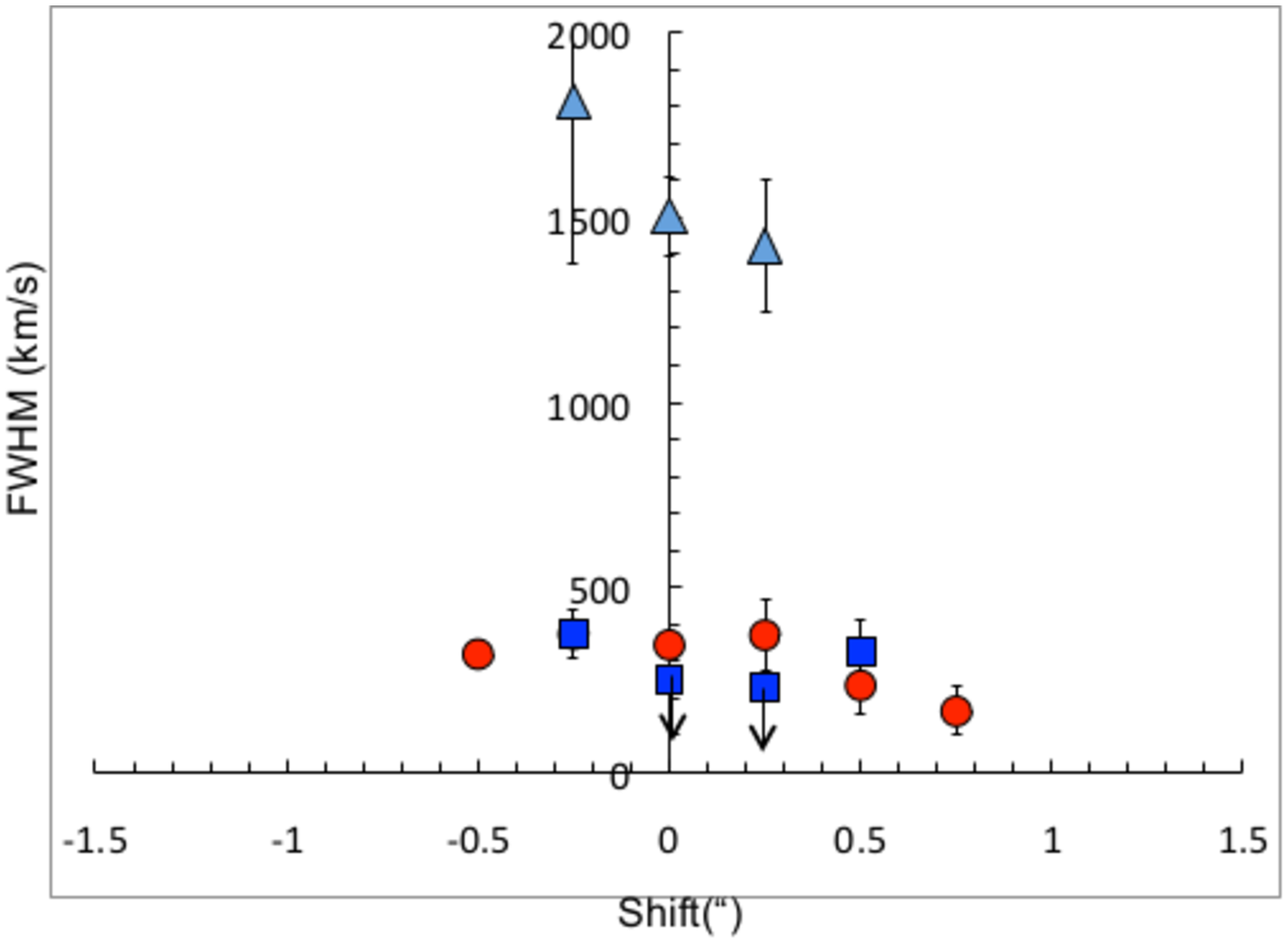}
\vspace{2.2in}
\includegraphics{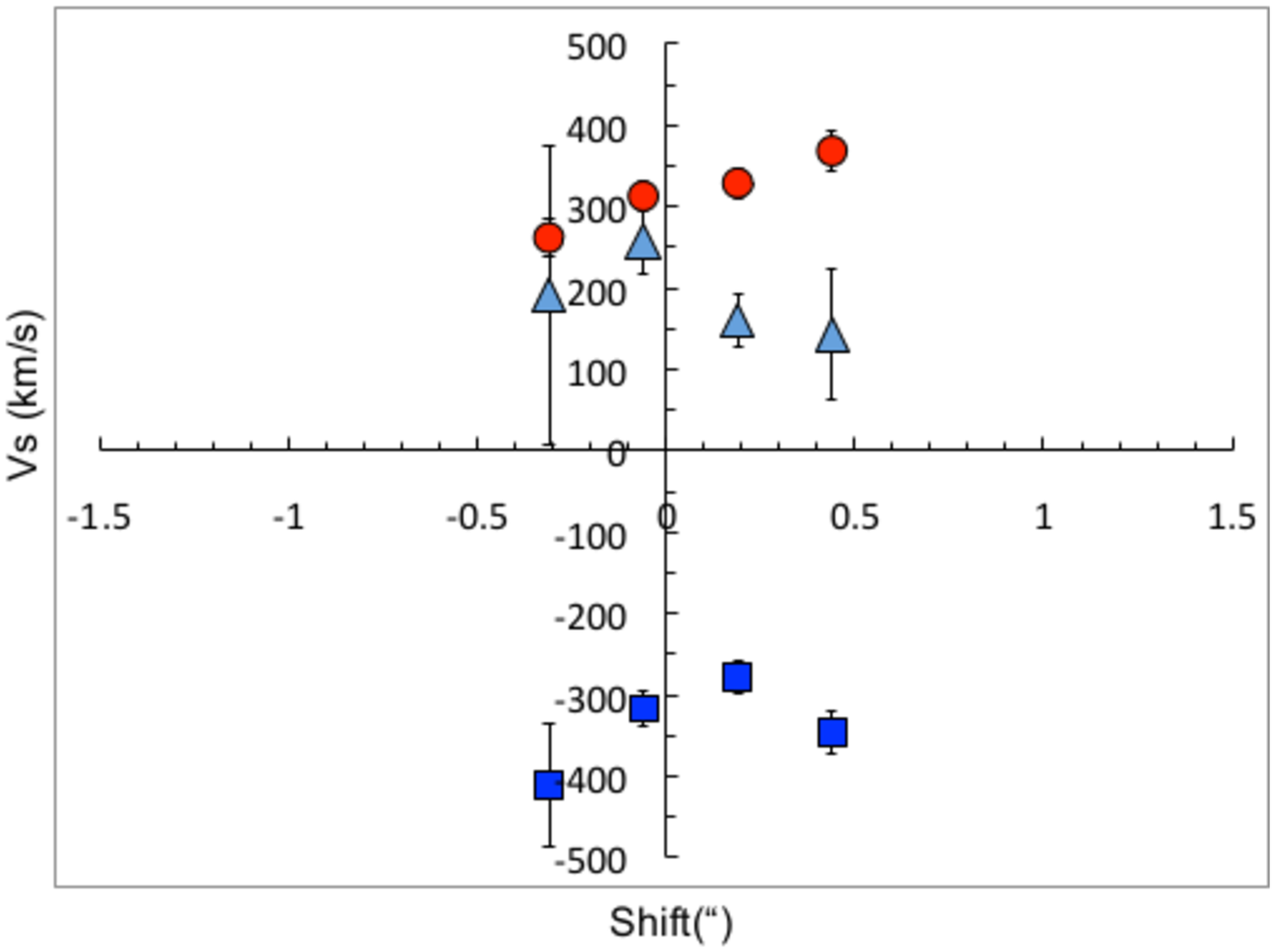}
\includegraphics{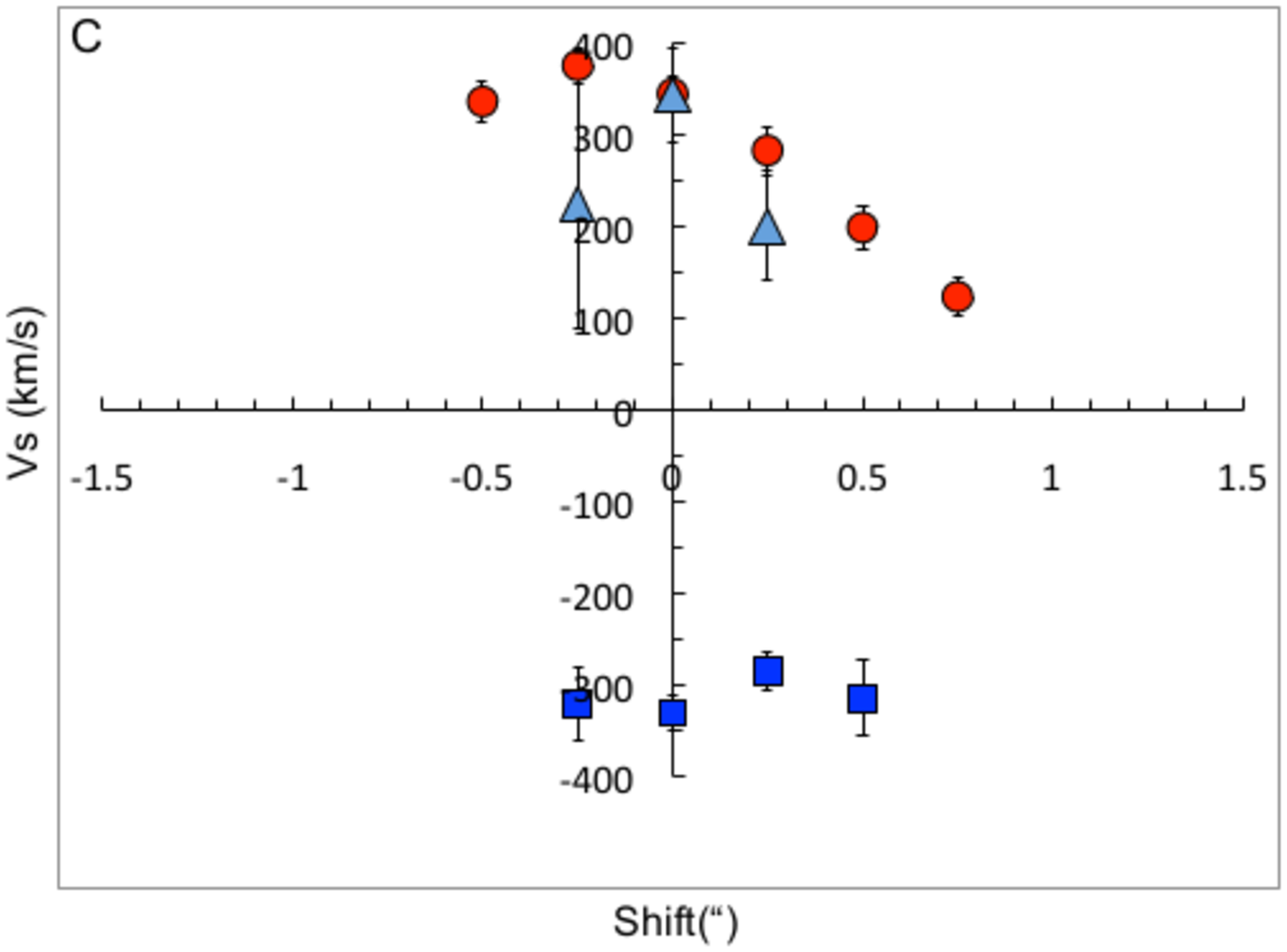}
\vspace{2.2in}
\caption {Spatially extended analysis of [OIII]$\lambda$5007 for SDSS J1014+02 along PA1 -5.9 (left) and PA2 +42.1 (right).  Because the [OIII] spatial profile (green solid line) is significantly narrower than the seeing (black solid line) along PA1, the spatial profiles of the individual kinematic components are compared
with both in panel B (left). Along PA2, the spatial centroids of the two narrow components (red circles and blue squares) are spatially shifted. This is highlighted by using solid blue and red lines. Line, symbol and color codes as in Fig. 5. The spatial scale of the B, C and D panels is zoomed relative to panel A for clarity.}
\label{spat1014}
\end{figure*}

\begin{figure*}
\includegraphics{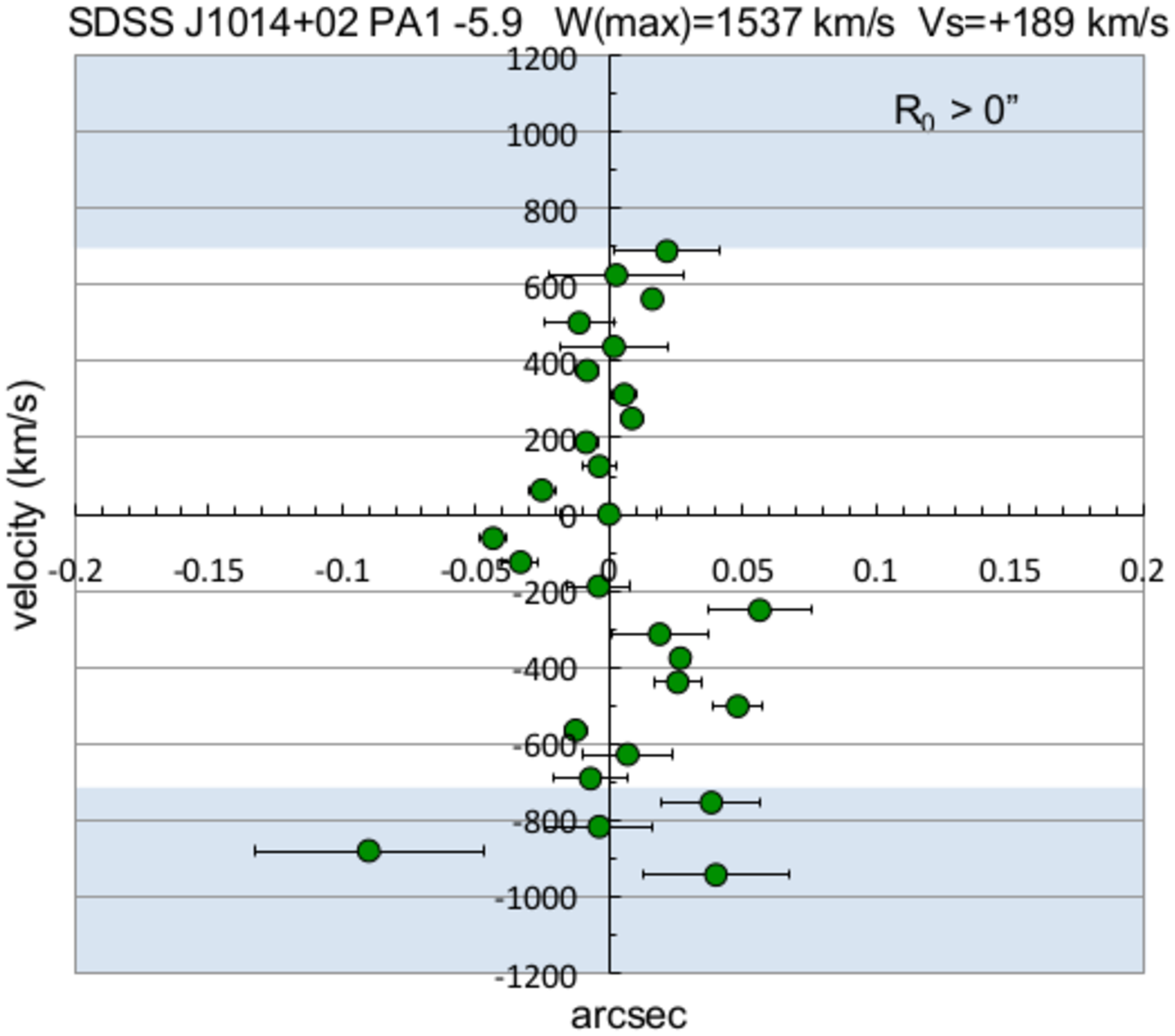}
\includegraphics{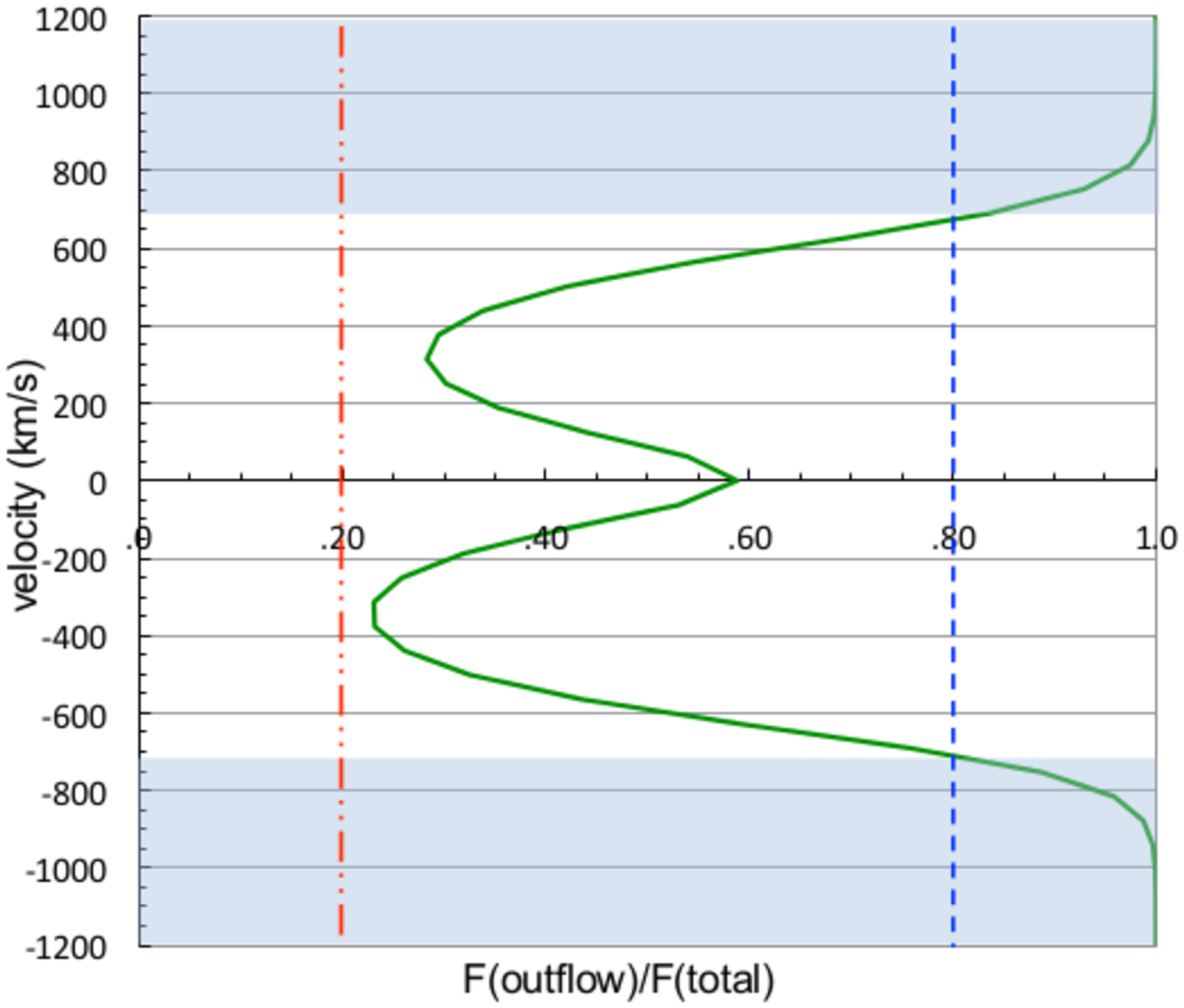}
\vspace{2.85in}
\includegraphics{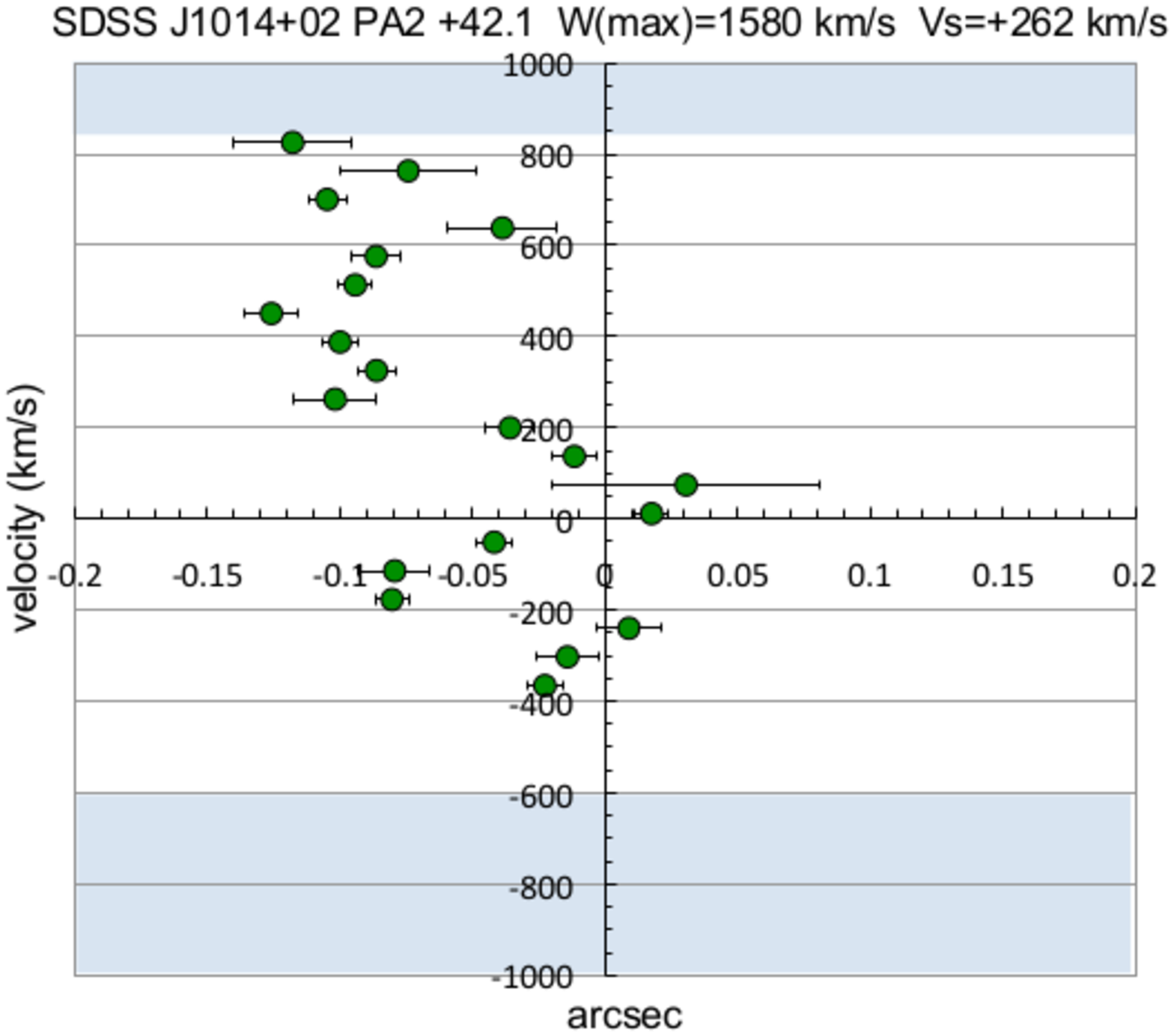}
\includegraphics{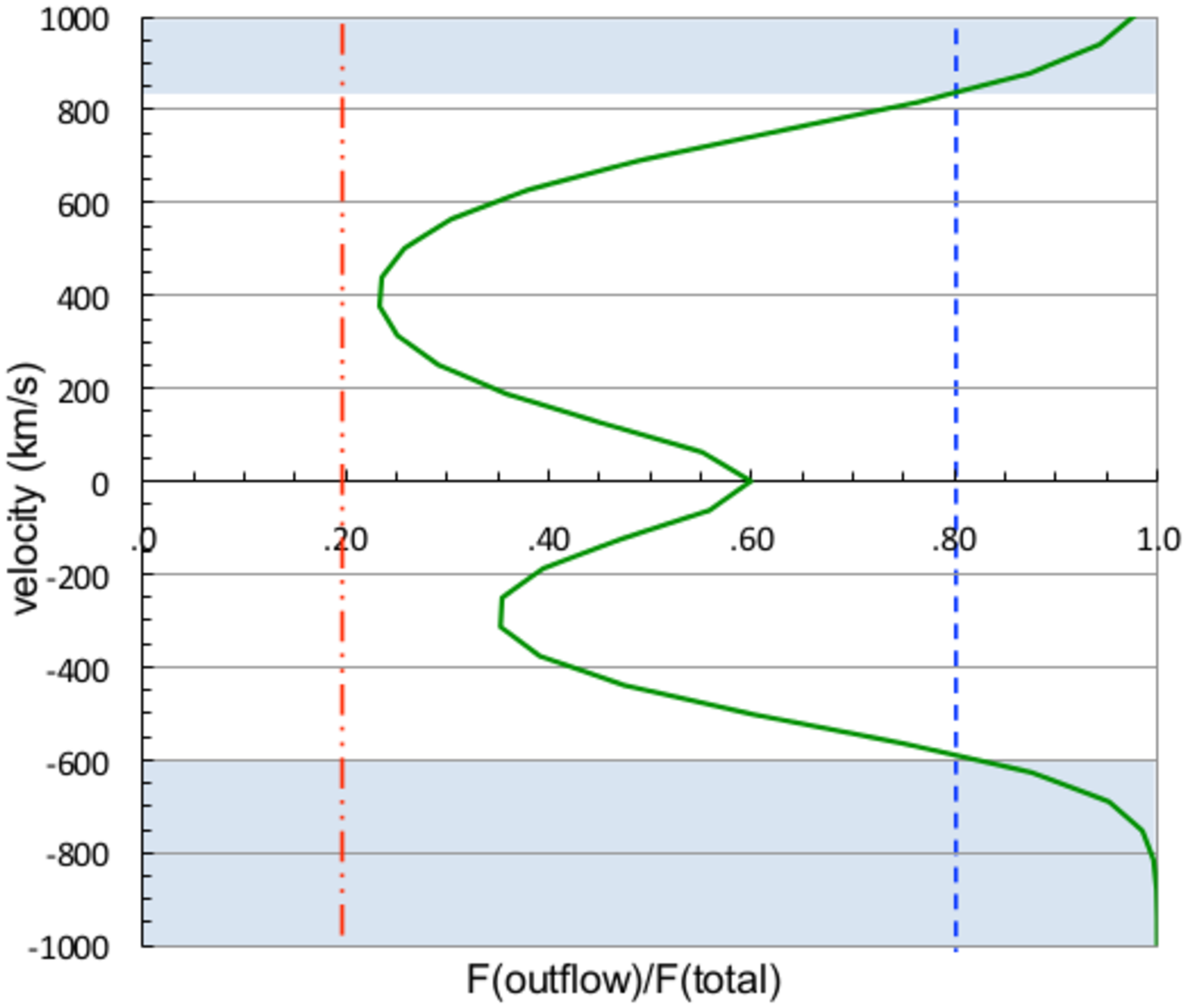}
\vspace{2.85in}
\caption{Spectroastrometry analysis for SDSS J1014+02 along PA1 -5.9 (top) and PA2 +42.1 (bottom). Line, symbol and color codes as in Fig.  6.}
\label{astrom1014}
\end{figure*}

As shown in \cite{hum15},  the [OIII] spatial profile   is dominated for both PA by the central unresolved source (Fig. \ref{spat1014}).
In fact, the object seems narrower than the seeing (FWHM=0.65$\pm$0.04 arcsec) along PA1. Only [OII] appears slightly extended
along PA2 according to \cite{hum15}. In addition, the authors reported very low surface brightness [OII] emission along PA2
up to $\sim$2.8 arcsec or 18 kpc from the continuum centroid toward the South-West. The lines
are narrow at this location with FWHM$\la$211 km s$^{-1}$.

  The pixel to pixel analysis   (Fig. \ref{spat1014}) shows that the two narrow components  are
  spatially unresolved along both PA. They are   shifted spatially along PA2 (panel B left). This is not apparent along PA1 (panel B right).  The broadest component (the outflow) is not sufficiently sampled, although it is consistent with being spatially unresolved. It shows no  clear kinematic substructure along any of the two slit PA. 
Thus, according to methods (i) and (ii) we find no evidence for the  ionized outflow to be extended. We estimate FWHM$_{\rm int}\la$0.31 arcsec or 2.0 kpc along both PA.

There is no clear evidence for the ionized outflow  to be extended along any of the two slit PA according to the spectroastrometric analysis  (Fig. ~\ref{astrom1014}). 

\vspace{0.2cm}

{\it SDSS J1017+03}
%z=0.454, 5.76 kpc/arcsec
\vspace{0.2cm}

\begin{figure*}
\includegraphics{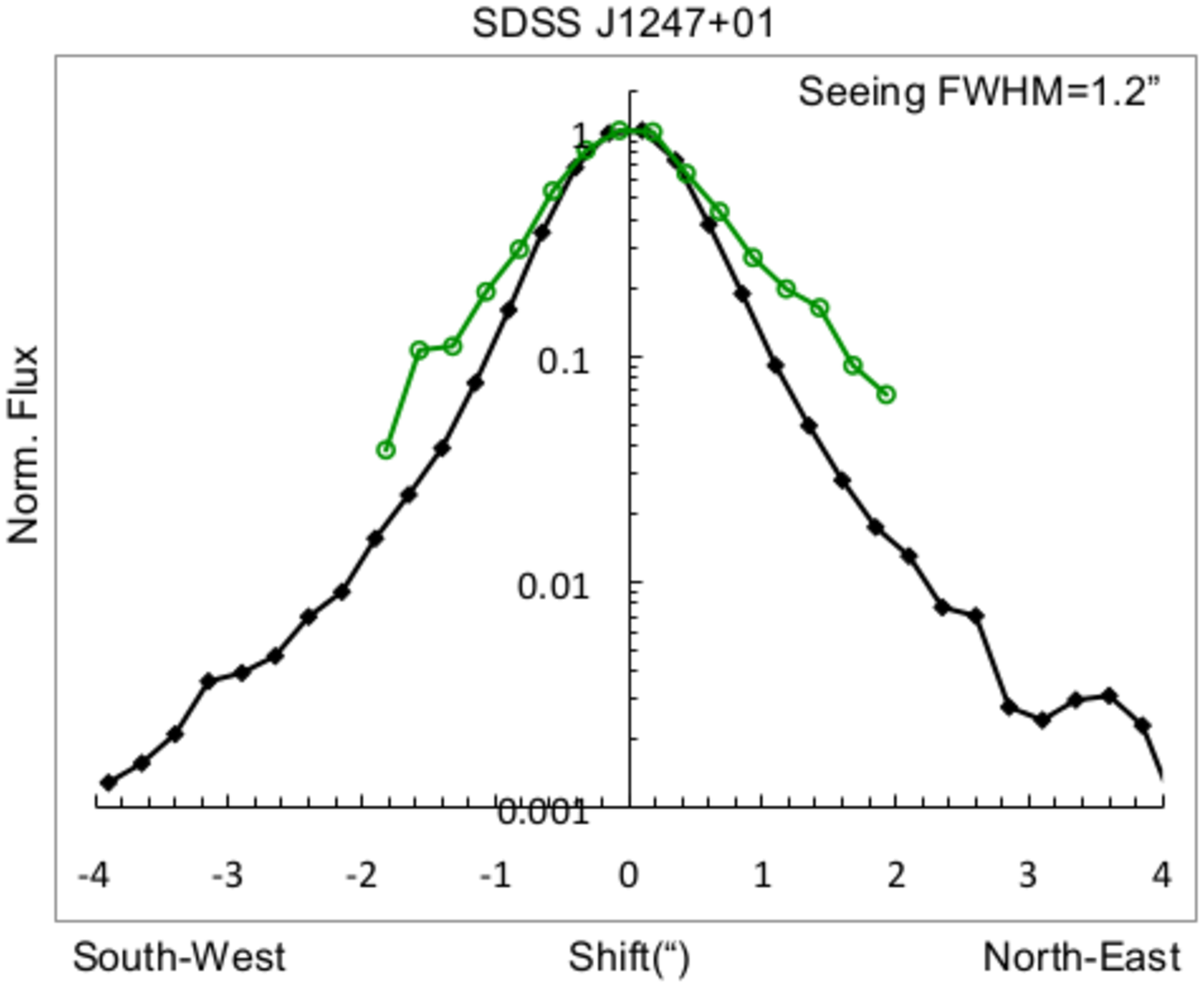}
\includegraphics{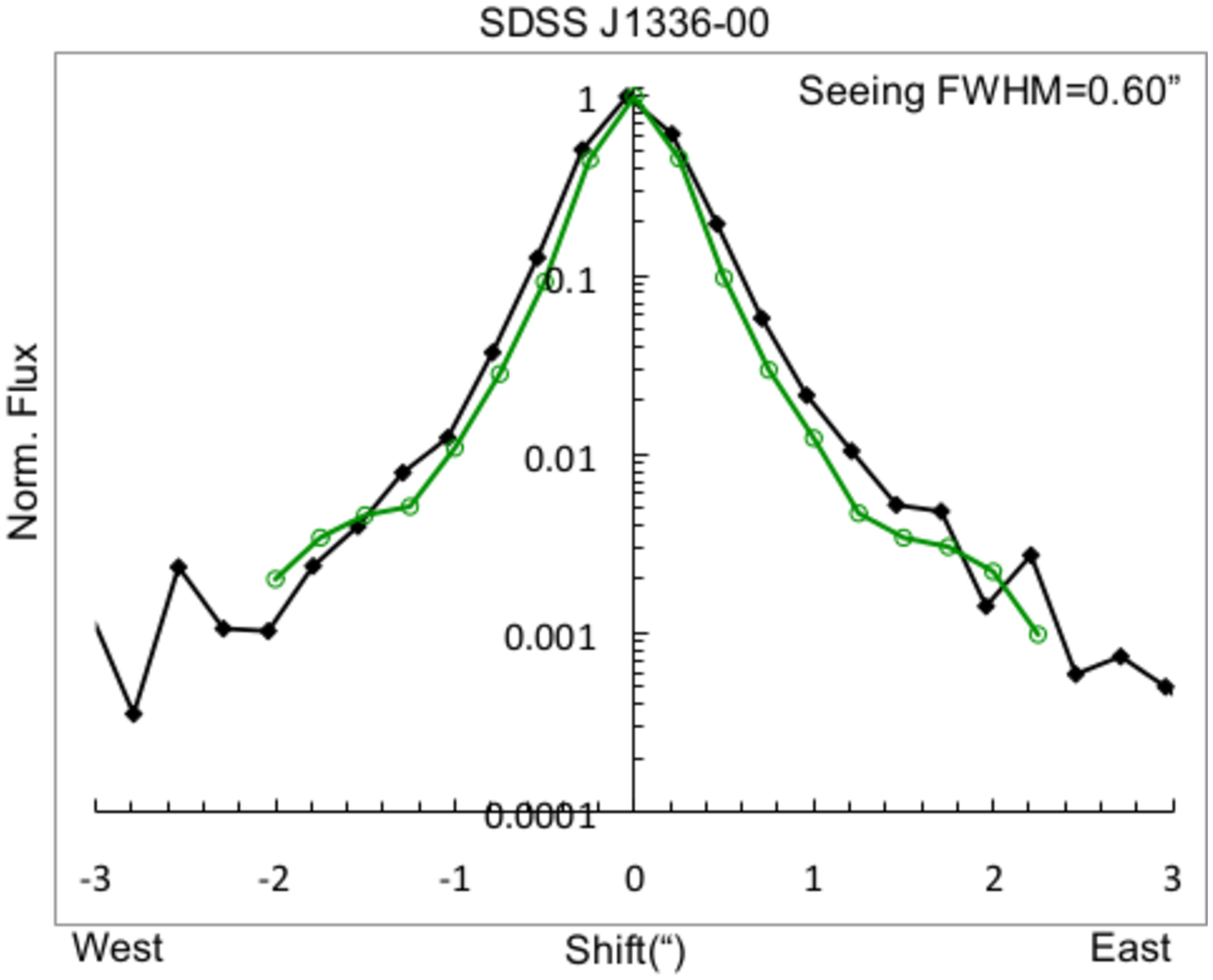}
\vspace{2.2in}
\includegraphics{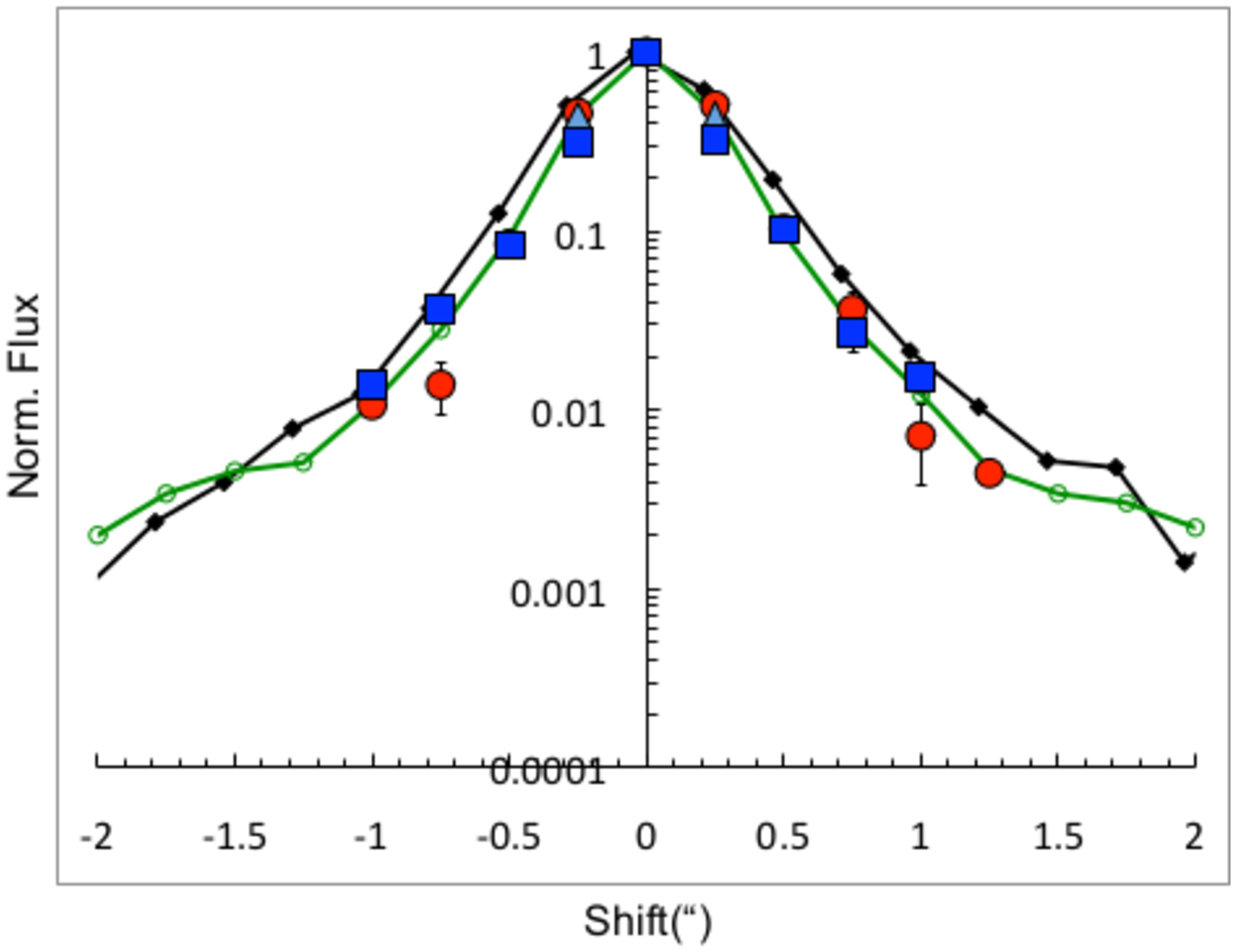}
\vspace{2.2in}
\includegraphics{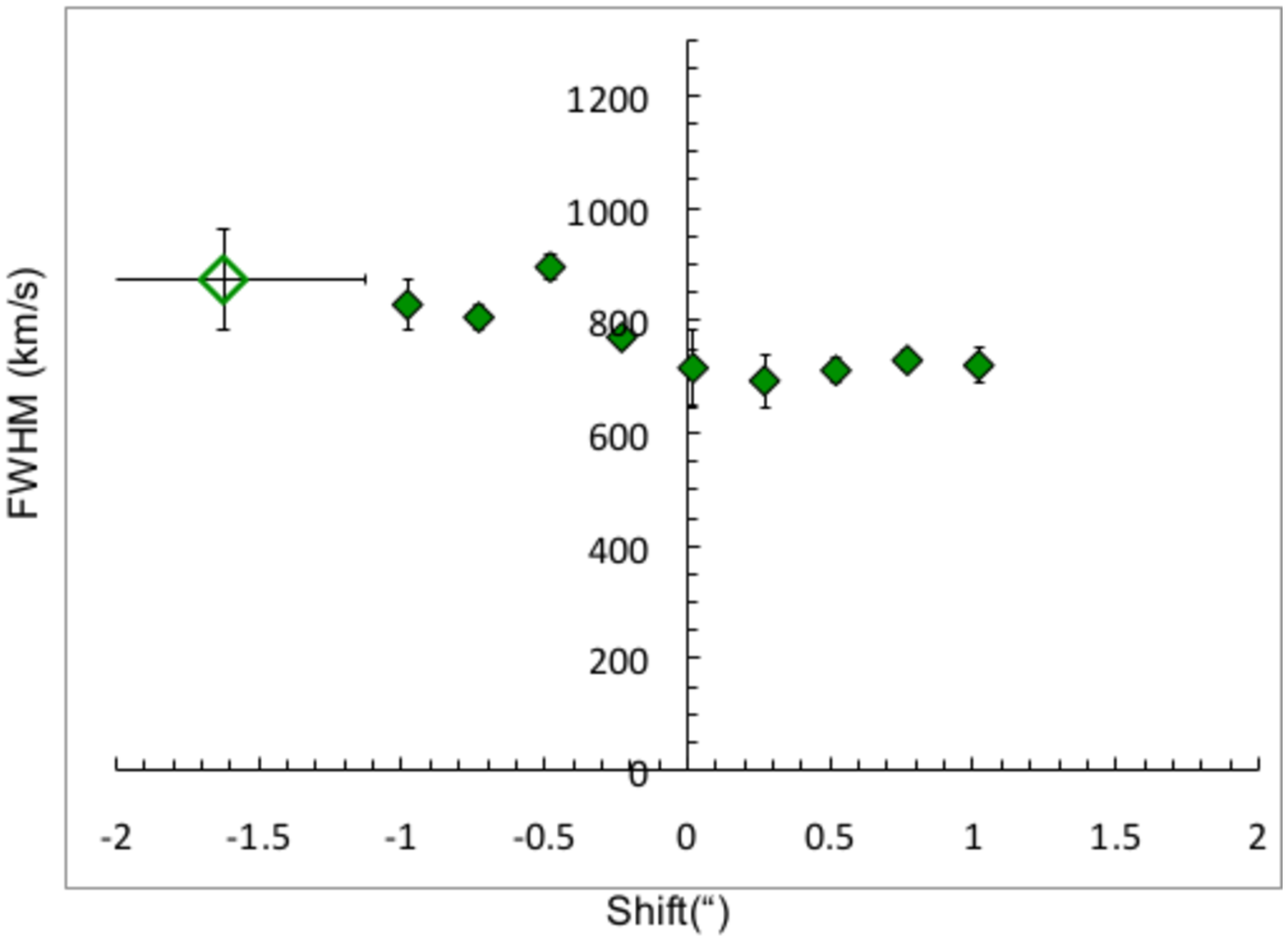}
\includegraphics{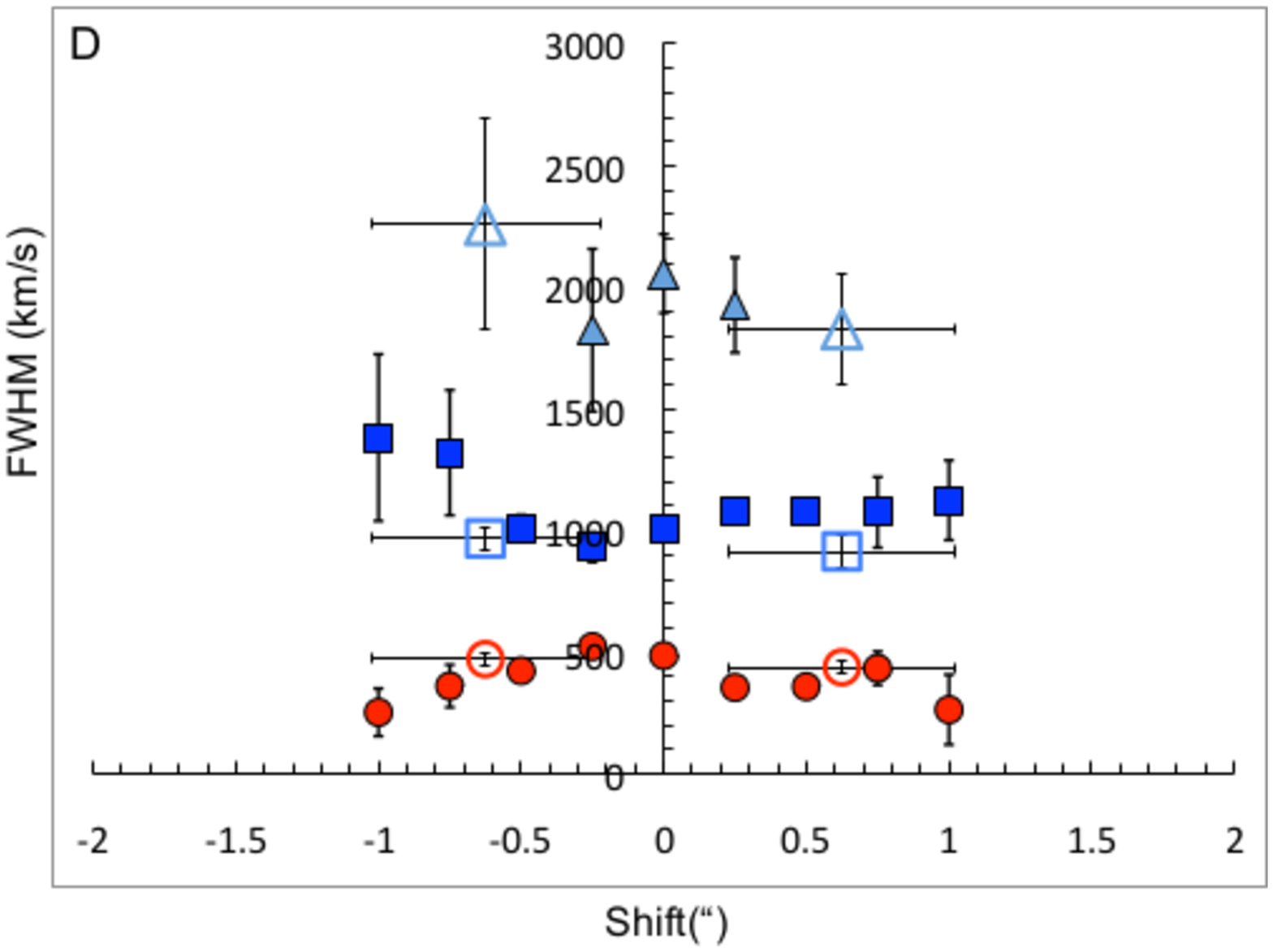}
\vspace{2.2in}
\includegraphics{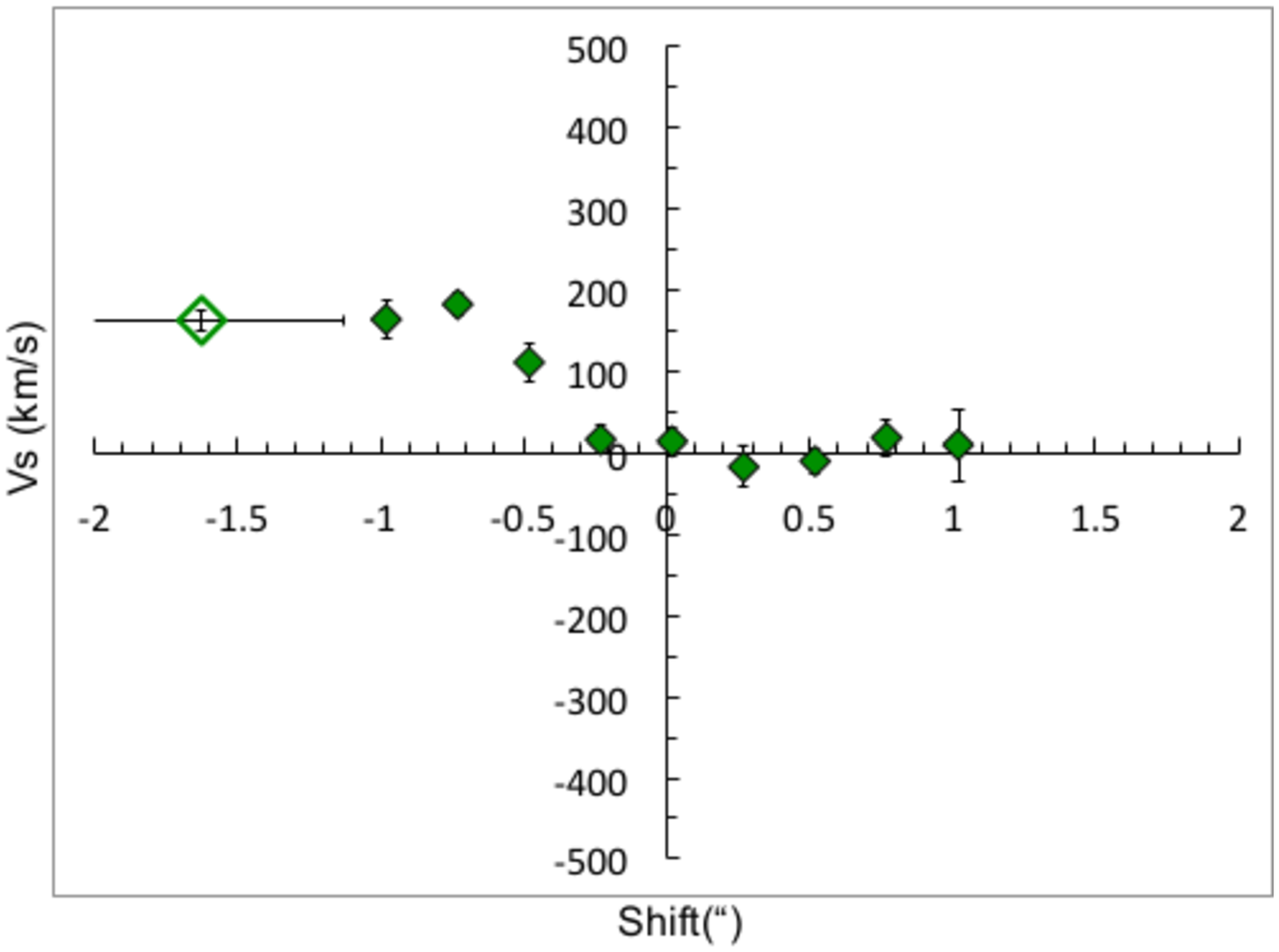}
\includegraphics{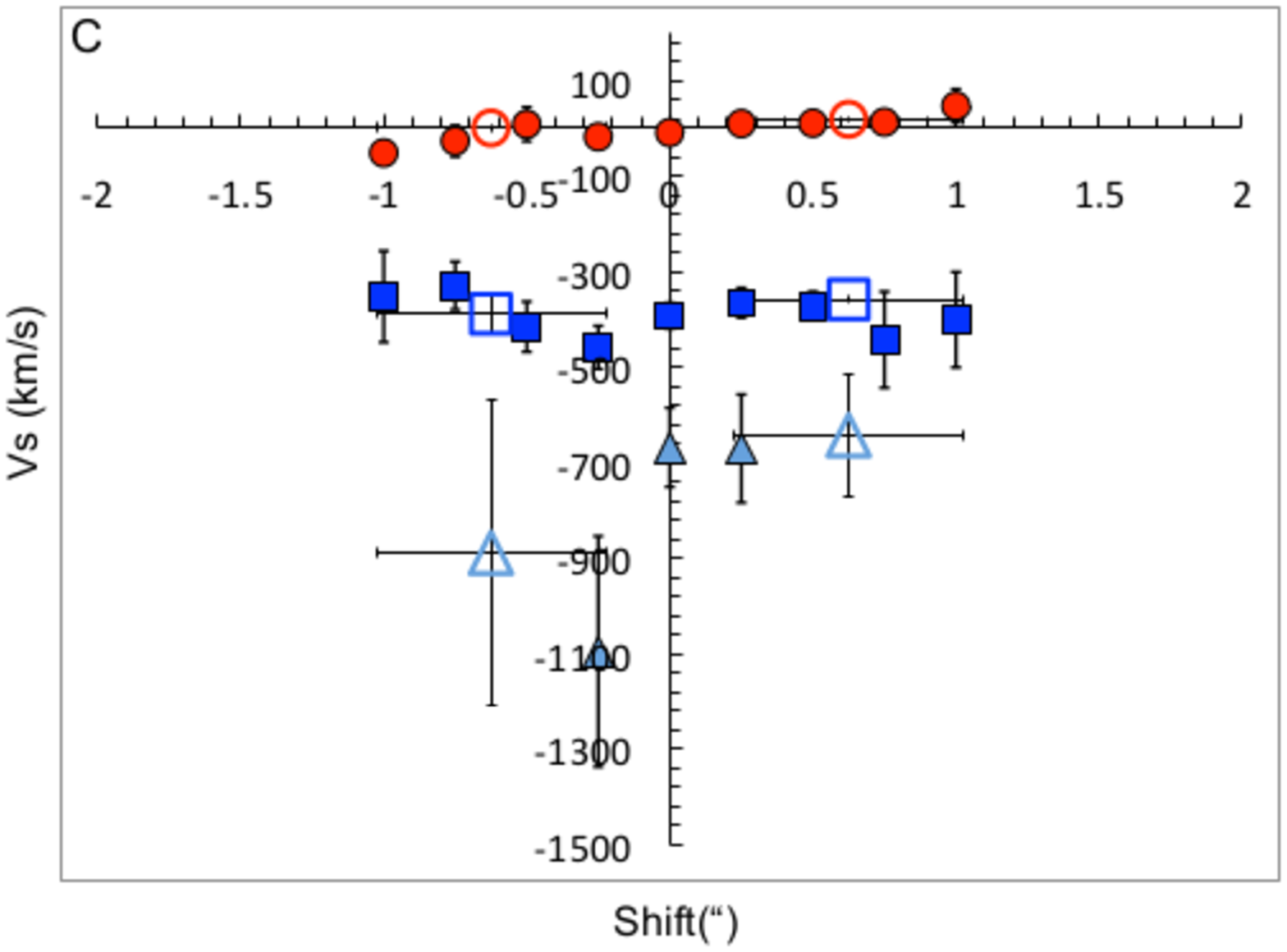}
\vspace{2.2in}
\caption{Spatially extended analysis  of [OIII]$\lambda$5007  for SDSS J1247+01 (left) and SDSS J1336-00 (right). 
The spatially extended kinematic line decomposition could not be applied in SDSS J1247+01. The $V_{\rm s}$ and FWHM values are derived from fitting single Gaussians. The [OIII] spatial profile for SDSS J1336-00 (top right panel) is narrower than the seeing disk.  The spatial profiles of the individual kinematic components are compared
with both  in panel B (left). Line, symbol and color codes as in Fig. 5. The spatial scale of the B, C and D panels is zoomed relative to panel A for clarity.}
\label{spat1247v1336}
\end{figure*}

\begin{figure*}
\includegraphics{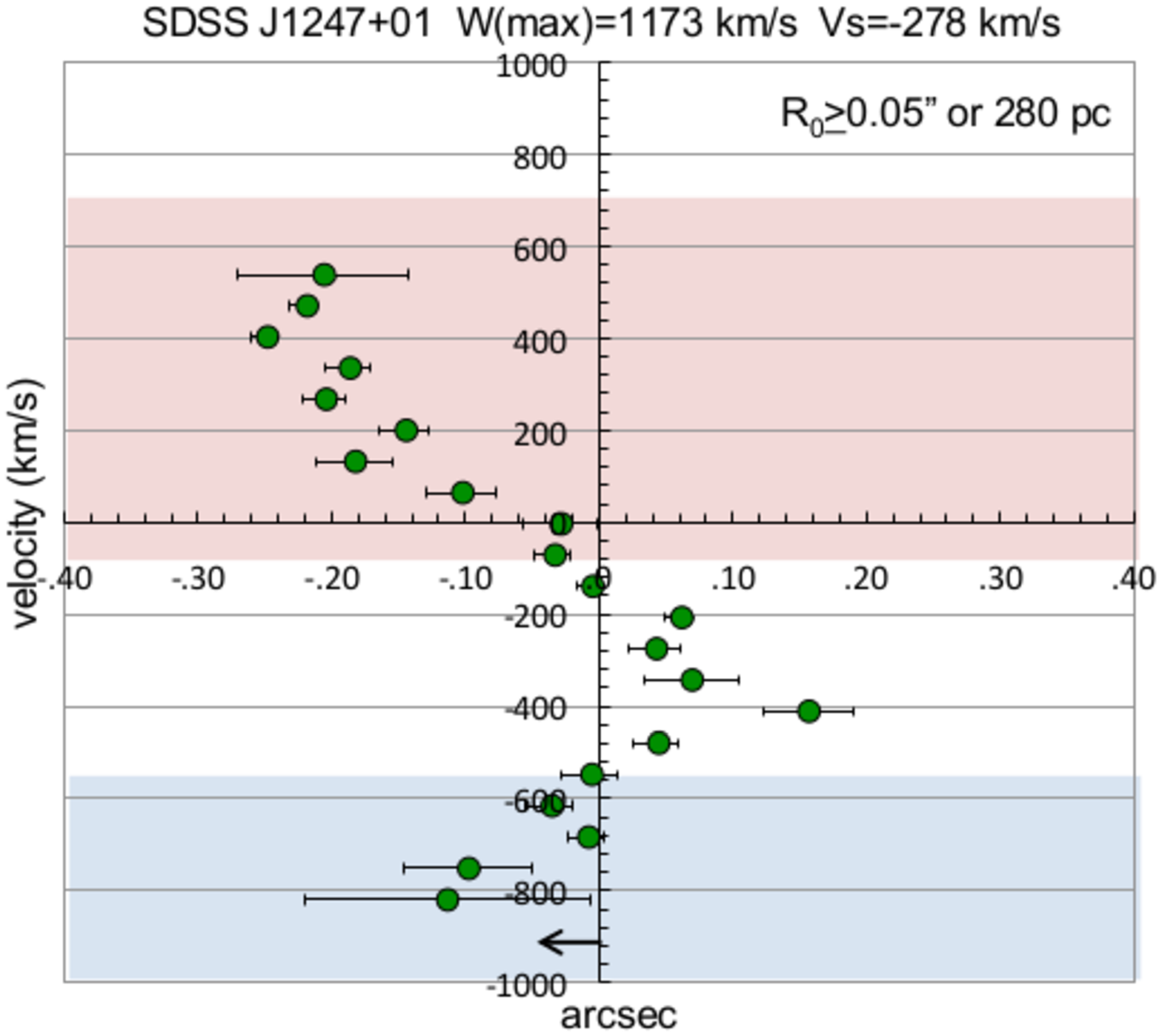}
\includegraphics{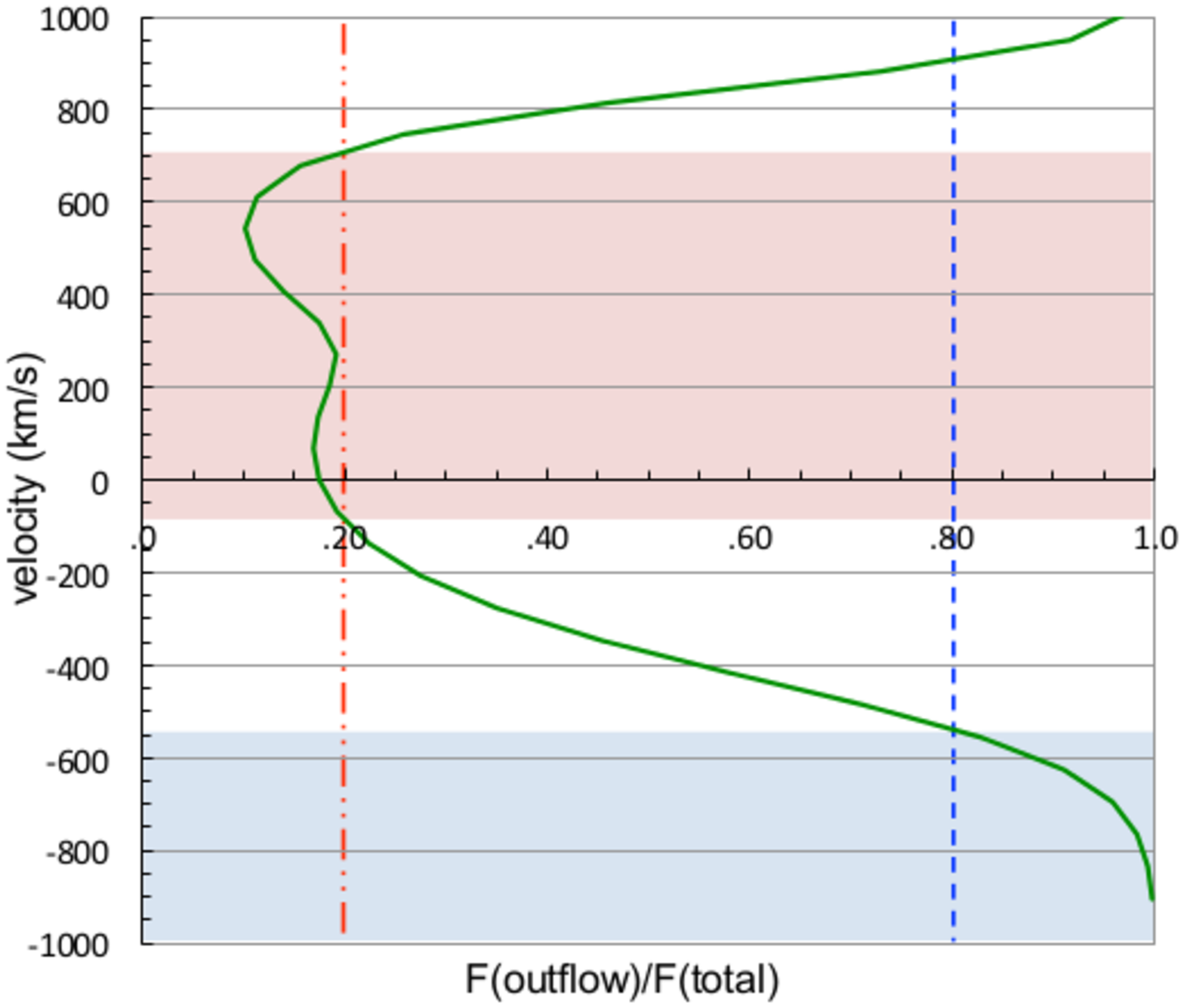}
\vspace{2.85in}
\includegraphics{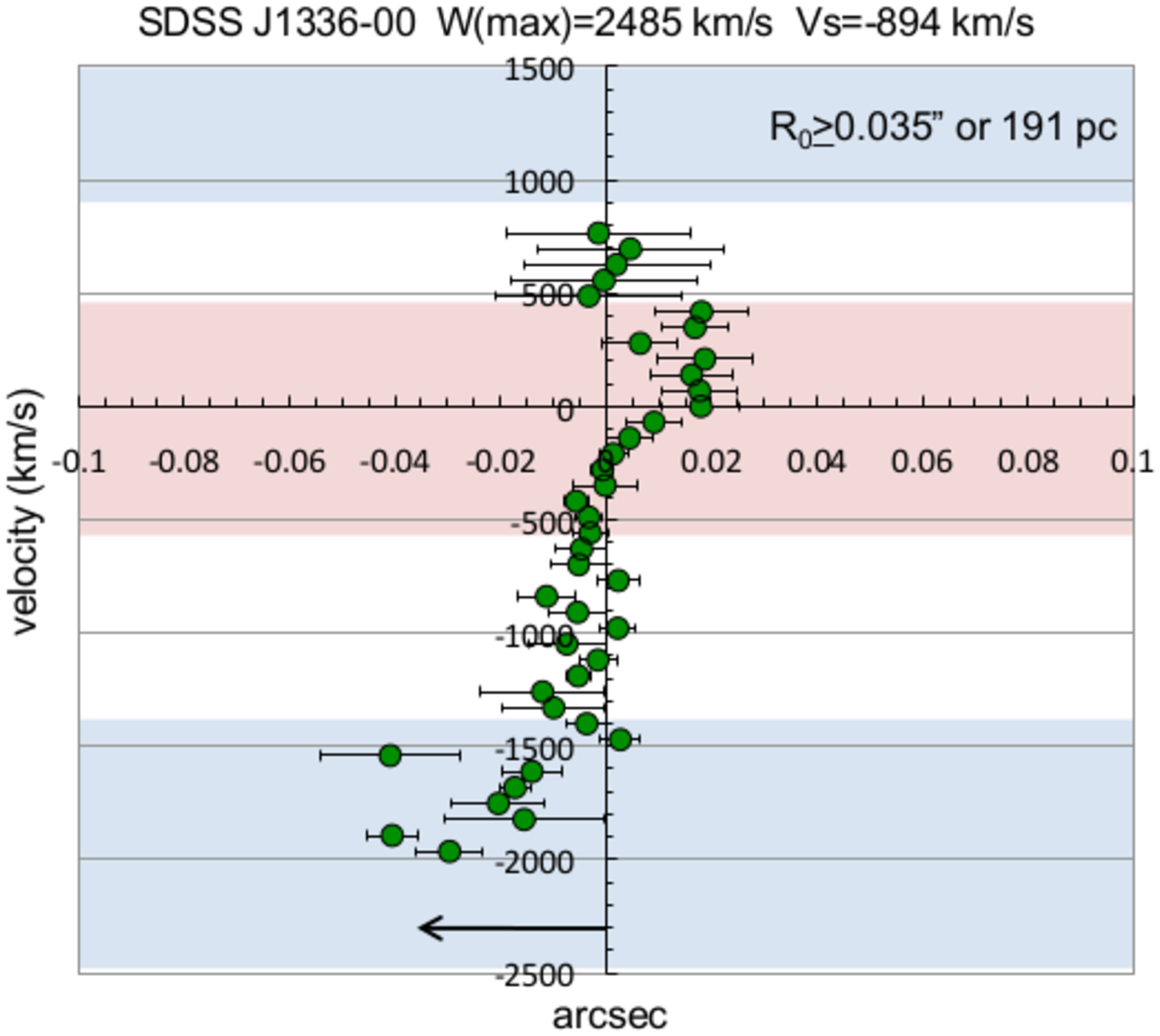}
\includegraphics{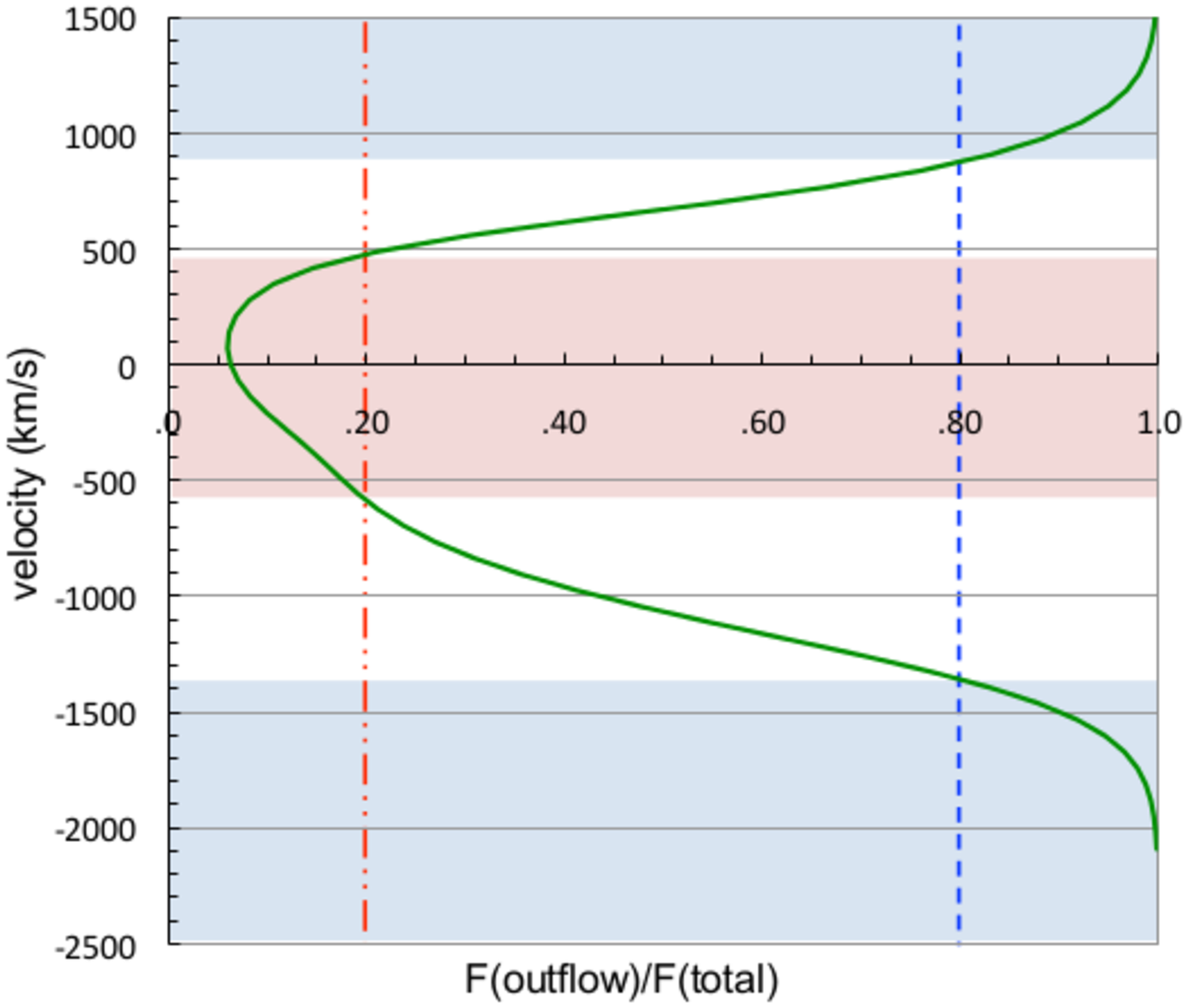}
\vspace{2.85in}
\caption{Spectroastrometry analysis for SDSS J1247+01 (top) and SDSS J1336-00 (bottom). Line, symbol and color codes as in Fig. 6.}
\label{astrom1247}
\end{figure*}

This radio-intermediate HLSy2 shows no clear evidence for a nuclear outflow.  
The  [OIII] lines  (Fig.  \ref{nuc1017})  can be  reproduced by single Gaussians
of FWHM=398$\pm$15 km s$^{-1}$, but for a very faint excess of emission on the blue wing (Fig.  \ref{nuc1017}).   A 2-Gaussian fit would result in two components of FWHM 365$\pm$16 and 550$\pm$128 km s$^{-1}$, with this being blueshifted by -348$\pm$64 km s$^{-1}$ and contributing only $\sim$9\% of the total line flux.  

As shown in \cite{hum15}, the [OIII] profile is dominated by a central unresolved source (seeing FWHM=0.75$\pm$0.05 arcsec). Low surface brightness extended gas up to $\sim$3.5 arcsec
or 20 kpc from the AGN was also reported.  The lines emitted by this gas are very narrow   with FWHM$\la$140 km s$^{-1}$.

\vspace{0.2cm}

{\it SDSS J1247+02}
%z=0.427, 5.56 kpc/arcsec 
\vspace{0.2cm}

This is a double peaked HLSy2 (Table \ref{tab:fitsnuc}, Fig. \ref{nuclei2}). It is the only radio-loud object in our sample. The nuclear spectrum shows two narrow components, with FWHM=422$\pm$50 and $\la$170 km s$^{-1}$ respectively and  shifted by 503$\pm$30 km s$^{-1}$. The most redshifted has $\sim$3.5 times more flux. 
 A broad component  (the ionized outflow) is also present, with W$_{\rm max}$=1173$\pm$198 km s$^{-1}$ and blueshifted by $V_{\rm max}$=-278$\pm$173 km s$^{-1}$ relative to the dominant narrow peak. 

 [OIII] is spatially extended. It shows a clear excess  at both sides of
the spatial centroid above the seeing disk (FWHM=1.25$\pm$0.05 arcsec);  see \cite{hum15} and Fig.   \ref{spat1247v1336}, left panel A). As reported in \cite{hum15}, very low surface brightness [OII] extended emission is detected  
up to $\sim$3$\arcsec$ NE of the AGN. It emits very narrows lines with  FWHM$\la$154 km s$^{-1}$.

The complexity of the lines and the low S/N of the individual spectra 
prevent a detailed
spatially extended analysis. 
 The results of fitting [OIII] pixel by pixel 
 with a single gaussian  are shown with green solid diamonds (Fig.\ref{spat1247v1336}, left panels B and C).  The lines are rather broad 
(FWHM$\sim$800-900 km s$^{-1}$) at all pixels, up
  to $\sim$1.0-1.5$\arcsec$ from the spatial centroid, at locations where the line emission seems well in  excess above the seeing disk.  However, given the prominent profiles of the two nuclear narrow components and the large velocity shift between them (Fig. \ref{nuclei2}), these 
  broad widths are  most likely due to the blend of both components. This is confirmed by creating an artificial spectrum consisting of two Gaussians of equal kinematic properties and relative amplitudes as those in the nuclear spectrum of SDSS J1247+02. A single Gaussian fit produces FWHM$\sim$850 km s$^{-1}$.
 We thus estimate 
  FWHM$_{\rm int}\la$0.51 arcsec or $\la$2.8 kpc for the outflowing gas.   
 
The spectroastrometric analysis appears to resolve the outflow at the highest, most blueshifted velocities, although the error bars are large and the data points scarce (Fig. ~\ref{astrom1247} top).  We infer $R_{\rm o}\ga$=0.05 arcsec or 280 pc.

\vspace{0.2cm}

{\it SDSS J1336-00}
%z=0.416, 5.47 kpc/arcsec

 \vspace{0.2cm}

The nuclear kinematics in this radio-quiet QSO2 is rather extreme (Table \ref{tab:fitsnuc}, Fig.\ref{nuclei2}). Three kinematic components
are isolated with FWHM=479$\pm$14, 984$\pm$29 and 2485$\pm$173 km s$^{-1}$
respectively. The broadest component is blueshifted by  $V_s$=-894$\pm$86 km s$^{-1}$
relative to the narrow core of the [OIII] line.

The [OIII] spatial profile is unresolved. In fact, it is narrower than the seeing size  (FWHM=0.64$\pm$0.05 arcsec; Fig.\ref{spat1247v1336} right, panel A). The pixel by pixel kinematic analysis shows that the spatial distribution of the three kinematic components
is consistent with all being unresolved (panel B). No clear evidence for kinematic substructure within the seeing disk is appreciated (panels C and D). We estimate FWHM$_{\rm int}\la$0.37 arcsec or 2.0 kpc.

The spectroastrometric results shown in  Fig. ~\ref{astrom1247} (bottom) are based on the assumption that only the broadest component contributes to the outflow. Whether the intermediate component also contributes is not discarded, but this uncertainty does not affect the inferred value of $R_{\rm o}$. We estimate a lower limit $R_{\rm}\ga$0.035 arcsec or 191 pc.

\vspace{1cm}
{\it SDSS J1416-02}
% z=0.315, 4.57 (4.56 in PaperI) kpc/arcsec 

\vspace{0.2cm}

This radio-quiet HLSy2 shows no clear evidence for an ionized outflow. 
The shape of the [OIII] profiles  (Fig.  \ref{nuclei2}) are dominated by a narrow spectrally unresolved core.  A second Gaussian, contributing $\sim$54\% of the total line flux, produces a better fit. It has FWHM=370$\pm$37 km s$^{-1}$. The narrow unresolved core traces the extended gas, which seems to be  rotating. The broadest component is possibly the emission from the central  classical NLR. 

 Although the lines are clearly extended (\cite{hum15})
up to $\sim$3.25$\arcsec$ or 14.8 kpc from the AGN, the lines are very narrow everywhere. 
At some spatially extended locations, they lines are actually narrower (FWHM$\sim$5.5-6.0 \AA) than the instrumental
profile derived from the arc and sky lines (FWHM=7.2$\pm$0.2 \AA),   indicating  that the gas did not fill the slit.  
We estimate the FWHM$\la$200 km s$^{-1}$ in the extended regions.

\vspace{0.2cm}
{\it SDSS J1452+00}
% z=0.305, 4.47 kpc/arcsec s
\vspace{0.2cm}

This radio-quiet HLSy2 shows no clear evidence for an outflow.  The lines show a small deviation from Gaussianity  (Fig.  \ref{nuclei2}) with a faint excess on the blue side. In this case, the bulk of the lines is dominated by a broad component of FWHM=835$\pm$21 km s$^{-1}$ and the small excess can be reproduced by  a narrow (FWHM$\la$225 km s$^{-1}$) extra component blueshifted by -250$\pm$31 km s$^{-1}$  and contributing $\sim$15\% of the total
line flux.  The  narrowness of this component suggests  an origin other than an outflow  (e.g. Villar Mart\'\i n et al. \citeyear{vm14}). The core of the line is  broad, 
but still consistent with gravitational motions (although additional broadening mechanisms cannot be discarded).  Using the relation between  $\sigma_{\rm [OIII]}$ and $\sigma_{*}$ inferred for AGN by Greene \& Ho (2005) and taking
into account its large scatter, we infer $\sigma_{*}=$290$\pm$178 km s$^{-1}$, which is  consistent with values measured for QSO2 hosts at similar $z$  (Bian et al. \citeyear{bian06}).

The [OIII] spatial profile  is dominated by 
a compact core consistent with the seeing size (FWHM=0.60$\pm$0.04 arcsec; \cite{hum15}).  We also reported extended gas up to $\sim$3.5$\arcsec$
or 16 kpc, possibly associated with a tidal feature. Its kinematic properties could not be constrained.


\begin{thebibliography}{}


\bibitem[\protect\citeauthoryear{}{2015}]{car15} 

 \bibitem[\protect\citeauthoryear{Alatalo et al.}{2015}]{ala15} 
Alatalo K. et al. 2015, ApJ, 798, 31

 \bibitem[\protect\citeauthoryear{Alexander et al.}{2010}]{ale10} 
Alexander D. M., Swinbank A. M., Smail Ian, McDermid R., Nesvadba N. P. H., 2010, MNRAS, 402, 2211

 \bibitem[\protect\citeauthoryear{Appenzeller et al.}{1998}]{app98} 
Appenzeller I. et al., 1998, The Messenger, 94, 1

 \bibitem[\protect\citeauthoryear{Arribas et al.}{2014}]{arr14} 
Arribas S., Colina L., Bellocchi E., Maiolino R., Villar Mart\'\i n M., 2014, A\&A, 568, 14

 \bibitem[\protect\citeauthoryear{Baldwin, Philips \& Televich}{1981}]{bal81} 
Baldwin J., Philips M., Televich R., 1981, PASP, 93, 5

\bibitem[\protect\citeauthoryear{Baskin \& Laor}{2005}]{bas05} 
Baskin A., Laor A., 2005, MNRAS, 358, 1043

\bibitem[\protect\citeauthoryear{Bennert et al.}{2006a}]{ben06a} 
Bennert N., Jungwiert B., Komossa S., Haas  M., Chini R., 2006a, A\&A, 456, 953

\bibitem[\protect\citeauthoryear{Bennert et al.}{2006b}]{ben06b} 
Bennert N., Jungwiert B., Komossa S., Haas  M., Chini R., 2006b, A\&A, 459, 55


\bibitem[\protect\citeauthoryear{Bellocchi et al.}{2013}]{bel13} 
Bellocchi E., Arribas S., Colina L., Miralles-Caballero D., 2013, A\&A, 557, 59


\bibitem[\protect\citeauthoryear{{Bessiere}, {Tadhunter}, {Ramos Almeida} \&{Villar Martin}}{{Bessiere} et~al.}{2012}]{bes12}{Bessiere} P.~S.,  {Tadhunter} C.~N.,  {Ramos Almeida} C.,    {Villar Martin}  M.,  2012, MNRAS, 426, 276

 \bibitem[\protect\citeauthoryear{Bian et al.}{2006}]{bian06} 
Bian Z., Gu Q., Zhao Y., Chao L., Cui Q.,  2006, MNRAS, 372, 876


	
 
 \bibitem[\protect\citeauthoryear{Cano D\'\i az et al.}{2012}]{can12} 
Cano-D\'\i az M., Maiolino R., Marconi A., Netzer H., Shemmer O., Cresci G., 2012,
A\&A, 537, L8
	
	
 \bibitem[\protect\citeauthoryear{Capaccioli \& Vaucouleurs}{1983}]{cap83} 
Capaccioli M., de Vaucouleurs G. 1983, ApJS, 52, 465
	
\bibitem[\protect\citeauthoryear{Carniani et al.}{2015}]{car15} 
Carniani S., Marconi A., Maiolino R. et al. 2015, A\&A, 508, 102


\bibitem[\protect\citeauthoryear{Cicone et al.}{2014}]{cic14} 
Cicone C. et al. 2014, A\&A, 562, 21


\bibitem[\protect\citeauthoryear{Cicone et al.}{2015}]{cic15} 
Cicone C. et al. 2015, A\&A, 574, 14


\bibitem[\protect\citeauthoryear{Crenshaw et al.}{2010}]{cre10} 
Crenshaw D. M., Schmitt, H. R., Kraemer S. B., Mushotzky R. F., Dunn J. P., 2010, ApJ, 708, 419


\bibitem[\protect\citeauthoryear{Cresci et al.}{2015}]{cre15} 
Cresci G. et al., 2015, ApJ, 799, 82

\bibitem[\protect\citeauthoryear{de Robertis \& Osterbrock}{1984}]{derob84} 
de Robertis M.M., Osterbrock D.E., 1984, ApJ, 276, 181

\bibitem[\protect\citeauthoryear{di Matteo et al.}{2005}]{dim05} 
Di Matteo T., Springel V. Hernquist L., 2005, Nature, 433, 604


\bibitem[\protect\citeauthoryear{Fabian}{2012}]{fab12} 
Fabian A.C., 2012, ARA\&A, 50, 455

\bibitem[\protect\citeauthoryear{Falcke \& Biermann}{1995}]{fal95} 
Falcke H., Biermann P.L., 1995, A\&A, 293, 665


\bibitem[\protect\citeauthoryear{Ferrarese \& Merritt}{2000}]{fer00} 
Ferrarese L., Merritt D., 2000, ApJ, 539, L9


 \bibitem[\protect\citeauthoryear{F\"orster Schreiber  et al.}{2014}]{for14} 
 F\"orster Schreiber N.M. et al. 2014, ApJ, 787, 38

 \bibitem[\protect\citeauthoryear{Genzel  et al.}{2015}]{gen15} 
Genzel R. et al. 2015, ApJ, 800, 20

\bibitem[\protect\citeauthoryear{Ghisellini et al.}{2004}]{ghi04} 
Ghisellini G., Haardt F., Matt G., A\&A, 413, 535	


 \bibitem[\protect\citeauthoryear{Greene \& Ho}{2005}]{gree05} 
Greene J., Ho L., 2005, ApJ, 630, 122


 \bibitem[\protect\citeauthoryear{Greene et al.}{2011}]{gre11} 
Greene J. E., Zakamska N., Ho L., Barth A.,  2011, ApJ, 732, 9 

 \bibitem[\protect\citeauthoryear{Greene et al.}{2012}]{gre12} 
Greene J. E., Zakamska N. L., Smith P. S., 2012, ApJ, 746, 86



 \bibitem[\protect\citeauthoryear{Hainline et al.}{2014}]{hai14} 
Hainline K. N., Hickox R. C., Greene J. E., Myers A. D., Zakamska N. L., Liu G.,
 Liu X., 2014, ApJ, 787, 65



\bibitem[\protect\citeauthoryear{Harrison  et al.}{2012}]{har12} 
Harrison C.M. et al., 2012, MNRAS, 426, 1073

 \bibitem[\protect\citeauthoryear{Harrison et al.}{2014}]{har14} 
Harrison C. M., Alexander D. M., Mullaney J. R., Swinbank A. M., 2014, MNRAS, 441, 3306

 \bibitem[\protect\citeauthoryear{Harrison et al.}{2015a}]{har15a} 
Harrison C. M., Thomson A. P., Alexander D. M., Bauer F. E., Edge A. C., 
Hogan M. T., Mullaney J. R., Swinbank A. M., 2015a, ApJ, 800, 45

 \bibitem[\protect\citeauthoryear{Harrison et al.}{2015b}]{har15b} 
Harrison C. M., Thomson A.P., Alexander D. M., Bauer F.E., Edge A.C., Hogan M.T.,
 Mullaney J. R., Swinbank A. M., 2015b, ApJ, 800, 45	

\bibitem[\protect\citeauthoryear{Heckman et al.}{1981}]{hec81} 
Heckman T. M., Miley G. K., van Breugel W. J. M., Butcher H. R., 1981, ApJ, 247, 403
	

\bibitem[\protect\citeauthoryear{Holt et al.}{2011}]{holt11} 
Holt J., Tadhunter C. N., Morganti R., Emonts B., 2011, MNRAS, 410, 1527

\bibitem[\protect\citeauthoryear{Hopkins et al.}{2012}]{hop12}
Hopkins, Quataert E., Murray N., 2012, MNRAS, 421, 3522

 \bibitem[\protect\citeauthoryear{Humphrey et al.}{2006}]{hum06}
Humphrey  A., Villar Mart\'\i n  M., Fosbury R., Vernet J., di Serego Alighieri S., 2006, MNRAS, 369, 1103
	
\bibitem[\protect\citeauthoryear{Humphrey et al.}{2010}]{hum10}
Humphrey A. Villar Mart\'\i n M., S‡nchez S. F., Mart\'\i nez-Sansigre A., Gonz\'alez Delgado R., P\'erez E.,
 Tadhunter C., P\'erez-Torres M. A., 2010, MNRAS, 408, L1
	
 \bibitem[\protect\citeauthoryear{Paper I}{}]{hum15}
Humphrey A., Villar Mart\'\i n M., Ramos Almeida C., Tadhunter C., Arribas S., Bessiere P., Cabrera Lavers, A., 2015,
MNRAS, 454, 4452 (Paper I)

% \bibitem[\protect\citeauthoryear{Kewley et al.}{2001}]{kew01} 
%Kewley L., Dopita M., Sutherland R., Heisler C. Trevena J., 2001, AJ, 556, 121

  \bibitem[\protect\citeauthoryear{Karouzos et al.}{2016}]{kar16} 
Karouzos M., Woo J.H., Bae H.J., 2016, ApJ, 819, 148

\bibitem[\protect\citeauthoryear{Lal \& Ho}{2010}]{lal10} 
Lal D. V., Ho L. C., 2010, A\&A, 139, 1089


\bibitem[\protect\citeauthoryear{Liu et al.}{2013a}]{liu13a} 
Liu G., Zakamska N. L.,Greene J. E., Nesvadba N., Liu X., 2013a, MNRAS, 430, 2327

\bibitem[\protect\citeauthoryear{Liu et al.}{2013b}]{liu13b} 
Liu G., Zakamska N. L.,Greene J. E., Nesvadba N., Liu X., 2013b, MNRAS, 436, 2576

 \bibitem[\protect\citeauthoryear{McElroy et al.}{2015}]{mce15}
  McElroy R., Croom S., Pracy M., Sharp R.,  Ho I.T., Medling A., 2015, MNRAS, 446, 2186 


\bibitem[\protect\citeauthoryear{Morganti et al.}{2005}]{mor05}
 Morganti R., Tadhunter C. N., Oosterloo T. A., 2005, A\&A, 447, L9

  \bibitem[\protect\citeauthoryear{Mullaney et al.}{2013}]{mul13} 
Mullaney J.R., Alexandeer D.M., Fine S., Goulding A.D., Harrison C.M., Hickox R.C., 2013, MNRAS, 433, 622

 \bibitem[\protect\citeauthoryear{Nesvadba et al.}{2006}]{nes06}
Nesvadba N. P. H., Lehnert, M. D., Eisenhauer F., Gilbert A., Tecza M., Abuter R., 2006, ApJ, 650, 693

 \bibitem[\protect\citeauthoryear{Nesvadb et al.}{2008}]{nes08}
Nesvadba N. P. H., Lehnert M. D., De Breuck C., Gilbert A. M., van Breugel W., 2008, A\&A, 491, 407

 \bibitem[\protect\citeauthoryear{Nesvadb et al.}{2008}]{nes11}
Nesvadba N. P. H., Polletta M., Lehnert M. D., Bergeron J., De Breuck C., Lagache G., Omont A., 2011, MNRAS, 415, 2359


%\bibitem[\protect\citeauthoryear{Netzer}{1990}]{net90}  
%Netzer H., 1990, in 	{\it Saas-Fee Advanced Course of the Swiss Society for Astrophysics and Astronomy: Active galactic nucle},  1990,  eds. R.D. Blandford, H. Netzer and L. Woltjer, p. 57

 %\bibitem[\protect\citeauthoryear{Padovani \& Raffanelli}{1988}]{pad88}
%Padovani  P., Raffanelli P., 1988, A\&A, 205, 53

\bibitem[\protect\citeauthoryear{Page et al.}{2012}]{pag12} 
Page M.J. et al., 2012, Nature 485, 213

\bibitem[\protect\citeauthoryear{Perna et al.}{2015}]{per15} 
Perna M. et al. 2015, A\&A, 574, 82


\bibitem[\protect\citeauthoryear{Ramos Almeida et al.}{2011}]{ram11} Ramos Almeida C., Tadhunter C.~N., Inskip 
K.~J., Morganti R., Holt J., Dicken D., 2011, MNRAS, 410, 1550 

\bibitem[\protect\citeauthoryear{Ramos Almeida et al.}{2012}]{2012MNRAS.419..687R} 
Ramos Almeida C., et al., 2012, MNRAS,  419, 687 

\bibitem[\protect\citeauthoryear{Reyes et al.}{2008}]{rey08} 
Reyes R., Zakamska N., Strauss M. et al. 2008, AJ, 136, 2373

\bibitem[\protect\citeauthoryear{Redr\'\i guez Zaur\'\i n  et al.}{2013}]{rod13} 
Rodr\'\i guez Zaur\'\i n J., Tadhunter C. N., Rose M., Holt J., 2013, MNRAS, 432, 138

\bibitem[\protect\citeauthoryear{Rupke \& Veilleux}{2011}]{rup11} 
Rupke D., Veilleux S., 2011, ApJ, 729, L27

\bibitem[\protect\citeauthoryear{Shih et al.}{2013}]{shih13} 
Shih H.Y., Stockton A., Kewley L., 2013, ApJ, 772, 138

\bibitem[\protect\citeauthoryear{Sol\'orzano-I\~narrea et al.}{2002}]{sol01} 
Sol\'orzano-I\~narrea C., Tadhunter C. N., Axon D. J., 2002, MNRAS, 323, 965

\bibitem[\protect\citeauthoryear{Stern \& Laor}{2012}]{stern12} 
Stern J., Laor A., 2012, MNRAS, 426, 2703


\bibitem[\protect\citeauthoryear{Stern et al.}{2014}]{stern14} 
Stern J., Laor A., Baskin A., 2014, MNRAS, 438, 901


\bibitem[\protect\citeauthoryear{Tadhunter et  al.}{1994}]{tad94} 
Tadhunter C., Shaw M., Clark N., Morganti R., 1994, A\&A, 288, L21
	

\bibitem[\protect\citeauthoryear{Tadhunter et al.}{2014}]{tadh14} 
	Tadhunter C., Morganti R., Rose M., Oonk J. B. R., Oosterloo T., 2014, Nature 511, 440

\bibitem[\protect\citeauthoryear{Tremaine et  al.}{2002}]{tre02} 
Tremaine S. et al. 2002, ApJ, 574, 740

\bibitem[\protect\citeauthoryear{Villar Mart\'\i n et al.}{2003}]{vm03} 
Villar Mart\'\i n M., Vernet J., di Serego Alighieri S., Fosbury R., Humphrey A., Pentericci L.,
2003, MNRAS, 346, 273


\bibitem[\protect\citeauthoryear{VM11a}{}]{vm11a} 
Villar Mart\'\i n M., Tadhunter C., Humphrey A.,  Fraga Encina R., Gonz\'alez Delgado R., P\'erez Torres M.,  Mart\'\i nez-Sansigre A.,
2011a, MNRAS, 416, 262 (VM11a)

\bibitem[\protect\citeauthoryear{VM11b}{}]{vm11b} 
Villar Mart\'\i n M., Humphrey A., Gonz\'alez Delgado R., Colina L., Arribas S., 2011b, MNRAS, 418, 2032 (VM11b)
	
\bibitem[\protect\citeauthoryear{Villar Mart\'\i n et al.}{2012}]{vm12} 
Villar Mart\'\i n M., Cabrera Lavers A., Bessiere P., Tadhunter C., Rose M., de Breuck C., 2012, MNRAS, 423, 80
	
\bibitem[\protect\citeauthoryear{Villar Mart\'\i n et al.}{2014}]{vm14} 
Villar Mart\'\i n M., Emonts B., Humphrey A., Cabrera Lavers A., Binette  L., 2014, MNRAS, 440, 3202 	

\bibitem[\protect\citeauthoryear{Villar Mart\'\i n et al.}{2015}]{vm15} 
Villar  Mart\'\i n M.,  Bellocchi  E.,   Stern J.,  Tadhunter C.,  Gonz\'alez Delgado R., 2015, 454, 439

\bibitem[\protect\citeauthoryear{Whittle}{1985}]{whi85} 
Whittle M., 1985,  MNRAS, 213, 1

\bibitem[\protect\citeauthoryear{Whittle}{1992}]{whi92} 
Whittle M., 1992, ApJSS, 79, 49

\bibitem[\protect\citeauthoryear{Xu et al.}{1999}]{xu99} 
Xu C., Livio M., Baum S., 1999, AJ, 118, 1169

\bibitem[\protect\citeauthoryear{York et al.}{2000}]{york00} 
York D. G. et al., 2000, AJ, 120, 1579

\bibitem[\protect\citeauthoryear{Zakamska et al.}{2003}]{zak03} 
Zakamska N. et al. 2003, AJ, 126, 2125

\bibitem[\protect\citeauthoryear{Zakamska \& Greene}{2014}]{zak14} 
Zakamska N., Greene J., 2014, MNRAS, 442, 784

\bibitem[\protect\citeauthoryear{Zakamska et al.}{2016}]{zak16} 
Zakamska N. et al. 2016, MNRAS, 455, 4191

\end{thebibliography}
\end{document}